\documentclass[10pt]{article}
\usepackage[margin=1in]{geometry}
\usepackage{amsmath,amssymb,amsthm,mathtools}
\usepackage{booktabs}
\usepackage{graphicx}
\usepackage{array}
\usepackage{longtable}
\usepackage[colorlinks=true,linkcolor=blue,citecolor=blue,urlcolor=blue]{hyperref}
\let\origthebibliography\thebibliography
\renewcommand{\thebibliography}[1]{\origthebibliography{#1}\setlength{\itemsep}{0pt}\setlength{\parsep}{0pt}}

\newtheorem{theorem}{Theorem}[section]
\newtheorem{proposition}[theorem]{Proposition}
\newtheorem{lemma}[theorem]{Lemma}
\newtheorem{corollary}[theorem]{Corollary}

\theoremstyle{definition}
\newtheorem{definition}[theorem]{Definition}
\newtheorem{example}[theorem]{Example}
\newtheorem{examples}[theorem]{Examples}
\newtheorem{remark}[theorem]{Remark}
\newtheorem{hypothesis}[theorem]{Hypothesis}
\numberwithin{equation}{section}

\newcommand{\R}{\mathbb R}
\newcommand{\C}{\mathbb C}
\newcommand{\Z}{\mathbb Z}
\newcommand{\N}{\mathbb N}
\newcommand{\cS}{\mathcal S}
\newcommand{\Hh}{\widehat H}
\newcommand{\Tr}{\operatorname{Tr}}
\newcommand{\tr}{\operatorname{tr}}
\newcommand{\Ind}{\operatorname{Ind}}
\newcommand{\ind}{\operatorname{ind}}
\newcommand{\SF}{\operatorname{SF}}
\newcommand{\sfl}{\operatorname{sf}}
\newcommand{\sgn}{\operatorname{sgn}}
\newcommand{\spec}{\operatorname{spec}}
\newcommand{\Op}{\operatorname{Op}}
\newcommand{\OpA}{\operatorname{Op}^A}

\newcommand{\supp}{\operatorname{supp}}
\newcommand{\Ran}{\operatorname{Ran}}
\newcommand{\dist}{\operatorname{dist}}
\newcommand{\curl}{\operatorname{curl}}
\newcommand{\jp}[1]{\langle #1\rangle}
\newcommand{\norm}[1]{\|#1\|}
\newcommand{\Lm}{\Lambda}
\newcommand{\Gr}{\operatorname{Gr}}
\newcommand{\ab}{a}
\newcommand{\trarea}{\mathcal T}
\newcommand{\cP}{\mathcal P}
\newcommand{\cD}{\mathcal D}
\newcommand{\ip}[2]{\langle #1,#2\rangle}
\newcommand{\Fl}{\mathcal F}
\newcommand{\U}{\mathcal U}
\newcommand{\Meff}{M_{\rm eff}}
\newcommand{\Wt}{\mathcal W}
\newcommand{\Cy}{\mathsf C}
\newcommand{\sff}{{\rm f}}
\newcommand{\stt}{{\rm t}}
\newcommand{\Herm}{\operatorname{Herm}}
\newcommand{\lsym}{\mathfrak l}

\title{\bf Interface currents of magnetized differential systems:\\ quantization and bulk--interface correspondence}
\author{Guillaume Bal\thanks{Departments of Statistics and Mathematics and CCAM, University of Chicago, Chicago, IL 60637;
guillaumebal@uchicago.edu}}
\date{\today}

\begin{document}
\maketitle

\begin{abstract}
This paper concerns the asymmetric transport along an interface separating two insulators modeled by self-adjoint differential systems $H$ of arbitrary
order on $L^2(\R^2;\C^n)$, coupled to a magnetic field either by Peierls substitution or, for operators of the form $L^*L+W$, in the factor $L$. Each bulk
carries a constant field, which may vanish. The principal symbol of $H$ is either elliptic or has a kernel of constant rank, in which case finitely many
bands remain bounded at large momentum, as in models of rotating shallow water, cold plasmas, Maxwell's equations and pseudospin-one materials. Ellipticity
is then replaced by a Douglis--Nirenberg structure and by a gap condition on an effective flat symbol, which persist in the field for factorized operators
of every order and for Peierls substitutions of order at most two.

We show that, for every energy $\alpha$ away from the flat-band set and in a spectral gap of both bulks, the interface current $\sigma_I=\Tr\,i[H,P]\varphi'(H)$
is well defined, with $2\pi\sigma_I$ a Fredholm index, and that it is given, for all walls in the elliptic case and under a freezing condition otherwise,
by a bulk--interface correspondence of the form
\[
  2\pi\sigma_I=J(a_+,B_+;\alpha)-J(a_-,B_-;\alpha)+\Cy .
\]
For a nonzero bulk field, $J=2\pi\sigma_H$ is the Kubo--Chern Hall conductivity, which we compute by an extension-corrected count of generalized Landau
levels. For a vanishing field, $J$ is the Chern integral of the field-free symbol, which is also, under a weak-field condition, the weak-field limit of the
Hall conductivity. The term $\Cy$ is the integral, over the transition region and the directions of the momentum, of the Chern form of the occupied projector
at infinite momentum. The current is stable under large perturbations, and boundaries of the applied field inside a homogeneous material carry
quantized currents that change at field-induced transitions. Near flat-band energies, where generalized Landau levels may accumulate, the current may
diverge.
\end{abstract}



\section{Introduction}\label{sec:intro}
This paper considers the asymmetric transport along interfaces separating two-dimensional insulators coupled to a magnetic field. At an interface between
insulators in different topological phases, modes localized near the interface carry a nonzero net flow of signal along it, which is robust against defects.
The first instance of such topologically protected transport is the integer quantum Hall effect, in which a two-dimensional electron gas in a strong
perpendicular magnetic field carries a quantized current along its edge \cite{Halperin,TKNN}. Interface modes of the same nature appear without applied
field in many continuous models of wave propagation, such as rotating shallow water \cite{DMV,TDV}, cold plasmas and Maxwell's equations in gyrotropic media
\cite{Silv}, and pseudospin-one materials such as the dice lattice \cite{MN16,KDDC17}. Classical waves do not couple to magnetic fields, but slowly
varying strains of honeycomb structures act on waves near a Dirac point as a \emph{pseudo-magnetic} field \cite{GRW21}. The resulting photonic
Landau levels and interface modes have been observed \cite{BGZ24}, and the effective equations have been derived \cite{BCMQ26}. In this
paper, $B$ denotes either a magnetic or a pseudo-magnetic field.

The asymmetric transport along an interface $\{y=0\}$ is quantified by the interface current observable
\begin{equation}\label{eq:sigmaI}
  \sigma_I=\Tr\,i[H,P]\,\varphi'(H).
\end{equation}
Here, $H$ is the Hamiltonian, $P=P(x)$ is a smooth function equal to $0$ near $x=-\infty$ and to $1$ near $x=+\infty$, so that $i[H,P]$ physically models the current through a line
transverse to the interface, and $\varphi'$ is a density of states supported in a common spectral gap of the two bulks \cite{Bal22,QB21,BEC}. This observable is the edge conductance of the integer quantum Hall effect
\cite{Halperin,Hatsugai93}, whose quantization and equality with the bulk Hall conductivity are established in \cite{SKR00,KRS,EG,EGS} for discrete models and
in \cite{KS04,CG05,Taarabt} for continuous magnetic operators; see \cite{GP,PSB} for topological insulators. For elliptic
differential operators without magnetic field, it was shown in \cite{Bal22,QB21,BEC} that $2\pi\sigma_I$ is an integer given by bulk data. For the magnetic Dirac
operator, $2\pi\sigma_I$ was computed in \cite{QB22} as a spectral flow.

For the Landau Hamiltonian, the bulk Hall conductivity and its quantization go back to \cite{Streda,ASS}, and interface currents between regions of different
fields are studied in \cite{DGR}. We extend these results to differential systems of arbitrary order and matrix size that need not be elliptic, which raises
two difficulties. First, in a field the bulk spectrum consists of generalized Landau levels of infinite multiplicity, which may accumulate at flat-band
energies. The bulk invariant is then the Hall conductivity, defined by an operator formula \cite{ASS,BES}, whose dependence on the energy and
on the field, which may differ or vanish in the two bulks, must be determined. Second, for rotating fluids, plasmas and Maxwell's equations \cite{DMV,TDV,Silv,GJT}, the principal
symbol has a kernel of constant rank $r\ge1$. Then $r$ bands remain bounded at infinite momentum, converge to the eigenvalues of an effective flat symbol
$\Meff$ depending on position and on the direction of the momentum, and produce essential spectrum on the flat-band set $\Fl$ of its values, so that the
analysis must be carried out away from $\Fl$ without ellipticity. The elliptic case is $r=0$.

We consider operators $H=a(u,\Pi)$, where $u=(x,y)\in\R^2$, $\Pi=D-A$ with $D=-i\nabla$ is the kinetic momentum, and the Hermitian matrix symbol $a$ is
polynomial of degree $m$ in the momentum and quantized by Weyl symmetrization, which is the Peierls substitution \cite{Peierls33,Lut51,IMP}, as well as factorized operators
$L_B^*L_B+W$, in which the field enters through the factor $L_B$. Outside a strip $\{|y|<y_0\}$, the \emph{domain wall}, or wall, the coefficients and the field are
constant, equal to $a_\pm(k)$ and $B_\pm$, and either field may vanish. In the non-elliptic case, the main assumption is a Douglis--Nirenberg structure \cite{DN}: conjugated by a frame
of the kernel bundle of $a_m$, the symbol has blocks of orders $0$, $\frac m2$, $\frac m2$ and $m$. This structure defines $\Meff$. It holds at every field
for first-order systems and for factorized operators, and for Peierls substitutions of order two if it holds without field, but it may fail in a field at
higher order. Both classes are defined in section~\ref{sec:operators}, and the hypotheses in section~\ref{sec:hypotheses}.

Our main result (Theorem~\ref{thm:BIC}) is that, for every energy $\alpha$ away from $\Fl$ and in a spectral gap of both bulks, $H$ is self-adjoint, $2\pi\sigma_I$ is the index of a
Fredholm operator (Theorems~\ref{thm:analysis} and \ref{thm:quant}), and, for all walls in the elliptic case and under a freezing condition otherwise,
\begin{equation}\label{eq:mainformula}
  2\pi\sigma_I=J(a_+,B_+;\alpha)-J(a_-,B_-;\alpha)+\Cy .
\end{equation}
For a nonzero bulk field, $J(\ab,B;\alpha)=2\pi\sigma_H(\ab,B;\alpha)$ is the Kubo--Chern Hall conductivity of the bulk \cite{TKNN,ASS,BES}. For a vanishing field,
$J(\ab,0;\alpha)=I(\ab;\alpha)$ is the Chern integral of the occupied projector of the field-free symbol \cite{Bal19,QB21}, which is also, under the weak-field condition (W), the weak-field limit of the Hall conductivity. 
The term $\Cy$ is the integral over the cylinder $[-y_0,y_0]\times S^1$ of the first Chern form of the occupied
projector $E^\infty(y,\theta)$ at infinite momentum in the direction $\theta$, at the position $y$ across the wall. It depends on the field only through $\Meff$. The quantities $J$ are not integers in general,
but the right-hand side is. The formula is stable under large perturbations that preserve the flat-band structure and the bulk gaps (Theorem~\ref{thm:slab}), and it covers
interfaces between two magnetized bulks, between a magnetized and a field-free bulk, and between two field-free bulks.

Each term of \eqref{eq:mainformula} is computable. Without field, $I_+-I_-+\Cy$ is a Chern number on a sphere in the variables $(y,\xi,\zeta)$, as in the
Fedosov--H\"ormander formula \cite{Fed96,Hormander3,Bal19,Bal22,QB21,Faure}, $\Cy$ being the contribution at infinite momenta. With field, $J$ is obtained from the eigenvalues of the one-dimensional operator
$K_B(\ab)$, the Weyl quantization of $\ab(By,\zeta)$ on $L^2(\R_y)$. Compared with the field-free invariant, $J$ changes by
the signed number of these generalized Landau levels that cross the energy as the field is switched on \eqref{eq:genFH}, as for magnetic barriers
\cite{DGR} and magnetic Dirac operators \cite{QB22}. This change is invisible to the symbol, so that no formula of Fedosov--H\"ormander type holds at strong
field.

In particular, at energies in the field-free gaps of both bulks, the field-free current is recovered at weak fields (Corollary~\ref{cor:trans}(c)). Conversely, a boundary of the applied
field inside a homogeneous material carries a quantized current, which for instance equals $-1$ beyond a field-induced transition in a Bernevig--Hughes--Zhang (BHZ) type
insulator \cite{BHZ06} (Corollary~\ref{cor:trans}(b)), and which diverges in a dice-lattice model as the energy approaches a flat-band energy carrying a Landau fan \cite{MN16} (Corollary~\ref{cor:main}(c)). The term $\Cy$ is in general nonzero
when $E^\infty$ depends on $y$. The choice of coupling to the field matters for flat bands: for a curl operator with a
gyrotropic potential, Peierls substitution and factorization give different transition fields (Examples~\ref{ex:examples}), and beyond order four Peierls substitution may destroy a
bounded band (Proposition~\ref{prop:Meffcomp}(e)), in line with the sensitivity of flat bands to the field observed in tight-binding models \cite{AAM96,RKY20}.

The lack of ellipticity is addressed by conjugating $H$ by a frame of the kernel bundle, which yields a Douglis--Nirenberg system with fixed blocks, and a
parametrix near the energy gives self-adjointness, resolvent estimates and continuity along deformations. The bulk Hall conductivity reduces to a pairing \eqref{eq:tau} for
the reduced one-dimensional operator, which is invariant along deformations to reference symbols, changes by the spectral flow when levels cross, and is
computed for the references by a semiclassical limit. The local dependence of interface currents allows us to reduce general walls to translation-invariant ones, where the current is a spectral flow of the fibers \cite{Phillips,QB22}. No regularization at infinite momentum is needed, in contrast with the regularizations used in \cite{Silv,Bal19,TDV,FB25} to obtain integer bulk
invariants: here the bulk invariants are in general not integers, their fractional parts are fixed by $E^\infty$, and $\Cy$ restores integrality; in a
field, they also have an operator definition by the Kubo formula.

\smallskip
\noindent\textbf{Related work and outline.} Bulk--edge correspondences for discrete and continuous models have a long history \cite{KRS,EG,EGS,GP,PSB,Drouot}.
For continuum magnetic Hamiltonians, they are obtained for Landau-type Schr\"odinger operators, possibly with disorder or magnetic
barriers \cite{Taarabt,DGR,CMT}, and magnetic Dirac operators \cite{CMS,QB22,BCLRS24,BCLRS25}; edge states along curved interfaces in a field are studied in
\cite{BBD}. Topological waves in fluids and continuous media, which motivate the non-elliptic setting, are discussed in
\cite{DMV,TDV,GJT,Silv}, where the lack of compactness at infinite momentum is known to affect the correspondence \cite{GJT,FB25}. The Fredholm and
cyclic-cohomology framework goes back to \cite{ASS,Connes}, and field-free correspondences for elliptic operators are proved in \cite{Bal22,QB21,BEC}. Unbounded domain walls \cite{Bal23,CB25} are not considered here. The present results recover these correspondences in the smooth setting considered here, including the cylinder term. They are new already for general magnetized elliptic systems, and are extended to non-elliptic systems with moving kernel bundles, to fields that may differ or vanish in the two bulks, and to walls depending on both variables. Two of the main mathematical tools we use are the magnetic Weyl calculus of \cite{IMP,IMP2} and a magnetic version of Douglis--Nirenberg theory \cite{DN}.

The rest of the paper is as follows. Section~\ref{sec:results} introduces the operators, defines the bulk and interface invariants, and states the main
results: the spectrum and the invariant of the magnetized bulk (Theorems~\ref{thm:bulkspec} and \ref{thm:bulk}), the quantization of the interface current and
its spectral-flow form (Theorems~\ref{thm:quant} and \ref{thm:TI}), the bulk--interface correspondence (Theorem~\ref{thm:BIC}), and stability under large perturbations (Theorem~\ref{thm:slab}). Section~\ref{sec:compute} computes the effective flat
symbol for Peierls and factorized operators and applies the results to Landau, Dirac, BHZ, shallow-water, dice and curl operators. Section~\ref{sec:engine}
constructs the graded parametrix and derives self-adjointness and resolvent estimates, section~\ref{sec:pbulk} reduces the magnetized bulk to a one-dimensional
operator and computes its invariant through the Kubo pairing, and section~\ref{sec:pBIC} proves the correspondence by locality, freezing and a spectral-flow
argument. Appendix~\ref{s:sec:models} contains further examples and numerical tests, Appendix~\ref{s:sec:tools} the symbol calculi, and
Appendices~\ref{s:sec:routine} and \ref{s:sec:routine2} the proofs of the more standard results. Appendix~\ref{s:sec:intro} summarizes the main techniques and the dependencies between the results, and lists the notation.

\smallskip
\noindent\textbf{Notation.} Points of the plane are denoted by $u=(x,y)$. The interface is the $x$-axis, and the two bulks are located at $y\to\pm\infty$. The
dual variables are $k=(\xi,\zeta)$, with $\rho=|k|$ and $\theta=k/|k|\in S^1$. We write $D=-i\nabla$ and $\jp\cdot=(1+|\cdot|^2)^{1/2}$. We denote by $S^\nu$ the
class of matrix symbols such that $|\partial^\gamma_u\partial^\beta_ka|\le C\jp k^{\nu-|\beta|}$; derivatives in $u$ are not required to gain decay. The
symbols used below are classical, that is, they have asymptotic expansions in terms homogeneous in $k$ of decreasing degree on $\{|k|\ge1\}$, and the
degree-$j$ part of a symbol refers to this expansion. We write $a=b+S^\nu$ when $a-b\in S^\nu$. $C_b^\infty$ is the space of smooth functions with bounded derivatives, $\mathfrak S_1$ and $\mathfrak S_2$ are the
trace and Hilbert--Schmidt classes, and $\Herm(n)$ is the space of Hermitian $n\times n$ matrices. We denote by $\Op^w$ the Weyl quantization on $\R^d$ without
field, by $\Op^w_y$ the Weyl quantization in the single variable $y$, with dual variable $\zeta$, by $\Op^w_h$ the semiclassical Weyl quantization with
parameter $h$, and by $\#$ (or $\#_h$) the corresponding Moyal products.
\section{Setting and main results}\label{sec:results}
Section~\ref{sec:operators} introduces the operators, and, in the non-elliptic setting where the principal symbol has a kernel of rank $r\ge1$ (Hypothesis~\ref{hyp:rank}), the frames of the kernel bundle and the effective flat symbol, and characterizes the flat-band set.
Section~\ref{sec:hypotheses} defines the bulk invariants and the cylinder term and states our main hypotheses. The last three subsections present our main results. Section~\ref{sec:magbulk} describes the spectrum
and the invariant of the magnetized bulk in terms of one-dimensional spectral data. Section~\ref{sec:interface} states the quantization of the interface current,
its spectral-flow form for translation-invariant walls, and the bulk--interface correspondence with its consequences for magnetic transitions and Landau fans.
Finally, section~\ref{sec:stability} shows that the current is stable under large perturbations.

\subsection{Operators and flat bands}\label{sec:operators}
\noindent\textbf{Operators and frames.} A (magnetic or pseudo-magnetic) field is a function $B\in C_b^\infty(\R^2)$, and a potential is a smooth vector field $A$ with
$\partial_xA_2-\partial_yA_1=B$ and $|\partial^\beta A|\le C_\beta\jp u$. The kinetic momentum is $\Pi=D-A$, and we set $\Lm=(1+\Pi^2)^{1/2}$. The magnetic
Sobolev space $H^s_A$, $s\in\N$, consists of the $f\in L^2$ with $\Pi^\beta f\in L^2$ for $|\beta|\le s$, and $H^\infty_A=\bigcap_sH^s_A$. We use the
gauge-covariant magnetic Weyl quantization \cite{IMP}
\[
  (\OpA(a)f)(u)=(2\pi)^{-2}\iint e^{i(u-u')\cdot k}e^{-i\int_{[u,u']}A}\,a\bigl(\tfrac{u+u'}2,k\bigr)f(u')\,du'\,dk .
\]
The magnetic Moyal product $\sharp^B$ is defined by $\OpA(a)\OpA(b)=\OpA(a\sharp^Bb)$, and it depends on $B$ only \cite{MP04,IMP}; its asymptotic
expansion is recalled in Appendix~\ref{s:sec:tools}. The operators considered in this paper are
of the form $H=\OpA(a)$, where $a(u,k)=\sum_{j=0}^ma_j(u,k)$ is Hermitian, $m\ge1$, and each $a_j$ is a homogeneous polynomial of degree $j$ in $k$ with
coefficients in $C_b^\infty(\R^2;\C^{n\times n})$. Thus $H$ is the Weyl-symmetrized polynomial $a(u,\Pi)$, which corresponds to the Peierls substitution. For
photonic and other classical-wave structures, $A$ can be the gauge potential induced by strain in the effective model near a Dirac point
\cite{GRW21,BCMQ26}, in which case the Peierls substitution describes the strained medium. We
call $a_m$ the principal symbol and write $W:=a_0$. A constant (bulk) symbol is written $a(k)$, and $H_B(a)=\OpA(a)$ for a constant field $B$.
Gauge transformations act by multiplication by $e^{i\lambda}$, which commutes with $P(x)$.

\smallskip
\noindent\textbf{Peierls and factorized operators.} A field-free symbol can be coupled to the field in two ways, both used below. We call
$\OpA(a)$, with $a$ independent of the field, a \emph{Peierls operator} \cite{Peierls33,Lut51,PST03,CIP19}: the field enters only through $\Pi$. 

We call \emph{factorized} an operator $H=L_B^*L_B+W$, where $L_B=\OpA(\lsym)$ is a differential operator whose symbol $\lsym(u,k)$ is an $(n-r)\times n$ polynomial of degree $p$ in $k$ with
coefficients in $C_b^\infty$, $1\le r<n$, whose principal part satisfies $\lsym_p\lsym_p^*\ge c\jp k^{2p}$ for $|k|\ge1$ and some $c>0$, and where $W\in C_b^\infty$ is Hermitian and acts by multiplication.
Then $m=2p$, $\ker a_m=\ker\lsym_p$ has rank $r$, and $H=\OpA(\lsym^*\sharp^B\lsym+W)$ is of the above form, with a symbol that depends on the field.
It differs from the Peierls operator with the same field-free symbol $\lsym^*\sharp^0\lsym+W$, which is $\lsym^*\lsym+W$ when $\lsym$ is constant, by
$\OpA(\lsym^*\sharp^B\lsym-\lsym^*\sharp^0\lsym)$, whose symbol is a polynomial of degree $m-2$ in $k$ with coefficients of size $O(B)$ in the field and its
derivatives. 

For the curl operator, $\curl_B u=i(\Pi_1u_2-\Pi_2u_1)$,
\begin{equation}\label{eq:curlPF}
  \curl_B^*\curl_B=\OpA\bigl(k^\perp(k^\perp)^T\bigr)-\tfrac B2\,\sigma_y ,\qquad k^\perp=(-\zeta,\xi),\quad \sigma_y=\bigl(\begin{smallmatrix}0&-i\\ i&0\end{smallmatrix}\bigr),
\end{equation}
for every field, since $[\Pi_1,\Pi_2]=iB$. Curl operators, Maxwell's equations and linear elasticity are naturally
factorized \cite{Silv}, whereas effective models of electrons in crystals and of strained photonic structures are usually written as Peierls operators
\cite{Lut56,GRW21}. The results of
this paper apply to both classes. For the same field-free symbol, the two operators coincide without field, but in a field they satisfy our hypotheses in different cases
(section~\ref{sec:compute}, before Proposition~\ref{prop:Meffcomp}) and have different generalized Landau levels (Examples~\ref{ex:examples} and
Remark~\ref{rem:quant}).
\begin{hypothesis}[constant kernel rank]\label{hyp:rank}
$\dim\ker a_m(u,\theta)=r\ge0$ for all $(u,\theta)\in\R^2\times S^1$, and $|a_m(u,\theta)v|\ge c_0|v|$ for $v\perp\ker a_m(u,\theta)$ for some $c_0>0$.
\end{hypothesis}
For $r=0$, this is uniform ellipticity. For $r\ge1$, we call $(u,\theta)\mapsto\ker a_m(u,\theta)$ the \emph{kernel bundle}, denote by $E_{\ker}(u,\theta)$ the projector onto $\ker a_m$, and
set $Q=1-E_{\ker}$. A Hermitian matrix $V$
is an \emph{elliptic direction} at $u$ if $E_{\ker}(u,\theta)VE_{\ker}(u,\theta)=0$ for every $\theta$, and $\mathcal V_0(u)$ denotes the space of such matrices.

A \emph{homogeneous frame} is a smooth unitary $U(u,\theta)=[V_\sff,V_\stt]$ with bounded derivatives whose first $r$ columns span $\ker a_m$. It diagonalizes
the principal symbol, $U^*a_mU=\mathrm{diag}(0,\tilde a_m)$, where the \emph{transverse principal symbol} $\tilde a_m$ satisfies $|\tilde a_mv|\ge c_0|v|$, and it is unique up to $U\mapsto U\,\mathrm{diag}(g_\sff,g_\stt)$. A
\emph{frame symbol} is a classical symbol $U\in S^0$, in the sense of the notation of section~\ref{sec:intro}, equal to $U(u,\theta(k))+S^{-1}$ for $|k|\ge R$, for some $R>0$, such that
$U^*\sharp^BU-1$ and $U\sharp^BU^*-1$ belong to $S^{-1}$ and have quantizations of norm at most $\frac12$. It is \emph{homogeneous} if it equals
$U(u,\theta(k))$ for $|k|\ge R$, that is, if it has no correction in $S^{-1}$ at large momentum. The quantization $\U:=\OpA(U)$ is then invertible, with $\U^*\U=\OpA(1+\upsilon)$ and $\upsilon\in S^{-1}$
(Lemma~\ref{lem:globalframe}(a)). The \emph{conjugated symbol} $\tilde a:=U^*\sharp^Ba\sharp^BU$ has blocks $\tilde a_{ij}$, $i,j\in\{\sff,\stt\}$, in the
splitting $\C^r\oplus\C^{n-r}$, which, unlike $\Ran E_{\ker}\oplus\Ran Q$, does not move with $(u,\theta)$. We call $\tilde a_{\sff\sff}$ the
\emph{flat block}, $\tilde a_{\sff\stt}$ and $\tilde a_{\stt\sff}$ the \emph{off-diagonal blocks}, and $\tilde a_{\stt\stt}$ the \emph{transverse block}. For $r=0$, $U=1$ and the objects attached to the kernel
are empty.
\begin{definition}[class (DN$_0$), effective flat symbol $\Meff$, flat-band set $\Fl$]\label{def:Meff}
$H$ is in class (DN$_0$) at the field $B$ if, for some frame symbol and on $\{|k|\ge R\}$, $\tilde a_{\sff\sff}\in S^0$, $\tilde a_{\sff\stt},\tilde
a_{\stt\sff}\in S^{m/2}$, and $\tilde a_{\stt\stt}-\tilde a_m\in S^{m-1}$. The \emph{effective flat symbol} $\Meff(u,\theta)$ is then the degree-$0$ part of
the Schur complement $\tilde a_{\sff\sff}-\tilde a_{\sff\stt}\tilde a_{\stt\stt}^{-1}\tilde a_{\stt\sff}$ of the transverse block, in which
$\tilde a_{\sff\stt}\tilde a_{\stt\stt}^{-1}\tilde a_{\stt\sff}$ is the \emph{Schur term}, and the \emph{flat-band set} is
$\Fl=\overline{\bigcup_{u,\theta}\spec\Meff(u,\theta)}$, with $\Fl=\emptyset$ for $r=0$.
\end{definition}
The set $\Fl$ is compact. For a constant symbol, it is a union of at most $r$ closed intervals, possibly reduced to points, and in general it may
have nonempty interior (Examples~\ref{ex:examples}).

A family $t\mapsto H_t$ of such operators is \emph{continuous} if its coefficients and the differences $A_t-A_0$ of its magnetic potentials are continuous in $C_b^\infty$, its frame symbols
are continuous in $S^0$, and its conjugated symbols are continuous in the graded class with blocks of orders $0$, $\frac m2$, $\frac m2$ and $m$.
The class (DN$_0$) consists of the operators for which $\tilde a$ is a Douglis--Nirenberg system \cite{DN} with weights $0$ and $\frac m2$ (classes $S^\nu$ are defined for real $\nu$). The bands of $a(u,\cdot)$ that remain bounded at large momentum converge
along each direction to the eigenvalues of $\Meff$, and we call them \emph{asymptotically flat bands}. Flat-band sets are computed for several examples in section~\ref{sec:compute}
(Examples~\ref{ex:examples}).
\begin{proposition}[flat bands]\label{prop:flat}
Assume Hypothesis~\ref{hyp:rank}.
\\[1mm]
(a) If $a_m$ does not depend on $u$ on $\{\pm y\ge y_0\}$, as under (N2) below, homogeneous frames exist. If (DN$_0$) holds for two frame symbols $U$ and $U'$, whose homogeneous parts are
related by $U'(u,\theta)=U(u,\theta)\,\mathrm{diag}(g_\sff,g_\stt)(u,\theta)$, then $\Meff'=g_\sff^*\Meff g_\sff$. In particular, $\Fl$ and the projection
$E^\infty$ defined below do not depend on the frame symbol for which (DN$_0$) holds. For $m\le2$, (DN$_0$) holds for all frame symbols if it holds for one.
For $m\ge3$, we do not claim that (DN$_0$) itself is independent of the frame symbol, and (N1) below requires it for one frame symbol.
\\[1mm]
(b) Assume (DN$_0$). Then $\alpha\in\Fl$ if and only if there are $\psi_n\in\cD(H)$ with $\norm{\psi_n}=1$, $\norm{(H-\alpha)\psi_n}\to0$ and
$\norm{\Lm^{-1}\psi_n}\to0$. If $\alpha\in\spec\Meff(u_0,\theta_0)$, then $\alpha\in\spec_{\rm ess}H$, with Weyl sequences localized near $u_0$, of kinetic
momentum tending to infinity along $\theta_0$, and with $\langle\psi_n,i[H,P]\psi_n\rangle\to0$ for the switches $P$ of section~\ref{sec:interface}.
\end{proposition}
The proof of the proposition is in section~\ref{sec:engine}. By (b),
every energy in $\Fl$ is essential spectrum of $H$, carried by states of arbitrarily large momentum located near the wall or in the bulks.
For $\varphi'$ supported near such an energy, $\varphi'(H)$ is then neither localized near the wall nor smoothing in momentum, and $\sigma_I$ is in general not
defined, which is why the flat-band gap (N3) below cannot be relaxed. Part (b)
also shows that $\Fl$ is determined by $H$, and not by the choice of frame.

Explicit formulas for $\Meff$ are given in Proposition~\ref{prop:Meffcomp}. In particular, (DN$_0$) holds
for $r=0$, and for $m=1$ when homogeneous frames exist. Moreover, $\Meff$ does not depend on the field for $m\le2$ and for factorized operators, and the field may enter $\Meff$ for $m\ge3$.

\medskip

We call \emph{occupied projector}, by analogy with the electronic case, the spectral projector below $\alpha$ at infinite momentum. For $\alpha\notin\Fl$ and
$(u,\theta)\in\R^2\times S^1$, it is
\[
  E^\infty(u,\theta):=U\bigl(E^\infty_{\rm fl}\oplus E^\infty_{\rm tr}\bigr)U^*,\qquad E^\infty_{\rm fl}:=\mathbf 1_{(-\infty,\alpha)}(\Meff(u,\theta)),\quad E^\infty_{\rm
  tr}:=\mathbf 1_{(-\infty,0)}(\tilde a_m(u,\theta)),
\]
of constant rank $q$. For $r=0$, it reduces to $E^\infty(u,\theta)=\mathbf 1_{(-\infty,0)}(a_m(u,\theta))$. For $x_0\in\R$, the \emph{cylinder term} is
\begin{equation}\label{eq:cyl}
  \Cy(x_0):=\int_{[-y_0,y_0]\times S^1}c_1\bigl(E^\infty(x_0,\cdot,\cdot)\bigr),\qquad\text{oriented by }dy\wedge d\theta ,
\end{equation}
where $c_1(E):=\frac i{2\pi}\tr(E\,dE\wedge dE)$ is the first Chern form of a smooth map $E$ with values in orthogonal projections. We write $\Cy$ for
$\Cy(x_0)$ when it does not depend on $x_0$.

\subsection{Bulk quantities and hypotheses}\label{sec:hypotheses}
For a constant symbol $\ab$ without field, $\spec H_0(\ab)=\overline{\bigcup_k\spec\ab(k)}$, and we set
$\mathcal G(\ab):=\R\setminus\spec H_0(\ab)$. For $\alpha\in\mathcal G(\ab)$ and $E_\alpha=\mathbf 1_{(-\infty,\alpha)}(\ab)$, the Chern integral
\begin{equation}\label{eq:I}
  I(\ab;\alpha):=-\int_{\R^2}c_1(E_\alpha),\qquad\R^2\text{ oriented by }d\xi\wedge d\zeta ,
\end{equation}
converges. It is not an integer in general. For a constant field $B\ne0$, the Landau gauge $A=(-By,0)$ yields fibers, the operators on $L^2(\R_y)$ obtained by Fourier transform in $x$, that are unitarily equivalent, by
translation to the guiding center $y_c=-\xi/B$, to the \emph{reduced bulk operator}
\begin{equation}\label{eq:KB}
  K_B(\ab):=\Op_y^w\bigl(\ab(By,\zeta)\bigr)\ \text{on }L^2(\R_y;\C^n).
\end{equation}
For $B=0$, we set $\spec K_0(\ab):=\spec H_0(\ab)$. We call the points of $\spec K_B(\ab)$, $B\ne0$, the \emph{generalized Landau levels}, since $\spec H_B(\ab)=\spec K_B(\ab)$ by Theorem~\ref{thm:bulkspec}, and
we denote by $\Fl_B(\ab)$ the flat-band set of $H_B(\ab)$. For $B\ne0$ and $\alpha\notin\spec K_B(\ab)\cup\Fl_B(\ab)$, with $P_\alpha=\mathbf 1_{(-\infty,\alpha)}(H_B(\ab))$ and $\hat
P_\alpha=\mathbf 1_{(-\infty,\alpha)}(K_B(\ab))$, we define
\begin{equation}\label{eq:sigmaH}
  \sigma_H(\ab,B;\alpha):=i\,\trarea\bigl(P_\alpha[[P_\alpha,x],[P_\alpha,y]]\bigr),\qquad J(\ab,B;\alpha):=2\pi\sigma_H(\ab,B;\alpha),
\end{equation}
where $\trarea X:=\lim_{L\to\infty}(2L)^{-2}\Tr(\mathbf 1_{\Lambda_L}X\mathbf 1_{\Lambda_L})$, $\Lambda_L=[-L,L]^2$, is the trace per unit area \cite{BES}. With this sign convention, a single Landau level of $|k|^2$ at $B>0$ gives
$J=+1$. The definition \eqref{eq:sigmaH} also applies at $B=0$, and it then gives $J=I$ for every
constant symbol and $\alpha\in\mathcal G(\ab)$, including the unregularized Dirac symbol (Lemma~\ref{lem:kubo}(b)). For $B=0$ and $\alpha\in\mathcal G(\ab)$, we therefore set
\begin{equation}\label{eq:J0}
  J(\ab,0;\alpha):=I(\ab;\alpha).
\end{equation}

We now state our main standing hypothesis on the operator, in particular that it has the  block structure of Definition~\ref{def:Meff}, that its coefficients and the field are constant outside a strip, and that the energy (or frequency depending on the application) of interest is in a spectral gap of both bulks and away from flat bands. It is under these conditions that the interface current is defined and quantized.
\begin{hypothesis}[{\bf N}]\label{hyp:N}
(N1) Hypothesis~\ref{hyp:rank} holds, and $H$ is in class (DN$_0$) at the given field.
\\[2pt]
(N2) There is $y_0>0$ such that $a=a_\pm(k)$ and $B=B_\pm$ are constant on $\{\pm y\ge y_0\}$. The values $B_\pm=0$ are allowed. We call the strip
$\{|y|<y_0\}$ the (domain) wall.
\\[2pt]
(N3) The flat-band gap holds, namely $\dist(\alpha,\Fl)>0$. This is vacuous for $r=0$.
\\[2pt]
(N4) The energy lies in a gap of both bulks, $\alpha\notin\spec K_{B_+}(a_+)\cup\spec K_{B_-}(a_-)$.
\end{hypothesis}
Some results hold under the weaker hypothesis \textbf{(N$'$)} (N1) and (N3) hold, and there are operators $H_\pm$ satisfying (N1) and (N3), agreeing with $H$ on
$\{\pm y\ge y_0\}$, and such that $\alpha\notin\spec H_\pm$. Hypothesis (N) implies (N$'$): there are potentials $A_\pm$ with
$\curl A_\pm=B_\pm$ and $A_\pm=A$ on $\{\pm y\ge y_0\}$, and the \emph{bulk operators} $H_\pm:=\Op^{A_\pm}(a_\pm)$ satisfy (N1), (N3) and
$\spec H_\pm=\spec K_{B_\pm}(a_\pm)$ (Lemma~\ref{s:lem:bulks}). 

We define the {\em margin} $\delta:=\min\{\dist(\alpha,\Fl),\dist(\alpha,\spec H_+\cup\spec H_-)\}$, and the {\em energy window}
$\mathcal E_\alpha:=(\alpha-\frac\delta2,\alpha+\frac\delta2)$. Unless stated otherwise, constants depend only on $n$, $m$, $c_0$, $y_0$, $|\alpha|$, $\delta$, the frame, and
finitely many seminorms of the coefficients and of $B$.
Under (N1) and (N3), $H$ is essentially self-adjoint on
$C_c^\infty$, with resolvent estimates in weighted spaces adapted to the Douglis--Nirenberg structure (Theorem~\ref{thm:analysis}).

\smallskip 
\noindent\textbf{Conditions (TI), (F), and (W).} The following conditions are required only by some of the results. Sufficient conditions for (F) and (W)
are given in Lemma~\ref{s:lem:Fsufficient} and Proposition~\ref{s:prop:Wholds}. The wall is \emph{translation invariant} (TI), when $a=a(y,k)$ and $B=B(y)$ do not depend on $x$. General walls are reduced to
translation-invariant ones by freezing the coefficients along the interface. For $x_0\in\R$, $L\ge1$ and $t\in[0,1]$, let $H^{x_0}_{t,L}$ be the operator with
coefficients $a(x_0+t\chi_L(x-x_0),y,k)$ and field $B(x_0+t\chi_L(x-x_0),y)$, where $\chi_L$ is odd, equal to $s$ for $|s|\le L$ and constant for $|s|\ge2L$,
with $0\le\chi_L'\le1$ and $|\chi_L^{(j)}|\le C_jL^{1-j}$. For factorized operators $L_B^*L_B+W$, the coefficients of the symbol $\lsym$ of $L_B=\OpA(\lsym)$ and those of $W$ are frozen instead of
those of $a$. 

We say that $H$ satisfies the \emph{freezing condition} (F) if there are $x_0$ and $L_0$ such that $H^{x_0}_{t,L}$ satisfies (N) with uniform
constants for $t\in[0,1]$ and $L\ge L_0$, continuous in $t$ in the sense of section~\ref{sec:operators}. Condition (F) holds under (TI),
for $r=0$, for first-order systems whose kernel bundle does not depend on $x$, and for factorized operators, under the conditions of Lemma~\ref{s:lem:Fsufficient}, which is proved in Appendix~\ref{s:sec:routine}. It can fail when
the kernel depends on $x$, since $\Meff$ contains $x$-derivatives of the frame.

When a bulk field vanishes, we require the Douglis--Nirenberg structure of the bulk symbol to persist under weak fields. Let $\ab$ be a constant symbol, or a
field-dependent family $\ab^{(B)}$ with $\ab^{(B)}-\ab^{(0)}\in S^{m-2}$ of seminorms $O(\norm B_{C^{N_0}})$, as for factorized operators, in which case $J(\ab,B;\alpha)$ is computed from
$\ab^{(B)}$ and $I(\ab;\alpha)$ from $\ab^{(0)}$. We say that $\ab$
satisfies (W) at the energy $\alpha$ if, for some $b_1>0$ and integer $N_0$, every field with $\norm B_{C^{N_0}}\le b_1$ admits a frame symbol $U^{(B)}(k)$, independent of $u$,
such that:
\\[1mm]
(W1) $U^{(0)}$ is unitary-valued and $U^{(B)}-U^{(0)}=O(\norm B_{C^{N_0}})$ in $S^{-1}$;
\\[1mm]
(W2) the conjugated symbol $\tilde a^{(B)}:=U^{(B)*}\sharp^B\ab^{(B)}\sharp^BU^{(B)}$ satisfies (DN$_0$);
\\[1mm]
(W3) its blocks differ from those of $U^{(0)*}\ab^{(0)}U^{(0)}$ by $O(\norm B_{C^{N_0}})$ in the classes $S^0$, $S^{m/2}$, $S^{m/2}$ and $S^m$ of (DN$_0$);
\\[1mm]
(W4) when such a field is added to $H$ in the bulk region where its symbol is $\ab$, the resulting operator still satisfies (N1), with frame and conjugated
symbols continuous in the field in the sense of section~\ref{sec:operators};
\\[1mm]
(W5) at the energy $\alpha$ under consideration, the occupied flat projector does not depend on the field:
$\mathbf 1_{(-\infty,\alpha)}(\Meff^{(B)}(u,\theta))=\mathbf 1_{(-\infty,\alpha)}(\Meff^{(0)}(\theta))$ for all $(u,\theta)$, where $\Meff^{(B)}$ is the effective
flat symbol of $\tilde a^{(B)}$.
\\[1mm]
Conditions (W1)--(W4) keep $\dist(\alpha,\Fl)$ bounded below under such fields, and (W5) means that the projection $E^\infty$ at $\alpha$ does not depend on
the field. It is needed because $J$ is determined modulo $\Z$ by $E^\infty$ (Theorem~\ref{thm:bulk}(b)), so that a field-dependent $E^\infty$ would give a
field-dependent fractional part of $J$, incompatible with the weak-field limit of Theorem~\ref{thm:bulk}(c). It holds automatically, after decreasing $b_1$,
when $r\le1$ and when $\Meff$ does not depend on the field.

Hypothesis (N1) at $B\ne0$ and condition (W) concern the operator. The cases in which they
can be read off the field-free symbol, which include elliptic, factorized and low-order Peierls operators, are listed in section~\ref{sec:compute}, before Proposition~\ref{prop:Meffcomp}.
\subsection{The magnetized bulk}\label{sec:magbulk}
\begin{theorem}[spectrum of the magnetized bulk]\label{thm:bulkspec}
Let $\ab$ satisfy (N1) at $B\ne0$. Then $\Fl_B(\ab)=\bigcup_\theta\spec\Meff(\theta)$, $K_B(\ab)$ is self-adjoint, $\spec_{\rm ess}K_B(\ab)=\Fl_B(\ab)$, and
$\spec H_B(\ab)=\spec K_B(\ab)$, each eigenvalue of $K_B(\ab)$ being an eigenvalue of infinite multiplicity of $H_B(\ab)$.
\end{theorem}
The proof is given in section~\ref{sec:pbulk}. For $r\ge1$, the flat-band energies are therefore essential spectrum of $K_B(\ab)$, for instance infinitely degenerate levels or accumulation points of
\emph{Landau fans}. 

\smallskip
The fractional part of $J$ comes from infinite momenta, where the occupied projector is the loop $\theta\mapsto E^\infty(\theta)$. To compare a bulk with
reference symbols whose invariant can be computed, we extend this loop to a disk, which plays the role of a compactified momentum plane. A \emph{bulk
extension} is a smooth map $E^{\rm ext}:\mathbb D\to\Gr_q(\C^n)$, where $\Gr_q(\C^n)$ is the Grassmannian of $q$-dimensional subspaces of $\C^n$, identified with
the corresponding orthogonal projections, and $\mathbb D$ is the unit disk oriented by $dw_1\wedge dw_2$, such that
$E^{\rm ext}(w)=E^\infty(w/|w|)$ for $\frac12\le|w|\le1$. Its Chern integral is
\begin{equation}\label{eq:kappa}
  \kappa(E^{\rm ext}):=\int_{\mathbb D}(E^{\rm ext})^*c_1=\frac i{2\pi}\int_{\mathbb D}\tr\bigl(E^{\rm ext}\,[\partial_{w_1}E^{\rm ext},\partial_{w_2}E^{\rm ext}]\bigr)\,dw_1\,dw_2 .
\end{equation}
For $0<q<n$, the Grassmannian $\Gr_q(\C^n)$ is simply connected and $\pi_2(\Gr_q(\C^n))\cong\Z$, the isomorphism being given by the integral of $c_1$ \cite[\S4.2]{Hatcher}. Hence bulk extensions exist. Two
extensions that agree on the annulus glue along it into a map from the sphere to $\Gr_q(\C^n)$, whose class measures their difference, and every class is
realized by a connected sum. The homotopy classes of extensions relative to the annulus thus form a torsor under $\pi_2(\Gr_q(\C^n))\cong\Z$, and $\kappa$
distinguishes them. For $q\in\{0,n\}$, $\Gr_q(\C^n)$ is a point and $\kappa=0$. The integer $\mathcal I_{E^{\rm ext}}$ of \eqref{eq:Ihat}, used below, is $-\sgn(B)$
times the spectral flow of $K_B$ along a gapped path from a reference symbol built on $E^{\rm ext}$ to the conjugated symbol of $\ab-\alpha$.
\begin{theorem}[the bulk invariant]\label{thm:bulk}
Let $\ab$ satisfy (N1) at $B\ne0$, and let $\alpha,\alpha'\notin\spec K_B(\ab)$ lie in the same connected component of $\R\setminus\Fl_B(\ab)$.
\\[1mm]
(a) For $\alpha<\alpha'$, $J(\alpha')-J(\alpha)=\sgn(B)\#(\spec K_B(\ab)\cap(\alpha,\alpha'))$, where $\#$ counts multiplicities. In particular, if the transverse
principal symbol $\tilde a_m$ is positive definite and $\alpha<\inf\Fl_B(\ab)$, then $J(\ab,B;\alpha)=\sgn(B)\#(\spec K_B(\ab)\cap(-\infty,\alpha))$.
\\[1mm]
(b) For every bulk extension $E^{\rm ext}$, $J=\mathcal I_{E^{\rm ext}}-\kappa(E^{\rm ext})$, where the integer $\mathcal I_{E^{\rm ext}}$ of \eqref{eq:Ihat} is a signed count of
the levels created along a deformation from a reference symbol built on $E^{\rm ext}$ to $\ab-\alpha$. In particular, $J\equiv-\kappa(E^{\rm ext})\pmod\Z$.
\\[1mm]
(c) Under (W) at $\alpha$, if $\alpha\in\mathcal G(\ab)$, then $J(\ab,B;\alpha)=I(\ab;\alpha)$ for $0<|B|\le b_0$, where $b_0$ is the constant of
Proposition~\ref{prop:weak} with $K=\{\alpha\}$.
\end{theorem}
The proof is given in section~\ref{sec:pbulk}. By (a), a generalized Landau level of multiplicity $d$ carries the Chern number $d\,\sgn(B)$, whatever the
symbol. By (b), non-integer values of $J$, such as $-\frac12\sgn M$ for the Dirac symbol, come from infinite momenta and not from the levels. When
$E^\infty$ depends on the field, which may happen for $r\ge2$ and $m\ge3$, so does the fractional part of $J$, which is why (c) requires (W5). By (c), $J\to I$
as $B\to0$ in a field-free gap, which justifies \eqref{eq:J0}; inside a field-free band $J$ has no limit, since $J\approx\sgn(B)\,\alpha/(2|B|)$ for $|k|^2$. Hall conductivities of
magnetized bulks are related to Landau levels for the Landau Hamiltonian in \cite{Streda,ASS,BES} and for magnetic Dirac operators in \cite{QB22,CMS}, and
St\v{r}eda-type formulas are discussed in \cite{CMT}.

For orthogonal projections $P$ and $Q$ with $P-Q$ compact, $\ind(P,Q):=\dim\ker(P-Q-1)-\dim\ker(Q-P-1)$ is the index of the pair. It is additive, and
it equals $\Tr(P-Q)$ when $P-Q\in\mathfrak S_1$ \cite{ASS}. We write $\sfl$ for the spectral flow, which counts the net number of eigenvalues crossing $0$ upward
\cite{Phillips}, and $\Fl_{tB}(\ab)$ is the flat-band set of $H_{tB}(\ab)$. The following consequences of Theorem~\ref{thm:bulk} describe the dependence of $J$
on the symbol at a fixed field and on the field.
\begin{corollary}[pair index and field dependence]\label{cor:bulk}
Let $\ab$ be as in Theorem~\ref{thm:bulk}.
\\[1mm]
(a) If $\ab'$ satisfies (N1) at $B$, with $E^\infty(\ab')$ at $\alpha'$ equal to $E^\infty(\ab)$ at $\alpha$, then
$J(\ab,B;\alpha)-J(\ab',B;\alpha')=\sgn(B)\ind(\hat P_\alpha(\ab),\hat P_{\alpha'}(\ab'))$. If (N1) and this equality hold at all fields near $B$, then
$\sigma_H(\ab)-\sigma_H(\ab')=\partial_Bn_{\rm rel}$ with $n_{\rm rel}=\frac{|B|}{2\pi}\ind$. If the transverse principal symbol $\tilde a_m$ is positive definite and $\alpha<\inf\Fl_B(\ab)$, then
$\sigma_H=\partial_Bn(\alpha,B)$ with $n=\frac{|B|}{2\pi}\#(\spec K_B(\ab)\cap(-\infty,\alpha))$.
\\[1mm]
(b) Let $0<t_1<1$, and assume (N1) and $\dist(\alpha,\Fl_{tB}(\ab))\ge c$ for some $c>0$ and all $t\in[t_1,1]$, and that $E^\infty$ at $\alpha$ does not depend on $t$.
Then $J(\ab,tB;\alpha)$ changes only when a level crosses $\alpha$, by $-\sgn(B)$ per upward crossing. If these assumptions hold on $[t_0,1]$ with
$0<t_0\le b_0/|B|$, where $b_0$ is the constant of Theorem~\ref{thm:bulk}(c) at $\alpha_0$, and (W) holds at some
$\alpha_0\in\mathcal G(\ab)\setminus\spec K_B(\ab)$, $\alpha_0<\alpha$, in the connected component of $\alpha$ in $\R\setminus\Fl_{tB}(\ab)$ for every $t\in[t_0,1]$, then
\[
  J(\ab,B;\alpha)=I(\ab;\alpha_0)-\sgn(B)\,\sfl\bigl(t\mapsto K_{tB}(\ab)-\alpha_0;\,t\in[t_0,1]\bigr)+\sgn(B)\,\#\bigl(\spec K_B(\ab)\cap(\alpha_0,\alpha)\bigr).
\]
\end{corollary}
The proof is given in section~\ref{sec:pbulk}. The second and third formulas of (a) are St\v{r}eda formulas. The assumptions on $E^\infty$ in (a) and (b) hold when
$r\le1$, when $m\le2$ and for factorized symbols. For $\alpha<\alpha_0$ in (b), the last term is replaced by $-\sgn(B)\,\#(\spec K_B(\ab)\cap(\alpha,\alpha_0))$.

\subsection{The interface current and the main result}\label{sec:interface}
A \emph{switch} is a smooth real-valued function equal to $0$ near $-\infty$ and to $1$ near $+\infty$, and
$\mathsf P_{x_0}:=\mathbf 1_{[x_0,\infty)}(x)$ is the sharp switch at $x_0$. Transport is counted positively in the direction of increasing $x$, so that a
single branch $E=\xi$ gives $2\pi\sigma_I=+1$. The current $\sigma_I$ is the index $\Ind T=\dim\ker T-\dim\ker T^*$ of a Fredholm operator $T$, which gives
its quantization and its independence of the switches. For translation-invariant walls between magnetized bulks, it is also a spectral flow, through which the bulk invariants enter.
\begin{theorem}[quantization]\label{thm:quant}
Assume (N$'$). For $\Phi\in C_c^\infty(\mathcal E_\alpha)$, the Schwartz kernel $K_\Phi$ of $\Phi(H)$ satisfies $|K_\Phi(u,u')|\le
C_Ne^{-c(|y|+|y'|)}\jp{x-x'}^{-N}$. The operator $i[H,P]\varphi'(H)$ is trace-class for every switch $P$ and every switch $\varphi$ with $\varphi'\in
C_c^\infty(\mathcal E_\alpha)$, and
\[
  2\pi\sigma_I=\Ind\bigl(\mathsf P_{x_0}U_\varphi\mathsf P_{x_0}|_{\Ran\mathsf P_{x_0}}\bigr)=-\Tr\bigl(U_\varphi^*PU_\varphi-P\bigr)\in\Z,\qquad U_\varphi=e^{2\pi i\varphi(H)},
\]
independently of $P$, of $x_0$, of $\varphi$, and of the gauge.
\end{theorem}
We write $\sigma_I(H)$. For translation-invariant walls, we denote by $\Hh(\xi)$, $\xi\in\R$, the fibers of $H$ under the partial Fourier transform in $x$, in
a gauge $A=(A_1(y),0)$, and set $\SF(\alpha):=\sfl(\xi\mapsto\Hh(\xi)-\alpha)$.
\begin{theorem}[translation-invariant walls]\label{thm:TI}
Assume (N), (TI) and $B_\pm\ne0$. The fibers $\Hh(\xi)$ have discrete spectrum in $\mathcal E_\alpha$ and are norm-resolvent analytic. Moreover, $\spec\Hh(\xi)\cap\mathcal E_\alpha=\emptyset$
for $|\xi|\ge Z_\alpha$, with $Z_\alpha\le C_0(1+\delta^{-1})$, and $2\pi\sigma_I=\SF(\alpha)=\int\tr(\partial_\xi\Hh\,\varphi'(\Hh))\,d\xi$.
Finally, for $\alpha_1<\alpha_2$ satisfying (N3) and (N4) with $[\alpha_1,\alpha_2]\cap\Fl=\emptyset$,
\[
  \SF(\alpha_2)-\SF(\alpha_1)=\sgn(B_+)\,n_+-\sgn(B_-)\,n_-,\qquad n_\pm=\#\bigl(\spec K_{B_\pm}(a_\pm)\cap(\alpha_1,\alpha_2)\bigr),
\]
counted with multiplicity.
\end{theorem}
The proofs of Theorems~\ref{thm:quant} and \ref{thm:TI} are given in Appendix~\ref{s:sec:routine2}. Once $\Phi(H)$ is localized near the wall
(Proposition~\ref{prop:loc}), they follow known arguments. The identification of $2\pi\sigma_I$ with a Fredholm index is a Fedosov-type computation as in
\cite{ASS,EG,EGS,BEC}, and its identification with the spectral flow of the fibers uses the fiber decomposition and Phillips' formula as in
\cite{Phillips,QB21,QB22}. For $B_\pm\ne0$, Theorem~\ref{thm:TI} extends \cite{QB22}.

\bigskip
\noindent\textbf{Bulk--interface correspondence.} We now present the main result of this paper. For an orthonormal frame $\psi$ of
$\Ran E^\infty(x_0,y,\cdot)$ depending smoothly on $y\in[-y_0,y_0]$, let $\Theta(y):=i\int_{S^1}\tr(\psi^*\partial_\theta\psi)\,d\theta$ be its Berry phase.
\begin{theorem}[bulk--interface correspondence]\label{thm:BIC}
Assume (N) and either (TI) or (F). If $B_+=0$ or $B_-=0$, assume (W) at $\alpha$ for the corresponding bulk symbol. Then $\Cy(x_0)$ does not depend on $x_0$, and
\[
  2\pi\sigma_I=J_+-J_-+\Cy ,\qquad J_\pm:=J(a_\pm,B_\pm;\alpha),
\]
with the convention \eqref{eq:J0}. Equivalently, $2\pi\sigma_I=\tilde J_+-\tilde J_-$ with $\tilde J_\pm:=J_\pm+\Theta(\pm y_0)/2\pi\in\Z$.
\end{theorem}
Theorem~\ref{thm:BIC} generalizes the correspondences obtained for magnetic barriers of the Landau Hamiltonian in \cite{DGR} and for magnetic Dirac
operators in \cite{QB22}, and, for $B_\pm=0$ and $r=0$, those of \cite{Bal19,Bal22,QB21,BEC} for elliptic operators; for the latter, see also
\cite{Drouot,Faure}. Changing $\psi$ by a unitary gauge $g(\theta)$ whose determinant has winding $w$ shifts $\tilde J_+$ and $\tilde J_-$ by the
same integer $-w$, so that only their difference is intrinsic. The term $\Cy$ depends on the field only through $\Meff$, and it vanishes if $E^\infty$ does not
depend on $y$. When $B_\pm\ne0$, the right-hand side is also the difference of the numbers of levels created in the two bulks along deformations to references
built on a common extension of $E^\infty$ (Theorem~\ref{thm:relcount}), which explains why it is an integer.
The proof is given in section~\ref{sec:pBIC}.

The theorem covers interfaces between two Landau insulators, \emph{magnetic transitions} between a Landau
insulator and a band insulator, and field-free interfaces. The last two cases and the weak-field limit are made explicit as follows.
\begin{corollary}[magnetic transitions]\label{cor:trans}
Assume the hypotheses of Theorem~\ref{thm:BIC}.
\\[1mm]
(a) If $B_+=B_-=0$, then $2\pi\sigma_I=I(a_+;\alpha)-I(a_-;\alpha)+\Cy$.
\\[1mm]
(b) Let $a\equiv\ab$ be a constant symbol, and assume that only the field varies across the wall, with $B_-=0\ne B_+$. Then
$2\pi\sigma_I=J(\ab,B_+;\alpha)-I(\ab;\alpha)+\Cy$, and $\Cy=0$ when $r\le1$, when $m\le2$, and for factorized operators. Under (W) at $\alpha$, if moreover (N1) and $\dist(\alpha,\Fl_{tB_+}(\ab))\ge c$ hold for some $c>0$ and all $t\in(0,1]$ and $E^\infty$ at $\alpha$ does not depend on $t$, then $J(\ab,B_+;\alpha)-I(\ab;\alpha)$ is $-\sgn(B_+)$ times the net number of upward crossings
of $\alpha$ by levels of $K_{tB_+}(\ab)$ for $t\in(0,1]$.
\\[1mm]
(c) Let $H_t$ have the coefficients of $H$ and the field $tB$. Assume that $H_t$ satisfies the hypotheses of Theorem~\ref{thm:BIC} for $|t|\le t_1$, for some $t_1>0$, with
continuous in $t$ in the sense of section~\ref{sec:operators}, that $a_+$ and $a_-$ satisfy (W) at $\alpha$, and that
$\alpha\in\mathcal G(a_+)\cap\mathcal G(a_-)$. Then there is $t_0\in(0,t_1]$ such that $\sigma_I(H_t)=\sigma_I(H_0)$ for $|t|\le t_0$, namely
$2\pi\sigma_I(H_t)=I(a_+;\alpha)-I(a_-;\alpha)+\Cy$.
\end{corollary}
By (b), a boundary of the applied field in a homogeneous material carries a quantized current, which vanishes at weak field. Part (c) states that the
field-free current is attained, and not merely approached, at weak field. It does not follow from continuity in $t$, since rescaling the field changes the
bulk fields by unbounded potentials, but from Theorem~\ref{thm:BIC} at both ends. Without (W), the weak-field limit may fail: for
$\rho^4k^\perp(k^\perp)^T+g\sigma_y$, every field $B\ne0$ makes the bounded band unbounded (Proposition~\ref{prop:Meffcomp}(e)), and infinitely many levels cross an energy below all field-free bands as $B\to0$ (Proposition~\ref{s:prop:exactB}(c) and Example~\ref{s:ex:higher}).
\begin{corollary}\label{cor:main}
Assume the hypotheses of Theorem~\ref{thm:BIC}.
\\[1mm]
(a) The current $\sigma_I$ depends only on $(a_\pm,B_\pm,\alpha)$ and on the homotopy class of $E^\infty(x_0,\cdot,\cdot)$ relative to $\{y=\pm y_0\}$. It is
therefore unchanged by modifications inside the strip, of arbitrary size, that preserve the hypotheses and this homotopy class.
\\[1mm]
(b) If the transverse principal symbols of $a_\pm$ are positive definite and $\alpha<\inf\Fl$, then $\Cy=0$ and $2\pi\sigma_I=\sgn(B_+)N_+(\alpha)-\sgn(B_-)N_-(\alpha)$, where $N_\pm$ are the finite
numbers of levels of $K_{B_\pm}(a_\pm)$ below $\alpha$, and where the term $\sgn(B_\pm)N_\pm$ is replaced by $I(a_\pm;\alpha)$ if $B_\pm=0$.
\\[1mm]
(c) Assume that $B_+\ne0$, that $\spec K_{B_+}(a_+)$ accumulates at $f\in\Fl$ from above, and that, for some $\varepsilon>0$, $(f,f+\varepsilon)$ contains finitely many points of
$\spec K_{B_-}(a_-)$, or that $B_-=0$ and $(f,f+\varepsilon)\subset\mathcal G(a_-)$, and that $(f,f+\varepsilon)\cap\Fl=\emptyset$. Then $|2\pi\sigma_I(\alpha)|\to\infty$ as
$\alpha\downarrow f$ through energies outside the levels of the bulks. The same holds with the roles of the two bulks exchanged, and, for levels accumulating at $f$
from below, as $\alpha\uparrow f$.
\end{corollary}
The proofs of the corollaries are given in section~\ref{sec:pBIC}. In Corollary~\ref{cor:trans}(b), the assumptions on $E^\infty$ hold when $r\le1$, when
$m\le2$ and for factorized operators. In Corollary~\ref{cor:main}(a), the homotopy class is preserved along every path that satisfies the hypotheses, in particular
by perturbations of degree $0$ in the elliptic directions and, for $r\le1$, by those in the non-elliptic directions that keep (N3). Part (b) gives quantized
currents inside field-free bands, since for $B_\pm\ne0$ only (N4) is required, and (c) shows that $\sigma_I$ may diverge at a Landau fan.

\smallskip
\noindent\textbf{The elliptic case.} When $r=0$, the setting simplifies as follows. No frame is
needed, $\Fl=\emptyset$, and (N1), (N3) and (F) hold, as does (W) for Peierls and factorized couplings, so that (N) reduces to coefficients and field constant outside a
strip and an energy in a gap of both bulks. The spectrum of $K_B(\ab)$ is discrete (Theorem~\ref{thm:bulkspec}), and $E^\infty=\mathbf 1_{(-\infty,0)}(a_m)$
involves only the principal symbol. Theorem~\ref{thm:BIC} then gives $2\pi\sigma_I=J_+-J_-+\Cy$ for walls depending on $(x,y)$, for all bulk fields including
vanishing ones, and at every energy in a common gap of the bulks. When $a_m$ is positive definite, $\Cy=0$ and $2\pi\sigma_I=\sgn(B_+)N_+-\sgn(B_-)N_-$, with $\sgn(B_\pm)N_\pm$ replaced by
$I(a_\pm;\alpha)$ when $B_\pm=0$ (Corollary~\ref{cor:main}(b)). Every perturbation of order at most $m-1$, whatever its size, preserves (N1) and (N3), so that only the bulk gaps need to be controlled along a
deformation, as in section~\ref{sec:stability}. These results generalize to arbitrary magnetized elliptic systems the non-magnetic setting of \cite{Bal22,QB21} and the results obtained for specific
models, Landau-type Schr\"odinger operators \cite{Taarabt,DGR,CMT} and magnetic Dirac operators \cite{CMS,QB22,BCLRS24,BCLRS25}.

\subsection{Stability}\label{sec:stability}
A perturbation consists of a Hermitian polynomial symbol $v(u,k)$ of degree at most $m$ in $k$ with
coefficients in $C_b^\infty(\R^2)$, and of a vector potential $\beta\in C_b^\infty(\R^2;\R^2)$, which changes the field by $\curl\beta$. We set
$H_s:=\Op^{A+s\beta}(a+sv)$, $s\in[0,1]$, and, for a factorized operator, $H_s:=L_{B,s}^*L_{B,s}+W+s\Op^{A+s\beta}(v)$ with $L_{B,s}=\Op^{A+s\beta}(\lsym)$. Two conditions
control such paths, one for each way a gap can close: at infinite momentum through the flat bands, and at finite momentum through the bulk spectra.
\begin{definition}[admissible perturbations]\label{def:admissible}
The perturbation $(v,\beta)$ is \emph{structurally admissible} if the operators $H_s$, $s\in[0,1]$, satisfy (N1) and (N3) with uniform constants, with
frame symbols and conjugated symbols depending continuously on $s$ in the sense of section~\ref{sec:operators}. It is \emph{gap preserving} if, moreover, there
are $y_1\ge y_0$ and operators $H_{\pm,s}$, $s\in[0,1]$, agreeing with $H_s$ on $\{\pm y\ge y_1\}$, satisfying (N1) and (N3) with uniform constants and
frame and conjugated symbols continuous in $s$, and such that $\alpha\notin\spec H_{\pm,s}$ for all $s$.
\end{definition}
Structural admissibility holds, whatever the size of the perturbation and its dependence on $(x,y)$, in three cases. (i) For $r=0$, every $v$ of degree
at most $m-1$ and every $\beta$ are admissible. (ii) Perturbations $v$ of degree $0$ with values in the elliptic directions $\mathcal V_0(u)$, which leave
$\Meff$ and $\Fl$ unchanged, are admissible with any $\beta$ if $m\le2$ or if the operator is factorized, and with $\beta=0$ otherwise. (iii) Perturbations
whose conjugated symbol belongs to the class $\mathcal S^{-1}_+$ of section~\ref{sec:engine}, with flat, off-diagonal and transverse blocks of orders $-1$,
$\frac m2-1$ and $m-1$, are admissible if $\beta=0$ or $m\le2$. A perturbation of
degree $0$ outside $\mathcal V_0(u)$ moves $\Meff$ by $V_\sff^*vV_\sff$ at leading order, and it is admissible as long as $\Fl$ stays at positive distance from
$\alpha$.

Gap preservation holds with $H_{\pm,s}=H_\pm$ for $v$ and $\beta$ supported in a strip $\{|y|\le y_1-1\}$, and, with $H_{\pm,s}$ the corresponding
perturbed bulk operators, for perturbations that are small relative to the bulk gaps in the weighted norm of Theorem~\ref{thm:analysis}(b), for instance
periodic or random lower-order terms and periodic field modulations of zero mean. Stability results of this kind are obtained for the Landau Hamiltonian in
\cite{DGR} and for magnetic Dirac operators in \cite{QB22}.
\begin{theorem}[stability]\label{thm:slab}
Let $H$ satisfy the hypotheses of Theorem~\ref{thm:quant}.
\\[1mm]
(a) If $(v,\beta)$ is structurally admissible and gap preserving, then $\sigma_I(H_1)=\sigma_I(H)$. If $H$ satisfies the hypotheses of
Theorem~\ref{thm:BIC}, then $2\pi\sigma_I(H_1)=J(a_+,B_+;\alpha)-J(a_-,B_-;\alpha)+\Cy$, with the bulk data and the cylinder term of $H$, although $H_1$
need not satisfy (TI), (F) or (N2). If, moreover, $v$ and $\beta$ are supported in $\{|y|\le y_1-1\}$, then $\Cy=\Cy_1(x_0)$ for every $x_0$, where $\Cy_1(x_0)$ is
the cylinder term of $H_1$ on $[-y_1,y_1]\times S^1$.
\\[1mm]
(b) If $v$ and $\beta$ are supported in a slab $\{|x-x_1|\le L\}$, no path is needed. It suffices that $H_1$ satisfies (N1) and (N3) with a frame symbol
equal to that of $H$ on $\{x\ge x_1+L+1\}$, and that
$\alpha\notin\spec\Op^{A_\pm+\beta}(a_\pm+v)$. Then $\sigma_I(H_1)=\sigma_I(H)$, whatever the order and the size of the perturbation.
\end{theorem}
In (b), a frame symbol of $H_1$ equal to that of $H$ on $\{x\ge x_1+L+1\}$ can always be chosen for $m\le2$ when the frame symbols are homogeneous.
Part (a) gives $2\pi\sigma_I(H_1)$ from the data of the unperturbed operator $H$ in general, and from the data of $H_1$ alone for perturbations supported
near the wall; without field, $I_+-I_-+\Cy_1(x_0)$ is then a formula of Fedosov--H\"ormander type for $H_1$ at every $x_0$, although $H_1$ need not be
translation invariant. For perturbations that do not vanish in the bulks, the bulks of $H_1$ are no longer translation invariant, and their Hall
conductivities, defined through a trace per unit area for periodic or ergodic perturbations \cite{BES}, are not covered by Theorem~\ref{thm:bulk}. The proof,
in Appendix~\ref{s:sec:routine2}, follows the continuity argument of \cite{QB21,BEC} for (a) and uses the locality of Theorem~\ref{thm:local} for (b).

Neither condition can be dropped: without structural admissibility, a flat-band energy may reach $\alpha$, which is then essential spectrum
(Proposition~\ref{prop:flat}(b)), and without gap preservation, a perturbed bulk may carry states at $\alpha$.
\section{Computation of the invariants and examples}\label{sec:compute}
By Theorem~\ref{thm:BIC}, $2\pi\sigma_I=J_+-J_-+\Cy$ at every energy $\alpha$ outside the flat-band set and outside the spectra of the two bulks. Computing
$\sigma_I$ therefore requires the invariants $J_\pm$ and $\Cy$, and the set of energies at which the formula applies, namely the complement of $\Fl$ and of the
generalized Landau levels of the bulks. This section computes the invariants, then the flat-band set, and illustrates the results on examples.

Without field, $I$ is computed by diagonalizing $\ab(k)$ pointwise, and the cylinder term is
$\Cy=(\Theta(y_0)-\Theta(-y_0))/2\pi$, with the Berry phases $\Theta$ of Theorem~\ref{thm:BIC}. The sum $I(a_+;\alpha)-I(a_-;\alpha)+\Cy$ is the integer
$-d_\alpha$, where $d_\alpha$ is the Chern number of the occupied projector of $a(x_0,y,k)$ on the boundary of $[-y_0,y_0]\times\bar{\mathbb B}$ in
$(y,\xi,\zeta)$-space, with $\bar{\mathbb B}$ the radially compactified momentum plane, $[-y_0,y_0]\times\bar{\mathbb B}$ oriented by $dy\wedge d\xi\wedge d\zeta$, and
its boundary oriented as a boundary. This is the integer $d$ of Lemma~\ref{lem:degree} for the bulk extensions $E_\alpha(a_\pm)$, for which
$\kappa=-I(a_\pm;\alpha)$. This is the Fedosov--H\"ormander formula in
the form of \cite{Faure}, in which $\Cy$ is the contribution of infinite momenta.

With field, $J$ is obtained from the spectrum of $K_B(\ab)$ by Theorem~\ref{thm:bulk}(a) and Corollary~\ref{cor:bulk}(b), starting from $J=0$ below the spectrum when the transverse principal symbol of $\ab$ is positive definite,
or from $J=I$ in a field-free gap at weak field under (W) at $\alpha$. For rotation-covariant symbols, $K_B(\ab)$ is block diagonal in the Hermite basis, with blocks of size at most $n$, and at most two
for $2\times2$ symbols (Example~\ref{s:ex:levels}). Assume (W) at $\alpha$, with $\alpha$ in the field-free gaps of both bulks, and assume that $E^\infty$ does not depend on
the field and that the flat-band gap holds along $t\mapsto tB_\pm$, as for $m\le2$ and for factorized operators. This gives
\begin{equation}\label{eq:genFH}
  2\pi\sigma_I=-d_\alpha-\sgn(B_+)\,s_++\sgn(B_-)\,s_-,
\end{equation}
where $s_\pm$ is the net number of generalized Landau levels of the bulk $\pm$ crossing $\alpha$ upward along $t\mapsto tB_\pm$, $t\in(0,1]$.

\noindent\textbf{Peierls and factorized operators in a field.} Hypothesis (N1) at $B\ne0$ and condition (W) can be read off the field-free symbol in
the following cases, by Propositions~\ref{prop:Meffcomp} and \ref{s:prop:Wholds}.
\\[1mm]
(i) For elliptic operators, $r=0$, they hold for every coupling.
\\[1mm]
(ii) For factorized operators of every order, (DN$_0$) holds at every field with $\Meff=V_\sff^*WV_\sff$ (Proposition~\ref{prop:Meffcomp}(d)), and (W)
holds (Proposition~\ref{s:prop:Wholds}). Condition (F) holds under the conditions of Lemma~\ref{s:lem:Fsufficient}.
\\[1mm]
(iii) When $\ker a_m(u,\theta)$ does not depend on $(u,\theta)$, a constant frame conjugates every coupling exactly, since a constant matrix commutes with
$\OpA$. Then (DN$_0$) only requires the flat block of $a$ to have order $0$ and its off-diagonal blocks order at most $\frac m2$, and it holds at every field,
with $\Meff$ independent of the field, together with (W) and (F), for every $m$. This covers a localized degree of freedom hybridized with a dispersive band.
\\[1mm]
(iv) For Peierls operators of order $m\le2$, (DN$_0$) and $\Meff$ do not depend on the field (Proposition~\ref{prop:Meffcomp}(c)), and (W) holds
(Proposition~\ref{s:prop:Wholds}).
\\[1mm]
(v) For Peierls operators of higher order with constant symbol and field, the block orders of (DN$_0$) persist for $m\le4$, $\Meff$ does not depend on the
field for $m=3$ and may depend on it for $m=4$, and (DN$_0$) fails for $m\ge5$ when the kernel rotates with $\theta$ (Proposition~\ref{prop:Meffcomp}(e)).
When the kernel depends on $u$, (DN$_0$) may fail for Peierls operators already without field (Remark~\ref{s:rem:disp}).

The generalized Landau levels are the eigenvalues of $K_B(\ab)$. The flat-band set is obtained from the effective flat symbol of
Definition~\ref{def:Meff}, which is computed in the following cases. Parts (a)--(c) and (e) concern Peierls operators, and part (d) concerns factorized
operators, both defined in section~\ref{sec:operators}.
\begin{proposition}[computation of the effective flat symbol]\label{prop:Meffcomp}
Let $U$ be a homogeneous frame symbol, i.e., equal to a homogeneous frame $U(u,\theta(k))$ for $|k|\ge R$ (section~\ref{sec:operators}). For every $m$, the degree-$0$ part of $\tilde a_{\sff\sff}$ consists of $V_\sff^*WV_\sff$ and, for $1\le j\le m$, of
the terms of the expansion of $V_\sff^*\sharp^Ba_j\sharp^BV_\sff$ that carry exactly $j$ derivatives in $k$. Under (DN$_0$), $\Meff$ is obtained by
subtracting $(\tilde a_{\sff\stt})_{m/2}\tilde a_m^{-1}(\tilde a_{\stt\sff})_{m/2}$ from that degree-$0$ part, where the subscript denotes the part of degree $\frac m2$. In particular, we have:
\\[1mm]
(a) For a first-order operator $a=k\cdot\gamma(u)+W(u)$, (DN$_0$) holds and $\Meff=V_\sff^*WV_\sff+T$, with the \emph{curvature correction}
\begin{equation}\label{eq:T}
  T=\frac i2\sum_j\Bigl(\partial_{k_j}V_\sff^*\,a_1\,\partial_{u_j}V_\sff-\partial_{u_j}V_\sff^*\,a_1\,\partial_{k_j}V_\sff\Bigr).
\end{equation}
The correction $T$ is homogeneous of degree $0$ and covariant, and it vanishes for constant $\gamma$, in which case $\Meff$ is the compression of $W$ to
the kernel of $a_1$.
\\[1mm]
(b) Let $m=2$ and $B=0$. The flat block contains at degree $1$ the term
$f_1=V_\sff^*a_1V_\sff+\frac i2\sum_j(\partial_{k_j}V_\sff^*a_2\partial_{u_j}V_\sff-\partial_{u_j}V_\sff^*a_2\partial_{k_j}V_\sff)$, and (DN$_0$) holds if and only if
$f_1\equiv0$. If $\ker a_2(u,\theta)$ does not depend on $u$, this condition reads $V_\sff^*a_1V_\sff=0$, and $\Meff$ is given by the explicit formula \eqref{s:eq:Meff2}.
\\[1mm]
(c) For Peierls operators of order $m\le2$, (DN$_0$) and $\Meff$ do not depend on $B$, so that (N3) is a condition on the field-free symbol.
\\[1mm]
(d) Let $H=L_B^*L_B+W$ be a factorized operator, with $L_B=\OpA(\lsym)$ of order $p$. Then $R_K:=(1+L_B^*L_B)^{-1}=1-L_B^*(1+L_BL_B^*)^{-1}L_B$ belongs to
$\OpA(S^0)$, with principal symbol $E_{\ker}$, and its symbol $r_K$ satisfies $\lsym\sharp^Br_K\in S^{-p}$. With the frame symbol built from
$[r_K\sharp^BV_\sff,V_\stt]$ in the proof, $H$ is in (DN$_0$) for every field, with off-diagonal blocks of order $0$, and $\Meff=V_\sff^*WV_\sff$.
\\[1mm]
(e) For Peierls operators of order $m\ge3$, the field may enter $\Meff$. Let $m\ge5$, and let a constant symbol satisfy (DN$_0$) at $B=0$ in a homogeneous
frame with $(\partial_\theta V_\sff)^*V_\stt\not\equiv0$. Then (DN$_0$) fails in homogeneous frames at every $B\ne0$. For the symbols
$|k|^{m-2}k^\perp(k^\perp)^T+g\sigma_y$, $m=6,8$, $g\in\R$, (N1) fails at every $B\ne0$ in every frame.
\end{proposition}
The proof, a computation with the Moyal product, is given below, except for (b), which is proved in Appendix~\ref{s:sec:routine}.

In (d), no invertibility of $L_BL_B^*$ is required. If
$L_BL_B^*\ge c>0$, the orthogonal projection $P_K=1-L_B^*(L_BL_B^*)^{-1}L_B$ onto $\ker L_B$ may be used in place of $R_K$, and then $\lsym\sharp^Bp_K=0$. In
(e), the field enters $\Meff$ for $m=4$ in Proposition~\ref{s:prop:exactB}(a), and not for $m=3$ with constant symbol and field
(Remark~\ref{s:rem:m3}). For $m\ge5$, the first-order magnetic term creates off-diagonal blocks of degree $m-2>\frac m2$ (Proposition~\ref{s:prop:hier}(c)).
For $m=6,8$, every field turns the bounded band into a branch of levels $\mu_j=|B|^{m/2}\nu_j\to-\infty$, infinitely many of which cross every negative energy
as $B\to0$ (Proposition~\ref{s:prop:exactB}), so that (W) and the weak-field limit of Theorem~\ref{thm:bulk}(c) also fail. Factorized operators are not
affected, by (d).
\begin{proof}[Proof of Proposition~\ref{prop:Meffcomp}(a),(c)--(e)]
A term of the expansion of $V_\sff^*\sharp^Ba_j\sharp^BV_\sff$ that carries $\nu$ derivatives in $k$ is
homogeneous of degree $j-\nu$ at large momentum, since $V_\sff$ is homogeneous of degree $0$, and derivatives in $u$ and factors of $B$ do not change the
degree. Hence the degree-$0$ part of the flat block consists of the terms with $\nu=j$, as stated before the proposition. Since $\tilde a_{\stt\stt}^{-1}=\tilde a_m^{-1}+S^{-m-1}$, the degree-$0$ part of the Schur term is
$(\tilde a_{\sff\stt})_{m/2}\tilde a_m^{-1}(\tilde a_{\stt\sff})_{m/2}$. 

For (a), let $m=1$. Without field, $f\sharp g=fg+\frac i2\{f,g\}+\cdots$ with $\{f,g\}:=\sum_j(\partial_{u_j}f\,\partial_{k_j}g-\partial_{k_j}f\,\partial_{u_j}g)$,
so that the first-order term in the flat block is $\frac i2(\{V_\sff^*,a\}V_\sff+\{V_\sff^*a,V_\sff\})$. Since $V_\sff^*a_1\equiv0$, the second bracket
receives no contribution from $a_1$, and since $\partial a_1V_\sff=-a_1\partial V_\sff$, the first one becomes
$\partial_kV_\sff^*a_1\partial_uV_\sff-\partial_uV_\sff^*a_1\partial_kV_\sff$, summed over $j$. With the factor $\frac i2$, this is $T$ of \eqref{eq:T}, which is
homogeneous of degree $0$. The second-order terms of $a_1$, the
first-order terms of $W$ and the magnetic terms have degree $\le-1$. The off-diagonal blocks are of degree $0<\frac12$ since $a_1V_\sff=0$, so that the Schur correction has degree
$\le-1$. Under $V_\sff\mapsto V_\sff g$, the terms involving $\partial g$ carry a factor $a_1V_\sff=0$, which gives covariance. For constant $\gamma$, we have
$\partial_uV_\sff=0$ and therefore $T=0$.

For (c), consider $m\le2$ and a homogeneous frame. The first-order magnetic term of the Moyal product is
$\frac i2B\sum_{p<q}(\partial^{(p)}_{k_1}\partial^{(q)}_{k_2}-\partial^{(p)}_{k_2}\partial^{(q)}_{k_1})$ \cite{MP04,IMP}, and for the factors $V_\sff^*$, $a_m$, $V_\sff$ it
vanishes in the flat block. Indeed, differentiating $a_mV_\sff=0$ and $V_\sff^*a_m=0$ gives $\partial a_m\,V_\sff=-a_m\partial V_\sff$ and
$V_\sff^*\partial a_m=-\partial V_\sff^*a_m$, so that each of the three pairs of factors contributes a multiple of
$\partial_{k_1}V_\sff^*\,a_m\,\partial_{k_2}V_\sff-\partial_{k_2}V_\sff^*\,a_m\,\partial_{k_1}V_\sff$, which vanishes since $\partial_{k_i}V_\sff=(\partial_{k_i}\theta)\,\partial_\theta
V_\sff$. For $m=2$, the first-order magnetic term of $a_1$ has degree $-1$, and that of $W$ has degree $-2$. The remaining magnetic terms, which contain a
factor $B$ and three derivatives, or a factor $\partial B$, or a factor $B^2$, have degree $\le m-3<0$. In the off-diagonal blocks, the magnetic terms have
degree $\le m-2\le\frac m2-1$, so that they change the Schur complement at degree $\le-1$. Hence neither the block orders nor $\Meff$ depend on $B$. For
$m\le2$ and a general frame symbol, we use Proposition~\ref{prop:flat}(a).

For (d), the operator $1+L_BL_B^*$ is elliptic of order $2p$, since $\lsym_p\lsym_p^*\ge c\jp k^{2p}$, and it is self-adjoint, by
Theorem~\ref{thm:analysis}(a) with $r=0$, and bounded below by $1$. Its inverse
therefore belongs to $\OpA(S^{-2p})$, by the parametrix and the inverse-closedness of the magnetic calculus \cite{IMP,IMP2,CHP18}. Let $g\in S^{-2p}$ be its
symbol, so that $(1+\lsym\sharp^B\lsym^*)\sharp^Bg=1$. The identity $(1+L_B^*L_B)L_B^*=L_B^*(1+L_BL_B^*)$ gives $R_K=1-L_B^*(1+L_BL_B^*)^{-1}L_B$, so that
$r_K=1-\lsym^*\sharp^Bg\sharp^B\lsym\in S^0$, with principal symbol $1-\lsym_p^*(\lsym_p\lsym_p^*)^{-1}\lsym_p=E_{\ker}$. Moreover,
$\lsym\sharp^Br_K=\lsym-(1-g)\sharp^B\lsym=g\sharp^B\lsym\in S^{-p}$. Let $U_R=[V_{\sff,R},V_{\stt,R}]$ be the unitary-valued homogeneous frame symbol of
Lemma~\ref{lem:globalframe}(a), and set
\[
  \tilde V_\sff:=V_{\sff,R}+\chi_R\,(r_K\sharp^BV_{\sff,R}-V_{\sff,R}),\qquad U_K:=[\tilde V_\sff,V_{\stt,R}] ,
\]
with $\chi_R$ the cut-off introduced before Proposition~\ref{prop:paramalpha}. Since $r_K=E_{\ker}(u,\theta(k))+S^{-1}$ and $E_{\ker}V_\sff=V_\sff$, the
correction $\chi_R(r_K\sharp^BV_{\sff,R}-V_{\sff,R})$ belongs to $S^{-1}$ and is $O(R^{-1})$ in $S^0$, so that $U_K$ is a frame symbol for $R$ large. Since
$(1-\chi_R)(V_{\sff,R}-r_K\sharp^BV_{\sff,R})$ has compact support in $k$, we have $\lsym\sharp^B\tilde V_\sff=\lsym\sharp^Br_K\sharp^BV_{\sff,R}+S^{-\infty}\in
S^{-p}$. With $a=\lsym^*\sharp^B\lsym+W$ and $(\lsym\sharp^Bf)^*=f^*\sharp^B\lsym^*$,
\[
  \tilde V_\sff^*\sharp^Ba\sharp^B\tilde V_\sff=(\lsym\sharp^B\tilde V_\sff)^*\sharp^B(\lsym\sharp^B\tilde V_\sff)+\tilde V_\sff^*\sharp^BW\sharp^B\tilde V_\sff,\
  V_{\stt,R}^*\sharp^Ba\sharp^B\tilde V_\sff=(\lsym\sharp^BV_{\stt,R})^*\sharp^B(\lsym\sharp^B\tilde V_\sff)+V_{\stt,R}^*\sharp^BW\sharp^B\tilde V_\sff .
\]
The first right-hand side is the sum of a symbol in $S^{-2p}$ and of a symbol in $S^0$ whose degree-$0$ part is $V_\sff^*WV_\sff$. The second one belongs
to $S^p\sharp^BS^{-p}+S^0\subset S^0$. The transverse block is $V_\stt^*\lsym_p^*\lsym_pV_\stt$ up to order $2p-1$, and it is elliptic since $\lsym_p$ is injective on
$\Ran V_\stt(\theta)$, with $|\lsym_pV_\stt v|\ge c|k|^p|v|$. Therefore (DN$_0$) holds with off-diagonal blocks of order $0$, the Schur correction has order
$-2p$, and $\Meff=V_\sff^*WV_\sff$ for every field. If $L_BL_B^*\ge c>0$, then $(L_BL_B^*)^{-1}\in\OpA(S^{-2p})$ in the same way, $P_K$ is the orthogonal
projection onto $\ker L_B$ by linear algebra, its symbol $p_K$ satisfies $\lsym\sharp^Bp_K=0$, and the same construction with $p_K$ in place of $r_K$
applies.

For (e), the first statement is Remark~\ref{s:rem:m3} and Proposition~\ref{s:prop:exactB}(a), and the second one is Proposition~\ref{s:prop:hier}(c).
For the third one, assume that (DN$_0$) holds at some $B\ne0$ for $K_B(\ab)$ in some frame, which we may take with $K_B(U)$ unitary by Lemma~\ref{lem:globalframe}(b), with $\ab=|k|^{m-2}k^\perp(k^\perp)^T+g\sigma_y$. The transverse
block is then positive at large momentum, and the flat block and the Schur correction are of order $0$, hence bounded. By the isotropic parametrix of
Lemma~\ref{lem:reduction}(c), in its form uniform in the spectral parameter (Proposition~\ref{s:prop:largeparam}), $K_B(\ab)-z$ is then invertible for $z$ real and sufficiently negative, through the Schur complement with respect to the
transverse block, so that $K_B(\ab)$ is bounded below. This contradicts Proposition~\ref{s:prop:exactB}(b).
\end{proof}

The following examples treat the Landau Hamiltonian, Dirac operators, a magnetic transition, two operators with a flat band at $0$, and a flat band
carrying a Landau fan. Further examples are collected in Appendix~\ref{s:sec:models} 
\begin{examples}\label{ex:examples}
\emph{The Landau Hamiltonian.} For $\ab=|k|^2$, we have $r=0$, $\Fl=\emptyset$, and $\spec K_B(\ab)=\{|B|(2j+1):j\ge0\}$ with simple levels, so that
$J(\ab,B;\alpha)=\sgn(B)\,\#\{j:|B|(2j+1)<\alpha\}$ by Theorem~\ref{thm:bulk}(a). For an interface between the fields $B_-$ and $B_+$,
Corollary~\ref{cor:main}(b) gives $2\pi\sigma_I=\sgn(B_+)N_+(\alpha)-\sgn(B_-)N_-(\alpha)$, where $N_\pm$ is the number of levels of the bulk $\pm$ below
$\alpha$. For $B_-<0<B_+$, this is $N_++N_-$, the number of snake states of the Iwatsuka model \cite{DGR}.
\\[2mm]
\emph{Dirac operators.} For the Dirac symbol $\ab=k\cdot\sigma+M\sigma_3$, $M\ne0$, we have $r=0$ and $\Fl=\emptyset$. The generalized Landau levels are
$\{M\sgn B\}\cup\{\pm\sqrt{2j|B|+M^2}:j\ge1\}$ \cite{QB22}. None of them lies in $(-|M|,|M|)$, for any field, so that
$J(\ab,B;\alpha)=I(\ab;\alpha)=-\frac12\sgn M$ for $|\alpha|<|M|$ and every $B$. Since $E^\infty=\mathbf 1_{(-\infty,0)}(\theta\cdot\sigma)$ does not depend on
the mass, $\Cy=0$ for a mass wall, and $2\pi\sigma_I=-\frac12\sgn M_++\frac12\sgn M_-$ at every field, including vanishing ones. For $\alpha$ between generalized Landau levels
of a magnetized bulk, Theorem~\ref{thm:bulk}(a) gives $J(\ab,B;\alpha)=-\frac12\sgn M+\sgn(B)\,n_B(\alpha)$, where $n_B(\alpha)$ is the number of levels in
$(0,\alpha)$ for $\alpha>0$, and minus the number of levels in $(\alpha,0)$ for $\alpha<0$. This is the bulk quantity $I(H_\pm;\alpha)$ of
\cite[Thm.~2.1]{QB22}, which is thereby identified with the Hall conductivity of the bulk. Theorem~\ref{thm:BIC} thus recovers \cite[Thm.~2.1]{QB22}, where
the roles of $x$ and $y$ are exchanged, at every energy outside the generalized Landau levels of the two bulks, including energies inside the field-free
bands. Constant electric potentials $V_\pm$ in the bulks shift the levels, and they are included by replacing $\alpha$ by $\alpha-V_\pm$. The bulk invariants of the
unregularized Dirac operator are thus defined at every field, including $B=0$, with fractional part fixed by $E^\infty$, whereas a regularization changes
$E^\infty$ and shifts $J$ to an integer (Example~\ref{s:ex:Dirac}).
\\[2mm]
\emph{A magnetic transition.} For the BHZ-type symbol $\ab=(|k|^2-\mu)\sigma_3+k\cdot\sigma$ with $\mu>0$, we have $r=0$, $\Fl=\emptyset$ and $I(\ab;0)=1$. The unpaired generalized Landau level
$\sgn(B)(|B|-\mu)$ crosses $0$ at $|B|=\mu$, and for $0<\mu<2+\sqrt3$ the paired levels do not vanish (Example~\ref{s:ex:levels}), so that
$J(\ab,B;0)=1$ for $|B|<\mu$ and $0$ for $|B|>\mu$. For $\mu>2+\sqrt3$, a paired level crosses $0$ twice below $\mu$, and $J$ also vanishes on a window of
fields below $\mu$. Across a boundary of the applied field in this
material, Corollary~\ref{cor:trans}(b) gives $2\pi\sigma_I=J(\ab,B;0)-I(\ab;0)$, which, for $0<\mu<2+\sqrt3$, is $0$ for $|B|<\mu$ and $-1$ beyond the transition field. This
creates a chiral mode at the edge of the magnetized region of a homogeneous insulator.
\\[2mm]
\emph{Curl operators.} For the factorized operator $H=\curl_B^*\curl_B+W(u)$ acting on vector fields in the plane, the field-free symbol is
$a=k^\perp(k^\perp)^T+W$, so that $m=2$, $r=1$ and $V_\sff=\theta$. Proposition~\ref{prop:Meffcomp}(d) gives $\Meff(u,\theta)=\theta^TW(u)\theta$, the
compression of $W$ to the kernel, for every field, and the flat-band set is the closure of the set of these values. By \eqref{eq:curlPF}, the Peierls operator
$\OpA(a)$ is $H+\frac B2\sigma_y$, and it has the same $\Meff$, since $\theta^T\sigma_y\theta=0$. For the gyrotropic potential $W=g(y)\sigma_y$, we have
$\Fl=\{0\}$, and the field-free bands are $\frac{\rho^2}2\pm\sqrt{\rho^4/4+g^2}$, $\rho=|k|$. The lower band is asymptotically flat and fills $[-|g|,0)$. For
$g=\tanh(\frac32y)$, $\mathcal G(a_\pm)=(-\infty,-1)\cup(0,1)$, and the field-free value above the flat-band energy, at $\alpha=\frac12$, is
$2\pi\sigma_I=I_+-I_-+\Cy=-2$.

The levels show the role of the coupling. For constant $g$ and $B>0$, the generalized Landau levels of $\curl_B^*\curl_B+g\sigma_y$ are $g$, $g+B$ and
\[
  \lambda_\pm(n)=\tfrac12\Bigl(B(2n+1)\pm\sqrt{B^2(2n+1)^2+4g(g+B)}\Bigr),\qquad n\ge1
\]
(Example~\ref{s:ex:curl}). The level $g$ is carried by $\ker\curl_B$, which contains zero modes of $\curl_B$ analogous to the Aharonov--Casher zero modes of
the Pauli operator \cite{AC79}, so that it does not move with the field. Complex conjugation maps $(B,g)$ to $(-B,-g)$, and the levels of the Peierls
operator are obtained by replacing $g$ by $g+\frac B2$. For $g>0$ and $B>0$, no level enters $(0,g)$, while for $g<0$ and $\alpha\in(\frac{|g|}3,|g|)$ the only level crossing
$\alpha$ is $g+B$, upward at $B=|g|+\alpha$; for smaller $\alpha$, the fan levels $\lambda_-(n)$, which increase to $\frac{|g|}{2n+1}$ for $B>|g|$, also cross $\alpha$. At $\alpha=\frac12$ and $B_+=B_-=B$, Theorem~\ref{thm:BIC} and
Corollary~\ref{cor:bulk}(b) give $2\pi\sigma_I=-2$ for $|B|<\frac32$ and $-1$ for $|B|>\frac32$ for the factorized operator, and $2\pi\sigma_I=-2$ for $|B|<1$,
$-1$ for $1<|B|<3$ and $0$ for $|B|>3$ for the Peierls operator. The two couplings have the same flat-band set and the same field-free current, and they
differ in the fields at which the current jumps.

Below the flat-band energy, for $\alpha\in(-1,0)$, the energy lies in a field-free band and $\sigma_I$ is
defined only with field. For $B=0.5$, the levels of both bulks inside the band accumulate at $0$ from below, and Corollary~\ref{cor:main}(b) gives
$2\pi\sigma_I=N_+(\alpha)-N_-(\alpha)$, which equals $+1$ and $+2$ at $\alpha=-0.35$ and $-0.25$ for the factorized operator (Example~\ref{s:ex:curl},
Figure~\ref{s:fig:curl}).
\\[2mm]
\emph{Spin-one operators: rotating shallow water and the dice lattice.} The rotating shallow-water symbol with Coriolis parameter $f(y)$,
\[
  a(y,k)=\begin{pmatrix}0&\xi&\zeta\\ \xi&0&-if(y)\\ \zeta&if(y)&0\end{pmatrix}=\xi A_1+\zeta A_2+f(y)A_3 ,
\]
is unitarily equivalent to $\xi S_x+\zeta S_y+f(y)S_z$, since the $A_j$ satisfy the spin-one commutation relations and generate an irreducible
representation. This is also the dice-lattice symbol
(Example~\ref{s:ex:aT3}) with mass $f$. Here $m=r=1$, $V_\sff=(0,\theta^\perp)$ with $\theta^\perp=(\hat\zeta,-\hat\xi)$, $\Meff=0$ and $\Fl=\{0\}$, the flat band of
geostrophic modes, and the other bands are $\pm\sqrt{f^2+|k|^2}$. Since $\Cy=0$ and $I(a_\pm;\alpha)=-\sgn f_\pm$, Theorem~\ref{thm:BIC} gives
$2\pi\sigma_I=\sgn f_--\sgn f_+$, which is $\pm2$ at the equator \cite{DMV}.

Writing a Hermitian perturbation of degree $0$ as
$v=\bigl(\begin{smallmatrix}v_0&p^*\\ p&S+b\sigma_y\end{smallmatrix}\bigr)$ in the variables (height, velocity), with $S$ real symmetric, we have
$V_\sff^*vV_\sff=(\theta^\perp)^TS\,\theta^\perp$. The elliptic directions are therefore the perturbations with $S=0$, which include the Coriolis term, and
Theorem~\ref{thm:slab} gives stability under such perturbations of arbitrary size, as long as the bulk gaps stay open. A non-elliptic perturbation
$S\ne0$ moves $\Fl$, and $\sigma_I$ is no longer defined once $\Fl$ reaches $\alpha$ (Proposition~\ref{prop:flat}(b)). Since $\Meff$ does not depend on the field
for $m=1$, this holds at every field, magnetic for the dice lattice or pseudo-magnetic in photonic realizations.

With field, the flat band carries a Landau
fan, and Appendix~\ref{s:sec:models} details the elliptic directions of this and the other examples and the effect of the field
(Figures~\ref{s:fig:spin1} and \ref{s:fig:curl}). For $f_\pm=\pm1$ and $B=0.5$, Theorems~\ref{thm:BIC} and \ref{thm:bulk}(a) and Corollary~\ref{cor:bulk}(a), combined with the numerically computed levels and pair indices of
Example~\ref{s:ex:aT3}, give $2\pi\sigma_I(\alpha)=-2+n(\alpha)$ for $0<\alpha<1$, where $n(\alpha)$ is the number of fan levels of the $f=-1$ bulk in $(\alpha,1)$. Independently of
the numerics, Corollary~\ref{cor:main}(c) shows that the current diverges as $\alpha\downarrow0$.

For electrons in $\alpha$--$\mathcal T_3$ materials, such as
HgCdTe near critical doping \cite{MN16,Teppe16}, $B$ is a genuine magnetic field, and the Peierls substitution is the minimal coupling of the effective
Kane model. For photonic realizations of the dice lattice, $B$ would be a pseudo-magnetic field induced by strain.
\end{examples}

\begin{remark}[other couplings]\label{rem:quant}
For constant fields, gauge-covariant couplings of a field-free symbol $a$ are of the form $\OpA(a+Bb_1+B^2b_2+\cdots)$, with $b_j$ of order
$m-2j$, and Peierls and factorized operators are two such choices; for variable fields, terms with derivatives of $B$ also appear. For $r=0$ and for $m=1$, the corrections are of lower order relative to every block of the symbol and only move
the generalized Landau levels, as the Zeeman term of the Pauli operator $(\sigma\cdot\Pi)^2=\Pi^2-B\sigma_3$. With flat bands, the correction $Bb_1$ acts on the
flat block, which has order $0$: it may move $\Meff$ for $m\ge2$, and for $m\ge5$ and a rotating kernel it must cancel the off-diagonal term of degree $m-2$
of Proposition~\ref{s:prop:hier}(c), which factorization achieves. When the kernel depends on $u$, a Peierls operator may violate (DN$_0$) already without
field (Remark~\ref{s:rem:disp}). In models derived from a microscopic structure, the corrections are part of the model, as for the parameter $\kappa$ of the
Luttinger Hamiltonian \cite{Lut56} and for the flat bands of tight-binding models
\cite{AAM96,RKY20}.
\end{remark}
\section{Frames, parametrix and resolvent estimates}\label{sec:engine}
The analysis of $H$ rests on a block decomposition. Conjugated by a frame of the kernel bundle of $a_m$ (Lemma~\ref{lem:globalframe}), $H$ becomes
$\U^*H\U=\OpA(\tilde a)$, with flat, off-diagonal and transverse blocks of orders $0$, $\frac m2$ and $m$. The Schur complement of $\tilde a-z$ with respect to
its transverse block is $\Meff-z+S^{-1}$, so that $\tilde a-z$ is invertible at large momentum when $\Re z$ stays away from $\Fl$, with an inverse whose blocks
have orders $0$, $-\frac m2$ and $-m$. A Neumann series in graded symbol classes then gives a parametrix (Proposition~\ref{prop:paramalpha}), and the rest of
the section follows from it: the identification of $\Meff$ for two frames and the Weyl sequences of Proposition~\ref{prop:flat}, whose proofs come next, and
self-adjointness, weighted resolvent and Combes--Thomas estimates (Theorem~\ref{thm:analysis}).

The frame must be fixed. In the moving splitting $\Ran E_{\ker}\oplus\Ran Q$, derivatives of the moving projector mix the blocks: the first composition error
contains $-a_m(\partial_kE_{\ker})E_{\ker}\,\partial_u(a-z)^{-1}_{\sff\sff}$, of order $m-1$, which exceeds the order $\frac m2-1$ allowed from the flat to the
transverse components. In a fixed frame, $\partial_u$ preserves the order of each block while $\partial_k$ lowers it by one.

The quantization $\U$ of a frame symbol is invertible but in general not unitary, since $U$ is at most unitary pointwise and $\sharp^B$ adds corrections in
$S^{-1}$. Two refinements are used below: frame symbols analytic in a strip in $k$, which give the Combes--Thomas estimate of Theorem~\ref{thm:analysis}(b),
and frame symbols for which $\U$ is unitary, which make $\U^*H\U$ unitarily equivalent to $H$ and are used for the functional calculus of
section~\ref{sec:pbulk}. The latter are obtained by adding to the transverse columns $V_\stt$ multiples of the flat columns $V_\sff$ and normalizing each
block, which keeps the block orders of (DN$_0$).
\begin{lemma}[frames]\label{lem:globalframe}
(a) Under Hypothesis~\ref{hyp:rank} and (N2), there is a homogeneous frame $U(u,\theta)$ with bounded derivatives, independent of $y$ on
$\{\pm y\ge y_0\}$, and independent of $x$ under (TI). This frame is modified by $V_\sff\mapsto V_\sff\,\mathrm{diag}(e^{-iw\theta},1,\dots,1)$ so that $\det U(u,\cdot)$ has winding zero. It
then extends to a unitary-valued symbol $U_R\in S^0$ equal to $U(u,\theta(k))$ for $|k|\ge R$, with $|\partial_k^\beta U_R|\le C_\beta R^{-|\beta|}$ on $|k|\le R$, which is a homogeneous frame symbol for $R$ large, and it may be
replaced by a frame symbol that is analytic in $k$ in the strip $\{|\Im k|\le\varrho_0\}$, with symbol bounds there. For every frame symbol $U$, $\U=\OpA(U)$ is invertible with
$\U^{-1}\in\OpA(S^0)$, and $\U^*\U=1+\OpA(\upsilon)$ with $\upsilon\in S^{-1}$.
\\[1mm]
(b) Let $U$ be a frame symbol for which (DN$_0$) holds, with $R$ large. There is a frame symbol $U^{\rm u}$ with $U^{\rm u}-U\in S^{-1}$ such that $\OpA(U^{\rm u})$ is
unitary. With $U^{\rm u}$ in place of $U$, $H$ is still in class (DN$_0$), with the same $\Meff$. The same holds for the reduced bulk operators $K_B(\ab)$ of
\eqref{eq:KB}, with the isotropic symbol classes of section~\ref{sec:pbulk} in place of $S^\nu$, and for the fibers $\Hh(\xi)$ of the translation-invariant walls
of Theorem~\ref{thm:TI}.
\end{lemma}
The proof, by parallel transport with Kato's connection \cite{Kato}, is given in Appendix~\ref{s:sec:routine}. 

Throughout the rest of this section, $U$ is a frame symbol for which (DN$_0$) holds,
$\U=\OpA(U)$, $s=(s_\sff,s_\stt)=(0,\frac m2)$, $\mathcal D:=\mathrm{diag}(1,\Lm^{m/2})$, $\Wt:=(\U^*)^{-1}\mathcal D\,\U^{-1}$, and
\[
  \tilde A_z:=\U^*(H-z)\U=\OpA(\tilde a_z),\qquad \tilde a_z:=\tilde a-z(1+\upsilon).
\]
For $r=0$, we have $U=1$, $\upsilon=0$ and $\mathcal D=\Lm^{m/2}$, and all statements reduce to the elliptic theory.

The block orders of (DN$_0$) are conveniently encoded in the following graded classes. A matrix symbol $c$ belongs to $\mathcal S^\mu_\epsilon$, with
$\epsilon\in\{+,0,-\}$, if on $\{|k|\ge R\}$ we have $c_{ij}\in S^{\epsilon s_i+\epsilon s_j+\mu}$ for $\epsilon=\pm$, and $c_{ij}\in S^{s_i-s_j+\mu}$ for $\epsilon=0$.
Thus $\tilde a_z\in\mathcal S^0_+$ by (DN$_0$) and since $\upsilon\in S^{-1}$. These classes encode the Schur-complement structure: $\mathcal S^0_+$ contains the conjugated symbols, $\mathcal S^0_-$ their inverses, and
$\mathcal S^0_0$ the products of the two. Let $Z=\{z:\Re z\in\mathbb J,|\Im z|\le1\}$, where $\mathbb J$ is a compact interval with $\dist(\mathbb J,\Fl)\ge\frac14\dist(\alpha,\Fl)$. Let $\chi\in C^\infty(\R)$ with $\chi=0$ on $(-\infty,1]$ and $\chi=1$ on $[2,\infty)$, and set
$\chi_R(k)=\chi(|k|/R)$.
\begin{proposition}[graded parametrix]\label{prop:paramalpha}
(a) We have $\mathcal S^\mu_+\sharp^B\mathcal S^\nu_-\subset\mathcal S^{\mu+\nu}_0$, $\mathcal S^\mu_-\sharp^B\mathcal S^\nu_+\subset\mathcal S^{\mu+\nu}_0$,
$\mathcal S^\mu_0\sharp^B\mathcal S^\nu_0\subset\mathcal S^{\mu+\nu}_0$ and $\mathcal S^\mu_-\sharp^B\mathcal S^\nu_0\subset\mathcal S^{\mu+\nu}_-$, and the term of order
$j$ of each product lies in the class of index $\mu+\nu-j$. The operators $\mathcal D^{-1}\OpA(c)\mathcal D^{-1}$ for $c\in\mathcal S^0_+$, $\mathcal
D\OpA(c)\mathcal D$ for $c\in\mathcal S^0_-$, and $\mathcal D^{\mp1}\OpA(c)\mathcal D^{\pm1}$ for $c\in\mathcal S^0_0$ are bounded.
\\[1mm]
(b) There is $R_1$ such that, for $|k|\ge R_1$, $u\in\R^2$ and $z\in Z$, $\tilde a_z(u,k)$ is invertible. Moreover, $q_z:=\tilde a_z^{-1}$ is holomorphic in $z$
and belongs to $\mathcal S^0_-$ on $\{|k|\ge R_1\}$ with seminorms bounded uniformly on $Z$, and $(q_z)_{\sff\sff}=(\Meff-z)^{-1}+S^{-1}$.
\\[1mm]
(c) There are operators $G_z$ and $G'_z$, holomorphic on $Z$, and $K_z,K'_z\in\OpA(S^{-\infty})$ such that $(H-z)G_z=1+K_z$ and $G'_z(H-z)=1+K'_z$ on $\cS'$.
Moreover, $\U^{-1}G_z(\U^*)^{-1}=\OpA(g_z)$ with $g_z\in\mathcal S^0_-$, and $\Wt G_z\Wt$ and $\Wt G'_z\Wt$ are bounded uniformly on $Z$.
\end{proposition}
The proof is given below. The constants of Proposition~\ref{prop:paramalpha}, and hence those of Theorem~\ref{thm:analysis} and of sections~\ref{sec:pbulk} and
\ref{sec:pBIC}, depend on finitely many derivatives of the coefficients. This is not an artifact of the proof: for $r\ge1$, the bulk--interface
correspondence may fail for non-smooth coefficients, as for a discontinuous Coriolis parameter in the shallow-water model \cite{BY24,RT24}.
\begin{proof}[Proof of Proposition~\ref{prop:paramalpha}]
(a) In the fixed splitting, the product of blocks is $(c\sharp d)_{ij}=\sum_lc_{il}\sharp d_{lj}$, which is a sum of scalar-type products of symbols in
$S^{\epsilon s_i+\epsilon s_l+\mu}$ and $S^{\epsilon' s_l+\epsilon's_j+\nu}$. The orders add, and each order of the expansion carries one more $k$-derivative,
which lowers every block by one. Boundedness follows from the Calder\'on--Vaillancourt theorem of the magnetic calculus \cite{IMP,CHP24} after composition with $\Lm^{\pm s_i}$.

(b) By Hypothesis~\ref{hyp:rank}, $|(\tilde a_z)_{\stt\stt}v|\ge\frac{c_0}2|k|^m|v|$ for $|k|$ large. The Schur complement
$s_z=(\tilde a_z)_{\sff\sff}-(\tilde a_z)_{\sff\stt}(\tilde a_z)_{\stt\stt}^{-1}(\tilde a_z)_{\stt\sff}$ equals $\Meff-z+S^{-1}$, and
$\dist(\Re z,\spec\Meff)\ge\frac14\dist(\alpha,\Fl)$ on $Z$, so that $\norm{s_z^{-1}}\le8/\dist(\alpha,\Fl)$ for $|k|\ge R_1$. Block inversion gives
$q_{\sff\sff}=s_z^{-1}\in S^0$ and $q_{\sff\stt}=-s_z^{-1}\tilde a_{\sff\stt}\tilde a_{\stt\stt}^{-1}\in S^{-m/2}$, similarly for $q_{\stt\sff}$, and $q_{\stt\stt}\in
S^{-m}$. For the derivatives, we use $\partial q=-q(\partial\tilde a_z)q$. Since $\partial_u$ preserves the type of $\tilde a_z$ while $\partial_k$ lowers it,
Part (a) applied pointwise gives $q(\partial_u\tilde a_z)q\in\mathcal S^0_-$ and $q(\partial_k\tilde a_z)q\in\mathcal S^{-1}_-$, and an induction gives all
derivatives.

(c) Let $g^{(0)}=\chi_{R_1}q_z\in\mathcal S^0_-$. By (a), $\tilde a_z\sharp^Bg^{(0)}=1+e_z$ with $e_z\in\mathcal S^{-1}_0$, up to a term supported in
$\{|k|\le2R_1\}$, which is in $S^{-\infty}$. Since $\mathcal S^{\mu}_0$ is a filtered algebra, the asymptotic Neumann series $\sum_j(-e_z)^{\sharp j}$ has an
asymptotic sum $f_z\in\mathcal S^0_0$, that is, $f_z-\sum_{j<N}(-e_z)^{\sharp j}\in\mathcal S^{-N}_0$ for every $N$, which exists by Borel's lemma
\cite[Prop.~18.1.3]{Hormander3} applied block by block and can be chosen holomorphic in $z$, and $g_z:=g^{(0)}\sharp^Bf_z\in\mathcal S^0_-$ satisfies $\tilde a_z\sharp^Bg_z-1\in S^{-\infty}$, holomorphically
in $z$. We set $G_z:=\U\OpA(g_z)\U^*$. Then
\[
  (H-z)G_z=(\U^*)^{-1}\OpA(\tilde a_z)\OpA(g_z)\U^*=1+K_z,\ K_z:=(\U^*)^{-1}\OpA(\tilde a_z\sharp^Bg_z-1)\U^*\in\OpA(S^{-\infty}).
\]
The left parametrix is built in the same way. Finally, $\Wt G_z\Wt=(\U^*)^{-1}\mathcal D\OpA(g_z)\mathcal D\,\U^{-1}$ is bounded by Proposition~\ref{prop:paramalpha}(a), and
the positivity of $\Wt$ follows from $\mathcal D\ge1$ and the invertibility of $\U$.
\end{proof}
\begin{proof}[Proof of Proposition~\ref{prop:flat}(a)]
Existence follows from Lemma~\ref{lem:globalframe}(a). Let $U$ and $U'$ be frame symbols. Then
$\U'=\U\,\OpA(y)$, where $y$ is the symbol of $\U^{-1}\U'\in\OpA(S^0)$ (Lemma~\ref{lem:globalframe}(a)), and $\tilde a'=y^*\sharp^B\tilde a\sharp^By$,
$U'^*\sharp^BU'=y^*\sharp^B(1+\upsilon)\sharp^By$. Since $\U^{-1}=(\U^*\U)^{-1}\U^*$ and $(\U^*\U)^{-1}-1\in\OpA(S^{-1})$, the symbol $y$ equals $U^*U'$ up to $S^{-1}$, so that
$y=y_{\rm d}+y_{\rm o}$, with $y_{\rm d}$ block diagonal, equal to $\mathrm{diag}(g_\sff,g_\stt)$ up to $S^{-1}$, and $y_{\rm o}\in S^{-1}$ off-diagonal. Let
$m\le2$, and assume (DN$_0$) for $U$. The term $y_{\rm d}^*\sharp^B\tilde a\sharp^By_{\rm d}$ has blocks $(y_{\rm d})_i^*\sharp^B\tilde a_{ij}\sharp^B(y_{\rm d})_j$,
of the orders of (DN$_0$). The terms containing $y_{\rm o}$ contribute to the flat block at orders $\frac m2-1\le0$ and $m-2\le0$, and to the off-diagonal
blocks at orders $\le m-1\le\frac m2$. Hence (DN$_0$) holds for $U'$. A change of cut-off changes $U$ by a term in $S^{-\infty}$.

For the relation between the effective flat symbols, let $U$ and $U'$ be frame symbols for which (DN$_0$) holds, with conjugated symbols $\tilde a$ and
$\tilde a'=y^*\sharp^B\tilde a\sharp^By$. Since $\Fl$ and $\Fl'$ are bounded, we may fix $z=\lambda+i$ with $\lambda\notin\Fl\cup\Fl'$, and apply
Proposition~\ref{prop:paramalpha}(b),(c) in both frames, with $\mathbb J$ a small interval around $\lambda$. In the frame $U$, the
symbol $g_z$ of Proposition~\ref{prop:paramalpha}(c) satisfies $\tilde a_z\sharp^Bg_z-1\in S^{-\infty}$ and, by its construction, $g_z-\chi_{R_1}q_z\in
\mathcal S^{-1}_-$, so that $(g_z)_{\sff\sff}=(\Meff-z)^{-1}+S^{-1}$ by Proposition~\ref{prop:paramalpha}(b). The operator $\OpA(y)=\U^{-1}\U'$ is invertible, with
inverse $\OpA(w)$, $w\in S^0$, and $w=w_{\rm d}+w_{\rm o}$ with $w_{\rm d}=\mathrm{diag}(g_\sff^*,g_\stt^*)+S^{-1}$ at large momentum and $w_{\rm o}\in S^{-1}$
off-diagonal. Since $\tilde a'_z=y^*\sharp^B\tilde a_z\sharp^By$, the symbol $g'':=w\sharp^Bg_z\sharp^Bw^*$ satisfies $\tilde a'_z\sharp^Bg''-1\in
S^{-\infty}$. Its flat block is $w_{\sff\sff}\sharp^B(g_z)_{\sff\sff}\sharp^Bw_{\sff\sff}^*$ plus terms containing a factor $w_{\rm o}$, of orders at most $-1$,
$-1-\frac m2$ and $-2-m$, so that $g''_{\sff\sff}=g_\sff^*(\Meff-z)^{-1}g_\sff+S^{-1}$. In the frame $U'$, Proposition~\ref{prop:paramalpha}(c) gives a right
parametrix $g'_z$ with $(g'_z)_{\sff\sff}=(\Meff'-z)^{-1}+S^{-1}$, and a left parametrix. Composing $\tilde a'_z\sharp^B(g''-g'_z)\in S^{-\infty}$ with the
left parametrix gives $g''-g'_z\in S^{-\infty}$. Comparing the parts of degree $0$ of the flat blocks, $(\Meff'-z)^{-1}=g_\sff^*(\Meff-z)^{-1}g_\sff$, and
hence $\Meff'=g_\sff^*\Meff g_\sff$, since $g_\sff$ is unitary. Consequently, $E^\infty_{\rm fl}$ is conjugated by $g_\sff$, $E^\infty_{\rm tr}$ by $g_\stt$, and
$E^\infty=U(E^\infty_{\rm fl}\oplus E^\infty_{\rm tr})U^*$ is unchanged.
\end{proof}
In particular, every $f$ in the maximal domain $\cD(H_{\max})=\{f\in L^2:Hf\in L^2\}$ satisfies $f=G'_\alpha(H-\alpha)f-K'_\alpha f$, where $\Wt G'_\alpha$ is bounded
and $K'_\alpha$ is smoothing, so that $\cD(H_{\max})\subset\Wt^{-1}L^2$.

\begin{proof}[Proof of Proposition~\ref{prop:flat}(b), first statement]
Let $\alpha\notin\Fl$, and take $\mathbb J\ni\alpha$ and $z=\alpha\in Z$, with the parametrix of Proposition~\ref{prop:paramalpha}. For $\psi_n$ as in the statement, $\psi_n=G'_\alpha(H-\alpha)\psi_n-K'_\alpha\psi_n$. The first
term tends to $0$, since $G'_\alpha$ is bounded, and $\norm{K'_\alpha\psi_n}\le\norm{K'_\alpha\Lm}\norm{\Lm^{-1}\psi_n}\to0$, which contradicts $\norm{\psi_n}=1$. Conversely, the Weyl sequences of the second statement of
Proposition~\ref{prop:flat}(b), constructed below, provide such sequences for every $\alpha\in\spec\Meff(u,\theta)$, and the set of energies admitting them is closed by a diagonal
argument.
\end{proof}
\begin{proof}[Proof of Proposition~\ref{prop:flat}(b), Weyl sequences]
Let $w$ be a unit eigenvector of $\Meff(u_0,\theta_0)$ for the eigenvalue $\alpha$. By gauge covariance, we may use the transversal gauge centered at
$u_0$, for which $|A(u)|\le C|u-u_0|$. We work with the conjugated operator $\tilde A:=\U^*(H-\alpha)\U=\OpA(\tilde a-\alpha(1+\upsilon))$, it suffices to construct
$\phi_n$ with $\norm{\tilde A\phi_n}\to0$, $\norm{\phi_n}\to1$, $\phi_n$ weakly null and $\norm{\Lm^{-1}\phi_n}\to0$, since $\psi_n:=\U\phi_n$ then satisfies
$\norm{(H-\alpha)\psi_n}\le\norm{(\U^*)^{-1}}\norm{\tilde A\phi_n}\to0$ and has the same properties up to constants. Let $G'\in\OpA(S^{-m})$ be a parametrix of
the elliptic transverse block, $\OpA(\tilde a_{\stt\stt}-\alpha(1+\upsilon_{\stt\stt}))G'=1+K$ with $K\in\OpA(S^{-\infty})$ \cite{IMP}, and let $s\in S^0$ be the
symbol of $\OpA(\tilde a_{\sff\sff}-\alpha(1+\upsilon_{\sff\sff}))-\OpA(\tilde a_{\sff\stt}-\alpha\upsilon_{\sff\stt})G'\OpA(\tilde a_{\stt\sff}-\alpha\upsilon_{\stt\sff})$. Its
degree-$0$ part is $\Meff-\alpha$, so that $s(u,k)=\Meff(u,\theta(k))-\alpha+S^{-1}$.

For $\rho\ge1$, let $\varepsilon=\rho^{-1/2}$, $\eta\in C_c^\infty(\R^2)$ with $\norm\eta_2=1$, $\eta_\varepsilon(u)=\varepsilon^{-1}\eta((u-u_0)/\varepsilon)$, and
$f_\rho=e^{i\rho\theta_0\cdot u}\eta_\varepsilon w$. We set $\phi_\rho=(f_\rho,g_\rho)$ in the fixed splitting, with
$g_\rho=-G'\OpA(\tilde a_{\stt\sff}-\alpha\upsilon_{\stt\sff})f_\rho$. Then the transverse component of $\tilde A\phi_\rho$ is $-K\OpA(\tilde a_{\stt\sff}-\alpha\upsilon_{\stt\sff})f_\rho$,
and its flat component is $\OpA(s)f_\rho$. Conjugation by $e^{i\rho\theta_0\cdot u}$ shifts the momentum, $e^{-i\rho\theta_0\cdot u}\OpA(p)e^{i\rho\theta_0\cdot
u}=\OpA(p(\cdot,\cdot+\rho\theta_0))$ exactly, since it amounts to replacing $A$ by $A-\rho\theta_0$. Hence $\norm{\OpA(s)f_\rho}=\norm{\OpA(s_\rho)\eta_\varepsilon w}$
with $s_\rho(u,k)=s(u,k+\rho\theta_0)$. Let $s_0=s(u_0,\rho\theta_0)$, so that $s_0w=O(\rho^{-1})$. On $\{|k|\le\rho/2\}$, we write
$s_\rho(u,k)-s_0=(u-u_0)\cdot d_1+k\cdot d_2$ with $d_1$ bounded in $S^0$ and $d_2=O(\rho^{-1})$ in $S^0$, by the homogeneity of $\Meff\circ\theta$ and the decay of the
remainder, and the region $|k|\ge\rho/2$ contributes $O(\rho^{-N})$ by the rapid decay of the Fourier transform of $\eta_\varepsilon$ at momenta larger than
$\varepsilon^{-1}$. By the magnetic calculus, $\OpA((u-u_0)_jb)=(u-u_0)_j\OpA(b)+\OpA(S^{-1})$ and $\OpA(k_jb)=\OpA(b)\Pi_j+\OpA(S^{-1})$, while
$\norm{(u-u_0)\eta_\varepsilon}=O(\varepsilon)$ and $\norm{\Pi\eta_\varepsilon}=O(\varepsilon^{-1})$, since $|A|=O(\varepsilon)$ on the support. Hence
$\norm{\OpA(s)f_\rho}\le C(\varepsilon+\rho^{-1}\varepsilon^{-1}+\rho^{-1})=O(\rho^{-1/2})$. Similarly, $\norm{\Lm^{-N}f_\rho}=O(\rho^{-N})$, so that the transverse
component of $\tilde A\phi_\rho$ is $O(\rho^{-N})$, $\norm{g_\rho}=O(\rho^{-m/2})$ since $G'\OpA(\tilde a_{\stt\sff})$ has order $-\frac m2$, and
$\norm{\Lm^{-1}\phi_\rho}=O(\rho^{-1})$. For $\rho_n\to\infty$ sufficiently fast, the $\phi_{\rho_n}$ are asymptotically orthonormal and weakly null, which proves
that $\alpha\in\spec_{\rm ess}H$. Finally, since $\psi_n\in H^\infty_A$ and $P$ is a bounded smooth function,
$\ip{\psi_n}{i[H,P]\psi_n}=-2\Im\ip{(H-\alpha)\psi_n}{P\psi_n}\to0$.
\end{proof}

\begin{theorem}[self-adjointness and resolvent estimates]\label{thm:analysis}
Assume (N1) and (N3).
\\[1mm]
(a) $H$ is essentially self-adjoint on $C_c^\infty$, and $\cD(H)\subset\Wt^{-1}L^2$. For $r=0$,
$\cD(H)=H^m_A$.
\\[1mm]
(b) Let $\mathbb J$ be a compact interval at distance at least $\frac14\dist(\alpha,\Fl)$ from $\Fl$, let $\Re z\in\mathbb J$ and $|\Im z|\le1$, and let
$d=\dist(z,\spec H)$. Then $\norm{\Wt(H-z)^{-1}\Wt}\le C(1+d^{-1})$. Moreover, $\Lm^N\Phi(H)$ is bounded for $\Phi\in C_c^\infty(\mathbb J)$ and all $N$, and,
for $d\le1$, the Combes--Thomas estimate $\norm{\Wt e^{\mu\phi}(H-z)^{-1}e^{-\mu\phi}\Wt}\le C/d$ holds for $|\nabla\phi|\le1$ with bounded higher
derivatives and $|\mu|\le cd$.
\\[1mm]
(c) Let $H_s=\Op^{A_s}(a_s)$, $s\in[0,1]$, satisfy (N1) and (N3) with uniform constants. Assume that the coefficients of $a_s$ depend continuously on $s$ in
$C_b^\infty$, that $A_s-A_0$ is bounded with bounded derivatives and depends continuously on $s$ in $C_b^\infty$, that the frame symbols $U_s$ depend
continuously on $s$ in $S^0$, and that the conjugated symbols $\tilde a_s$ depend continuously on $s$ in the graded class $\mathcal S^0_+$. Then
$s\mapsto(H_s-z)^{-1}$ is norm continuous for $z\notin\R$.
\end{theorem}
It follows from Proposition~\ref{prop:paramalpha} as in the elliptic theory \cite{DN}, and the proof is given in Appendix~\ref{s:sec:routine}.

\section{The magnetized bulk and its invariant}\label{sec:pbulk}
This section proves Theorems~\ref{thm:bulkspec} and \ref{thm:bulk}. Throughout, $\ab$ is a constant symbol satisfying (N1) at a constant field $B\ne0$,
and we use the Landau gauge $A=(-By,0)$. A change of gauge conjugates $H_B(\ab)$ by a multiplication operator $e^{i\lambda}$, which commutes with $x$ and $y$
and preserves $\trarea$, so that $\sigma_H$ does not depend on the gauge. Three unitarily equivalent forms of the reduced operator are used: $K_B(\ab)$ in
the variables $(y,\zeta)$, the semiclassical form $\Op^w_h(\ab\circ\varsigma_B)$ with $h=|B|$, used for the semiclassical limit of
Proposition~\ref{prop:semiclassical}, and the dilated form of Lemma~\ref{lem:reduction}(d), in which the classes $\cP_\varepsilon$ do not depend on the field and
Lemma~\ref{lem:tau} applies along families in $B$.

The strategy is as follows. In the Landau gauge, $H_B(\ab)$ is a direct integral of translates of the one-dimensional operator $K_B(\ab)$, and the Kubo
formula reduces to the pairing $\tau$ of \eqref{eq:tau} of the spectral projection $\hat P_\alpha$ of $K_B(\ab)$ with $y$ and $D_y$ (Lemma~\ref{lem:kubo}). This pairing is defined
for projections whose symbols are, up to decaying terms, the projection $E^\infty$ along each direction at infinity, it is invariant along deformations in that
class, and it changes by the index of a pair, or by the spectral flow, when levels cross (Lemma~\ref{lem:tau}). We deform $\ab-\alpha$ to a reference symbol
whose negative projection is an extension of $E^\infty$ to the disk, and the levels created along the deformation give the integer part of $J$. For the
reference, $\tau$ does not depend on the size $R$ of the disk in momentum, and its semiclassical limit as $R\to\infty$ is the Chern integral of the
extension (Proposition~\ref{prop:semiclassical}), which gives the fractional part. At weak field, the same semiclassical limit, with the field as the small
parameter, gives the field-free Chern integral. All proofs are given in this section, except those of Propositions~\ref{prop:weak}, \ref{prop:gradedFC},
\ref{prop:semiclassical} and \ref{prop:gap}, which are given in Appendix~\ref{s:sec:routine}. The last three are statements of the isotropic and
semiclassical symbol calculus, and the first is a perturbative inversion. 

\smallskip
\noindent\textbf{Reduction to one dimension.} Let
\begin{equation}\label{eq:ell}
  \ell(y,\zeta):=(By,\zeta),\qquad X=(y,\zeta)\in\R^2 .
\end{equation}
The reduced bulk operator \eqref{eq:KB} is $K_B(c)=\Op^w_y(c\circ\ell)$ for a constant-coefficient symbol $c(k)$. We use the isotropic (Shubin) classes
$\Gamma^\mu$ on the phase plane, defined by $|\partial_X^\beta b|\le C_\beta\jp X^{\mu-|\beta|}$ \cite{Shubin}. In these classes all derivatives gain,
$b_1\#b_2-b_1b_2\in\Gamma^{\mu_1+\mu_2-2}$, and $\Op^w(\Gamma^{-\varepsilon})$ consists of compact operators for $\varepsilon>0$. Since $\ell$ is linear and
invertible, $c\circ\ell\in\Gamma^\mu$ for every constant-coefficient $c\in S^\mu$.
\begin{lemma}\label{lem:reduction}
Let $c$ and $d$ be constant-coefficient symbols. (a) The partial Fourier transform $\mathsf F_x$ in $x$ maps $H_B(c)$ to the direct integral of the fibers
$\Op^w_y(c(\xi+By,\zeta))=e^{-iy_cD_y}K_B(c)e^{iy_cD_y}$, $y_c=-\xi/B$. (b) $K_B(c)K_B(d)=K_B(c\sharp^Bd)$. (c) The graded classes and Proposition~\ref{prop:paramalpha} hold for the operators $\Op^w(b)$, $b$ an isotropic symbol on $\R^2_X$, with $\jp X$ in place of
$\jp k$ and the Moyal product $\#$ in place of $\sharp^B$. In particular, $K_B(\ab)$ is self-adjoint on its maximal domain. (d) For $h=|B|$, $K_B(\ab)$ is unitarily equivalent to
$\Op^w_h(\ab\circ\varsigma_B)$, where $\varsigma_B$ is the identity if $B>0$ and the swap of the two phase-space variables if $B<0$, and, after the
isotropic dilation $X\mapsto\sqrt hX$, to $\Op^w(\ab\circ\varsigma_B(\sqrt h\,\cdot))$. In this last form, the direction at infinity $\theta(X)$ of the principal
symbol does not depend on $|B|$ for fields of a fixed sign, and the resolvents remain norm continuous in $B$.
\end{lemma}
\begin{proof}
For a linear potential, the line integral in the magnetic quantization is $A(\frac{u+u'}2)\cdot(u'-u)$, so that $\OpA(c)$ is the Weyl quantization of
$c(k-A(u))=c(\xi+By,\zeta)$. This symbol does not depend on $x$, so that $\OpA(c)$ commutes with translations in $x$, and $\mathsf F_x$ maps it to the family of
operators $\Op^w_y(c(\xi+By,\zeta))$. Since $c(\xi+By,\zeta)=(c\circ\ell)(y-y_c,\zeta)$, this fiber is $e^{-iy_cD_y}K_B(c)e^{iy_cD_y}$, which is (a). The magnetic
calculus gives $H_B(c)H_B(d)=H_B(c\sharp^Bd)$ \cite{MP04,IMP}, where $c\sharp^Bd$ is again a constant-coefficient symbol. Taking fibers at $\xi$ and conjugating by $e^{-iy_cD_y}$
gives (b). For (c), the proofs of these statements only use composition and inversion in the graded classes and the boundedness of operators of order $0$,
which hold in the isotropic calculus with $\jp X$ in place of $\jp k$, since all derivatives gain there. Self-adjointness then follows as in
Theorem~\ref{thm:analysis}(a), the parametrix at $z=\pm i$ being available since $\pm i\notin\Fl_B(\ab)$. Part (d) is a metaplectic change of variables,
proved in Proposition~\ref{s:prop:shubin}, followed by the dilation, which is also metaplectic; norm continuity in $B$ holds because $X\cdot\nabla_X$ of the
dilated symbol has the same order.
\end{proof}

\begin{proof}[Proof of Theorem~\ref{thm:bulkspec}]
\emph{Step 1: $\spec_{\rm ess}K_B(\ab)\subset\Fl_B(\ab)$.} Let $U(k)$ be a frame symbol of $\ab$, independent of $u$, and let $\tilde a=U^*\sharp^B\ab\sharp^BU$ be the
conjugated symbol, which satisfies (DN$_0$) by (N1), with $U^*\sharp^BU=1+\upsilon$. By Lemma~\ref{lem:reduction}(b),
$K_B(U)^*(K_B(\ab)-E)K_B(U)=K_B(\tilde a-E(1+\upsilon))$, and $K_B(U)$ is invertible. The blocks of $\tilde a\circ\ell$ lie in $\Gamma^0$, $\Gamma^{m/2}$,
$\Gamma^{m/2}$ and $\Gamma^m$. For $E\notin\Fl_B(\ab)$, the isotropic versions of Proposition~\ref{prop:paramalpha}(b),(c)
(Lemma~\ref{lem:reduction}(c)) give right and left parametrices of $K_B(\tilde a-E(1+\upsilon))$, bounded between the weighted spaces, with remainders in
$\Op^w(\Gamma^{-\infty})$, which are compact. Hence $K_B(\ab)-E$ is Fredholm, and $E\notin\spec_{\rm ess}K_B(\ab)$.

\emph{Step 2: $\Fl_B(\ab)\subset\spec_{\rm ess}K_B(\ab)$.} Let $f\in\spec\Meff(\theta_0)$ for some $\theta_0\in S^1$, and let $w\in\C^r$ be a unit eigenvector,
$\Meff(\theta_0)w=fw$. Let $T:=\Op^w(\tilde a_{\stt\stt}\circ\ell)$, which is elliptic of order $m$ in $\Gamma$, let $G'$ be a parametrix of $T$ modulo
$\Op^w(\Gamma^{-\infty})$, and let
\[
  S:=\Op^w(\tilde a_{\sff\sff}\circ\ell)-\Op^w(\tilde a_{\sff\stt}\circ\ell)\,G'\,\Op^w(\tilde a_{\stt\sff}\circ\ell).
\]
By the isotropic calculus, $S=\Op^w(s)$ with $s-(\tilde a_{\sff\sff}-\tilde a_{\sff\stt}\tilde a_{\stt\stt}^{-1}\tilde a_{\stt\sff})\circ\ell\in\Gamma^{-1}$, so
that $s=\Meff\circ\theta\circ\ell+\Gamma^{-1}$ by Definition~\ref{def:Meff}, where $\theta(k)=k/|k|$ and $\ell$ is the map \eqref{eq:ell}. We use the quasimodes of the proof of Proposition~\ref{prop:flat}(b), with widths adapted to the isotropic calculus. For $\rho>0$, let
$X_\rho=(y_\rho,\zeta_\rho):=\ell^{-1}(\rho\theta_0)$, $L=\rho^{1/2}$, and
\[
  \varphi_{\rho}(y):=(\pi L^2)^{-1/4}\,e^{i\zeta_\rho y}\,e^{-(y-y_\rho)^2/(2L^2)}\,w,
\]
the normalized Gaussian centered at $X_\rho$, with widths $L$ in $y$ and $L^{-1}$ in $\zeta$. Since $\Meff\circ\theta\circ\ell$ is homogeneous of degree $0$,
$s(X_\rho)w=fw+O(\rho^{-1})$ and, on $\{|X-X_\rho|\le|X_\rho|/2\}$, $s(X)-s(X_\rho)=(y-y_\rho)d_1+(\zeta-\zeta_\rho)d_2$ with $d_1,d_2$ of size $O(\rho^{-1})$ in
$\Gamma^0$, while the complementary region contributes $O(\rho^{-N})$. In the isotropic calculus \cite{Shubin}, $\Op^w((y-y_\rho)d_1)=(y-y_\rho)\Op^w(d_1)$
and $\Op^w((\zeta-\zeta_\rho)d_2)=\Op^w(d_2)(D_y-\zeta_\rho)$ up to $O(\rho^{-1})$ in norm, and $\norm{(y-y_\rho)\varphi_\rho}=O(L)$,
$\norm{(D_y-\zeta_\rho)\varphi_\rho}=O(L^{-1})$. Hence
\[
  \norm{(S-f)\varphi_\rho}\le C\bigl(\rho^{-1}+L\rho^{-1}+L^{-1}\rho^{-1}\bigr)\le C\rho^{-1/2}.
\]
Let
$\rho_n\to\infty$, $\varphi_n:=\varphi_{\rho_n}$, and
\[
  \psi_n:=K_B(U)\bigl(\varphi_n\oplus(-G'\Op^w(\tilde a_{\stt\sff}\circ\ell)\varphi_n)\bigr).
\]
Since the centers $X_{\rho_n}$ tend to infinity, $\varphi_n\rightharpoonup0$, and $\psi_n\rightharpoonup0$. Since $K_B(U)$ is invertible, $\norm{\psi_n}\ge c>0$.
Moreover, $K_B(U)^*(K_B(\ab)-f)\psi_n=K_B(\tilde a-f(1+\upsilon))(\varphi_n\oplus(-G'\Op^w(\tilde a_{\stt\sff}\circ\ell)\varphi_n))$. Its flat component is
$(S-f)\varphi_n$ plus terms of the form $\Op^w(\upsilon\circ\ell)$ or $\Op^w(\Gamma^{-m/2})$ applied to $\varphi_n$, and its transverse component is
$(1-TG')\Op^w(\tilde a_{\stt\sff}\circ\ell)\varphi_n$ plus terms of the same form. These operators are compact, hence map the weakly null sequence $\varphi_n$ to a
sequence tending to $0$ in norm. Therefore $\norm{(K_B(\ab)-f)\psi_n}\to0$, and $f\in\spec_{\rm ess}K_B(\ab)$ by Weyl's criterion. Since $S^1$ is compact,
$\Fl_B(\ab)=\bigcup_\theta\spec\Meff(\theta)$ is closed, and the inclusion follows.

\emph{Step 3: the spectrum of $H_B(\ab)$.} By Lemma~\ref{lem:reduction}(a), $H_B(\ab)$ is unitarily equivalent to the direct integral over $\xi\in\R$ of the
operators $e^{-iy_c(\xi)D_y}K_B(\ab)e^{iy_c(\xi)D_y}$, which are all unitarily equivalent to $K_B(\ab)$. Hence $\spec H_B(\ab)=\spec K_B(\ab)$. If
$K_B(\ab)\psi=E\psi$, then $\mathsf F_x^{-1}(g(\xi)e^{-iy_c(\xi)D_y}\psi)$ is an eigenvector of $H_B(\ab)$ for every $g\in L^2(\R)$, so that $E$ is an eigenvalue of
infinite multiplicity.
\end{proof}

\smallskip
\noindent\textbf{Spectral projections and the Kubo pairing.} Recall that $\hat P_\alpha=\mathbf 1_{(-\infty,\alpha)}(K_B(\ab))$ is the spectral projection of the
reduced bulk operator, and $P_\alpha=\mathbf 1_{(-\infty,\alpha)}(H_B(\ab))$ that of the bulk operator, as in \eqref{eq:sigmaH}. These projections belong to the
following classes.
\begin{definition}\label{def:Peps}
Let $E:S^1\to\Gr_q(\C^n)$ be smooth, let $\varepsilon\in(0,\frac13)$, and let $\Gamma^\mu_{3/4}$ be the isotropic class of symbols on $\R^2_X$ with
$|\partial^\beta b|\le C_\beta\jp X^{\mu-\frac34|\beta|}$ \cite{Shubin}. Let $\tilde\theta$ be smooth with $\tilde\theta(X)=X/|X|$ for $|X|\ge1$. We denote by
$\cP_\varepsilon(E)$ the set of orthogonal projections $\Op^w(p)$ on $L^2(\R;\C^n)$ with $p-E\circ\tilde\theta\in\Gamma^{-1+\varepsilon}_{3/4}$. A family $s\mapsto
P_s\in\cP_\varepsilon(E)$ is called $C^1$ if $s\mapsto p_s-E\circ\tilde\theta$ is $C^1$ in $\Gamma^{-1+\varepsilon}_{3/4}$.
\end{definition}
\begin{proposition}\label{prop:gradedFC}
Let $\alpha\notin\spec K_B(\ab)\cup\Fl_B(\ab)$, and let $E^\infty_\ell:=E^\infty\circ\theta\circ\ell$. Then $\hat P_\alpha\in\cP_\varepsilon(E^\infty_\ell)$ for every
$\varepsilon\in(0,\frac13)$. If $\ab_s$ is a $C^1$ family satisfying (N1) uniformly, with fixed $E^\infty$ at $\alpha$ and $\alpha\notin\spec K_B(\ab_s)$, then $s\mapsto\hat
P_\alpha(\ab_s)$ is $C^1$ in $\cP_\varepsilon(E^\infty_\ell)$. The same holds for symbols of the same graded type that are not of the form $\ab\circ\ell$, and in
particular for the reference and deformation symbols of Lemma~\ref{lem:path} below.
\end{proposition}
The proof, given in Appendix~\ref{s:sec:routine}, follows the functional calculus of pseudodifferential operators \cite{DS99,Shubin}: it writes $\hat
P_\alpha$ as an integral of resolvents along a vertical line and describes the symbol of the resolvent in the graded classes for all values of the spectral
parameter, which is the new ingredient.

For a projection $\hat P$ on $L^2(\R;\C^n)$ such that $\hat P[[\hat P,D_y],[\hat P,y]]$ is trace-class, we set
\begin{equation}\label{eq:tau}
  \tau(\hat P):=-i\Tr\bigl(\hat P[[\hat P,D_y],[\hat P,y]]\bigr).
\end{equation}
This is the pairing of $\hat P$ with the cyclic $2$-cocycle defined by the derivations $[D_y,\cdot]$ and $[y,\cdot]$, of the type used for the quantum Hall
effect in \cite{Connes,BES}, and we call it the Kubo pairing. Lemma~\ref{lem:kubo} below shows that $J=\sgn(B)\tau(\hat P_\alpha)$. With
$\delta_1=[D_y,\cdot]$ and $\delta_2=[y,\cdot]$, which commute since $[D_y,y]=-i$ is central, $\tau(P)=-i\Tr(P[\delta_1P,\delta_2P])$, and for
$P\in\cP_\varepsilon(E)$ the symbols of $\delta_1P$ and $\delta_2P$ are $-i\partial_yp$ and $i\partial_\zeta p$.
\begin{lemma}\label{lem:tau}
(i) For $P\in\cP_\varepsilon(E)$, $P[[P,D_y],[P,y]]\in\mathfrak S_1$, so that $\tau(P)$ is defined, and $P-P'$ is compact for $P'\in\cP_\varepsilon(E)$.
(ii) For $P,P'\in\cP_\varepsilon(E)$, $\tau(P)-\tau(P')=\ind(P,P')$. In particular, $\tau(P)-\tau(P')=\Tr(P-P')$ if $P-P'$ is trace-class.
(iii) The pairing $\tau$ is constant along $C^1$ families in $\cP_\varepsilon(E)$.
(iv) Let $\lambda\mapsto c_\lambda$, $\lambda\in[0,1]$, be polynomial in $\lambda$ and of Douglis--Nirenberg type with a common frame and uniform constants, let
$\Op^w(c_0)$ and $\Op^w(c_1)$ be invertible, let $0$ be at positive distance from the essential spectra, and let $P_\lambda:=\mathbf
1_{(-\infty,0)}(\Op^wc_\lambda)\in\cP_\varepsilon(E)$ whenever $0\notin\spec\Op^w(c_\lambda)$. Then $\tau(P_1)-\tau(P_0)=\ind(P_1,P_0)=-\sfl(\lambda\mapsto\Op^wc_\lambda)$.
(v) The pairing $\tau$ is constant along norm-continuous families in $\cP_\varepsilon(E)$.
\end{lemma}
\begin{proof}
We use the following rules of the isotropic calculus (Appendix~\ref{s:sec:tools}): compositions gain $\frac32$ in the first
correction; $\Op^w(c)$ is trace-class with $\Tr\Op^w(c)=\frac1{2\pi}\int\tr c$ if $c\in\Gamma^\mu_{3/4}$ with $\mu<-2$; $\Tr(ST)=\Tr(TS)$ if $S\in\mathfrak S_1$ and $T$ is
bounded, or if $S\in\Op\Gamma^\mu_{3/4}$ and $T\in\Op\Gamma^\nu_{3/4}$ with $\mu+\nu<-2$; and $\Tr\delta_j(Z)=0$ for $Z\in\Op\Gamma^\mu_{3/4}$ with $\mu<0$ and
$\mu-\frac34<-2$, since the symbol of $\delta_jZ$ is a partial derivative of that of $Z$ up to a constant, which is integrable with integral $0$.

(i) Write $p=E\circ\tilde\theta+w$ with $w\in\Gamma^{-1+\varepsilon}_{3/4}$. Since $\partial(E\circ\tilde\theta)\in\Gamma^{-1}$ and $\partial w\in\Gamma^{-7/4+\varepsilon}_{3/4}$,
we have $\delta_jP\in\Op\Gamma^{-1}_{3/4}$. The symbol of $P[\delta_1P,\delta_2P]$ is $p[\partial_yp,\partial_\zeta p]$ up to $\Gamma^{-7/2}_{3/4}$. The
derivatives $\partial_y(E\circ\theta)$ and $\partial_\zeta(E\circ\theta)$ are both scalar multiples of $(\partial_\theta E)\circ\theta$, so that the commutator of the
leading parts vanishes. The remaining terms contain $w$ or a derivative of $w$, and belong to $\Gamma^{-3+\varepsilon}_{3/4}$ or $\Gamma^{-11/4+\varepsilon}_{3/4}$, with
$-\frac{11}4+\varepsilon<-2$. Hence $P[\delta_1P,\delta_2P]\in\Op\Gamma^{-11/4+\varepsilon}_{3/4}$ is trace-class. Finally, $P-P'=\Op^w(p-p')$ with
$p-p'\in\Gamma^{-1+\varepsilon}_{3/4}$, which is compact.

(iii) The traces defining $\tau(P_s)$ are traces of operators in $\Op\Gamma^{-11/4+\varepsilon}_{3/4}$ depending $C^1$ on $s$ in that class, so that we may differentiate
under the trace. We write $P=P_s$, $\bar P=1-P$ and $\dot P=\partial_sP$. Since $P^2=P$, we have $\dot P=P\dot P\bar P+\bar P\dot PP$ and
$\delta_jP=P\delta_jP\bar P+\bar P\delta_jPP$. In particular, $\dot P=[X,P]$ with $X:=[\dot P,P]\in\Op\Gamma^{-1+\varepsilon}_{3/4}$, and
$C:=[\delta_1P,\delta_2P]\in\Op\Gamma^{-2}_{3/4}$ commutes with $P$, being a product of two operators that are off-diagonal with respect to $P$. Then
\[
  i\,\partial_s\tau=\Tr(\dot PC)+\Tr\bigl(P[\delta_1\dot P,\delta_2P]+P[\delta_1P,\delta_2\dot P]\bigr).
\]
The first term vanishes, since $\Tr(P\dot P\bar PC)=\Tr(P\dot PC\bar P)=\Tr(\bar PP\dot PC)=0$, and likewise for $\bar P\dot PP$. For the second term, we use
$\delta_j\dot P=[\delta_jX,P]+[X,\delta_jP]$. The terms containing $X$ undifferentiated sum to $\Tr(P[X,C])=\Tr(PXC)-\Tr(CPX)=0$, by $PC=CP$ and
cyclicity. For $Y\in\Op\Gamma^{-7/4+\varepsilon}_{3/4}$, expanding with $P\delta_jPP=0=\bar P\delta_jP\bar P$ and using cyclicity gives
$\Tr(P[[Y,P],\delta_2P])=-\Tr(Y\delta_2P)$ and $\Tr(P[\delta_1P,[Y,P]])=\Tr(Y\delta_1P)$. Hence the remaining terms are
\[
  -\Tr(\delta_1X\,\delta_2P)+\Tr(\delta_2X\,\delta_1P)=\Tr(X\,\delta_1\delta_2P)-\Tr(X\,\delta_2\delta_1P)=0,
\]
where we used $\Tr\delta_1(X\delta_2P)=0=\Tr\delta_2(X\delta_1P)$ and $[\delta_1,\delta_2]=0$. Therefore $\partial_s\tau=0$.

(ii) We first assume that $P-P'\in\Op\Gamma^{-\infty}$. Let $X=PD_yP$ and $Y=PyP$, with primes for the corresponding operators for $P'$. The identity
$P[[P,D_y],[P,y]]P=[X,Y]-P[D_y,y]P$, the relation $P[D_y,y]P=-iP$ and the cyclicity of the trace give
$\tau(P)-\tau(P')=-i\Tr\bigl([X-X',Y]+[X',Y-Y']\bigr)+\Tr(P-P')$. Here $X-X'$ and $Y-Y'$ are smoothing, and $\Tr[S,T]=0$ for $S\in\Op\Gamma^{-\infty}$ and
$T\in\Op\Gamma^1$, so that $\tau(P)-\tau(P')=\Tr(P-P')=\ind(P,P')$ \cite{ASS}. In general, let $c_\lambda:=(1-\lambda)(1-2p)+\lambda(1-2p')$, a family of
order-$0$ symbols equal to $1-2E\circ\tilde\theta$ modulo $\Gamma^{-1+\varepsilon}_{3/4}$. The operators $\Op^w(c_\lambda)$ are bounded, self-adjoint and
analytic in $\lambda$, and equal to $1-2\Op^w(E\circ\tilde\theta)$ up to compact operators, so that their essential spectra are contained in $\{-1,1\}$. Their
spectrum near $0$ thus consists of finitely many analytic eigenvalue branches, with eigenvectors in $\Op\Gamma^{-\infty}$-ranges, and $\Op^w(c_0)=1-2P$ and
$\Op^w(c_1)=1-2P'$ are invertible. For $\lambda$ outside the finitely many crossing points, $\mathbf 1_{(-\infty,0)}(\Op^wc_\lambda)\in\cP_\varepsilon(E)$ by the
functional calculus of order-$0$ isotropic operators \cite{Shubin}, and it is $C^1$ in $\lambda$ between crossings, so that $\tau$ is constant there by
(iii). At a crossing, the spectral projection changes by a finite-rank projection with range in the Schwartz class, so that $\tau$ and the index of the pair
with $P$ change by the same amount by the smoothing case \cite{ASS94,Phillips}. Hence $\tau(P')-\tau(P)=\ind(P',P)$.

(iv) Let $\mathcal D=\mathrm{diag}(1,(1+y^2+D_y^2)^{m/4})$ and $z_0=ic$, with $0<c\le1$ smaller than the distance from $0$ to the degree-$0$ Schur
complements. By the isotropic version of Proposition~\ref{prop:paramalpha}, uniform in $\lambda$, $T_\lambda:=\mathcal D^{-1}(\Op^w(c_\lambda)-z_0)\mathcal
D^{-1}$ is a polynomial in $\lambda$ with bounded coefficients and uniformly bounded inverse on $[0,1]$, so that $(\Op^w(c_\lambda)-z_0)^{-1}=\mathcal
D^{-1}T_\lambda^{-1}\mathcal D^{-1}$ is holomorphic near $[0,1]$. The family is thus holomorphic in the generalized sense \cite[\S VII.1.2]{Kato}, and its
isolated eigenvalues near $0$ are real-analytic in $\lambda$ \cite[Thm.~VII.3.9]{Kato}. They vanish at finitely many points. Near any $\lambda$, we choose $g>0$ with
$\pm g\notin\spec$ and write $P_\lambda=P'_\lambda+\Pi^-_\lambda$, where $P'_\lambda=\mathbf 1_{(-\infty,-g]}(\Op^wc_\lambda)\in\cP_\varepsilon(E)$ is $C^1$ in
$\lambda$ by Proposition~\ref{prop:gradedFC}, and $\Pi^-_\lambda$ is the finite-rank projection onto the eigenvalues in $(-g,0)$, whose eigenvectors are in
the Schwartz class. By (iii), $\tau(P'_\lambda)$ is locally constant, and by (ii), $\tau(P_\lambda)-\tau(P'_\lambda)=\operatorname{rank}\Pi^-_\lambda$. Hence
$\tau(P_\lambda)$ changes only at the crossing points, by minus the net number of upward crossings, and so does $\ind(P_\lambda,P_0)$, by additivity of the
index and since $\ind=\Tr$ for finite-rank differences.

(v) If $P,P'\in\cP_\varepsilon(E)$ and $\norm{P-P'}<1$, then $\ind(P,P')=0$ \cite{ASS94}, and hence $\tau(P)=\tau(P')$ by (ii). A norm-continuous family is
therefore locally constant, hence constant.
\end{proof}
\begin{lemma}[Kubo formula]\label{lem:kubo}
(a) $J(\ab,B;\alpha)=\sgn(B)\,\tau(\hat P_\alpha)$, and $\tau(\hat P)=\operatorname{rank}\hat P$ for finite-rank projections in $\Op\Gamma^{-\infty}$.
\\[1mm]
(b) For a constant Hermitian symbol $\ab$ and $\alpha\in\mathcal G(\ab)$, let $P_\alpha=\mathbf 1_{(-\infty,\alpha)}(\ab(D))$, a Fourier multiplier with symbol
$E_\alpha(k)$. Then $i\,\trarea\bigl(P_\alpha[[P_\alpha,x],[P_\alpha,y]]\bigr)=\frac1{2\pi}I(\ab;\alpha)$. In particular, formula \eqref{eq:sigmaH} extends to
$B=0$ and gives directly $J(\ab,0;\alpha)=I(\ab;\alpha)$, the convention \eqref{eq:J0}.
\end{lemma}
\begin{proof}
(a) By Lemma~\ref{lem:reduction}(a), the fiber of $P=P_\alpha$ at $\xi$ is $\hat P(\xi)=e^{-iy_cD_y}\hat P_\alpha e^{iy_cD_y}$ with $y_c=-\xi/B$. With
$\mathsf F_xf(\xi)=\int e^{-ix\xi}f\,dx$, multiplication by $x$ acts on fibers as $i\partial_\xi$, so that the fiber of $[P,x]$ is $-i\partial_\xi\hat P(\xi)$.
Since $\partial_\xi=-\frac1B\partial_{y_c}$ and $\partial_s(e^{-isD_y}Ze^{isD_y})=-i[D_y,e^{-isD_y}Ze^{isD_y}]$, this fiber is $\frac1B[D_y,\hat P(\xi)]$. The fiber of
$[P,y]$ is $[\hat P(\xi),y]=e^{-iy_cD_y}[\hat P_\alpha,y]e^{iy_cD_y}$, since $e^{isD_y}ye^{-isD_y}=y+s$. The fibers of $Z:=P[[P,x],[P,y]]$ are therefore $e^{-iy_cD_y}\hat
Z_0e^{iy_cD_y}$ with $\hat Z_0=-\frac1B\hat P_\alpha[[\hat P_\alpha,D_y],[\hat P_\alpha,y]]$, which is trace-class with a continuous kernel by
Proposition~\ref{prop:gradedFC} and Lemma~\ref{lem:tau}(i). The kernel of $Z$ on the diagonal is therefore
continuous and constant, equal to $\frac1{2\pi}\int\hat Z_0(y-y_c,y-y_c)\,d\xi=\frac{|B|}{2\pi}\Tr\hat Z_0$ after the change of variables $\xi=-By_c$, and
$\trarea Z$ is this constant. Hence
$\trarea Z=-\frac{\sgn B}{2\pi}\Tr\bigl(\hat P_\alpha[[\hat P_\alpha,D_y],[\hat P_\alpha,y]]\bigr)=-\frac{\sgn B}{2\pi}\,i\tau(\hat P_\alpha)$, and
$\sigma_H=i\trarea Z=\frac{\sgn B}{2\pi}\tau(\hat P_\alpha)$. For a finite-rank projection $\hat P$ in $\Op\Gamma^{-\infty}$, both $\hat P$ and $0$ belong to $\cP_\varepsilon(0)$, so that
$\tau(\hat P)=\tau(0)+\Tr\hat P=\operatorname{rank}\hat P$ by Lemma~\ref{lem:tau}(ii). In particular, a single
Landau level of $|k|^2$ at $B>0$ gives $J=+1$, consistently with the sign convention of \eqref{eq:sigmaH}.

(b) Here $P_\alpha$ is the Fourier multiplier with symbol $E_\alpha=\mathbf 1_{(-\infty,\alpha)}(\ab)$ of \eqref{eq:I}. Since $x_j$ acts as $i\partial_{k_j}$ on the Fourier side, $[P_\alpha,x_j]$ is the Fourier multiplier with symbol $-i\partial_{k_j}E_\alpha$, and the
commutator of two Fourier multipliers is the Fourier multiplier with the matrix commutator of their symbols, so $[[P_\alpha,x],[P_\alpha,y]]$ has symbol
$(-i\partial_\xi E_\alpha)(-i\partial_\zeta E_\alpha)-(-i\partial_\zeta E_\alpha)(-i\partial_\xi E_\alpha)=-[\partial_\xi E_\alpha,\partial_\zeta E_\alpha]$. Hence
$Z:=P_\alpha[[P_\alpha,x],[P_\alpha,y]]$ is the Fourier multiplier with symbol $-E_\alpha[\partial_\xi E_\alpha,\partial_\zeta E_\alpha]$, its kernel is constant on the diagonal, and
$\trarea Z=(2\pi)^{-2}\int\tr\bigl(-E_\alpha[\partial_\xi E_\alpha,\partial_\zeta E_\alpha]\bigr)\,d^2k$. Since $c_1(E_\alpha)=\frac i{2\pi}\tr(E_\alpha[\partial_\xi
E_\alpha,\partial_\zeta E_\alpha])\,d\xi\wedge d\zeta$ and $I(\ab;\alpha)=-\int c_1(E_\alpha)$, this gives $\trarea Z=-\frac i{2\pi}I(\ab;\alpha)$, and hence
$i\,\trarea Z=\frac1{2\pi}I(\ab;\alpha)$.
\end{proof}

The pairing $\tau$ of the projections of the references below is computed in the semiclassical limit.
In the following statement, $Y=(s,\sigma)\in\R^2$ is the phase-space variable of operators on $L^2(\R_s;\C^n)$, in the applications a rescaling of $X$,
$\Op^w_h$ is the semiclassical Weyl
quantization on $L^2(\R;\C^n)$, and the graded types refer to the weights $\jp Y$ in place of $\jp k$.
\begin{proposition}\label{prop:semiclassical}
Let $c_h(Y)$, $0<h\le1$, be Hermitian symbols of graded type $\mathcal S^0_+$, uniformly in $h$, invertible for all $Y$ with inverses of graded type
$\mathcal S^0_-$, uniformly in $h$. Let $Q_h$ be the pointwise negative projection of $c_h$, let $\hat\U_h=\Op^w_h(\hat u_h)$ be unitary with $\hat u_h$
bounded in $\Gamma^0$, and let $P_h=\hat\U_h\mathbf 1_{(-\infty,0)}(\Op^w_h c_h)\hat\U_h^*$. Then, for $h\le h_0$, $\Op^w_h(c_h)$ is invertible and
\[
  \tau(P_h)=-\int_{\R^2}c_1(\hat u_hQ_h\hat u_h^*)+O(h)\qquad(h\to0).
\]
\end{proposition}
The proof is given in Appendix~\ref{s:sec:routine}. It expands the trace defining $\tau$ by the semiclassical calculus, the leading term being
$-\frac1{2\pi}\int\tr(q[\partial_sq,\partial_\sigma q])$ for the pointwise projection $q$, which is $-\int c_1$.

\smallskip
\smallskip
\noindent\textbf{References and deformations.} The bulk invariant is computed by comparison with \emph{reference} operators, whose invariant is given by an
extension of $E^\infty$ (defined before \eqref{eq:kappa}), through a deformation along which the spectral flow counts the levels that are created. Both are constructed for symbols
written in the frame, that is, in the fixed splitting $\C^r\oplus\C^{n-r}$ of section~\ref{sec:results}. A symbol $c$ in the frame corresponds to the
operator $(\U^*)^{-1}\OpA(c)\U^{-1}$, which is $H-z$ for $c=\tilde a_z$. All spectral statements below are made in frames for which $\U$ is unitary, which can
be arranged by Lemma~\ref{lem:globalframe}(b), and this operator is then unitarily equivalent to $\OpA(c)$.

Let $\tilde E^\infty:=\mathrm{diag}(\mathbf 1_{(-\infty,\alpha)}(\Meff),\mathbf 1_{(-\infty,0)}(\tilde a_m))=U^*E^\infty U$, a projection-valued function of $(u,\theta)$ in
the fixed splitting. Let $E^{\rm ext}$ be a bulk extension of $E^\infty$, and let $U_R$ be the unitary-valued homogeneous frame symbol of
Lemma~\ref{lem:globalframe}(a), which equals $U(u,\theta(k))$ for $|k|$ near $R$ and beyond. We set
\[
  \tilde E_R(u,k):=U_R^*E^{\rm ext}(k/R)U_R\quad(|k|\le R),\qquad \tilde E_R(u,k):=\tilde E^\infty(u,\theta(k))\quad(|k|\ge R).
\]
The two definitions agree near $|k|=R$, since $E^{\rm ext}(w)=E^\infty(w/|w|)$ for $\frac12\le|w|\le1$, so that $\tilde E_R$ is a smooth projection-valued symbol.
With $\Sigma_R=1-2\tilde E_R$, $\Lm_R=(R^2+\jp k^2)^{1/2}$ and $D_R=\mathrm{diag}(1,\Lm_R^{m/2})$, the \emph{reference symbol} is
\[
  b_{\rm ref}:=D_R\Sigma_RD_R ,
\]
and the \emph{congruence reference} is the corresponding operator $H_{\rm ref}:=(\U^*)^{-1}\OpA(b_{\rm ref})\U^{-1}$. Since $\Sigma_R^2=1$, the
symbol $b_{\rm ref}$ is invertible with inverse $D_R^{-1}\Sigma_RD_R^{-1}$. For $|k|\ge R$, it is block diagonal and equal to
$\mathrm{diag}(1-2E^\infty_{\rm fl},\Lm_R^m(1-2E^\infty_{\rm tr}))$, which is of Douglis--Nirenberg type, with degree-$0$ Schur complement $1-2E^\infty_{\rm fl}$ of
spectrum $\{\pm1\}$ and an elliptic transverse block. Let $Q_{\rm ref}$ be the pointwise negative projection of $(U_R^*)^{-1}b_{\rm ref}U_R^{-1}=U_Rb_{\rm ref}U_R^*$. For $|k|\ge R$, $D_R$ commutes with $\Sigma_R$, so that $Q_{\rm ref}=U_R\tilde E^\infty U_R^*=E^\infty$
there, and $Q_{\rm ref}$ restricted to $\{|k|\le R\}$ and rescaled to the unit disk is a bulk extension $E^{\rm ext}_{\rm ref}$ of $E^\infty$. It is homotopic to
$E^{\rm ext}$ relative to the annulus, through the negative projections of $U_RD_{R,s}\Sigma_RD_{R,s}U_R^*$ with $D_{R,s}=(1-s)+sD_R$, $s\in[0,1]$, since at
$s=0$ this negative projection is $U_R\tilde E_RU_R^*=E^{\rm ext}(\cdot/R)$. The references are uniformly gapped.
\begin{proposition}\label{prop:gap}
For $R$ large, the reduced bulk references $K_B(b_{\rm ref})$ and, under (TI), the fibers of the interface reference have no spectrum in $(-c,c)$ for some $c>0$,
uniformly in $\xi$.
\end{proposition}
The proof is given in Appendix~\ref{s:sec:routine}. It inverts $b_{\rm ref}$ in the graded classes, the inverse $D_R^{-1}\Sigma_RD_R^{-1}$ being
of norm $O(1)$ in the flat components and $O(R^{-m})$ in the transverse ones.

A path $s\mapsto c_s$ of symbols written in a common frame is called \emph{gapped} if it is piecewise polynomial in $s$ and if, for all $s$ and with
uniform constants, $c_s$ is of Douglis--Nirenberg type $\mathcal S^0_+$ with elliptic transverse block, its degree-$0$ Schur complement at large $|k|$ has
spectrum at distance at least $c>0$ from $0$, and its occupied projector at infinity is $E^\infty$. Along a gapped path of bulk symbols,
Lemma~\ref{lem:tau}(iv) applies on each polynomial piece, by Proposition~\ref{prop:gradedFC}. The straight path from the reference symbol $b_{\rm ref}$ to $\tilde a-\alpha(1+\upsilon)$ is in general not gapped: its Schur complement is not the convex combination of the Schur complements, and
the off-diagonal blocks of $\tilde a$ contribute a term of indefinite sign that can close the flat-band gap. We use instead the following two-step path:
step (A) removes the off-diagonal blocks at large momentum without changing the pointwise Schur complement, and step (B) interpolates linearly between two
block-diagonal symbols whose blocks have the same signs at large momentum, which keeps the flat-band gap.
\begin{lemma}\label{lem:path}
Let $\tilde c=\tilde a-\alpha(1+\upsilon)$, let $R$ be large enough for $\tilde c_{\stt\stt}$ to be invertible on $\{|k|\ge R\}$, let
$N:=\tilde c_{\sff\stt}\tilde c_{\stt\stt}^{-1}\tilde c_{\stt\sff}$ there, and let $\chi_R=\chi(|k|/R)$ with $\chi=0$ on $(-\infty,1]$ and $\chi=1$ on $[2,\infty)$.
\\[1mm]
(A) For $s\in[0,1]$, let $c^A_s$ have blocks $\tilde c_{\sff\sff}+((1-s\chi_R)^2-1)N$, $(1-s\chi_R)\tilde c_{\sff\stt}$, $(1-s\chi_R)\tilde c_{\stt\sff}$ and
$\tilde c_{\stt\stt}$.
\\[1mm]
(B) For $\lambda\in[0,1]$, let $c^B_\lambda=(1-\lambda)\,b_{\rm ref}+\lambda\,c^A_1$.
\\[1mm]
Then the path from $b_{\rm ref}$ to $c^A_1$ along (B), followed by the path from $c^A_1$ to $c^A_0=\tilde c$ along (A), is gapped. The same
construction applies to interface symbols, uniformly in $u$, and it is constant in $u$ for $\pm y\ge y_0$, where it reduces to the bulk paths.
\end{lemma}
\begin{proof}
Both families are polynomial in their parameter, of degree $2$ and $1$, and they are written in the common frame $U$.

\emph{Step (A).} Since $\chi_R=0$ on $\{|k|\le R\}$, the symbol $((1-s\chi_R)^2-1)N$ is defined everywhere. On $\{|k|\ge R\}$, the pointwise Schur complement of
$c^A_s$ is
\[
  \tilde c_{\sff\sff}+((1-s\chi_R)^2-1)N-(1-s\chi_R)^2\,\tilde c_{\sff\stt}\tilde c_{\stt\stt}^{-1}\tilde c_{\stt\sff}=\tilde c_{\sff\sff}-N,
\]
which is the Schur complement of $\tilde c$, independently of $s$. Its degree-$0$ part is $\Meff-\alpha$, of spectrum at distance at least $\delta$ from $0$.
Since $\tilde c_{\sff\stt},\tilde c_{\stt\sff}\in S^{m/2}$ and $\tilde c_{\stt\stt}^{-1}\in S^{-m}$ on $\{|k|\ge R\}$, we have $N\in S^0$, and $\chi_R\in S^0$, so that
$c^A_s$ is of type $\mathcal S^0_+$ with seminorms uniform in $s$. The transverse block $\tilde c_{\stt\stt}$ does not depend on $s$ and is elliptic. The
occupied projector at infinity is determined by the degree-$0$ Schur complement and by the transverse principal symbol, which do not depend on $s$, and it is
therefore $E^\infty$.

\emph{Step (B).} On $\{|k|\ge2R\}$, $c^A_1$ is block diagonal, with flat block $\tilde c_{\sff\sff}-N=\Meff-\alpha+S^{-1}$ and transverse block $\tilde
c_{\stt\stt}=\tilde a_m+S^{m-1}$, and $b_{\rm ref}$ is block diagonal with blocks $1-2E^\infty_{\rm fl}$ and $\Lm_R^m(1-2E^\infty_{\rm tr})$. The
matrices $1-2E^\infty_{\rm fl}$ and $\Meff-\alpha$ commute and have the same sign on each eigenspace, and $|\Meff-\alpha|\ge\delta$, so that every eigenvalue
of $(1-\lambda)(1-2E^\infty_{\rm fl})+\lambda(\Meff-\alpha)$ has modulus at least $(1-\lambda)+\lambda\delta\ge\min(1,\delta)$. Likewise, $1-2E^\infty_{\rm tr}$ and
$\tilde a_m$ commute and have the same sign, with $|\tilde a_m|\ge c_0|k|^m$, so that $(1-\lambda)\Lm_R^m(1-2E^\infty_{\rm tr})+\lambda\tilde a_m$ has modulus
at least $\min(1,c_0)|k|^m$. The remainders in $S^{-1}$ and $S^{m-1}$ are absorbed for $|k|\ge R'$, with $R'$ independent of $\lambda$. Hence, for all
$\lambda$, $c^B_\lambda$ is block diagonal at large momentum, of type $\mathcal S^0_+$, with flat block of spectrum at distance at least $\min(1,\delta)/2$ from $0$
and elliptic transverse block, and its negative projection at infinity is $\mathrm{diag}(E^\infty_{\rm fl},E^\infty_{\rm tr})$, that is, $U^*E^\infty U$.

For interface symbols, the same formulas apply pointwise in $u$, with constants uniform in $u$ by the uniform hypotheses. On $\{\pm y\ge y_0\}$, the symbol
$\tilde c$ and the extension do not depend on $u$, so that the paths coincide there with the bulk paths.
\end{proof}
Lemma~\ref{lem:tau}(iv) is applied to the reduced operators $K_B(c_s)$, in a frame for which $\U$ is unitary (Lemma~\ref{lem:globalframe}(b)). In section~\ref{sec:pBIC}, the same paths are used
for interface operators, which must agree with the bulk operators away from the wall. Since the symbols $c_s$ are not polynomial in $k$, this agreement holds
up to pseudolocal errors, which are smoothing and decay away from the wall (Lemma~\ref{s:lem:pseudoloc}, a standard property of the magnetic calculus).

For $B\ne0$ and a bulk extension $E^{\rm ext}$, the \emph{relative level count} and its signed version are
\begin{equation}\label{eq:Ihat}
  \mathcal N_{E^{\rm ext}}(\ab,B;\alpha):=-\sfl\bigl(s\mapsto K_B(c_s)\bigr),\qquad \mathcal I_{E^{\rm ext}}:=\sgn(B)\,\mathcal N_{E^{\rm ext}},
\end{equation}
computed along the gapped path $c_s$ of Lemma~\ref{lem:path} from the reference symbol of an extension in the class of $E^{\rm ext}$ relative to the annulus
to $\tilde a-\alpha(1+\upsilon)$. For homogeneous frame symbols, the reference of $E^{\rm ext}$ itself may be used, since $E^{\rm ext}_{\rm ref}$ is homotopic to $E^{\rm ext}$.
In general, replacing $E^{\rm ext}$ by its connected sum with a map $S^2\to\Gr_q$ of degree $k$ changes the class of $E^{\rm ext}_{\rm ref}$ by $k$, so that every class is
realized.

\smallskip
\noindent\textbf{Weak fields.} The following proposition shows that weak fields create no level in field-free gaps.
\begin{proposition}\label{prop:weak}
Let $\ab$ satisfy (W1)--(W4), and let $K\subset\mathcal G(\ab)$, the field-free gaps of \eqref{eq:I}, be compact. There is $b_0\in(0,b_1]$, with $b_1$ and $N_0$ as in (W), such that $K\cap\spec K_B(\ab)=\emptyset$ for $0<|B|\le b_0$,
and $\alpha\notin\spec\OpA(\ab^{(B)})$ for every field with $\norm B_{C^{N_0}}\le b_0$ and every $\alpha\in K$, with $\ab^{(B)}=\ab$ except in the
factorized case.
\end{proposition}
The proof, given in Appendix~\ref{s:sec:routine}, inverts $H-\alpha$ perturbatively around the field-free pointwise inverse $(\tilde a^{(0)}-\alpha)^{-1}$, the error being $O(\norm B_{C^{N_0}})$ in
the weighted norms of section~\ref{sec:engine} by (W). It applies verbatim to the reduced operators $K_B(\ab)$, by Lemma~\ref{lem:reduction}(b).

\begin{proof}[Proof of Theorem~\ref{thm:bulk}]
By Lemma~\ref{lem:kubo}, $J=\sgn(B)\tau(\hat P_\alpha)$, the trace-class property being given by Proposition~\ref{prop:gradedFC} and Lemma~\ref{lem:tau}(i).

(a) For $\alpha<\alpha'$ in the same component of $\R\setminus\Fl_B(\ab)$, the spectrum of $K_B(\ab)$ in $[\alpha,\alpha']$ is discrete by
Theorem~\ref{thm:bulkspec}, and $\hat P_{\alpha'}-\hat P_\alpha$ is the finite-rank projection onto the eigenvalues in $(\alpha,\alpha')$, whose eigenvectors are
in the Schwartz class by the parametrix of Theorem~\ref{thm:bulkspec}. Lemma~\ref{lem:tau}(ii) gives $\tau(\hat P_{\alpha'})-\tau(\hat
P_\alpha)=\operatorname{rank}(\hat P_{\alpha'}-\hat P_\alpha)$, and Lemma~\ref{lem:kubo} concludes. If the transverse principal symbol $\tilde a_m$ is positive definite and
$\alpha<\inf\Fl_B(\ab)$, then $E^\infty_{\rm fl}=0$ and $E^\infty_{\rm tr}=0$. Thus $E^\infty=0$,
$\hat P_\alpha$ is a compact projection, hence of finite rank with range in the Schwartz class, and $\tau(\hat P_\alpha)=\operatorname{rank}\hat
P_\alpha=\#(\spec K_B(\ab)\cap(-\infty,\alpha))$ by Lemma~\ref{lem:kubo}.

(b) The levels created along a deformation to a reference give $\mathcal I_{E^{\rm ext}}$, and the semiclassical limit for the reference gives
$-\kappa(E^{\rm ext})$. We use a frame for which $\U$ is unitary (Lemma~\ref{lem:globalframe}(b)), so that $\mathcal V:=K_B(U)$ is unitary and $K_B(\ab)-\alpha=\mathcal V(K_B(\tilde
a-\alpha))\mathcal V^*$. Let $E^{\rm ext}$ be a bulk extension, let $c_s$ be the gapped path of Lemma~\ref{lem:path} from the reference symbol $b_{\rm ref}$ of an extension in the class of $E^{\rm ext}$ to $\tilde c=\tilde a-\alpha$, and let $Q_0=\mathbf 1_{(-\infty,0)}(K_B(b_{\rm ref}))$ and
$Q_1=\mathbf 1_{(-\infty,0)}(K_B(\tilde c))$. Then $\hat P_\alpha=\mathcal VQ_1\mathcal V^*$, and the reduced reference projection is $\hat P_{\rm ref}:=\mathcal VQ_0\mathcal V^*$. All four
projections belong to classes $\cP_\varepsilon$ by Proposition~\ref{prop:gradedFC}, with $E^\infty_\ell$ for $\hat P_\alpha$ and $\hat P_{\rm ref}$. By
Lemma~\ref{lem:tau}(ii), the unitary invariance of the index of a pair, and Lemma~\ref{lem:tau}(iv) applied on each polynomial piece of the path,
\[
  \tau(\hat P_\alpha)-\tau(\hat P_{\rm ref})=\ind(\hat P_\alpha,\hat P_{\rm ref})=\ind(Q_1,Q_0)=-\sfl\bigl(s\mapsto K_B(c_s)\bigr)=\mathcal N_{E^{\rm ext}}.
\]
In particular, $\mathcal N_{E^{\rm ext}}$ does not depend on the gapped path, and $J=\sgn(B)\tau(\hat P_{\rm ref})+\mathcal I_{E^{\rm ext}}$ by Lemma~\ref{lem:kubo}. It remains to
show that $\sgn(B)\tau(\hat P_{\rm ref})=-\kappa(E^{\rm ext})$. The value $\tau(\hat P_{\rm ref})$ does not depend on $R$ large, by Lemma~\ref{lem:tau}(iii), since
the references form a gapped family (Proposition~\ref{prop:gap}) that is $C^1$ in $R$ by Proposition~\ref{prop:gradedFC}. By
Lemma~\ref{lem:reduction}(d), $K_B(c)$ is unitarily equivalent to $\Op^w_{|B|}(c\circ\varsigma_B)$, and after the dilation $Y\mapsto Y/R$ to
$\Op^w_\hbar(c(R\,\varsigma_B\cdot))$ with $\hbar=|B|/R^2$. The unitary equivalence and the dilation are implemented by symplectic linear maps of the phase plane, which preserve $\tau$; the change of orientation
for $B<0$ enters only through $\varsigma_B$ in the curvature integral below. The rescaled reference symbol is of graded type with normalized bounds uniform in $R$,
and pointwise invertible, so that Proposition~\ref{prop:semiclassical} gives $\tau(\hat P_{\rm ref})=-\sgn(B)\int c_1(Q_R)+O(R^{-2})$, where
$Q_R(Y)=Q_{\rm ref}(RY)$. On $\{|Y|\le1\}$, $Q_R$ is the extension $E^{\rm ext}_{\rm ref}$, and on $\{|Y|\ge1\}$, $Q_R=E^\infty\circ\theta$, whose Chern form
vanishes since it depends only on the angle. Hence $\sgn(B)\tau(\hat P_{\rm ref})=-\kappa(E^{\rm ext}_{\rm ref})=-\kappa(E^{\rm ext})$, the last equality since $E^{\rm ext}_{\rm ref}$ is in the class of $E^{\rm ext}$ and $\kappa$ depends only on the class relative to the annulus. Therefore $J=\mathcal I_{E^{\rm ext}}-\kappa(E^{\rm ext})$, and
$J\equiv-\kappa(E^{\rm ext})\pmod\Z$.

(c) Let $\alpha\in\mathcal G(\ab)$, let $b_0$ be the constant of Proposition~\ref{prop:weak} with $K=\{\alpha\}$, and let $B_h=h\sgn(B)$ for $0<h\le b_0$. By
Proposition~\ref{prop:weak}, $\alpha\notin\spec K_{B_h}(\ab)$. By Lemma~\ref{lem:reduction}(d), and in the unitary frames of (W) corrected as in
Lemma~\ref{lem:globalframe}(b), $K_{B_h}(\ab)-\alpha$ is unitarily equivalent to $\Op^w_h(c_h)$, where $c_h=(\tilde a^{(B_h)}-\alpha)\circ\varsigma_B$ is the
conjugated symbol. By (W) and the proof of Proposition~\ref{prop:weak}, $c_h$ is of graded type uniformly in $h$, pointwise invertible with uniform
bounds, and continuous in $h$ in the graded classes, by Lemma~\ref{s:lem:Bcont}. Let $\hat\U_h=\Op^w_h(\hat u_h)$, $\hat u_h=U\circ\varsigma_B$, be the corresponding unitary frame and $\Pi_h=\hat\U_h\mathbf 1_{(-\infty,0)}(\Op^w_h c_h)\hat\U_h^*$, which is
unitarily equivalent to $\hat P_\alpha$ at the field $B_h$. Its symbol class
$\cP_\varepsilon(E)$ has $E$ independent of $h$: after the dilated normalization of Lemma~\ref{lem:reduction}(d), $E$ is determined by the homogeneous part of
$U^{(B_h)}$, which is that of $U^{(0)}$ by (W1), by the transverse principal symbol, which does not depend on the field, and by
$\mathbf 1_{(-\infty,\alpha)}(\Meff^{(B_h)})$, which does not depend on $h$ by (W5). In the dilated form of Lemma~\ref{lem:reduction}(d), the projections $D_h\Pi_hD_h^*$, with $D_h$ the dilation, lie in a class $\cP_\varepsilon(E)$
independent of $h$ and are norm continuous in $h$, so that $\tau(D_h\Pi_hD_h^*)$ does not depend on $h\in(0,b_0]$ by Lemma~\ref{lem:tau}(v). Since the
dilation is symplectic, $\tau(\Pi_h)=\tau(D_h\Pi_hD_h^*)$ does not depend on $h$ either. As $h\to0$, Proposition~\ref{prop:semiclassical} gives $\tau(\Pi_h)=-\int
c_1(\hat u_hQ_h\hat u_h^*)+O(h)$, with $Q_h$ the pointwise negative projection of $c_h$. By (W1) and (W3), $\tilde
a^{(B_h)}\to U^*\ab U$ in the graded classes, with $U$ unitary-valued, so that $\hat u_hQ_h\hat u_h^*$ converges to $E_\alpha\circ\varsigma_B$ with uniform decay
of the curvature, and $\int c_1(\hat u_hQ_h\hat u_h^*)\to\sgn(B)\int c_1(E_\alpha)$, the sign accounting for the orientation of $\varsigma_B$. Hence
$\tau(\Pi_h)=-\sgn(B)\int c_1(E_\alpha)=\sgn(B)I(\ab;\alpha)$. Since the unitary equivalence is symplectic, $\tau(\hat P_\alpha)=\tau(\Pi_h)$, and
$J(\ab,B_h;\alpha)=I(\ab;\alpha)$ by Lemma~\ref{lem:kubo}.
\end{proof}

\begin{proof}[Proof of Corollary~\ref{cor:bulk}]
(a) By Proposition~\ref{prop:gradedFC}, $\hat P_\alpha(\ab)$ and $\hat P_{\alpha'}(\ab')$ belong to $\cP_\varepsilon(E^\infty_\ell)$ with the same $E^\infty$, and
Lemma~\ref{lem:tau}(ii) gives $\tau(\hat P_\alpha(\ab))-\tau(\hat P_{\alpha'}(\ab'))=\ind(\hat P_\alpha(\ab),\hat P_{\alpha'}(\ab'))$, so that $J(\ab)-J(\ab')=\sgn(B)\ind$ by Lemma~\ref{lem:kubo}, that is, $\sigma_H(\ab)-\sigma_H(\ab')=\frac{\sgn B}{2\pi}\ind$. Under the additional assumption of (a), the two reduced projections at the field $B'$ belong to the
same class $\cP_\varepsilon$ for all $B'$ near $B$ after the dilated normalization of Lemma~\ref{lem:reduction}(d), in which the direction at
infinity does not depend on the field. When $r\le1$, this assumption follows from the equality at $B$, since $E^\infty_{\rm fl}\in\{0,1\}$ is locally
constant in the field, while the homogeneous frame and the transverse principal symbol do not depend on the field; when $m\le2$ and for factorized symbols,
it follows from the field independence of $\Meff$ (Proposition~\ref{prop:Meffcomp}(c),(d)). For $B'$ near $B$, the operators $K_{B'}(\ab)$ and $K_{B'}(\ab')$ are unitarily equivalent to
operators that, in the dilated form of Lemma~\ref{lem:reduction}(d), lie in fixed classes and depend continuously on $B'$ in norm-resolvent sense, so that $\alpha$ and $\alpha'$ remain outside their spectra, the two pairings are constant by Lemma~\ref{lem:tau}(v), and so is the index, which is
their difference. Hence $\sigma_H(\ab)-\sigma_H(\ab')=\partial_Bn_{\rm rel}$ with $n_{\rm rel}=\frac{|B|}{2\pi}\ind$. If the transverse principal symbol $\tilde a_m$ is positive definite and $\alpha<\inf\Fl_B(\ab)$, Theorem~\ref{thm:bulk}(a) gives
$\sigma_H=\frac{\sgn B}{2\pi}\#(\spec K_B(\ab)\cap(-\infty,\alpha))$, and since this number is locally constant in $B$, $\sigma_H=\partial_Bn(\alpha,B)$.

(b) Let $t_*$ be a value at which a level of $K_{t_*B}(\ab)$ equals $\alpha$, and choose $\alpha'$ close to $\alpha$, in the same component of $\R\setminus\Fl_{tB}(\ab)$
for $t$ near $t_*$, such that no level equals $\alpha'$ for $t$ near $t_*$. Since $E^\infty$ at $\alpha$, hence at $\alpha'$, does not depend on $t$, the
projections $\hat P_{\alpha'}(tB)$, after the dilated normalization of Lemma~\ref{lem:reduction}(d), belong to a class $\cP_\varepsilon(E)$ independent of $t$ and are norm continuous in $t$, so that
$\tau(\hat P_{\alpha'}(tB))$ is constant for $t$ near $t_*$ by Lemma~\ref{lem:tau}(v), and by Theorem~\ref{thm:bulk}(a), $J(\ab,tB;\alpha)$ differs from $J(\ab,tB;\alpha')$ by $\sgn(B)$ times the signed number of levels between
$\alpha$ and $\alpha'$. The same argument shows that $J(\ab,tB;\alpha)$ is constant on intervals of $t$ without crossings. Hence $J$ changes only at crossings,
by $-\sgn(B)$ per upward crossing. The formula follows by combining Theorem~\ref{thm:bulk}(c) at $t_0$, where $\alpha_0\notin\spec K_{t_0B}(\ab)$ by Proposition~\ref{prop:weak}, this statement on $[t_0,1]$ at the
energy $\alpha_0$, for which $E^\infty$ is that at $\alpha$ since both lie in the same component of $\R\setminus\Fl_{tB}(\ab)$, and Theorem~\ref{thm:bulk}(a) between $\alpha_0$ and $\alpha$ at
$t=1$, which uses $\alpha_0\notin\spec K_B(\ab)$.
\end{proof}
\section{The interface current and the bulk--interface correspondence}\label{sec:locality}\label{sec:pBIC}
This section proves Theorem~\ref{thm:BIC} and its corollaries. The current depends only on the operator near the switch and near each wall
(Proposition~\ref{prop:loc} and Theorem~\ref{thm:local}), which allows walls to be modified and, under (F), frozen to translation-invariant ones. For a
translation-invariant wall between magnetized bulks, the current is the spectral flow of the fibers $\Hh(\xi)$ at $\alpha$ (Theorem~\ref{thm:TI}), and at
large $|\xi|$ the fibers decouple into translates of the bulk operators (Lemma~\ref{lem:decouple}). Deforming $H-\alpha$ to an interface reference built on
an extension of $E^\infty$ to the solid cylinder $[-y_0,y_0]\times\mathbb D$ (Lemma~\ref{lem:degree}), and following the spectral flow around a rectangle in
$(\xi,s)$, the current becomes the difference of the numbers of levels created in the two bulks (Theorem~\ref{thm:relcount}). Theorem~\ref{thm:bulk}(b)
identifies these numbers with $J_\pm$ up to the Chern integrals of the bulk extensions, whose difference is $\Cy$. Field-free bulks are reached by switching on
a weak field on a wall that carries no current. The localization, locality and fiber properties adapt standard arguments
\cite{ASS,EG,EGS,BEC,QB21,QB22}, and their proofs, as well as those of Theorems~\ref{thm:quant}, \ref{thm:TI} and \ref{thm:slab}, are given in
Appendix~\ref{s:sec:routine2}.

\begin{proposition}\label{prop:loc}
Assume (N$'$), and let $\Phi\in C_c^\infty(\mathcal E_\alpha)$, with the energy window $\mathcal E_\alpha$ of section~\ref{sec:hypotheses}. There are constants $c>0$ and $C_N$ such that the following holds.
(a) The operator $\Lm^Ne^{c\jp y}\Phi(H)e^{c\jp y}\Lm^N$ is bounded for every $N$. (b) For $\chi_1,\chi_2\in C_b^\infty(\R^2)$ with supports at distance $L$,
$\norm{\Lm^Ne^{c\jp y}\chi_1\Phi(H)\chi_2e^{c\jp y}\Lm^N}\le C_NL^{-N}$. (c) The kernel of $\Phi(H)$ satisfies $|K_\Phi(u,u')|\le C_Ne^{-c(|y|+|y'|)}\jp{u-u'}^{-N}$.
More generally, if $H$ agrees on a region $\mathcal O$ with an operator $H'$ satisfying (N1) and (N3) with $\alpha\notin\spec H'$, then $\Phi(H)$ decays like
$e^{-c\,\dist(\cdot,\R^2\setminus\mathcal O)}$ inside $\mathcal O$, in the sense of (a).
\end{proposition}
The proof combines geometric resolvent identities with the bulk operators, the weighted Combes--Thomas estimate of
Theorem~\ref{thm:analysis}(b) and the diamagnetic inequality $|e^{-t\Pi^2}g|\le e^{t\Delta}|g|$ (Lemma~\ref{s:lem:diamag}).
\begin{theorem}[locality and freezing]\label{thm:local}
Let $H$ and $H'$ satisfy the hypotheses of Theorem~\ref{thm:quant} with common constants.
\\[1mm]
(a) If $H=H'$ on a slab $\{|x-x_0|<L\}$, with equal
potentials and frames there, then $|\sigma_I(H)-\sigma_I(H')|\le C_NL^{-N}$, so that $\sigma_I(H)=\sigma_I(H')$ for $L$ large.
\\[1mm]
(b) Under (F),
$\sigma_I(H)=\sigma_I(H^{[x_0]})$, where $H^{[x_0]}$ has the coefficients and field of $H$ frozen at $x=x_0$. The operator $H^{[x_0]}$ satisfies (TI), and the cylinder term $\Cy(x_0)$ of Theorem~\ref{thm:BIC} is the same for both operators and does not depend on $x_0$.
\\[1mm]
(c) Assume that $H$ consists of two walls separated by a region of width $L$ in $y$ on which it agrees
with a gapped operator, and let $H^{(1)}$ and $H^{(2)}$ be obtained by replacing that operator beyond the midline on one side or the other. Then
$\sigma_I(H)=\sigma_I(H^{(1)})+\sigma_I(H^{(2)})$ for $L$ large.
\\[1mm]
(d) Under (W1)--(W4) and for $\alpha\in\mathcal G(\ab)$, a wall across which only the field changes, from a
constant $\varepsilon$ to $0$, with the same symbol $\ab$ on both sides and a field $B$ with $\norm B_{C^{N_0}}\le b_0$, has $\sigma_I=0$, where $b_0$ is the constant of
Proposition~\ref{prop:weak} with $K=\{\alpha\}$.
\end{theorem}
The proof is given in Appendix~\ref{s:sec:routine2}. Parts (a) and (b) reduce general walls to translation-invariant ones, while (c) and (d) are the main tools to analyze magnetic transitions.

For translation-invariant walls with $B_\pm\ne0$, we use the gauge $A=(A_1(y),0)$, in which $A_1(y)=-B_\pm y+c_\pm$ on $\{\pm y\ge y_0\}$, and a frame that
does not depend on $x$. The fibers of $\U$ and of $\U^*(H-z)\U$ are $\hat\U(\xi)=\Op_y^w(U(y,\xi-A_1(y),\zeta))$ and
$\tilde A_z(\xi):=\hat\U(\xi)^*(\Hh(\xi)-z)\hat\U(\xi)=\Op_y^w(\tilde a_z(y,\xi-A_1(y),\zeta))$, and Proposition~\ref{prop:paramalpha} and Theorem~\ref{thm:analysis} hold for them
uniformly in $\xi$ (Appendix~\ref{s:sec:tools}). On the side $\pm$, the kinetic momentum is $\xi-A_1(y)=B_\pm(y-y_c^\pm(\xi))$, with the \emph{guiding center}
$y_c^\pm(\xi):=(c_\pm-\xi)/B_\pm$, so that the fiber is a translate of the reduced bulk operator $K_{B_\pm}(a_\pm)$ near $y_c^\pm(\xi)$ when $y_c^\pm(\xi)$ lies in the
bulk $\pm$, which happens for $|\xi|$ large and one sign of $\xi$.
\begin{lemma}\label{lem:decouple}
Let $K\subset\C$ be compact and disjoint from $\Fl\cup\spec K_{B_+}(a_+)\cup\spec K_{B_-}(a_-)$, and let $P(\xi)$ and $P_\pm$ be the Riesz projections of $\Hh(\xi)$
and $K_{B_\pm}(a_\pm)$ for a contour in $K$. For $|\xi|$ large, $K$ lies in the resolvent set of $\Hh(\xi)$, and
\[
  \Bigl\|P(\xi)-\sum_\pm e^{-iy_c^\pm(\xi)D_y}P_\pm e^{iy_c^\pm(\xi)D_y}\Bigr\|\to0\qquad\text{as }\xi\to+\infty\text{ and as }\xi\to-\infty,
\]
where the sum runs over the guiding centers that lie in their bulk.
\end{lemma}
The proof is given in Appendix~\ref{s:sec:routine2}, together with that of Theorem~\ref{thm:TI}.

\smallskip
\noindent\textbf{Proof of the bulk--interface correspondence.} We first prove Theorem~\ref{thm:BIC} when both bulk fields are nonzero, and assume
$B_\pm\ne0$ until the end of that proof.
\begin{lemma}\label{lem:degree}
Let $E^{\rm ext}_\pm$ be bulk extensions of $E^\infty$ for the bulks $\pm$, as defined before \eqref{eq:kappa}. The maps $E^{\rm ext}_-$ on $\{-y_0\}\times\mathbb D$, $E^\infty(x_0,\cdot,\cdot)$ on $[-y_0,y_0]\times\partial\mathbb D$ and $E^{\rm ext}_+$ on
$\{y_0\}\times\mathbb D$ define a map $F:\partial\mathcal C\to\Gr_q$, where $\mathcal C=[-y_0,y_0]\times\mathbb D$, and an \emph{interface extension}, a smooth extension of $F$ to $\mathcal C$, exists if and only if
$d:=\int_{\partial\mathcal C}F^*c_1=0$. The \emph{interface reference} built on it is defined as the reference symbol of section~\ref{sec:pbulk}, with
$E^{\rm ext}$ replaced by the interface extension at each $y$. With $\mathcal C$ oriented by $dy\wedge dw_1\wedge dw_2$ and $\partial\mathcal C$ oriented as its boundary, $d=\kappa(E^{\rm ext}_+)-\kappa(E^{\rm ext}_-)-\Cy(x_0)$. For $0<q<n$, replacing $E^{\rm ext}_+$ by its
connected sum with a map $S^2\to\Gr_q$ of degree $k$ changes $d$ by $k$. For $q\in\{0,n\}$, $d=0$.
\end{lemma}
\begin{proof}
For $0<q<n$, since $\Gr_q(\C^n)$ is simply connected with $\pi_2\cong\Z$ detected by $c_1$ (section~\ref{sec:magbulk}), a map $\partial\mathcal C\cong S^2\to\Gr_q$
extends to the ball if and only if its degree vanishes. The face $\{y_0\}\times\mathbb D$ carries the orientation $+dw_1\wedge dw_2$, the face $\{-y_0\}\times\mathbb D$ carries
$-dw_1\wedge dw_2$, and the lateral face carries $d\theta\wedge dy=-dy\wedge d\theta$, which gives the formula for $d$.
\end{proof}
With compatible extensions, the current is a difference of relative level counts.
\begin{theorem}\label{thm:relcount}
Assume (N), (TI) and $B_\pm\ne0$, and let $E^{\rm ext}_\pm$ be bulk extensions with $d=0$. Then $2\pi\sigma_I=\mathcal I_{E^{\rm ext}_+}(a_+,B_+;\alpha)-\mathcal
I_{E^{\rm ext}_-}(a_-,B_-;\alpha)$.
\end{theorem}
\begin{proof}
We follow the spectral flow around the rectangle $[-Z,Z]\times[0,1]$ in $(\xi,s)$: the side $s=0$ carries the spectral flow $\SF(\alpha)=2\pi\sigma_I$ of Theorem~\ref{thm:TI}, the side $s=1$, where
the fibers are those of an interface reference, carries nothing, and the sides $\xi=\pm Z$ carry the level counts $\mathcal N_\pm$ of the bulks selected at
each end.

\emph{The path.} We choose an interface extension
$E^{\rm ext}$, which exists by Lemma~\ref{lem:degree}, and a frame whose fibers are unitary (Lemma~\ref{lem:globalframe}(b)). Let $G(\xi,s)$, $(\xi,s)\in\R\times[0,1]$, be the fibers of the
gapped interface path of Lemma~\ref{lem:path}, run from $\tilde a-\alpha$ at $s=0$ to the interface reference at $s=1$. By Lemma~\ref{s:lem:pdofibers}, Theorem~\ref{thm:TI} and Lemma~\ref{lem:decouple} apply to these fibers uniformly in $s$, and the family is
norm-resolvent continuous in $(\xi,s)$. It is invertible at $s=1$ for all $\xi$ by Proposition~\ref{prop:gap}, and at $s=0$ its spectral flow
in $\xi$ is $\SF(\alpha)=2\pi\sigma_I$ by Theorem~\ref{thm:TI}.

\emph{The ends $\xi=\pm Z$.} Along the path the flat-band gap at $0$ holds uniformly. The bulk spectra near $0$ are discrete and depend continuously on $s$, so that there are a partition
$0=s_0<\dots<s_N=1$ and numbers $\delta_j\in(0,c/2)$, with $c$ the constant of the gapped path, and $\eta_j\in(0,\delta_j)$ such that $\pm\delta_j$ and $-\eta_j$ do not belong to the spectra of the bulk
paths for $s\in[s_{j-1},s_j]$, and such that $[-\eta_j,0)$ contains no point of these spectra at $s=s_{j-1}$ and $s=s_j$. By Lemma~\ref{lem:decouple}, in the
uniform form of Lemma~\ref{s:lem:pdofibers} applied to small circles around $\pm\delta_j$ and $-\eta_j$ and to contours enclosing $[0,\delta_j)$ and
$[-\eta_j,0)$, there is $Z\ge Z_\alpha$ such that, at $\xi=\pm Z$, the points $\pm\delta_j$ are not in the spectrum of $G(\xi,s)$ for $s\in[s_{j-1},s_j]$, and the
spectral projections $\mathbf 1_{[0,\delta_j)}(G(\pm Z,s))$, $s\in\{s_{j-1},s_j\}$, have the ranks of the corresponding projections of the direct sum of the
bulk paths selected at that end. Since $G(\xi,0)$ has no spectrum at $0$ for $|\xi|\ge Z_\alpha$, the spectral flow of $s=0$ on $[-Z,Z]$ is still
$\SF(\alpha)$. By Lemma~\ref{s:lem:sfcompare}, the spectral flows $f_{\rm R}$ and $f_{\rm L}$ along $s\in[0,1]$ at $\xi=+Z$ and $\xi=-Z$ therefore equal those of
the bulk paths selected at that end. These paths run from $a_\pm-\alpha$ to the references, the reverse of the paths defining $\mathcal N_{E^{\rm ext}_\pm}$ in \eqref{eq:Ihat}, so that each selected bulk $\pm$ contributes $\mathcal N_{E^{\rm ext}_\pm}$.

\emph{The count.} The total spectral flow around the boundary
of the rectangle vanishes, so that $\SF(\alpha)=f_{\rm L}-f_{\rm R}$. For $B_\pm>0$, $\xi=+Z$ selects $a_-$ and $\xi=-Z$ selects $a_+$, which gives
$\mathcal N_+-\mathcal N_-$. For $B_\pm<0$, the ends are exchanged, which gives $\mathcal N_--\mathcal N_+$. For $B_+<0<B_-$, $\xi=+Z$
carries both bulks and $\xi=-Z$ none, which gives $-\mathcal N_+-\mathcal N_-$, and for $B_-<0<B_+$ we obtain $\mathcal N_++\mathcal N_-$. In each case,
this is $\mathcal I_{E^{\rm ext}_+}-\mathcal I_{E^{\rm ext}_-}$.
\end{proof}

\begin{proof}[Proof of Theorem~\ref{thm:BIC} for $B_\pm\ne0$]
By Theorem~\ref{thm:local}(b), we may assume (TI) when (F) holds. We fix $E^{\rm ext}_-$ and, by Lemma~\ref{lem:degree}, choose $E^{\rm ext}_+$ with $d=0$, so that $\kappa(E^{\rm ext}_+)-\kappa(E^{\rm ext}_-)=\Cy$. Theorem~\ref{thm:relcount} and Theorem~\ref{thm:bulk}(b) then give
\[
  2\pi\sigma_I=\mathcal I_{E^{\rm ext}_+}-\mathcal I_{E^{\rm ext}_-}=(J_++\kappa(E^{\rm ext}_+))-(J_-+\kappa(E^{\rm ext}_-))=J_+-J_-+\Cy .
\]
The term $\Cy$ involves only $E^\infty$, which depends on the field only through $\Meff$. The integer form follows from Stokes' theorem on the
cylinder, since $\tr(E\,dE\wedge dE)=d\tr(\psi^*d\psi)$ for an orthonormal frame $\psi$ of $\Ran E$, which gives $\Cy=(\Theta(y_0)-\Theta(-y_0))/2\pi$ with the Berry phase $\Theta$ defined before Theorem~\ref{thm:BIC}, and from $J_\pm\equiv-\kappa(E^{\rm ext}_\pm)\equiv-\Theta(\pm y_0)/2\pi\pmod\Z$. For first-order operators
with constant $\gamma$, $E^\infty_{\rm tr}$ does not depend on $u$, so that only $E^\infty_{\rm fl}$ contributes.
\end{proof}

We now allow $B_+=0$ or $B_-=0$. Quantization still holds under (N$'$), since the field-free bulk $\Op^{A_\pm}(a_\pm)$ has spectrum $\spec H_0(a_\pm)$,
which avoids $\alpha$ by (N4). However, the fibers no longer have compact resolvent on a field-free side, and the spectral flow argument above is not
available. We reduce instead to magnetized bulks, by inserting beyond the wall a region of weak constant field $\varepsilon$, which changes neither the current
nor, by Theorem~\ref{thm:bulk}(c), the bulk invariant. Two facts are used: changes of the field on a strip do not
modify the current, by Theorem~\ref{thm:slab}(a) with $v=0$, and a wall across which a weak field is switched on carries no current, by
Theorem~\ref{thm:local}(d).

\begin{proof}[Proof of Theorem~\ref{thm:BIC} when $B_+=0$]
Let $L$ be large, to be chosen, let $\beta_L(y)$ be a smooth function equal to $1$ on $[y_0+1,y_0+L]$ and to $0$ outside
$[y_0,y_0+L+1]$, with derivatives bounded independently of $L$, and let $\varepsilon>0$ with $\varepsilon\norm{\beta_L}_{C^{N_0}}\le b_0$, also to be chosen small. We introduce three auxiliary operators. The operator $H^\natural$ has the coefficients of $H$ and the field $B+\varepsilon\beta_L$. The
operator $H^{(1)}$ has the coefficients of $H$ and a field equal to $B$ for $y\le y_0$ and to $\varepsilon$ for $y\ge y_0+1$, interpolated as
$B+\varepsilon\beta_L$ in between. Finally, $H^{(2)}=\OpA(a_+)$ has the field $\varepsilon$ for $y\le y_0+L$ and $\varepsilon\beta_L$ for $y\ge y_0+L$.

We first apply Theorem~\ref{thm:slab}(a) with $v=0$ and the bounded potential $\beta=(-\varepsilon\int_{y_0}^y\beta_L,0)$ of the field $\varepsilon\beta_L$, which vanishes for $y\le y_0$ and equals a constant $\beta_\infty$ for $y\ge y_1:=y_0+L+1$. This gives
$\sigma_I(H)=\sigma_I(H^\natural)$. The perturbation is structurally admissible by (W4), since the added field is small and supported in the region where the symbol is
$a_+$, and it is gap preserving with $H_{-,s}=H_-$ and $H_{+,s}=\Op^{A_++s\beta_\infty}(a_+)$, which is gauge equivalent to $H_+$. Next, $H^\natural=H^{(1)}$ on $\{y<y_0+L\}$ and $H^\natural=H^{(2)}$ on $\{y>y_0+1\}$, and on $\{y_0+1<y<y_0+L\}$ all three operators agree with
$H_{\varepsilon}(a_+)=\OpA(a_+)$ at the constant field $\varepsilon$, for which $\alpha\notin\spec K_\varepsilon(a_+)$ by Proposition~\ref{prop:weak}.
Theorem~\ref{thm:local}(c) therefore gives $\sigma_I(H^\natural)=\sigma_I(H^{(1)})+\sigma_I(H^{(2)})$ for $L$ large. Moreover, $\sigma_I(H^{(2)})=0$ by
Theorem~\ref{thm:local}(d). It remains to compute $\sigma_I(H^{(1)})$. The operator $H^{(1)}$ has bulks $(a_+,\varepsilon)$ and $(a_-,B_-)$, and it satisfies
(N), as well as (TI) or (F) if $H$ does, since the added region is translation invariant. If $B_-\ne0$, the case $B_\pm\ne0$ of Theorem~\ref{thm:BIC}, proved above,
gives $2\pi\sigma_I(H^{(1)})=J(a_+,\varepsilon;\alpha)-J(a_-,B_-;\alpha)+\Cy+\Cy'$, where $\Cy'$ is the integral of $c_1(E^\infty)$ over the added cylinder $[y_0,y_0+1]\times S^1$, on which the symbol is $a_+$ and the field
$\varepsilon\beta_L$ has $C^{N_0}$ norm at most $b_0\le b_1$. By (W5), (W1) and Proposition~\ref{prop:flat}(a), $E^\infty$ there is the field-free projection
$E^\infty(a_+)(\theta)$, which does not depend on $y$, so that $\Cy'=0$. By Theorem~\ref{thm:bulk}(c), $J(a_+,\varepsilon;\alpha)=I(a_+;\alpha)=J(a_+,0;\alpha)$,
which concludes the proof in this case.

The case $B_-=0\ne B_+$ is symmetric. If $B_+=B_-=0$, we perform the construction on both sides, with three walls separated by regions of field $\varepsilon$.
Theorem~\ref{thm:local}(c) applies twice, the two field walls carry no current, the added cylinders contribute nothing as above, and the middle wall gives $J(a_+,\varepsilon)-J(a_-,\varepsilon)+\Cy=I(a_+)-I(a_-)+\Cy$.
\end{proof}

We now complete the proofs of the corollaries of section~\ref{sec:results}.

\begin{proof}[Proof of Corollary~\ref{cor:trans}]
Part (a) is Theorem~\ref{thm:BIC} with the convention \eqref{eq:J0}. For (b), if the symbol is $\ab$ everywhere, then $E^\infty$ depends on $u$ only through $E^\infty_{\rm fl}$, hence only on $\theta$, and
$\Cy=0$, when $\Meff$ does not depend on the field ($m\le2$ and factorized operators) or when $r\le1$, since $E^\infty_{\rm fl}$ has constant rank. Theorem~\ref{thm:BIC} gives $2\pi\sigma_I=J(\ab,B_+;\alpha)-I(\ab;\alpha)+\Cy$, and Corollary~\ref{cor:bulk}(b) expresses it through the crossings along $t\mapsto
tB_+$. For (c), Theorem~\ref{thm:BIC} gives $2\pi\sigma_I(H_t)=J(a_+,tB_+;\alpha)-J(a_-,tB_-;\alpha)+\Cy_t$ for $|t|\le t_1$, where the cylinder term $\Cy_t$ at the field $tB$ depends continuously on $t$,
since $\Meff$, and hence $E^\infty$, does. Let $b_0$ be the smaller of the constants of Theorem~\ref{thm:bulk}(c) for $a_+$ and $a_-$, and let
$t_0=\min\{t_1,b_0/\max(|B_+|,|B_-|)\}$, with $t_0=t_1$ if $B_+=B_-=0$. For $|t|\le t_0$, we have $J(a_\pm,tB_\pm;\alpha)=I(a_\pm;\alpha)$, by
Theorem~\ref{thm:bulk}(c) if $tB_\pm\ne0$ and by the convention \eqref{eq:J0} if $tB_\pm=0$. The right-hand side is therefore an integer depending continuously on $t$, hence constant, on
$[-t_0,t_0]$, and at $t=0$ it is the value given by (a).
\end{proof}

\begin{proof}[Proof of Corollary~\ref{cor:main}]
Part (a) follows from Theorem~\ref{thm:BIC}, since the cylinder term $\Cy$ is a homotopy invariant of $E^\infty$ with fixed bulks. Along a path satisfying the hypotheses,
$\Meff$ and the frame depend continuously on the parameter and the flat-band gap holds uniformly, so that $E^\infty$ is deformed continuously relative to
$\{y=\pm y_0\}$. For $r\le1$, $E^\infty_{\rm fl}$ has constant rank $0$ or $r$ on the connected set $\R^2\times S^1$, so that $E^\infty$ is determined by the frame and by
$\tilde a_m$, hence by the principal symbol. For (b), the positive definiteness of the
transverse principal symbols and $\alpha<\inf\Fl$ give $E^\infty=0$, hence $\Cy=0$, and the second statement of Theorem~\ref{thm:bulk}(a) gives
$J_\pm=\sgn(B_\pm)N_\pm(\alpha)$ for $B_\pm\ne0$, where $N_\pm<\infty$ by Theorem~\ref{thm:bulkspec}, and Theorem~\ref{thm:BIC} concludes. For (c), Theorem~\ref{thm:bulk}(a)
gives $J_+(\alpha)=J_+(\alpha_0)-\sgn(B_+)\#(\spec K_{B_+}(a_+)\cap(\alpha,\alpha_0))$ for $f<\alpha<\alpha_0<f+\varepsilon$, which diverges as $\alpha\downarrow f$. On the
other hand, $J_-(\alpha)$ stays bounded, since there are finitely many levels or $J_-=I(a_-;\cdot)$ is constant on a field-free gap, and $\Cy$ is constant on the
component. Finally, the statements of the elliptic-case paragraph of section~\ref{sec:interface} follow from ellipticity, for (N1) and (N3), and from
Lemma~\ref{s:lem:Fsufficient} and Proposition~\ref{s:prop:Wholds}, for (F) and (W).
\end{proof}

\section*{Declarations}
\noindent\textbf{Funding.} This work was supported in part by NSF grant DMS-2606421 and ONR grant N00014-26-1-2017.

\smallskip
\noindent\textbf{Competing interests.} The author has no competing interests to declare.

\smallskip
\noindent\textbf{Data and code availability.} No experimental data were generated in this work. The scripts reproducing the numerical tests and figures of
Appendix~\ref{s:sec:examples} are available from the author upon request.

\smallskip
\noindent\textbf{Use of generative AI.} Generative AI (Claude Opus 5.5) was used in the preparation of this work: to carry out the numerical tests of
Appendix~\ref{s:sec:examples} and prepare the figures, and to write the scripts reproducing them; to help draft and revise the text and the proofs; to audit
the manuscript for internal consistency and gaps in its arguments and proofs; and to help with bibliographical data. All such output was checked by the
author, who is alone accountable for the correctness of all statements and proofs.

\appendix
\section{Examples and numerical tests}\label{s:sec:models}\label{s:sec:examples}
Recall that $I(\ab;\alpha)=-\int_{\R^2}c_1(E_\alpha)$ is the Chern integral of the
projector $E_\alpha$ onto the bands of $\ab(k)$ below $\alpha$, that $J=2\pi\sigma_H$ is the bulk Hall conductivity at the field $B$, with $J=I$ at $B=0$, that
the generalized Landau levels $\spec K_B(\ab)$ are the eigenvalues of $K_B(\ab)$, at each of which $J$ jumps by $\sgn(B)$ times the multiplicity, that $\Cy$
is the integral of $c_1(E^\infty)$ over $[-y_0,y_0]\times S^1$, and that $2\pi\sigma_I=J(a_+,B_+;\alpha)-J(a_-,B_-;\alpha)+\Cy$. Unless stated otherwise, the
values of $I$, $J$ and $\Cy$ are obtained from the rules of section~\ref{sec:compute}: curvature integrals for $I$ and $\Cy$, and counts of generalized Landau levels for $J$. The numerical
methods are described below, and the numerical tests are reported with the examples.

\begin{figure}[ht]
\centering
\includegraphics[width=\textwidth]{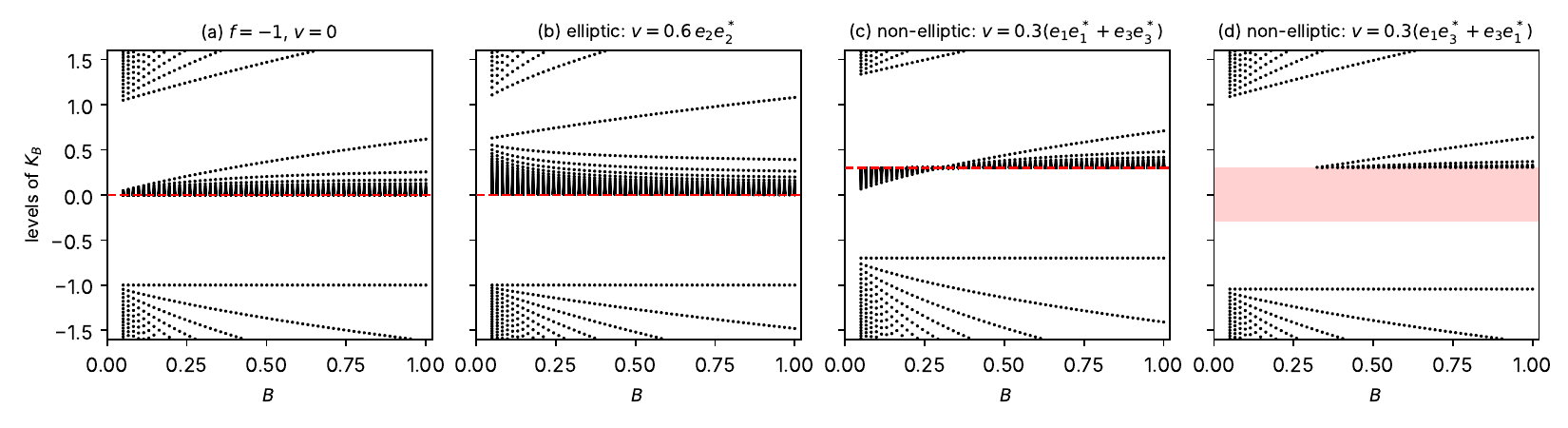}
\caption{Generalized Landau levels of the spin-one bulk symbol $\xi S_x+\zeta S_y+fS_z+v$ with mass $f=-1$, as functions of $B$. (a) $v=0$: a Landau fan
accumulates at $\Fl=\{0\}$ from above. (b) Elliptic perturbation $v=0.6\,e_2e_2^*$, a potential on the hub site, which is the height in shallow-water
variables: the levels move, and they still accumulate at $0$. (c) Non-elliptic perturbation $v=0.3(e_1e_1^*+e_3e_3^*)$: the flat
eigenvector $V_\sff$ has weight $\frac12$ on each of $e_1$ and $e_3$ and none on $e_2$, so that $\Meff=V_\sff^*vV_\sff=0.3$ in every direction, and the fan
accumulates at $\Fl=\{0.3\}$, from below for $B<0.3$ and from above for $B>0.3$, collapsing onto $\Fl$ at $B=0.3$, as the curl fan does at $B=|g|$. (d) Non-elliptic perturbation $v=0.3(e_1e_3^*+e_3e_1^*)$: $\Meff=-0.3\cos(2\theta+\theta_0)$ takes all values in $[-0.3,0.3]$, so
that $\Fl=[-0.3,0.3]$, which is essential spectrum of the magnetized bulk. 
\\
Red dashed lines (a,b,c) mark $\Fl$ when it is a point, and the red shaded band (d) marks it
when it is an interval. Levels whose eigenvectors are not resolved by the truncation are not shown (see the numerical methods above).}
\label{s:fig:spin1}
\end{figure}

\begin{figure}[ht]
\centering
\includegraphics[width=0.8\textwidth]{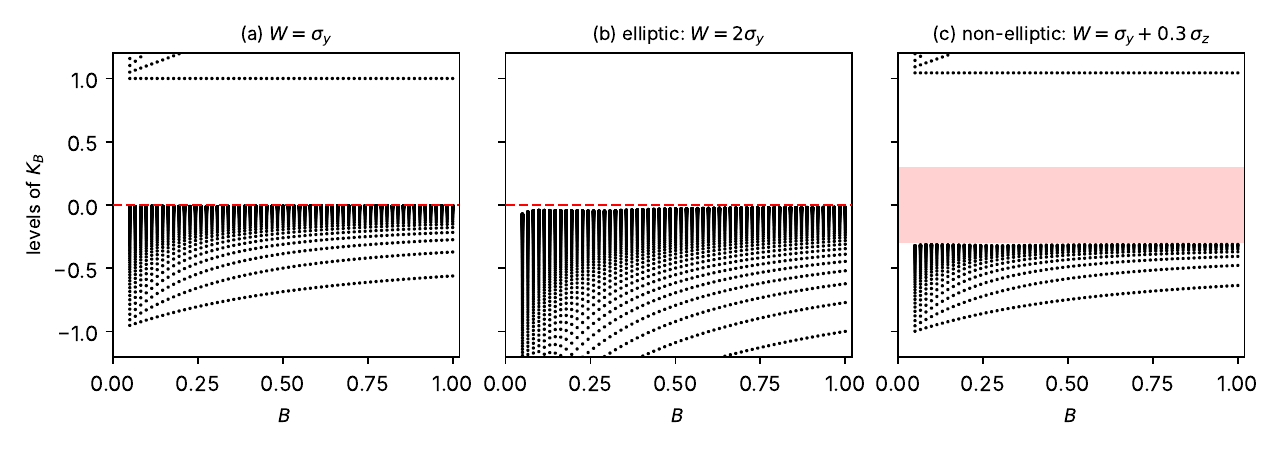}
\caption{Generalized Landau levels of the factorized operator $\curl_B^*\curl_B+W$ as functions of $B$; the levels of the Peierls operator
$\OpA(k^\perp(k^\perp)^T)+W=\curl_B^*\curl_B+W+\frac B2\sigma_y$ are obtained by replacing $W$ by $W+\frac B2\sigma_y$. (a) $W=\sigma_y$: the levels of the
asymptotically flat band accumulate at $\Fl=\{0\}$ from below, and the level $1$ above the gap, carried by $\ker\curl_B$, does not move with the field. (b) Elliptic perturbation $W=2\sigma_y$: $\Fl=\{0\}$ is unchanged. (c) Non-elliptic perturbation $W=\sigma_y+0.3\,\sigma_z$:
$\Meff=0.3\,\theta^T\sigma_z\theta$ takes all values in $[-0.3,0.3]$. Red dashed (a,b) lines mark $\Fl$ when it is a point, and the red shaded band (c) marks it when it is
an interval.}
\label{s:fig:curl}
\end{figure}

\smallskip
\noindent\textbf{Numerical methods.} Field-free fibers were diagonalized by
Fourier-spectral differentiation on a periodic cell with two walls, and the spectral flow was attributed to a wall by the localization of the eigenvectors.
Magnetized fibers were computed in a Hermite basis adapted to the field, with position-dependent coefficients in a discrete-variable representation, and the
crossings of $\alpha$ were counted exactly by matrix inertia ($LDL^*$ factorization). Rotation-covariant bulks were computed from exact Landau blocks. Two
kinds of truncation artifacts must be removed: crossings at large $|\xi|$ that move outward with the basis size, and pseudo-levels pinned at band edges whose
eigenvectors sit at the edge of the basis, such as the apparent level at $-f$ for the dice bulk with $B>0$.
Figures~\ref{s:fig:spin1} and \ref{s:fig:curl} are computed in the Hermite basis of $\mathfrak a=(\Pi_1+i\Pi_2)/(2|B|)^{1/2}$, truncated at
$N=\max(200,\min(700,60/|B|))$ levels per component, with the Weyl-symmetrized operators assembled from $\Pi_1$ and $\Pi_2$, that is, with the Peierls
operators of section~\ref{sec:results}, except for Figure~\ref{s:fig:curl} and the table of Example~\ref{s:ex:curl}, which are computed for
the factorized operator $\curl_B^*\curl_B+W$, assembled as $L^*L$ from the matrix of $L=(-\Pi_2,\Pi_1)$ in the same basis. Products are formed in a basis
enlarged by four levels before truncation. The dice and spin-one operators are of first order, for which the two couplings coincide, and the remaining
examples are Peierls operators unless stated otherwise. Eigenvectors with weight larger than
$10^{-6}$ on the upper fifth of the basis are discarded, which removes the truncation artifacts but also the levels that the truncation does not resolve, in
particular inside non-degenerate flat-band intervals.

\smallskip
\noindent\textbf{Elliptic directions and the field.} For $r=1$, $E_{\ker}(u,\theta)VE_{\ker}(u,\theta)=\tr(E_{\ker}(u,\theta)V)\,E_{\ker}(u,\theta)$. The space $\mathcal V_0(u)$
of elliptic directions is therefore the orthogonal complement, for the inner product $\tr(VV')$ on the $n^2$-dimensional real space of Hermitian matrices, of
the span of the projectors $E_{\ker}(u,\theta)$, $\theta\in S^1$. The number of non-elliptic directions at $u$ is the dimension of this span, and
$V_\sff^*VV_\sff=\tr(E_{\ker}V)$ is the component of $V$ that enters $\Meff$. A perturbation $v$ of degree $0$ with values in $\mathcal V_0$ changes the flat block by
Moyal terms of degree at most $-1$ and the Schur complement at degree at most $-\frac m2$, so that $\Meff$, $\Fl$ and (N3) are unchanged. Such perturbations
are structurally admissible for every $m$, by case (ii) of structural admissibility in section~\ref{sec:stability}, with $\beta=0$ if $m\ge3$.

The space $\mathcal V_0$ depends only on the principal symbol, and hence not on the field. For $m\le2$ and for factorized operators, $\Meff$ does not depend on
the field either, so that the classification and its consequences hold at every field. With field, the flat-band values form the essential spectrum of the
magnetized bulks, at which the generalized Landau levels accumulate (Theorem~\ref{thm:bulkspec}). A perturbation with values in
$\mathcal V_0$ moves the levels but not their accumulation set, while a non-elliptic perturbation moves the accumulation set together with $\Fl$
(Figures~\ref{s:fig:spin1} and \ref{s:fig:curl}). Stability under perturbations in $\mathcal V_0$ thus requires that no level of the perturbed bulks crosses
$\alpha$ along the path. Table~\ref{s:tab:elliptic} summarizes the examples below.
\begin{table}[ht]
\centering\small
\begin{tabular}{lccccl}
\toprule
symbol & $n$ & $r$ & elliptic & non-elliptic & $\Fl$ for constant $W$\\
\midrule
Dirac-type, spin $s\in\frac12+\N$ & $2s+1$ & $0$ & $(2s+1)^2$ & $0$ & $\emptyset$\\
spin one: shallow water, dice & $3$ & $1$ & $6$ & $3$ & $\{0\}$ (Coriolis term, $S_z$ mass)\\
$\alpha$--$\mathcal T_3$ (Example~\ref{s:ex:aT3}) & $3$ & $1$ & $6$ & $3$ & $\{s^2\Delta_A+c^2\Delta_C\}$\\
$k^\perp(k^\perp)^T+W$ (curl) & $2$ & $1$ & $1$ & $3$ & $\{\theta^TW\theta:\theta\in S^1\}$\\
\bottomrule
\end{tabular}
\caption{Real dimensions of the spaces of elliptic and non-elliptic directions among the $n^2$ real dimensions of Hermitian perturbations of degree $0$.}
\label{s:tab:elliptic}
\end{table}

\begin{example}[Dirac-type symbols]\label{s:ex:Dirac}
For the Dirac symbol $\ab=k\cdot\sigma+M\sigma_3$, $I=-\frac12\sgn M$ in the gap, and $J=-\frac12\sgn M$ for every field (Examples~\ref{ex:examples}).
For the $p$-fold Dirac symbol, with $p$ cones of the same chirality and mass $M>0$, $I=J=-\frac p2$ for every field. For the regularized Dirac symbol
$\ab=k\cdot\sigma+(M-\epsilon|k|^2)\sigma_3$ with $M,\epsilon>0$, the principal symbol is $-\epsilon|k|^2\sigma_3$, $E^\infty$ is constant, $I(\ab;0)=-1$ is an integer, and
$J(\ab,B;0)=-1$ for $|B|<M/\epsilon$ and $0$ beyond when $M\epsilon<2+\sqrt3$: a level crosses $0$ at the transition field. Indeed, the rescaling $k=\kappa/\epsilon$
maps this symbol to $\epsilon^{-1}$ times a BHZ-type symbol with $\mu=M\epsilon$ at the field $\epsilon^2B$, and for $M\epsilon>2+\sqrt3$ additional
crossings occur below $M/\epsilon$, as in Examples~\ref{ex:examples}. A boundary of the applied field in this material therefore
carries $2\pi\sigma_I=J-I$, which is $0$ for $|B|<M/\epsilon$ and $+1$ beyond when $M\epsilon<2+\sqrt3$. For the spin-$s$ symbols $k\cdot S+MS_z$ with half-integer $s$, which are elliptic,
$I=-\frac12(s+\frac12)^2\sgn M$. For $s=\frac32$, the span of $|\frac32\rangle\otimes|1\rangle$ and $|\frac12\rangle\otimes|0\rangle$ is invariant for $B>0$, with the block
$\bigl(\begin{smallmatrix}3M/2&\sqrt{3B/2}\\ \sqrt{3B/2}&M/2\end{smallmatrix}\bigr)$ of determinant $\frac34(M^2-2B)$, so that a level crosses $0$ at
$|B|=\frac{M^2}2$, and $J(\ab,B;0)=-2$ for $|B|<\frac{M^2}2$ and $-1$ for $|B|>\frac{M^2}2$; we checked numerically, for both signs of $B$, that no other level
crosses $0$ up to $|B|=8$ for $M=1$. Magnetic transitions thus occur for spin $\frac32$ but not for spin $\frac12$. For a mass wall between $M_\pm$ with $|M_\pm|$ large compared with the fields, $\Cy=0$ and
$2\pi\sigma_I=\frac12(s+\frac12)^2(\sgn M_--\sgn M_+)$. For the Dirac symbol, $E^\infty$ does not depend on $y$, so that the degree of Lemma~\ref{lem:degree}
vanishes with $E^{\rm ext}_-=E^{\rm ext}_+$, and the hemisphere extensions with $\kappa=\pm\frac12$, related by degree $\pm1$, correspond to the
signs of a reference mass.

\emph{Magnetization and regularization.} Integer bulk invariants of the Dirac operator are obtained without field only after a regularization at
infinite momentum \cite{Bal19}, and different regularizations give different integers. Here, $J(\ab,0;\alpha)=I=-\frac12\sgn M$ is defined without
regularization, its fractional part is fixed by $E^\infty$, and the cylinder term restores the integrality of interface currents; with a field, $J$ is in
addition given by the Kubo formula, with weak-field limit $I$. A regularization such as $-\epsilon|k|^2\sigma_3$ changes $E^\infty$ and
shifts $J$ by the Chern integral of an extension, here to $-1$ at weak field; it does not reproduce the magnetized value. The same holds for the coupled Dirac
model (1) of \cite{FB26} for $N_L$-layer rhombohedral graphene, with interlayer coupling $\gamma$ and gate potential $V$,, which is elliptic of order $1$ with $E^\infty$ independent of $V$. At weak field,
$J$ is the curvature integral, which is the normalized invariant of \cite{FB26}, equal to $\sgn(V)\frac{N_L}2$ for small $V$; we checked this numerically for
$N_L=2,3,4$ and $\gamma=1$. At larger fields, levels cross the gap: for $N_L=2$, the Landau block of index $1$ has the eigenvalue $0$ at
$B=\frac12(\gamma^2+\frac{V^2}4)$, beyond which $J$ drops from $1$ to $0$ for $V>0$.

For elliptic bulks, the values of $J(a,B;0)$ are collected in Table~\ref{s:tab:levelsJ}.
\begin{table}[ht]\centering\small
\begin{tabular}{@{}>{\raggedright\arraybackslash}p{0.31\textwidth}>{\raggedright\arraybackslash}p{0.14\textwidth}>{\raggedright\arraybackslash}p{0.14\textwidth}>{\raggedright\arraybackslash}p{0.32\textwidth}@{}}
\toprule
model & $I$ & $R$ & $J(a,B;0)$\\
\midrule
Dirac, $M>0$ & $-\frac12$ & $-\frac12$ & $-\frac12$ for all $B$\\
spin $s=\frac32,\frac52$, $M=1$ & $-\frac12(s+\frac12)^2$ & $-\frac12(s+\frac12)^2$ & $s=\frac32$: $-2$ ($|B|<\frac12$), $-1$ ($|B|>\frac12$)\\
$-\Delta$, any $\alpha$ & $0$ ($\alpha<0$) & $0$ & $\sgn(B)\#\{j:|B|(2j+1)<\alpha\}$\\
regularized Dirac, $M\epsilon<2+\sqrt3$ & $-1$ & $-1$ & $-1$ for $|B|<M/\epsilon$, $0$ beyond\\
$p$-fold Dirac, $M>0$ & $-\frac p2$ & $-\frac p2$ & $-\frac p2$ for all $B$\\
$d$-wave, $m=\mu=c_0=1$, $c=1$ & $-1.5995$ & $-2$ & $I$ at weak field\\
BHZ $(|k|^2-\mu)\sigma_3+k\cdot\sigma$, $\mu<2+\sqrt3$ & $1$ & & $1$ ($|B|<\mu$), $0$ ($|B|>\mu$)\\
twist bulks $\mp(|k|^2+M)\sigma_3$ & $0$ & & $0$; interface value $=\Cy=+2$\\
\bottomrule
\end{tabular}
\caption{Bulk invariants $J(a,B;0)$ of elliptic bulks, with the field-free Chern integral $I$ and the half-plane rotation $R$ of \cite{BEC}.}
\label{s:tab:levelsJ}
\end{table}
The $d$-wave entry differs from $R$ because the pairing is of principal order, and $I=-\sgn(c)(1+v)$ with $v\in(0,1)$ a solid-angle correction. The interface spectral flows agree with $J_+-J_-+\Cy$ in $12$ out of $12$ cases for
$(|k|^2-\mu)\sigma_3+k\cdot\sigma$, in $50$ out of $50$ cases for $(|k|^2+0.3|k|^4-\mu)\sigma_3+(1+0.5|k|^2)k\cdot\sigma$, and in $30$ out of $30$ cases for the
non-covariant symbol $(k_1^2+2k_2^2-\mu)\sigma_3+k\cdot\sigma$. They give $+2$ and $0$ for twisted and untwisted interfaces at $B=\pm1,0$.
\end{example}

\begin{example}[Closed-form spectra]\label{s:ex:Landau}
For $\ab=|k|^2$, each level $|B|(2j+1)$ is simple, and for any positive elliptic symbol the count of section~\ref{sec:compute} for an interface between
fields $B_\pm$ holds with $N_\pm$ the numbers of eigenvalues of $K_{B_\pm}(a_\pm)$ below $\alpha$.
\label{s:ex:levels}
Other spectra are available in closed form. For the Dirac symbol, $\spec K_B(\ab)=\{M\sgn B\}\cup\{\pm\sqrt{2j|B|+M^2}:j\ge1\}$ \cite{QB22}. For
$\ab=(|k|^2-\mu)\sigma_3+\Delta(k_1\sigma_1+k_2\sigma_2)$ and $B>0$, $\spec K_B(\ab)=\{B-\mu\}\cup\{B\pm\sqrt{(2Bj-\mu)^2+2B\Delta^2j}:j\ge1\}$, by pairing $e_1\otimes|j\rangle$
with $e_2\otimes|j-1\rangle$ in the Hermite basis. For rotation-covariant symbols
$\ab=\bigl(\begin{smallmatrix}f(|k|^2)&g(|k|^2)k_-^\nu\\ \bar g(|k|^2)k_+^\nu&h(|k|^2)\end{smallmatrix}\bigr)$, the operator $K_B$ splits into $\nu$ unpaired levels
and $2\times2$ blocks whose entries are Laguerre transforms, for instance $\langle j|\Op^w(f(|k|^2))|j\rangle=\frac{(-1)^j}{|B|}\int_0^\infty
f(s)e^{-s/|B|}L_j(2s/|B|)\,ds$ for $B>0$.
\end{example}
\begin{example}[The cylinder term]\label{s:ex:twist}
Let $a_m=|k|^2\bigl(\cos\beta(y)\sigma_3+\sin\beta(y)(\cos2\theta\,\sigma_1+\sin2\theta\,\sigma_2)\bigr)$, with $\beta$ increasing from $0$ to $\beta_+$ across the wall. At
infinite momentum the occupied projector depends on $y$, and
\[
  \Cy=\frac1{4\pi}\int_0^{2\pi}\!\!\int2\beta'(y)\sin\beta(y)\,dy\,d\theta=1-\cos\beta_+ .
\]
For $\beta_+=\pi$ and bulks $\mp(|k|^2+M)\sigma_3$, $J_\pm=0$ at $\alpha=0$, so that $2\pi\sigma_I=\Cy=2$ is carried entirely by the wall at infinite momentum,
whatever the fields; with constant bulk extensions, the boundary map of Lemma~\ref{lem:degree} covers $\Gr_1(\C^2)=S^2$ twice. For the $d$-wave entry of Table~\ref{s:tab:levelsJ}, with $c(y)$ increasing from $-1$ to $+1$,
$I_+-I_-\approx-3.1990$ is not an integer, $\Cy\approx-0.8010$, and the sum is $-4$ to quadrature accuracy, the value $2(\sgn c_--\sgn c_+)$ of
\cite[Prop.~2.8]{QB21}. Bulk-difference formulas without $\Cy$ are therefore correct only when $E^\infty$ does not depend on $y$.
\end{example}

\begin{example}[Rotating shallow water]\label{s:ex:sw}
We complete the shallow-water example of section~\ref{sec:compute}. The unitary equivalence with $\xi S_x+\zeta S_y+fS_z$ maps the height to the eigenvector
of $S_z$ with eigenvalue $0$, and (N3), (N4) hold without field for $\alpha\in(-\min|f_\pm|,\min|f_\pm|)\setminus\{0\}$. Neither $V_\sff$ nor the eigenvector
spanning $E^\infty_{\rm tr}$ depends on $y$, so that $\Cy=0$. The value $I(a_\pm;\alpha)=-\sgn f_\pm$ on both sides of $0$, the flat band carrying no
curvature, was obtained numerically: for $f=1$ and $\alpha=\pm0.5$, the Fukui--Hatsugai--Suzuki discretization on $[-K,K]^2$ gives $-0.9955$ at $K=200$, with a
tail of order $K^{-1}$, and $-0.4977$ for the Dirac symbol with $M=1$, which fixes the sign convention.

In the decomposition $v=\bigl(\begin{smallmatrix}v_0&p^*\\ p&S+b\sigma_y\end{smallmatrix}\bigr)$ of section~\ref{sec:compute}, with $v_0,b\in\R$, $p\in\C^2$ and
$\sigma_y$ acting on the velocity, the Coriolis term is $b=f$, and $V_\sff^*vV_\sff=(\theta^\perp)^TS\,\theta^\perp$ since the matrices
$\theta^\perp(\theta^\perp)^T$ span the real symmetric $2\times2$ matrices. The elliptic directions, of real dimension $6$, are a potential $v_0$ on the
height, arbitrary couplings $p$ between height and velocity, and Coriolis-type terms $b\sigma_y$; the non-elliptic directions are the real symmetric velocity
blocks $S$, of real dimension $3$. For constant $S$, $\Fl$ is the interval between the eigenvalues of $S$, reduced to $\{s\}$ for $S=s\,\mathrm{Id}$. Condition
(F) holds for Coriolis parameters $f(x,y)$, since the kernel does not depend on $u$. With field, the flat band carries a Landau fan, which an elliptic
perturbation deforms without moving its accumulation point, and which a non-elliptic perturbation moves with $\Fl$. Figure~\ref{s:fig:spin1} shows this in the
equivalent dice-lattice form, where the potential on the height becomes the potential on the hub site.
\end{example}

\begin{example}[Pseudospin-one materials]\label{s:ex:aT3}
For the $\alpha$--$\mathcal T_3$ symbol
\[
  a=\begin{pmatrix}\Delta_A&c\,k_-&0\\c\,k_+&\Delta_H&s\,k_-\\0&s\,k_+&\Delta_C\end{pmatrix},\qquad k_\pm=\xi\pm i\zeta,\ c=\cos\varphi,\ s=\sin\varphi,
\]
the dice lattice, $\varphi=\pi/4$, is unitarily equivalent to the shallow-water symbol of Example~\ref{s:ex:sw}, whose analysis carries over. We record what
depends on $\varphi$. We have $r=1$, $V_\sff=(se^{-i\theta},0,-ce^{i\theta})$, $\Meff=\Delta_As^2+\Delta_Cc^2$, and $iV_\sff^*\partial_\theta V_\sff=-\cos2\varphi$, so
that the flat band carries the Berry phase $-2\pi\cos2\varphi$ at infinity. The non-integrality of single-valley Hall conductivities observed in
\cite{KDDC17} is thus contained in $J_\pm$, and the corresponding integers are $\tilde J_\pm$. For a Hermitian perturbation $v=(v_{ij})$ of degree $0$,
\[
  V_\sff^*vV_\sff=s^2v_{11}+c^2v_{33}-2sc\,\Re\bigl(e^{2i\theta}v_{13}\bigr),
\]
so that, as in Example~\ref{s:ex:sw}, the elliptic directions have real dimension $6$, and the non-elliptic directions, $s^2v_{11}+c^2v_{33}$ and $v_{13}$,
dimension $3$. For constant $v$, $\Fl$ is the interval of center $s^2v_{11}+c^2v_{33}$ and half-width $2sc|v_{13}|$. The $S_z$ mass $f(y)(1,0,-1)$ of the dice
lattice is elliptic, so that $\Fl=\{0\}$ and $\mathcal G=(-1,0)\cup(0,1)$ for $|f|=1$, whereas the staggered mass is not: it moves $\Fl$ into the bulk gap, and
then (N3) fails throughout the gap, which is essential spectrum for every field by Proposition~\ref{prop:flat}(b). With the $S_z$ mass, a Landau
fan accumulates at $0$ (Examples~\ref{ex:examples}), and a boundary of the applied field in the bulk of mass $-1$ carries
$2\pi\sigma_I=-n(\alpha)$, the number of fan levels in $(\alpha,1)$.

For the dice mass wall at $B=0.5$, theory and numerics agree on $2\pi\sigma_I=-2,-1,0,+1$ at $\alpha=0.5,0.25,0.17,0.13$. The pair
index of the dice bulk with $f=-1$ with respect to the reference $f=+1$ takes the values $2,1,0,-1$ at the same energies, which gives $J=1-n(\alpha)$. HgCdTe near critical doping maps onto $\alpha$--$\mathcal T_3$ at $\tan\varphi=1/\sqrt3$ \cite{MN16}, with a mass that changes sign with temperature
and composition \cite{Teppe16}, and which mass type the mapping produces decides (N3).
\end{example}

\begin{example}[Variable leading coefficients]\label{s:ex:rot}
For rotating dice matrices $\gamma_j(y)=U(y)\gamma_j^0U(y)^*$, with $U=e^{i\beta(y)G}$ and $G=e_Ae_H^*+e_He_A^*$, the conjugation is exact, and the curvature
correction of Proposition~\ref{prop:Meffcomp}(a) is $T=-\beta'\cos\varphi\sin^2\varphi\sin2\theta$. Formula \eqref{eq:T} reproduces
it to five digits: at $\beta'=0.4$, it gives $-0.07985,-0.14142,-0.09552,+0.10703$ at $\theta=0.3,0.785,1.2,2.0$, for $|k|=5$ and $|k|=50$. The flat wall spectrum
then fills $[-\max|T|,\max|T|]$, although $E_{\ker}WE_{\ker}=0$, so that a criterion based on $E_{\ker}WE_{\ker}$ alone would miss it. Condition (F) holds for any
dependence of $W$ on $x$ (Lemma~\ref{s:lem:Fsufficient}).
\end{example}

\begin{example}[Gyrotropic and Maxwell operators]\label{s:ex:curl}
The factorized operator $H=\curl_B^*\curl_B+g(y)\sigma_y$ of section~\ref{sec:compute} satisfies (F) and (W). The Peierls operator $H+\frac B2\sigma_y$
satisfies (W) by Proposition~\ref{s:prop:Wholds} and needs no freezing condition under (TI). For the Peierls operator, Corollary~\ref{cor:main}(b) gives $+4$
and $+5$ at $\alpha=-0.35$ and $-0.3$, $\alpha=-0.25$ being a level of both bulks; the values for the factorized operator are in the table below. Maxwell
operators $\varepsilon^{-1/2}\curl^*\mu^{-1}\curl\,\varepsilon^{-1/2}$ are factorized and covered for arbitrary smooth $\varepsilon(x,y)$ and $\mu(x,y)$, with
$\Fl=\{0\}$. For photons, $B$ is a pseudo-magnetic field. Near Dirac points, the effective model is of first order and the two couplings coincide
\cite{GRW21}; in general, which coupling describes a strained structure depends on the derivation of the effective model.

For constant $g$ and $B>0$, the levels stated in Examples~\ref{ex:examples} are obtained in the Hermite basis $|n\rangle$ of
$\mathfrak a$ and the basis $e_\pm=(1,\pm i)/\sqrt2$ of $\C^2$, in which $\sigma_ye_\pm=\pm e_\pm$. The reduced operator of $\curl_B$ maps $e_+\otimes f$ to a
multiple of modulus $\sqrt B$ of $\mathfrak af$, and $e_-\otimes f$ to a multiple of modulus $\sqrt B$ of $\mathfrak a^*f$. Hence $\curl_B^*\curl_B$ annihilates
$e_+\otimes|0\rangle$, equals $B$ on $e_+\otimes|1\rangle$, and acts on $\mathrm{span}\{e_+\otimes|n+1\rangle,e_-\otimes|n-1\rangle\}$, $n\ge1$, by a matrix
unitarily equivalent to $B\bigl(\begin{smallmatrix}n+1&\sqrt{n(n+1)}\\ \sqrt{n(n+1)}&n\end{smallmatrix}\bigr)$, while $g\sigma_y=\mathrm{diag}(g,-g)$ on each
of these blocks. Its trace is $B(2n+1)$ and its determinant is $-g(g+B)$, which gives $\lambda_\pm(n)$. Complex conjugation maps $\curl_B^*\curl_B$ to
$\curl_{-B}^*\curl_{-B}$ and $\sigma_y$ to $-\sigma_y$, which gives $B<0$. Identity \eqref{eq:curlPF} gives the levels of the Peierls operator.
For $g<0$ and $B>0$, the levels $\lambda_-(n)$ cross the flat-band energy $0$ simultaneously at $B=|g|$, so that the Landau fan accumulates at $\Fl$ from
below for $B<|g|$ and from above for $B>|g|$; for the Peierls operator, the corresponding fields are $B=\frac23|g|$ and $B=2|g|$. We checked these levels and
the transition fields against the numerical diagonalization in the Hermite basis.

Here $E_{\ker}(\theta)=\theta\theta^T$, and these projectors span the real symmetric $2\times2$ matrices. The only elliptic direction is therefore $\sigma_y$, of real
dimension $1$, and the non-elliptic directions are $\mathrm{Id}$, $\sigma_z$ and $\sigma_x$. For $W=g\sigma_y+w_0\mathrm{Id}+w_3\sigma_z+w_1\sigma_x$ constant,
$\Meff=\theta^TW\theta=w_0+w_3\cos2\theta+w_1\sin2\theta$, so that $\Fl=[w_0-(w_1^2+w_3^2)^{1/2},w_0+(w_1^2+w_3^2)^{1/2}]$. The gyrotropic coefficient $g$ may
therefore be changed arbitrarily without moving $\Fl$, while any non-elliptic perturbation moves it. The operator is factorized, so that this holds at every
field (Figure~\ref{s:fig:curl}).

For the gyrotropic operator $\curl^*\curl$ without field and $g=\tanh(1.5y)$, the spectral flow at $\alpha=0.5$ is $-2$, as predicted by $I_+-I_-+\Cy=-2$. At
$B=0.5$, we obtain the following values.
\begin{center}\small
\begin{tabular}{@{}lcccccc@{}}
\toprule
$\alpha$ & $-1.2$ & $-0.6$ & $-0.45$ & $-0.35$ & $-0.25$ & $+0.5$\\
\midrule
$N_+(\alpha),N_-(\alpha)$ & $0,0$ & $1,1$ & $2,2$ & $3,2$ & $5,3$ & ---\\
prediction & $0$ & $0$ & $0$ & $+1$ & $+2$ & $-2$\\
numerics & $0$ & $0$ & $0$ & $+1$ & $+2$ & $-2$\\
\bottomrule
\end{tabular}
\end{center}
We also tested an energy that lies in a field-free band. Coupling the model to a scalar band $-\Pi^2+c$ with strength $0.3$ puts $\alpha=0.5$ in a
field-free band for $c=2.2$. Deforming $c$ to $-1$, where $\alpha$ lies in a field-free gap, moves two levels through $\alpha$ in each bulk, the crossings cancel, and the numerics give $-2$
at $c=2.2,1.2,-1$.

Without field, the difference between factorized and Peierls operators comes from the ordering of the position-dependent coefficients. It is visible on
$U(y)k^\perp(k^\perp)^TU(y)^*$ with $U=e^{i\beta(y)\sigma_z}$. The factorized operator $U\curl^*\curl U^*$ is unitarily equivalent to $\curl^*\curl$ and has
all its near-flat states exactly at $0$. For the Weyl quantization, the flat block contains $f_1=\beta'(y)\rho\cos\theta\sin2\theta$, which is indefinite on
the strip, so that $\spec_{\rm ess}=\R$ by Remark~\ref{s:rem:disp}; the spreading of the near-flat states over $(-0.89,0.89)$ observed numerically is limited by
the momentum cutoff of the discretization. A constant twist rate $\beta'=\beta_1$ makes this explicit for every $m$. Writing $s=(1,-1)$
for the eigenvalues of $\sigma_z$, the identity $\Op^w(e^{i\gamma y}p(\zeta))=e^{i\gamma y}p(D_y+\frac\gamma2)$ gives
$U^*\Op^w(Ua_0U^*)U=\Op^w(\tilde a_0)$ with $(\tilde a_0)_{ij}(\xi,\zeta)=(a_0)_{ij}(\xi,\zeta+\frac{\beta_1}2(s_i+s_j))$, a constant-coefficient operator. For
$a_0=\rho^{m-2}k^\perp(k^\perp)^T$, the lowest band of $\tilde a_0$ is $\beta_1\rho^{m-1}\cos\theta\sin2\theta+O(\rho^{m-2})$, as we checked numerically for
$m=2,4$, so that the bounded band becomes unbounded in both directions, while the factorized operator $U\curl^*\curl U^*$ keeps it at $0$.

The Lam\'e operator is elliptic for $\mu>0$ and $\lambda+2\mu>0$, and its non-elliptic limits
are factorized.
\end{example}


\begin{example}[Higher order]\label{s:ex:higher}
For non-factorized operators of order $m\ge3$, the field may enter $\Meff$ and the weak-field limit $J\to I$ may fail
(Proposition~\ref{prop:Meffcomp}(e), Appendix~\ref{s:sec:pframe}, and Proposition~\ref{s:prop:exactB} below).
Theorem~\ref{thm:BIC} still applies at every fixed nonzero field for which (N) holds, with $J_\pm$ computed by
Theorem~\ref{thm:bulk}(b) or by a pair index with respect to a reference. For $r=1$, as in the examples below, a field-dependent $\Meff$
does not change $E^\infty$ as long as it does not cross $\alpha$. For $r\ge2$, it may rotate $E^\infty_{\rm fl}$, in which case (W5) fails and the
fractional part of $J$ depends on the field.

The dependence of the flat block on the field at orders $4$ and $6$ was tested as follows. For $\rho^2k^\perp(k^\perp)^T+\lambda\rho^2\sigma_y+g\sigma_y$ with
$\lambda=0.3$ and $g=1$, the limit of the bounded branch at Landau index $2\cdot10^6$ gives $(\Meff+\lambda^2)/B=-0.614,-0.626,-0.663,-0.725,-0.850$ at
$B=0.01,0.02,0.05,0.1,0.2$, that is, $-2\lambda-\frac54B$. Whether an energy can be connected to a field-free gap then depends on $B$: at $B=0.2$, $\alpha=-0.2$ is in a magnetic gap and
connected to $\alpha_0=0.5$, while at $B=0$ it lies in the field-free flat band. For $\rho^4k^\perp(k^\perp)^T+g\sigma_y$, every $B\ne0$ turns the bounded band into a
branch $\approx-\frac74B^2\rho^2$. This branch crosses $\alpha_0=-1.5$, which lies in the field-free gap below all bands, at Landau index $\approx426,3426,53568$ for
$B=0.1,0.05,0.02$, with scaling $B^{-3}$, as predicted by Proposition~\ref{s:prop:exactB}(b),(c), so that infinitely many levels cross $\alpha_0$ as $B\to0$, and
Proposition~\ref{prop:weak} fails without (W).
\end{example}

The field terms of these examples are given explicitly by the following computation in the Hermite basis.
\begin{proposition}\label{s:prop:exactB}
Let $B\ne0$, $\rho_j^2:=2|B|j$, and let $K_B(a)$ be the reduced bulk operator of the symbol $a$.
(a) For $a=\rho^2k^\perp(k^\perp)^T+\lambda\rho^2\sigma_y+g\sigma_y$ ($m=4$), $K_B(a)$ has a sequence of eigenvalues $\mu_j$, $j\ge0$, with
$\mu_j\to-\lambda^2-2\lambda B-\frac54B^2$ as $j\to\infty$. When (N1) holds at $B$, this limit is $\Meff(B)$.
(b) For $a=\rho^{m-2}k^\perp(k^\perp)^T$ with $m=6,8$, $K_B(a)$ has a sequence of eigenvalues
$\mu_j=|B|^{m/2}\nu_j$, where the numbers $\nu_j$ do not depend on $B$ and $\nu_j=-\frac{m+1}4(2j)^{m/2-2}+O(j^{m/2-3})$ as $j\to\infty$, so that
$\mu_j=-\frac{m+1}4B^2\rho_j^{m-4}+O(\rho_j^{m-6})$. For $a+g\sigma_y$, $g\in\R$, the corresponding eigenvalues differ from $\mu_j$ by at most $|g|$.
(c) In particular, for $m=6,8$, $|g|<|\alpha_0|$ and $\alpha_0<0$, every $j$ with $\nu_j<0$ gives a level of $K_B(a+g\sigma_y)$ below $\alpha_0$ for
$|B|^{m/2}|\nu_j|>|\alpha_0|+|g|$ and above $\alpha_0$ for $|B|^{m/2}|\nu_j|<|\alpha_0|-|g|$, so that infinitely many levels cross $\alpha_0$ as $B\to0$.
\end{proposition}
\begin{proof}
The reflection $y\mapsto-y$ changes $B$ into $-B$ and, after conjugation by $R=\mathrm{diag}(1,-1)$, leaves $k^\perp(k^\perp)^T$ invariant and changes $\sigma_y$
into $-\sigma_y$. It thus maps the symbol with parameters $(B,\lambda,g)$ to the symbol with parameters $(-B,-\lambda,-g)$, and the statements are invariant
under this change. We may therefore assume $B>0$.

Write $\Pi=(By,D_y)$, so that $K_B(a)$ is the Weyl-symmetrized polynomial $a(\Pi)$ and $[\Pi_1,\Pi_2]=iB$. The operator $\mathfrak a=(\Pi_1+i\Pi_2)/\sqrt{2B}$
satisfies $[\mathfrak a,\mathfrak a^*]=1$, with Hermite basis $|n\rangle$, $n\ge0$, and number operator $\mathsf N=\mathfrak a^*\mathfrak a$. With $k_\pm=\xi\pm i\zeta$, we
have $\Op^w(\rho^2)=B(2\mathsf N+1)$ and $\Op^w(k_-)=\Pi_1-i\Pi_2=\sqrt{2B}\,\mathfrak a^*$. In the basis $e_\pm=(1,\pm i)/\sqrt2$ of $\C^2$, $\sigma_y=\mathrm{diag}(1,-1)$
and
\[
  k^\perp(k^\perp)^T=\frac12\begin{pmatrix}\rho^2&-k_-^2\\-k_+^2&\rho^2\end{pmatrix}.
\]
The product of symbols is $b\sharp c=b\exp\bigl(\frac{iB}2(\overleftarrow\partial_\xi\overrightarrow\partial_\zeta-\overleftarrow\partial_\zeta\overrightarrow\partial_\xi)\bigr)c$
\cite[\S4.3]{Zworski}, with arrows indicating the factor on which each derivative acts, so that $\xi\sharp\zeta-\zeta\sharp\xi=iB$. Since $\rho^2$ is quadratic, the expansion of $\rho^2\sharp c$ terminates, and
\[
  \rho^2\sharp c=\rho^2c+iB\,\partial_\theta c-\tfrac{B^2}4\Delta_kc,\qquad \partial_\theta:=\xi\partial_\zeta-\zeta\partial_\xi .
\]
For $c=\rho^{2p}$, we have $\partial_\theta c=0$ and $\Delta_kc=4p^2\rho^{2p-2}$. For $c=\rho^{2p}k_-^2$, we have $\partial_\theta c=-2ic$ and, since $k_-^2$ is
harmonic and homogeneous of degree $2$, $\Delta_kc=4p(p+2)\rho^{2p-2}k_-^2$. Hence
\[
  \Op^w(\rho^{2p+2})=\Op^w(\rho^2)\Op^w(\rho^{2p})+p^2B^2\Op^w(\rho^{2p-2}),
\]
\[
  \Op^w(\rho^{2p+2}k_-^2)=\bigl(\Op^w(\rho^2)-2B\bigr)\Op^w(\rho^{2p}k_-^2)+p(p+2)B^2\Op^w(\rho^{2p-2}k_-^2),
\]
with $\Op^w(k_-^2)=2B\,\mathfrak a^{*2}$. By induction, $\Op^w(\rho^{2p})=B^pf_p(2\mathsf N+1)$ and
$\Op^w(\rho^{2p}k_-^2)|n\rangle=2B^{p+1}h_p(2n+3)\sqrt{(n+1)(n+2)}\,|n+2\rangle$, where
\[
  f_0=1,\quad f_1(x)=x,\quad f_{p+1}=xf_p+p^2f_{p-1};\qquad h_0=1,\quad h_1(y)=y,\quad h_{p+1}=yh_p+p(p+2)h_{p-1}.
\]
Thus $f_2=x^2+1$, $f_3=x^3+5x$, $f_4=x^4+14x^2+9$, $h_2=y^2+3$ and $h_3=y^3+11y$. The operator $\Op^w(\rho^{2p}k_+^2)$ is the adjoint of
$\Op^w(\rho^{2p}k_-^2)$.

Let $p=\frac m2$. It follows that $K_B(a)$ leaves invariant the lines spanned by $e_+\otimes|0\rangle$ and $e_+\otimes|1\rangle$ and the planes
$V_j=\mathrm{span}\{e_+\otimes|j+2\rangle,e_-\otimes|j\rangle\}$, $j\ge0$, on which it is the Hermitian matrix with diagonal entries
\[
  d_+=\tfrac12B^pf_p(2j+5)+\lambda B(2j+5)+g,\qquad d_-=\tfrac12B^pf_p(2j+1)-\lambda B(2j+1)-g,
\]
where $\lambda=g=0$ for $m=6,8$, and off-diagonal entry $o=-B^ph_{p-1}(2j+3)\sqrt{(j+1)(j+2)}$. Let $\mu_j$ be the smaller eigenvalue on $V_j$, and set
$T_j=d_++d_-$ and $\Delta_j=d_+d_--o^2$. Then
\[
  \mu_j=\frac{\Delta_j}{T_j/2+\sqrt{T_j^2/4-\Delta_j}}=\frac{\Delta_j}{T_j}\Bigl(1+O\bigl(|\Delta_j|T_j^{-2}\bigr)\Bigr).
\]
A direct expansion of these polynomials gives, for $m=4$, $T_j=4B^2j^2+O(j)$ and $\Delta_j=-(4\lambda^2B^2+8\lambda B^3+5B^4)j^2+O(j)$; for $m=6$,
$T_j=8B^3j^3+O(j^2)$ and $\Delta_j=-28B^6j^4+O(j^3)$; and for $m=8$, $T_j=16B^4j^4+O(j^3)$ and $\Delta_j=-144B^8j^6+O(j^5)$. In each case $\Delta_jT_j^{-2}\to0$.
For $m=4$, $\mu_j\to-\lambda^2-2\lambda B-\frac54B^2$. For $m=6,8$, where $\lambda=g=0$, the entries $d_\pm$ and $o$ are $B^p$ times functions of $j$
only, so that $\mu_j=B^p\nu_j$ with $\nu_j$ independent of $B$. For $m=6$, $\mu_j=-\frac72B^3j+O(1)=-\frac74B^2\rho_j^2+O(1)$, and for $m=8$,
$\mu_j=-9B^4j^2+O(j)=-\frac94B^2\rho_j^4+O(\rho_j^2)$, which is (b). The term $g\sigma_y=\mathrm{diag}(g,-g)$ in the basis $e_\pm$ preserves the planes $V_j$
and has norm $|g|$, which gives the statement on $a+g\sigma_y$ by Weyl's inequality, and (c) follows. Finally, when (N1) holds at $B$, Theorem~\ref{thm:bulkspec} gives
$\spec_{\rm ess}K_B(a)=\Fl_B(\ab)$, and $\Fl_B(\ab)=\{\Meff(B)\}$ since $r=1$ and the symbol is rotation covariant, so that the limit in (a) is $\Meff(B)$.
\end{proof}

\begin{example}[Bulk invariants at a fixed field]\label{s:ex:fixed}
We next tested the change of extension class. We modified the standard extension $\frac12(1-\sigma_3)$ of $E^\infty$ on the disk $\{|k|\le R\}$ by a map
into the Grassmannian of nonzero degree, localized in that disk (a \emph{bubble}), which gives an extension $E^{\rm ext\prime}$ in another class, with the sign
matrix $\Sigma'=1-2E^{\rm ext\prime}(\cdot/R)$. For bubbles of degree $\pm1$ and $\pm2$, $m=2$ and $R=6$, the count $\#{\rm
neg}K_B(\Lm_R^m\Sigma')-\#{\rm neg}K_B(\Lm_R^m\sigma_3)$ of negative eigenvalues equals $-\sgn(B)\kappa(E^{\rm ext\prime})$ in all $16$ cases tested, namely $B=\pm1,\pm\frac12$ and the elliptic
skyrmion, antiskyrmion, two-skyrmion and skyrmion--antiskyrmion bubbles, with open gaps and stable truncations. With $R=3$, the gaps nearly close and the
counts fail, consistent with the semiclassical nature of Proposition~\ref{prop:semiclassical}, which requires the bubble to cover many
magnetic cells.

The St\v{r}eda formulas were tested on the pair index at a single field for the winding-two model $(|k|^2-1)\sigma_3+(k_-^2\sigma_++{\rm h.c.})$ with reference
$\mu\to-1$. The indices are $2,1,0$ at $B=0.2,0.5,1.5$, which gives $J=1.7071,0.7071,-0.2929$, that is, the field-free value $I=1+1/\sqrt2$ followed by two field-driven
crossings.
\end{example}

\section{Symbol calculi and kernel bounds}\label{s:sec:tools}
We recall that $D=-i\nabla$, $\Pi=D-A$, $\Lm=(1+\Pi^2)^{1/2}$, that $S^\nu$ is the class of matrix symbols with $|\partial^\gamma_u\partial^\beta_ka|\le C\jp k^{\nu-|\beta|}$, that $H^s_A$ is the space of $f\in L^2$ with $\Pi^\beta f\in L^2$ for $|\beta|\le s$, and that $\sharp^B$ is the product of the magnetic Weyl calculus $\OpA$.

We first recall the properties of the magnetic Weyl calculus \cite{IMP}. For $B\in C_b^\infty$ and potentials with polynomially bounded derivatives, the
calculus is gauge covariant. The product $\sharp^B$ maps $S^{\mu_1}\times S^{\mu_2}$ to $S^{\mu_1+\mu_2}$ with an asymptotic expansion whose $j$-th term lies in
$S^{\mu_1+\mu_2-j}$, and the terms containing $B$ or its derivatives carry at least two $k$-derivatives. At first order,
$a\sharp^Bb=ab+\frac i2\{a,b\}+\frac i2B\,(\partial_{k_1}a\,\partial_{k_2}b-\partial_{k_2}a\,\partial_{k_1}b)+\cdots$, with $\{a,b\}=\sum_j(\partial_{u_j}a\,\partial_{k_j}b-\partial_{k_j}a\,\partial_{u_j}b)$
and $B$ evaluated at the base point; for a constant field, the full expansion is the exponential of these two bilinear forms \cite{MP04}. The Calder\'on--Vaillancourt theorem and the
inverse-closedness of $\OpA(S^0)$ hold \cite{IMP,IMP2}; short proofs of these facts and of the Beals criterion, by a representation of magnetic
pseudo-differential operators as matrices localized near the diagonal in tight magnetic Gabor frames, are given in \cite{CHP18,CHP24}. Moreover, the kernels of $\OpA(a)$ equal $e^{-i\int_{[u,u']}A}$ times those of $\Op^0(a)$.

The calculus is pseudolocal in the following sense.
\begin{lemma}\label{s:lem:pseudoloc}
Let $c\in S^\nu$ vanish on an open set $\Omega$, and let $\chi_1,\chi_2\in C_b^\infty$ with $\supp\chi_1\subset\{u:\dist(u,\R^2\setminus\Omega)\ge L\}$. Then
$\norm{\Lm^{s}\chi_1\OpA(c)\chi_2\Lm^{s'}}\le C_{N,s,s'}L^{-N}$ for all $s,s',N$, and the same holds for $\OpA(c)\chi_1$. Consequently, if two symbols agree on
$\Omega$, compositions with them agree on $\supp\chi_1$ up to such terms.
\end{lemma}
\begin{proof}
This is the pseudolocality of the magnetic calculus \cite{IMP}. The kernel at $(u,u')$ is an oscillatory integral of $c(\frac{u+u'}2,\cdot)$, which vanishes unless $|u-u'|\ge2\dist(u,\R^2\setminus\Omega)\ge2L$ for
$u\in\supp\chi_1$. Integration by parts in $k$ gives decay $|u-u'|^{-N}$ with any number of derivatives, and the magnetic phase only adds polynomial factors.
\end{proof}
We also use the following facts on the Weyl calculus with order functions \cite{Zworski,DS99}, in which we write $\hbar$ for the semiclassical
parameter $h$, to distinguish it from the Planck function $h_g$ below and on the Weyl--H\"ormander calculus
\cite[Ch.~18]{Hormander3}, \cite{Lerner,BonyChemin}. An order function on $\R^{2d}$ is a positive function $M$ with $M(X)\le C\jp{X-Y}^NM(Y)$, and $S(M)$ is
the space of smooth functions $a$ with $|\partial^\beta a|\le C_\beta M$ for all $\beta$. (C1) For $a\in S(M_1)$, $b\in S(M_2)$ and $0<\hbar\le1$, the
semiclassical Moyal product satisfies $a\#_\hbar b=ab+\hbar R_\hbar(a,b)$, where
$R_\hbar(a,b)=i\int_0^1\bigl(e^{i\theta\hbar A(D)}A(D)(a\otimes b)\bigr)|_{\rm diag}\,d\theta$ and $A(D)=\frac12\sigma(D_X;D_{X'})$. Thus $R_\hbar(a,b)$ is bilinear in
the first derivatives of $a$ and $b$, each term containing a derivative in a position variable of one factor and in the conjugate momentum variable of the
other. Its seminorms in $S(M_1M_2)$ are bounded by sums of products of seminorms of these derivatives, with constants depending only on the
constants of $M_1$ and $M_2$ \cite[\S\S4.3--4.4]{Zworski}, \cite[Ch.~7]{DS99}. (C2) For a slowly varying temperate metric $g$ satisfying the uncertainty
principle with Planck function $h_g$, and for $g$-admissible weights, $a\#_\hbar b-ab\in S(M_1M_2\hbar h_g,g)$, with seminorms bounded by products of
seminorms of $a$ and $b$ and constants depending only on the structure constants of $g$ and of the weights \cite[Thm.~18.5.4]{Hormander3},
\cite[Ch.~2]{Lerner}. This applies to the isotropic classes $\Gamma^\mu_{3/4}=S(\jp X^\mu,|dX|^2/\jp X^{3/2})$ of \cite{Shubin}, with $h_g=\jp X^{-3/2}$, and to the metric
$|dX|^2/\jp X^2$, uniformly in $\hbar$. (C3) Operators in $\Op^w_\hbar S(1)$ are bounded on $L^2$, with norms bounded by finitely many seminorms, uniformly in $\hbar$
\cite[\S4.5]{Zworski}. (C4) If $T=\Op^w(a)$ with $a\in S(1,g)$ is invertible on $L^2$, then $T^{-1}=\Op^w(b)$ with $b\in S(1,g)$, whose seminorms are
bounded in terms of finitely many seminorms of $a$ and of $\norm{T^{-1}}$. This includes the flat metric, for which $S(1,g)=S(1)$ \cite{BonyChemin},
\cite[\S2.6]{Lerner}. (C5) If $\partial^\beta a\in L^1(\R^{2d})$ for $|\beta|\le2d+1$, then $\Op^w_\hbar(a)$ is trace-class and
$\Tr\Op^w_\hbar(a)=(2\pi\hbar)^{-d}\int\tr a$ \cite[Ch.~9]{DS99}. In particular, $\Op^w\Gamma^\mu_{3/4}\subset\mathfrak S_1$ for $\mu<-2d$. (C6) For the isotropic classes $\Gamma^\mu$ of symbols with
$|\partial^\beta b|\le C\jp X^{\mu-|\beta|}$, elliptic operators in $\Op^w\Gamma^\mu$ with $\mu>0$ have compact resolvent, their compactly supported functions lie
in $\Op^w\Gamma^{-\infty}$, and operators in $\Op^w\Gamma^{-\varepsilon}$ are compact \cite{Shubin}.

For the magnetic calculus, we use, besides the composition theorem, that magnetic pseudo-differential operators of order $\nu$ map the magnetic Sobolev
spaces $H^{s}_A$ into $H^{s-\nu}_A$, that $\OpA(S^{-\infty})$ maps $H^{-s}_A$ into $H^s_A$ for all $s$, and the Beals-type characterization of
$\OpA(S^0)$ by commutators \cite{IMP,IMP2,CHP18}. The following three lemmas describe the dependence of the magnetic calculus on the potential, on the field, and
on exponential weights.
\begin{lemma}\label{s:lem:gauge}
Let $A$ and $A'=A+\delta A$ be potentials with $\delta A\in C_b^\infty$. For every $c\in S^\nu$, there is a unique $c^\flat\in S^\nu$ with $\Op^{A'}(c)=\OpA(c^\flat)$.
Moreover, $c^\flat-c(u,k-\delta A(u))\in S^{\nu-1}$, the map $(c,\delta A)\mapsto c^\flat$ is continuous from $S^\nu\times C_b^\infty$ into $S^\nu$, and it acts
entrywise on matrix symbols.
\end{lemma}
\begin{proof}
The kernels of $\Op^{A'}(c)$ and $\OpA(c)$ differ by the factor $e^{-i\int_{[u,u']}\delta A}=e^{i(u-u')\cdot F(u,u')}$, where $F(u,u')=\int_0^1\delta
A(u'+t(u-u'))\,dt$ is smooth with bounded derivatives. Hence the kernel of $\Op^{A'}(c)$ is $e^{-i\int_{[u,u']}A}(2\pi)^{-2}\int e^{i(u-u')\cdot k}a(u,u',k)\,dk$
with the amplitude $a(u,u',k)=c(\frac{u+u'}2,k-F(u,u'))$, which satisfies $|\partial^\alpha_{u,u'}\partial^\beta_ka|\le C\jp k^{\nu-|\beta|}$ since $F$ is bounded
with bounded derivatives. The reduction of such amplitudes to Weyl symbols,
$c^\flat(v,k)=(2\pi)^{-2}\iint e^{-iz\cdot\eta}a(v+\frac z2,v-\frac z2,k+\eta)\,dz\,d\eta$, takes place on the kernel, where the common factor
$e^{-i\int_{[u,u']}A}$ plays no role. It gives $c^\flat\in S^\nu$ with $c^\flat-a(v,v,k)\in S^{\nu-1}$, continuously in $a$ \cite[\S18.1]{Hormander3}, and
$a(v,v,k)=c(v,k-\delta A(v))$. Uniqueness holds since the kernel determines the symbol.
\end{proof}
\begin{lemma}\label{s:lem:Bcont}
Let $f\in S^{\nu_1}$ and $g\in S^{\nu_2}$, and let $B,B'$ be fields with $\norm B_{C^N},\norm{B'}_{C^N}\le1$. Then $f\sharp^Bg-f\sharp^{B'}g\in
S^{\nu_1+\nu_2-2}$, with seminorms bounded by $C\norm{B-B'}_{C^{N'}}$ times products of seminorms of $f$ and $g$.
\end{lemma}
\begin{proof}
By \cite{IMP}, $(f\sharp^Bg)(X)=\pi^{-4}\iint e^{-2i\sigma(Y,Z)}\omega^B(x,y,z)f(X-Y)g(X-Z)\,dY\,dZ$ as an oscillatory integral, with $X=(x,\xi)$, $Y=(y,\eta)$,
$Z=(z,\zeta)$, $\omega^B=e^{-i\Gamma^B}$, and $\Gamma^B(x,y,z)$ the flux of $B$ through the triangle with vertices $x-y-z$, $x+y-z$ and $x-y+z$. This flux is a
constant multiple of $(y\wedge z)\gamma^B(x,y,z)$, where $\gamma^B$ is an average of $B$ over the triangle. It is linear in $B$ and smooth, with derivatives
bounded by $C\norm B_{C^N}$. Hence $\omega^B-\omega^{B'}$ is a constant multiple of $(y\wedge z)\gamma^{B-B'}\int_0^1\omega^{B_t}dt$ with $B_t=tB+(1-t)B'$. The
factors $y_j$ and $z_l$ are absorbed by integration by parts, since they are, up to constants, derivatives of the phase in the momenta $\zeta_j$ and $\eta_l$.
The derivatives then fall on $g$ and on $f$, because $\gamma$ and $\omega$ do not depend on the momenta. The resulting integral has the same structure, with
$f$ and $g$ replaced by first derivatives in $k$ and with an amplitude in the positions whose derivatives have polynomial growth and are bounded by
$C\norm{B-B'}_{C^{N'}}$. The proof of the composition theorem in \cite{IMP} applies to such integrals and gives the claim.
\end{proof}
\begin{lemma}\label{s:lem:expconj}
Let $0<\varrho\le1$. (a) If $f\in S^{\nu_1}$ and $g\in S^{\nu_2}$ extend analytically in $k$ to $\{|\Im k|<\varrho\}$ with symbol bounds uniform there, then so does
$f\sharp^Bg$. (b) Let $p\in S^\nu$ have this property, and let $\phi$ be real with $|\nabla\phi|\le1$ and bounded derivatives of all orders. For real
$|\mu|\le\varrho/2$, $e^{\mu\phi}\OpA(p)e^{-\mu\phi}=\OpA(p_\mu)$ on $C_c^\infty$, where $p_\mu-p=\mu p^{(1)}_\mu$ and $p^{(1)}_\mu$ is bounded in $S^{\nu-1}$
uniformly in $\mu$.
\end{lemma}
\begin{proof}
Such conjugation properties are standard \cite{CT,IMP2}; we include the argument for the graded symbols used here. For (a), in the integral representation of the previous proof, the momentum $\xi$ only enters through the arguments of $f$ and $g$. Replacing $\xi$ by
$\xi+i\kappa$ with $|\kappa|<\varrho$ gives an oscillatory integral of the same type, with symbols satisfying the same bounds, and analyticity follows by
differentiation under the integral. For (b), the kernel of $e^{\mu\phi}\OpA(p)e^{-\mu\phi}$ is that of $\OpA(p)$ multiplied by $e^{\mu(\phi(u)-\phi(u'))}
=e^{\mu(u-u')\cdot\Phi(u,u')}$, where $\Phi(u,u')=\int_0^1\nabla\phi(u'+t(u-u'))\,dt$ is real, bounded by $1$, with bounded derivatives. Shifting the contour of
integration in $k$ by $i\mu\Phi(u,u')$, which is allowed since $|\mu\Phi|\le\varrho/2$ and $p$ is analytic with symbol bounds in the strip, and which is
justified by inserting $\jp{u-u'}^{-2M}(1-\Delta_k)^M$ in the oscillatory integral, shows that the conjugated operator has the amplitude
$a_\mu(u,u',k)=p(\frac{u+u'}2,k+i\mu\Phi(u,u'))$ in the sense of the proof of Lemma~\ref{s:lem:gauge}. This amplitude satisfies
$|\partial^\alpha_{u,u'}\partial^\beta_ka_\mu|\le C\jp k^{\nu-|\beta|}$ uniformly in $|\mu|\le\varrho/2$, and
$a_\mu-a_0=i\mu\Phi\cdot\int_0^1(\partial_kp)(\frac{u+u'}2,k+is\mu\Phi)\,ds$ satisfies the same bounds with $\nu-1$ in place of $\nu$ and a factor $\mu$. Since
$a_0$ depends only on the midpoint, its reduction is $p$ itself, and the reduction of $a_\mu-a_0$ gives $p_\mu-p=\mu p^{(1)}_\mu$ with $p^{(1)}_\mu$ bounded in
$S^{\nu-1}$.
\end{proof}

The following diamagnetic bounds are used for kernel estimates.
\begin{lemma}\label{s:lem:diamag}
For $s>\frac12$, the operator $\Lm^{-2s}$ has a kernel bounded by $G_s(u-u')$, where $\widehat{G_s}(k)=\jp k^{-2s}$. In particular, $G_s\in L^2(\R^2)$,
$|G_s(v)|\le C_se^{-|v|/2}$ for $|v|\ge1$, and for $f\in L^2(\R^2)$,
\[
  \norm{f\,\Lm^{-2s}}_{\mathfrak S_2}\le\norm f_2\norm{G_s}_2 .
\]
If $T$ and $\Lm^{2s}T\Lm^{2s}$ are bounded, then $T$ has a continuous kernel with $|K_T(u,u')|\le\norm{G_s}_2^2\,\norm{\Lm^{2s}T\Lm^{2s}}$.
\end{lemma}
\begin{proof}
The diamagnetic inequality $|e^{-t\Pi^2}g|\le e^{t\Delta}|g|$, which holds pointwise \cite{AHS,Simon}, and $\Lm^{-2s}=\Gamma(s)^{-1}\int_0^\infty
t^{s-1}e^{-t}e^{-t\Pi^2}dt$ give $|\Lm^{-2s}g|\le(1-\Delta)^{-s}|g|$, and hence the kernel bound. The Hilbert--Schmidt bound follows. Since $\Lm^{-2s}$ maps $L^2$
into bounded continuous functions with $|(\Lm^{-2s}g)(u)|\le\norm{G_s}_2\norm g$, the vectors $e_u:=\Lm^{-2s}\delta_u$ lie in $L^2$ with $\norm{e_u}\le\norm{G_s}_2$
and depend continuously on $u$, and $K_T(u,u')=\ip{e_u}{(\Lm^{2s}T\Lm^{2s})e_{u'}}$.
\end{proof}
We next recall the Weyl--H\"ormander calculus, which we use for the fibers of translation-invariant walls. Let $g$ be a metric on $\R^{2d}$ that is slowly
varying, satisfies the uncertainty principle with Planck function $h_g\le1$, and is temperate, and let $M$, $M_1$ and $M_2$ be $g$-admissible weights
\cite[Ch.~18]{Hormander3}, \cite{Lerner,BonyChemin}. We use the following properties: (T1) the product $\#$ maps $S(M_1,g)\times S(M_2,g)$ to $S(M_1M_2,g)$,
and $a\#b-ab\in S(M_1M_2h_g,g)$, with the Moyal expansion to all orders; (T2) $\Op^w(S(1,g))$ is bounded on $L^2$; (T3) if $M\to0$ at infinity, then
$\Op^w(S(M,g))$ is compact, and trace-class if moreover $M\in L^1$; (T4) the calculus is inverse-closed (Beals' characterization), with seminorm bounds
polynomial in the commutator bounds; (T5) a Hermitian $a\in S(M,g)$ with $M\ge1$ that is elliptic for $M$ large is self-adjoint on $H(M,g)$, with a parametrix
in $S(M^{-1},g)$; (T6) for $f\in C_c^\infty$, $f(\Op^w a)\in\Op^wS(M^{-N},g)$ for all $N$, by the Helffer--Sj\"ostrand formula \cite{DS99}. All constants depend only
on the structure constants of $g$, on the weights and on finitely many seminorms. The fiber metric is admissible uniformly in the momentum along the wall.
\begin{lemma}\label{s:lem:admissible}
For $\Lm_\xi(y,\zeta)=\jp{\xi-A_1(y),\zeta}$, the metric $g_\xi=|dy|^2+|d\zeta|^2/\Lm_\xi^2$ is admissible with Planck function $h_g=\Lm_\xi^{-1}$ and structure
constants depending only on $\norm B_\infty$ and not on $\xi$, and every power of $\Lm_\xi$ is a $g_\xi$-admissible weight. The same holds for
$\Lm_{\xi,R}=(R^2+\Lm_\xi^2)^{1/2}$ uniformly in $R\ge1$, with $h_g\le R^{-1}$.
\end{lemma}
\begin{proof}
We verify the conditions of \cite[\S18.4]{Hormander3} and \cite[Ch.~2]{Lerner}. We have $|\partial_y\Lm_\xi|\le\norm B_\infty$ and $|\partial_\zeta\Lm_\xi|\le1$, so that $|\Lm_\xi(X)-\Lm_\xi(Y)|\le\norm B_\infty|y-y'|+|\zeta-\zeta'|$. If
$g_X(X-Y)\le r^2$, then $|\Lm(X)-\Lm(Y)|\le(\norm B_\infty+1)r\Lm(X)$, which gives slow variation for $r=(2\norm B_\infty+2)^{-1}$. The dual metric is
$g^\sigma_X=\Lm^2|dy|^2+|d\zeta|^2$, so that $h_g=\Lm^{-1}$. Temperance follows from $|\Lm(X)-\Lm(Y)|\le(\norm B_\infty+1)g^\sigma_Y(X-Y)^{1/2}$. The same
inequalities hold for $\Lm_{\xi,R}$.
\end{proof}
The fiber symbol $a(y,\xi-A_1(y),\zeta)$ of a differential operator of order $m$ lies in $S(\Lm_\xi^m,g_\xi)$ uniformly in $\xi$. Indeed, $\zeta$-derivatives lower
the order in $p=(\xi-A_1,\zeta)$, while $y$-derivatives produce bounded factors $A_1^{(j)}=-B^{(j-1)}$, and $g_\xi$ requires no gain in $y$. Combined with the
fixed frame, this gives the fiber versions of Proposition~\ref{prop:paramalpha}, Theorem~\ref{thm:analysis}, Proposition~\ref{prop:loc} and Theorem~\ref{thm:local},
with constants uniform in $\xi$.

\begin{proposition}\label{s:prop:shubin}
Let $h=|B|$. Then $K_B(\ab)$ is unitarily equivalent to $\Op_h^w(\ab\circ \varsigma_B)$ on $L^2(\R)$, where $\varsigma_B$ is the identity if $B>0$ and the swap of the two
phase-space variables if $B<0$.
\end{proposition}
\begin{proof}
Weyl quantization is covariant under linear symplectic maps $\varsigma$ of the phase plane: there is a unitary $M_\varsigma$, unique up to a phase, such that
$M_\varsigma^{-1}\Op^w(c)M_\varsigma=\Op^w(c\circ\varsigma)$ for all symbols $c$ \cite[Ch.~4]{Folland}. For the dilation $\varsigma(y,\zeta)=(y/h,h\zeta)$, $M_\varsigma$ is
$f\mapsto h^{1/2}f(h\,\cdot)$, and $\ab(By,\zeta)\circ\varsigma=\ab(\sgn(B)y,h\zeta)$, whose Weyl quantization is $\Op^w_h(\ab(\sgn(B)\,\cdot,\cdot))$. This proves the
claim for $B>0$. For $B<0$, the semiclassical Fourier transform, composed if necessary with the reflection $f\mapsto f(-\,\cdot)$, implements the symplectic map
$(y,\zeta)\mapsto(-\zeta,y)$, which transforms the symbol $\ab(-y,\zeta)$ into $\ab(\zeta,y)=\ab\circ\varsigma_B(y,\zeta)$.
\end{proof}

\section{Proofs for the operator and the bulk}\label{s:sec:pframe}\label{s:sec:routine}
Recall that $E_{\ker}(u,\theta)$ is the projector onto $\ker a_m(u,\theta)$, of rank $r$, and $Q=1-E_{\ker}$; that a homogeneous frame is a smooth unitary $U=[V_\sff,V_\stt]$ whose first $r$ columns span $\ker a_m$; that the conjugated symbol $\tilde a=U^*\sharp^Ba\sharp^BU$ has blocks $\tilde a_{ij}$, $i,j\in\{\sff,\stt\}$, in the fixed splitting $\C^r\oplus\C^{n-r}$; that (DN$_0$) requires $\tilde a_{\sff\sff}\in S^0$, $\tilde a_{\sff\stt},\tilde a_{\stt\sff}\in S^{m/2}$ and $\tilde a_{\stt\stt}-V_\stt^*a_mV_\stt\in S^{m-1}$ at large momentum; and that $\Meff$ is the degree-$0$ part of the Schur complement $\tilde a_{\sff\sff}-\tilde a_{\sff\stt}\tilde a_{\stt\stt}^{-1}\tilde a_{\stt\sff}$.

\smallskip
\noindent\textbf{Frames and resolvent estimates.} We give here the proofs of Lemma~\ref{lem:globalframe} and Theorem~\ref{thm:analysis}.
\begin{proof}[Proof of Lemma~\ref{lem:globalframe}(a)]
The construction uses parallel transport with Kato's connection \cite{Kato}. On $\{y\ge y_0\}$, $E_{\ker}(u,\theta)=E_{\ker}^+(\theta)$ depends only on $\theta$. Vector bundles over $S^1$ are trivial since $U(r)$ and $U(n-r)$ are connected,
so that there is a smooth unitary $U_+(\theta)$ whose first $r$ columns span $\Ran E_{\ker}^+(\theta)$, and multiplying one flat column by $e^{-iw\theta}$
removes the winding $w$ of $\det U_+$. Let $T(x,y,y',\theta)$ be Kato's transformation function, $\partial_yT=[\partial_yE_{\ker},E_{\ker}](x,y,\theta)\,T$ with
$T=1$ at $y=y'$ \cite{Kato}. It is unitary and maps $\Ran E_{\ker}$ and $\Ran Q$ at $y'$ onto $\Ran E_{\ker}$ and $\Ran Q$ at $y$. We set
$U(u,\theta)=T(x,\max(y,-y_0),y_0,\theta)\,U_+(\theta)$ for $y\le y_0$ and $U=U_+$ for $y\ge y_0$. The generator vanishes for $|y|\ge y_0$, so that $U$ is
smooth, and since the transport runs over an interval of length at most $2y_0$, $U$ has bounded derivatives, those of $E_{\ker}$ being bounded thanks to the gap
$c_0$. By construction, $U$ does not depend on $y$ on $\{\pm y\ge y_0\}$, it does not depend on $x$ under (TI), and $\det U(u,\cdot)$ has the winding of
$\det U_+$, namely zero.

For the extension, let $E_+$ be a smooth extension of $U_+$ to the unit disk with values in $U(n)$, equal to $U_+(w/|w|)$ for $|w|\ge\frac34$, which exists
since $\det U_+$ has winding zero, and let $\varrho\in C^\infty([0,1])$ be nondecreasing, equal to $0$ near $[0,\frac12]$ and to $1$ near $1$. With
$\theta_w=w/|w|$, we set $\tilde U(u,w)=E_+(2w)$ for $|w|\le\frac12$ and
$\tilde U(u,w)=T(x,y_0+\varrho(|w|)(\max(y,-y_0)-y_0),y_0,\theta_w)\,U_+(\theta_w)$ for $\frac12\le|w|\le1$, and $U_R(u,k)=\tilde U(u,k/R)$ for $|k|\le R$.
Then $U_R$ is smooth, unitary-valued, equal to $U(u,\theta(k))$ for $|k|\ge R$, with bounded $u$-derivatives and $|\partial^\beta_kU_R|\le C_\beta R^{-|\beta|}$ on
$|k|\le R$. Since $U_R^*U_R=U_RU_R^*=1$ pointwise, $U_R^*\sharp^BU_R-1$ and $U_R\sharp^BU_R^*-1$ consist of Moyal terms, which belong to $S^{-1}$ and are $O(R^{-1})$
in $S^0$, so that $U_R$ is a homogeneous frame symbol for $R$ large. Convolution in $k$ with $\check\chi$, where $\chi\in C_c^\infty$ equals $1$ near $0$,
produces a symbol that is analytic in every strip $\{|\Im k|\le\varrho_0\}$ with symbol bounds there, and it changes $U_R$ by a term in $S^{-\infty}$ that is
$O(R^{-1})$ in $S^0$, so that it is again a frame symbol for $R$ large.

Finally, for a frame symbol $U$, $\U^*\U=1+\OpA(U^*\sharp^BU-1)$ and $\U\U^*=1+\OpA(U\sharp^BU^*-1)$ are invertible since the perturbations have norm at most
$\frac12$, so that $\U$ is invertible, and $\U^{-1}=(\U^*\U)^{-1}\U^*\in\OpA(S^0)$ by inverse-closedness of the magnetic calculus \cite{IMP}.
\end{proof}

\begin{proof}[Proof of Lemma~\ref{lem:globalframe}(b)]
Recall that the blocks of a matrix symbol refer to the fixed splitting $\C^r\oplus\C^{n-r}$ of section~\ref{sec:results}, whose first $r$
components correspond to the flat columns $V_\sff$ of the frame and the last $n-r$ to the transverse columns $V_\stt$.
Let $Y=\U^*\U=1+\OpA(\upsilon)$, with blocks $Y_{ij}$ in the fixed splitting. By Lemma~\ref{lem:globalframe}(a), $\upsilon\in S^{-1}$ and $\norm{\OpA(\upsilon)}\le\frac12$. Let
$N:=-Y_{\sff\sff}^{-1}Y_{\sff\stt}$, which belongs to $\OpA(S^{-1})$ by the inverse-closedness of the magnetic calculus \cite{IMP}, and let $L$ be the
block upper-triangular operator with diagonal blocks equal to $1$ and with $N$ in the block $(\sff,\stt)$. A direct computation gives that $L^*YL$ is block
diagonal, with diagonal blocks $Y_{\sff\sff}$ and $Y'_{\stt\stt}:=Y_{\stt\stt}-Y_{\stt\sff}Y_{\sff\sff}^{-1}Y_{\sff\stt}$, both of the form $1+\OpA(S^{-1})$ and at
norm distance at most $\frac12$ from $1$, since $\frac12\le Y\le\frac32$. Let $G=\mathrm{diag}(Y_{\sff\sff}^{-1/2},(Y'_{\stt\stt})^{-1/2})$. The square roots are given by binomial series that converge in norm and,
by the Beals-type characterization of magnetic pseudo-differential operators \cite{IMP2,CHP18}, also in $\OpA(S^0)$, and they belong to $1+\OpA(S^{-1})$. Then
$\hat\U:=\U LG$ is unitary, and its symbol $U^{\rm u}$ satisfies $U^{\rm u}-U\in S^{-1}$. Let $\tilde A=\U^*H\U$ with blocks $\tilde A_{ij}$. Then
\[
  L^*\tilde AL=\begin{pmatrix}\tilde A_{\sff\sff}&\tilde A_{\sff\stt}+\tilde A_{\sff\sff}N\\ \tilde A_{\stt\sff}+N^*\tilde A_{\sff\sff}&\tilde A_{\stt\stt}+N^*\tilde A_{\sff\stt}+\tilde A_{\stt\sff}N+N^*\tilde A_{\sff\sff}N\end{pmatrix}.
\]
Since $N$ has order $-1$, the added terms have orders at most $-1$ in the off-diagonal blocks and $\frac m2-1$ in the transverse block, and the flat block is
unchanged. Conjugation by the block-diagonal operator $G$ preserves the orders of all blocks. Hence (DN$_0$) holds for $U^{\rm u}$. The flat block of the
inverse of the conjugated symbol, whose inverse is the Schur complement, changes by terms of degree at most $-1-\frac m2$, so that $\Meff$ is unchanged.
In the isotropic and fiber calculi, the same computation applies with the corresponding inverse-closedness and Beals-type characterization
\cite{Shubin,Lerner}.
\end{proof}

The proof of Theorem~\ref{thm:analysis} uses the parametrices $G_z$, $G'_z$ and the smoothing remainders $K_z$, $K'_z$ of
Proposition~\ref{prop:paramalpha}, with $R(z)=(H-z)^{-1}$ and $d=\dist(z,\spec H)$.
\begin{proof}[Proof of Theorem~\ref{thm:analysis}(a)]
The argument is that of the elliptic theory \cite{ADN,IMP2}, with the parametrix of Proposition~\ref{prop:paramalpha}. Since $H$ is formally symmetric, $(H|_{C_c^\infty})^*=H_{\max}$, and $H$ is essentially self-adjoint if and only if $\ker(H_{\max}-z)=\{0\}$ for one $z$ in each
open half-plane. Let $z=\alpha\pm i$, which belongs to $Z$ when $\alpha\in\mathbb J$, and let $f\in\ker(H_{\max}-z)$. Since $(H-z)f=0$ in $\cS'$, Proposition~\ref{prop:paramalpha} gives
$f=-K'_zf$. As $K'_z\in\OpA(S^{-\infty})$ maps $L^2$ into $H^\infty_A$ \cite{IMP}, $f\in H^m_A$. The space $C_c^\infty$ is dense in $H^m_A$, since cut-offs $\chi(u/R)$ commute with
$\Pi^\beta$ up to terms of order $R^{-1}$ and mollification converges in $H^m_A$ on compact sets, and $H$ is bounded from $H^m_A$ to $L^2$, being a polynomial of degree $m$ in $\Pi$ with
coefficients in $C_b^\infty$. Hence $f$ belongs to the domain of the closure $\bar H$ of $H|_{C_c^\infty}$, and $\bar Hf=zf$. Since $\bar H$ is symmetric and
$\Im z\ne0$, we obtain $f=0$. The inclusion $\cD(H)\subset\Wt^{-1}L^2$ was shown after the proof of Proposition~\ref{prop:paramalpha}, and for $r=0$, the parametrix $G'_z$ lies in $\OpA(S^{-m})$ and maps $L^2$ into $H^m_A$, so that $\cD(H)=H^m_A$.
\end{proof}

\begin{proof}[Proof of Theorem~\ref{thm:analysis}(b),(c)]
We start with (b). The identities $G'_z(H-z)=1+K'_z$ and $(H-z)G_z=1+K_z$ hold on $\cD(H)$ and on $L^2$, respectively, since $\cD(H)=\cD(H_{\max})$ by (a).
They give
\[
  R(z):=(H-z)^{-1}=G'_z-K'_zG_z+K'_zR(z)K_z .
\]
The operator $\Wt G'_z\Wt$ is bounded, $\Wt K'_zG_z\Wt=(\Wt K'_z\Wt)(\Wt^{-1}G_z\Wt)$ is bounded by Proposition~\ref{prop:paramalpha}(a), and $\Wt K'_z$ and $K_z\Wt$ are bounded
since $K_z$ and $K'_z$ are smoothing. With $\norm R\le d^{-1}$, this gives the weighted bound. For the momentum smoothing, let $\tilde\Phi$ be an almost analytic
extension of $\Phi$ supported in $Z$. Since $(H-z)^{-1}=G_z-(H-z)^{-1}K_z$ and $G_z$ is holomorphic on $Z$,
\[
  \Phi(H)=\frac1\pi\int\bar\partial\tilde\Phi(z)\,(H-z)^{-1}d^2z=-\frac1\pi\int\bar\partial\tilde\Phi(z)\,(H-z)^{-1}K_z\,d^2z ,
\]
where $K_z\Lm^N$ is bounded uniformly on $Z$ while $|\bar\partial\tilde\Phi|\norm{(H-z)^{-1}}\le C_M|\Im z|^{M-1}$. Taking adjoints gives the other side.

For the Combes--Thomas estimate, we first take $\phi$ bounded. The bounds will not depend on $\norm\phi_\infty$. By Lemma~\ref{lem:globalframe}(a), we may take the
frame symbol $U$ analytic in the strip $\{|\Im k|\le\varrho_0\}$. Then $U^*$ and the conjugated symbol $\tilde a_z$ are analytic in the same strip, with symbol
bounds there, by Lemma~\ref{s:lem:expconj}(a). Since $(H-z)^{-1}=\U\OpA(\tilde a_z)^{-1}\U^*$, we have
\[
  e^{\mu\phi}R(z)e^{-\mu\phi}=\bigl(e^{\mu\phi}\U e^{-\mu\phi}\bigr)\bigl(e^{\mu\phi}\OpA(\tilde a_z)e^{-\mu\phi}\bigr)^{-1}\bigl(e^{\mu\phi}\U^*e^{-\mu\phi}\bigr).
\]
By Lemma~\ref{s:lem:expconj}(b), $e^{\mu\phi}\OpA(p)e^{-\mu\phi}=\OpA(p)+\mu\OpA(p^{(1)}_\mu)$ for $|\mu|\le\varrho_0/2$ and every symbol $p$ analytic in the strip,
where each entry of $p^{(1)}_\mu$ has order one less than the corresponding entry of $p$, uniformly in $\mu$. Applied entrywise in the fixed splitting, this
gives $e^{\mu\phi}\OpA(\tilde a_z)e^{-\mu\phi}=\OpA(\tilde a_z)+\mu\OpA(e_\mu)$ with $e_\mu\in\mathcal S^{-1}_+$ uniformly, while the outer factors are bounded. We
write $\OpA(\tilde a_z)+\mu\OpA(e_\mu)=\mathcal D\bigl(S_0+\mu\mathcal D^{-1}\OpA(e_\mu)\mathcal D^{-1}\bigr)\mathcal D$, where $S_0=\mathcal D^{-1}\OpA(\tilde a_z)\mathcal D^{-1}$
satisfies $\norm{S_0^{-1}}=\norm{\mathcal D\U^{-1}R(z)(\U^*)^{-1}\mathcal D}\le C/d$ by the weighted bound, and $\mathcal D^{-1}\OpA(e_\mu)\mathcal D^{-1}$ is bounded by
Proposition~\ref{prop:paramalpha}(a). A Neumann series then gives the claim for $|\mu|\le cd$, with constants that do not depend on $\norm\phi_\infty$. For
a general $\phi$, we apply this bound to the bounded functions $\phi_n=\chi_n\circ\phi$, where $\chi_n$ is smooth with $\chi_n(t)=t$ for $|t|\le n$, $|\chi_n'|\le1$
and bounded higher derivatives, and let $n\to\infty$ on $C_c^\infty$, on which $e^{\pm\mu\phi_n}=e^{\pm\mu\phi}$ for $n$ large.

We now prove (c). By Lemma~\ref{s:lem:gauge}, $H_s=\Op^{A_0}(b_s)$ and $\U_s=\Op^{A_0}(U_s^\flat)$, where the transformed symbols depend continuously on $s$, and
the transformation acts entrywise and preserves the orders of the entries. The conjugated symbols in the quantization $\Op^{A_0}$ are therefore the
transforms of $\tilde a_s$, which depend continuously on $s$ in $\mathcal S^0_+$, and we may assume that $A_s=A_0$. Let $z=\alpha+i$, and let
$T_s:=\mathcal D^{-1}\OpA(\tilde a_{z,s})\mathcal D^{-1}$, which is bounded and depends continuously on $s$ in norm by Proposition~\ref{prop:paramalpha}(a), since the bounds
there depend linearly on finitely many seminorms in $\mathcal S^0_+$. Since $H_s$ is self-adjoint by (a), $H_s-z$ is invertible, and
$T_s^{-1}=\mathcal D\U_s^{-1}(H_s-z)^{-1}(\U_s^*)^{-1}\mathcal D=\U_s^*\Wt_s(H_s-z)^{-1}\Wt_s\U_s$, the identities being verified on $\cS$ and extended by
continuity. By (b), whose constants are uniform along the path, $\norm{T_s^{-1}}\le C$ uniformly in $s$. Hence $T_s^{-1}-T_{s'}^{-1}=T_s^{-1}(T_{s'}-T_s)T_{s'}^{-1}$
tends to $0$ as $s'\to s$. Since $(H_s-z)^{-1}=\U_s\mathcal D^{-1}T_s^{-1}\mathcal D^{-1}\U_s^*$, where $\U_s$ is norm continuous and $\mathcal D^{-1}$ is
bounded, the resolvent is norm continuous at $z$, and the resolvent identity extends this to all $z\notin\R$.
\end{proof}

\smallskip
\noindent\textbf{The freezing condition.} The following cases are those quoted after the definition of (F) in section~\ref{sec:hypotheses}.
\begin{lemma}\label{s:lem:Fsufficient}
Condition (F) holds in each of the following cases: (i) when (TI) holds; (ii) when $r=0$ and (N) holds; (iii) when $m=1$, $\partial_xE_{\ker}\equiv0$, and
(N) holds; (iv) when $H=L_B^*L_B+W$ with $L_B$ as in Proposition~\ref{prop:Meffcomp}(d), and (N2)--(N4) hold for the bulks.
\end{lemma}
\begin{proof}
In case (i), $H^{x_0}_{t,L}=H$. In case (ii), (N1) is ellipticity, which is preserved by freezing, and $\Fl=\emptyset$. In case (iii), (DN$_0$) is automatic
for $m=1$, and with a frame independent of $x$ the curvature correction $T$ of Proposition~\ref{prop:Meffcomp}(a) involves only $y$-derivatives of the frame and the value of $a_1$. Hence
$\Meff(H^{x_0}_{t,L})(x,y,\theta)=V_\sff^*W(x_0+t\chi_L(x-x_0),y)V_\sff+T(x_0+t\chi_L(x-x_0),y,\theta)=\Meff(H)(x_0+t\chi_L(x-x_0),y,\theta)$ takes values in $\Fl(H)$. In case (iv),
the operators $H^{x_0}_{t,L}$, obtained by freezing $\lsym$ and $W$, are factorized with the same structure, with $\lsym_p\lsym_p^*\ge c\jp k^{2p}$ uniformly,
and $\Meff=V_\sff^*WV_\sff$ is evaluated at the frozen point, so that its spectrum lies in $\Fl(H)$. The frame symbols of
Proposition~\ref{prop:Meffcomp}(d) depend continuously on $t$, since the symbols of $(1+L_tL_t^*)^{-1}$ do, by the resolvent identity
$(1+L_tL_t^*)^{-1}-(1+L_sL_s^*)^{-1}=(1+L_tL_t^*)^{-1}(L_sL_s^*-L_tL_t^*)(1+L_sL_s^*)^{-1}$ and the Beals-type characterization \cite{IMP2,CHP18}; no lower
bound on $L_tL_t^*$ is needed. In all cases the bulks are unchanged.
\end{proof}

\smallskip
\noindent\textbf{Weak fields and condition (W).} We prove Proposition~\ref{prop:weak} and give the cases in which (W) holds.
\begin{proof}[Proof of Proposition~\ref{prop:weak}]
Let $U^{(B)}$ be the frame symbols of (W), $U:=U^{(0)}$, and $\tilde a^{(0)}:=U^*\ab^{(0)}U$. Since $U$ is unitary-valued and $\alpha\in\mathcal G(\ab)$, the
pointwise inverse $q:=(\tilde a^{(0)}-\alpha)^{-1}=U^*(\ab^{(0)}-\alpha)^{-1}U$ exists for all $k$, uniformly for $\alpha\in K$, and it is of type $\mathcal S^0_-$
by Proposition~\ref{prop:paramalpha}(b) at large momentum. Let $\upsilon^{(B)}$ be defined by $U^{(B)*}\sharp^BU^{(B)}=1+\upsilon^{(B)}$. Since the symbols do not depend
on $u$, the field-free products are pointwise, so that $U^*\sharp^0U=1$, and Lemma~\ref{s:lem:Bcont} together with (W) gives $\upsilon^{(B)}=O(\norm
B_{C^{N_0}})$ in $S^{-1}$. We write
\[
  \bigl(\tilde a^{(B)}-\alpha(1+\upsilon^{(B)})\bigr)\sharp^Bq-1=\bigl(\tilde a^{(B)}-\tilde a^{(0)}-\alpha\upsilon^{(B)}\bigr)\sharp^Bq+\bigl((\tilde
  a^{(0)}-\alpha)\sharp^Bq-(\tilde a^{(0)}-\alpha)\sharp^0q\bigr)=:e^{(B)},
\]
where we used $(\tilde a^{(0)}-\alpha)\sharp^0q=(\tilde a^{(0)}-\alpha)q=1$. By (W) and Proposition~\ref{prop:paramalpha}(a), the first term belongs to $\mathcal S^0_0$ with
seminorms $O(\norm B_{C^{N_0}})$. By Lemma~\ref{s:lem:Bcont} applied entrywise, the second term belongs to $\mathcal S^{-2}_0$ with seminorms $O(\norm
B_{C^{N_0}})$. Hence $\mathcal D^{-1}\OpA(e^{(B)})\mathcal D$ is bounded with norm $O(\norm B_{C^{N_0}})$ by Proposition~\ref{prop:paramalpha}(a), and it has norm at most
$\frac12$ for $\norm B_{C^{N_0}}\le b_0$ small enough. Writing $1+\OpA(e^{(B)})=\mathcal D(1+\mathcal D^{-1}\OpA(e^{(B)})\mathcal D)\mathcal D^{-1}$, we obtain the
right inverse $G:=\OpA(q)\mathcal D(1+\mathcal D^{-1}\OpA(e^{(B)})\mathcal D)^{-1}\mathcal D^{-1}$ of $\tilde A_\alpha=\OpA(\tilde a^{(B)}-\alpha(1+\upsilon^{(B)}))$, with
$\mathcal DG\mathcal D$ bounded since $q\in\mathcal S^0_-$. A left inverse is obtained in the same way. Hence $\tilde A_\alpha$ is invertible between the weighted
spaces, and so is $H-\alpha=(\U^*)^{-1}\tilde A_\alpha\U^{-1}$, so that $\alpha\notin\spec\OpA(\ab^{(B)})$. The same holds for every energy in a neighborhood of
$K$ of fixed size. For $r=0$, the frame is trivial and the argument only uses $q=(\ab-\alpha)^{-1}\in S^{-m}$.

For a constant field $0<|B|\le b_0$, the same computation applies to the reduced operators. Indeed, $K_B$ is a homomorphism on constant-coefficient symbols,
so that $K_B(\tilde a^{(B)}-\alpha(1+\upsilon^{(B)}))K_B(q)=K_B(1+e^{(B)})$ with the same symbol $e^{(B)}$. By the metaplectic normalization of
Lemma~\ref{lem:reduction}(d), $K_B(c)$ is unitarily equivalent to $\Op^w_h(c\circ\varsigma_B)$ with $h=|B|$, and constant-coefficient symbols of order $\nu$ belong to
the flat class $S(\jp Y^\nu)$ uniformly. The semiclassical calculus with order functions \cite{Zworski} then gives the analogue of Proposition~\ref{prop:paramalpha}(a) for the
weights $\Op^w_h(\mathrm{diag}(1,\jp Y^{m/2}))$, with constants independent of $h$, and the argument above shows that $K_B(\ab^{(B)})-\alpha$ is invertible, that
is, $K\cap\spec K_B(\ab)=\emptyset$.
\end{proof}

\begin{proposition}\label{s:prop:Wholds}
Condition (W) holds when $r=0$, when $m\le2$, and for factorized operators $L_B^*L_B+W$ whose bulk symbols are of the form $\lsym^*\sharp^B\lsym+W$, where $L_B=\OpA(\lsym)$ with $\lsym(k)$ a constant
$(n-r)\times n$ polynomial symbol of order $p$, $1\le r<n$, whose principal part satisfies $\lsym_p\lsym_p^*\ge c|k|^{2p}$, and $W$ is a constant Hermitian matrix.
In all three cases, $\Meff$ does not depend on the field, so that (W5) holds at every $\alpha\notin\Fl$.
\end{proposition}
\begin{proof}
In all three cases, the symbols do not depend on $u$, so that the field-free products are pointwise. For $r=0$, we take $U^{(B)}=1$. Then $\tilde
a^{(B)}=\ab$ and there is nothing to prove. For $m\le2$, we take for $U^{(B)}=U$ a unitary-valued frame symbol, independent of $B$, equal to a homogeneous
frame of $\ker\ab_m(\theta)$ for $|k|\ge R$. By Lemma~\ref{s:lem:Bcont} applied twice, $\tilde a^{(B)}-U^*\ab U=U^*\sharp^B\ab\sharp^BU-U^*\sharp^0\ab\sharp^0U$
belongs to $S^{m-2}$ with seminorms $O(\norm B_{C^{N_0}})$. Since $m-2\le0$, $m-2\le\frac m2$ and $m-2\le m$, each block lies in the class required by (W), and
(DN$_0$) holds for $\tilde a^{(B)}$ since it holds for $U^*\ab U$. Condition (W4) holds for $r=0$ with the trivial frame, and for $m\le2$
because every frame symbol satisfies (DN$_0$) for every field and $\Meff$ does not depend on the field (Propositions~\ref{prop:flat}(a) and \ref{prop:Meffcomp}(c)), so
that the frame of $H$ can be kept fixed while the field varies. For the factorized case, $\ab^{(B)}=\lsym^*\sharp^B\lsym+W$, and
$\ab^{(B)}-\ab^{(0)}\in S^{2p-2}$ with seminorms $O(\norm B_{C^{N_0}})$ by Lemma~\ref{s:lem:Bcont}. Since $\lsym_p\lsym_p^*\ge c|k|^{2p}$, there is $R'$ such that
$\lsym(k)$ is onto for $|k|\ge R'$, so that $\ker\lsym(k)$ has rank $r$ there. We let $V_\sff(k)$ be an orthonormal frame of $\ker\lsym(k)$ and $V_\stt(k)$ one
of its orthogonal complement for $|k|\ge R'$, equal to a homogeneous frame of $\ker\lsym_p(\theta)$ up to $S^{-1}$, and we extend $U^{(0)}:=[V_\sff,V_\stt]$ to a
unitary-valued symbol on $\{|k|\le R'\}$, after removing the winding of the determinant as in Lemma~\ref{lem:globalframe}(a). Let $g^{(B)}$
be the symbol of $(1+L_BL_B^*)^{-1}\in\OpA(S^{-2p})$ and $r_K^{(B)}=1-\lsym^*\sharp^Bg^{(B)}\sharp^B\lsym$ that of $(1+L_B^*L_B)^{-1}$, as in the proof of
Proposition~\ref{prop:Meffcomp}(d). At $B=0$, all products are pointwise, $r_K^{(0)}=(1+\lsym^*\lsym)^{-1}$, and $r_K^{(0)}V_\sff=V_\sff$ for
$|k|\ge R'$. Writing
\[
  g^{(B)}-g^{(0)}=g^{(B)}\sharp^B\bigl((1+\lsym\lsym^*)\sharp^0g^{(0)}-(1+\lsym\sharp^B\lsym^*)\sharp^Bg^{(0)}\bigr),
\]
and using Lemma~\ref{s:lem:Bcont} for the difference between $\sharp^0$ and $\sharp^B$ and for $\lsym\sharp^B\lsym^*-\lsym\lsym^*$, together with the uniform bounds on
$g^{(B)}$ in $S^{-2p}$ given by the inverse-closedness of the calculus \cite{IMP,IMP2,CHP18}, we obtain $g^{(B)}-g^{(0)}=O(\norm B_{C^{N_0}})$ in $S^{-2p-2}$, and
hence $r_K^{(B)}-r_K^{(0)}=O(\norm B_{C^{N_0}})$ in $S^{-2}$. We set $\tilde V_\sff^{(B)}:=V_\sff+\chi_R(r_K^{(B)}\sharp^BV_\sff-V_\sff)$ with $R\ge R'$ and
$U^{(B)}:=[\tilde V_\sff^{(B)},V_\stt]$. Then $U^{(B)}$ is a frame symbol for $R$ large, $U^{(B)}$ equals $U^{(0)}$ at $B=0$, which is unitary-valued, and
$U^{(B)}-U^{(0)}=O(\norm B_{C^{N_0}})$ in $S^{-2}$, which is (W1). As in the proof of Proposition~\ref{prop:Meffcomp}(d),
$\lsym\sharp^B\tilde V_\sff^{(B)}\in S^{-p}$, and it vanishes for $|k|\ge2R$ at $B=0$, so that it is $O(\norm B_{C^{N_0}})$ in $S^{-p-2}$ there. The flat block of
$\tilde a^{(B)}$ is $(\lsym\sharp^B\tilde V_\sff^{(B)})^*\sharp^B(\lsym\sharp^B\tilde V_\sff^{(B)})+\tilde V_\sff^{(B)*}\sharp^BW\sharp^B\tilde V_\sff^{(B)}$, and the
off-diagonal ones are $(\lsym\sharp^BV_\stt)^*\sharp^B(\lsym\sharp^B\tilde V_\sff^{(B)})+V_\stt^*\sharp^BW\sharp^B\tilde V_\sff^{(B)}$ and its adjoint. They are of
order $0$, which gives (W2), and they differ from their values at $B=0$ by $O(\norm B_{C^{N_0}})$ in $S^0$. The transverse block differs from its value at
$B=0$ by $O(\norm B_{C^{N_0}})$ in $S^{2p-2}$, which gives (W3). The same argument, applied with two fields $B$ and $B'$ instead of $B$ and $0$, shows that
$U^{(B)}$ and $\tilde a^{(B)}$ depend on $B$ in a Lipschitz way in these classes, and (W4) follows from Proposition~\ref{prop:Meffcomp}(d)
applied to the factorized operator $H$, whose frame symbols depend continuously on the field by the same argument. Finally,
$\Meff^{(B)}=V_\sff^*WV_\sff$ does not depend on the field, which gives (W5). In the cases $r=0$ and $m\le2$, (W5) holds since $\Fl=\emptyset$, respectively
since $\Meff$ does not depend on the field (Proposition~\ref{prop:Meffcomp}(c)). This proves (W).
\end{proof}

\smallskip
\noindent\textbf{The flat block in a field.} We prove Proposition~\ref{prop:Meffcomp}(b) and describe how the field enters the flat block at
higher order. The computation of the conjugated symbol relies on the Moyal product of three factors, which we write in exponential form \cite[Ch.~18]{Hormander3}. For
$B=0$, we have $a_1\sharp a_2\sharp a_3=\exp\bigl(\frac i2\sum_{p<q}\sigma_{pq}\bigr)(a_1a_2a_3)$, where
$\sigma_{pq}=\partial_u^{(p)}\cdot\partial_k^{(q)}-\partial_k^{(p)}\cdot\partial_u^{(q)}$ acts on the $p$-th and $q$-th factors. If $V$ does not depend on $u$, then
$\sigma_{13}=0$ in $V^*\sharp a\sharp V$, the exponent is $X=\frac i2\sum_i\partial^{(2)}_{u_i}(\partial^{(3)}_{k_i}-\partial^{(1)}_{k_i})$, and
\begin{align}
  V^*\sharp a\sharp V&=V^*aV+\frac i2\sum_j\bigl(V^*\partial_{u_j}a\,\partial_{k_j}V-\partial_{k_j}V^*\,\partial_{u_j}a\,V\bigr)\notag\\
  &\quad-\frac18\sum_{i,j}\Bigl(V^*\partial_{u_iu_j}a\,\partial_{k_ik_j}V-2\,\partial_{k_i}V^*\,\partial_{u_iu_j}a\,\partial_{k_j}V+\partial_{k_ik_j}V^*\,\partial_{u_iu_j}a\,V\Bigr)+r_3,\label{s:eq:triple}
\end{align}
where $r_3$ contains three derivatives in $k$.

\begin{proof}[Proof of Proposition~\ref{prop:Meffcomp}(b)]
We first show that, for $m=2$ and $B=0$, when $\ker a_2(u,\theta)$ does not depend on $u$, (DN$_0$) holds if and only if $V_\sff^*a_1V_\sff=0$, and $\Meff$ is the
degree-$0$ part of
\begin{multline}\label{s:eq:Meff2}
  V_\sff^*WV_\sff+\frac i2\sum_j\Bigl(V_\sff^*\,\partial_{u_j}a_1\,\partial_{k_j}V_\sff-\partial_{k_j}V_\sff^*\,\partial_{u_j}a_1\,V_\sff\Bigr)\\
  +\frac14\sum_{i,j}\partial_{k_i}V_\sff^*\,\partial_{u_i}\partial_{u_j}a_2\,\partial_{k_j}V_\sff-\bigl(V_\sff^*\check a_1V_\stt\bigr)\tilde a_2^{-1}\bigl(V_\stt^*\check a_1V_\sff\bigr),
\end{multline}
with $\tilde a_2=V_\stt^*a_2V_\stt$ and $\check a_1=a_1+\frac i2\sum_j(\partial_{u_j}a_2\,\partial_{k_j}E_{\ker}-\partial_{k_j}E_{\ker}\,\partial_{u_j}a_2)$.
Here $\partial_uV_\sff=0$, and $a_2(u)V_\sff=0$ for all $u$ gives $\partial_u^\alpha a_2\,V_\sff=0$ and $V_\sff^*\partial_u^\alpha a_2=0$. We
apply \eqref{s:eq:triple} with $a=a_2+a_1+W$. The degree-$2$ flat term $V_\sff^*a_2V_\sff$ vanishes. At degree $1$, the flat block is $V_\sff^*a_1V_\sff$, since the
first-order term of $a_2$ contains the factors $\partial_ua_2V_\sff$ or $V_\sff^*\partial_ua_2$. Thus (DN$_0$) forces $V_\sff^*a_1V_\sff=0$, and all blocks then have
the required orders. At degree $0$, the flat block consists of $V_\sff^*WV_\sff$, of the first-order term of $a_1$, and of the second-order term of $a_2$, in
which only the middle term of \eqref{s:eq:triple} survives and gives $\frac14\sum\partial_{k_i}V_\sff^*\,\partial_{u_iu_j}a_2\,\partial_{k_j}V_\sff$. The off-diagonal
block at degree $1$ is $V_\sff^*a_1V_\stt$ plus the first-order term of $a_2$, namely $\frac i2\sum_j(\partial_{k_j}V_\sff^*)\,\cdot\,(-\partial_{u_j}a_2V_\stt)$ and its
transverse counterpart. Written in terms of $E_{\ker}$, this is $V_\sff^*\check a_1V_\stt$. The Schur correction at degree $0$ is therefore the last term of
\eqref{s:eq:Meff2}. In the general case, when $V_\sff$ depends on $u$, all six terms of the first-order product contribute. Using $\partial a_2V_\sff=-a_2\partial V_\sff$ and
$V_\sff^*\partial a_2=-\partial V_\sff^*a_2$, they reduce to $\frac i2(\partial_kV_\sff^*a_2\partial_uV_\sff-\partial_uV_\sff^*a_2\partial_kV_\sff)$, which is of degree
$1$. Together with $V_\sff^*a_1V_\sff$, this is the degree-$1$ flat block $f_1$, which must vanish for (DN$_0$) to hold. Conversely, if $f_1\equiv0$, the flat
block has order $0$, and the off-diagonal blocks have order at most $1=\frac m2$ since $a_2V_\sff=0$, so that (DN$_0$) holds.
\end{proof}

Proposition~\ref{prop:Meffcomp}(e) follows from the
results below and from Proposition~\ref{s:prop:exactB}. Throughout, $B$ is constant unless stated otherwise, and the kernel of $a_m$ does not
depend on $u$.

\begin{proposition}\label{s:prop:hier}
Assume (DN$_0$) at $B=0$ for a homogeneous frame. Then $a_jV_\sff=0$ for every $j>\frac m2$. Moreover, (a) the first-order magnetic terms of $a_j$, $j>\frac m2$,
vanish in the flat block; (b) those of $a_j$, $j\le\frac m2$, have degree $j-2$, and hence are absent from $\Meff$ for $m\le3$ and may be present for $m=4$,
through $j=2$; (c) for $a_m=\rho^mQ(\theta)$, the off-diagonal first-order magnetic terms of $a_m$ are $-\frac i2Bm\rho^{m-2}(\partial_\theta V_\sff)^*V_\stt\,\tilde Q$, with
$\tilde Q=V_\stt^*QV_\stt$ invertible, of degree $m-2$. They respect the block orders of (DN$_0$) in this frame if and only if $m\le4$ or
$(\partial_\theta V_\sff)^*V_\stt\equiv0$, and their Schur contribution has degree $m-4$. For a constant symbol, they are the only off-diagonal terms of degree
$m-2$ at $B\ne0$, since (DN$_0$) at $B=0$ forces $a_{m-2}V_\sff=0$ when $m-2>\frac m2$, and the conclusion holds in every homogeneous frame, since two such
frames differ by a block-diagonal $g(\theta)$ and the first-order magnetic term of two symbols that depend on $k$ only through $\theta$ vanishes.
\end{proposition}
\begin{proof}
We prove $a_jV_\sff=0$ by downward induction on $j$. If $a_{j'}V_\sff=0$ for all $j'>j$ and all $u$, then $\partial_u^\alpha a_{j'}V_\sff=0$, so that the Moyal
corrections of the $a_{j'}$ to the flat block and to the off-diagonal block at degree $j$ vanish, since each of them contains a factor $\partial^\alpha_ua_{j'}V_\sff$
or its adjoint, as $\partial_uV_\sff=0$. The flat block at degree $j\ge1$ is then $V_\sff^*a_jV_\sff$, and the off-diagonal one is $V_\stt^*a_jV_\sff$. Condition
(DN$_0$) forces both to vanish when $j>\frac m2$, and Hermiticity gives $a_jV_\sff=0$. Part (a) follows from the observation on the magnetic Moyal terms at the beginning of this section, and (b) is a degree count. For (c),
the terms of the pairs $(a_m,V_\stt)$ and $(V_\sff^*,V_\stt)$ cancel by $V_\sff^*\partial a_m=-\partial V_\sff^*a_m$, as in the proof of Proposition~\ref{prop:Meffcomp}(c) in
section~\ref{sec:engine}. In the remaining pair $(V_\sff^*,a_m)$, we write
$\partial_{k_i}a_m=m\rho^{m-1}\hat k_iQ+\rho^m(\partial_{k_i}\theta)\partial_\theta Q$ and antisymmetrize against
$\partial_{k_j}V_\sff=(\partial_{k_j}\theta)\partial_\theta V_\sff$. Only the radial part survives, with $\hat k_2\partial_{k_1}\theta-\hat k_1\partial_{k_2}\theta=-\rho^{-1}$,
and $QV_\stt=V_\stt\tilde Q$ since $QV_\sff=0$. For a constant symbol, the non-magnetic Moyal terms vanish, the first-order magnetic terms of $a_j$, $j<m$,
have degree at most $m-3$, and the pointwise term $V_\sff^*a_{m-2}V_\stt$ vanishes by the first statement.
\end{proof}
For third-order operators, the situation is intermediate.
\begin{remark}\label{s:rem:m3}
For $m=3$, the first-order magnetic terms do not enter $\Meff$, by Proposition~\ref{s:prop:hier}. The second-order magnetic terms have degree $m-3=0$, and they
contain a $u$-derivative of $a_3$ or a derivative of $B$. Hence $\Meff$ does not depend on the field for $m=3$ if $a_3$ does not depend on $u$ and $B$ is
constant, but it may in general. In particular, field independence does not extend to all third-order systems.
\end{remark}
The parity of the symbol under $k\mapsto-k$ constrains the flat terms of odd degree.
\begin{proposition}\label{s:prop:parity}
For differential operators, the frame can be chosen $\pi$-periodic in $\theta$, with zero determinant winding. Then the degree-$j$ part of the flat block, and
of its Schur complement, satisfies $f_j(u,\theta+\pi)=(-1)^jf_j(u,\theta)$ for every field. Nonzero flat terms of odd degree are therefore indefinite.
\end{proposition}
\begin{proof}
Since $\ker a_m(u,\theta)=\ker a_m(u,\theta+\pi)$, the kernel bundle descends to $\R^2\times\R P^1$, where it is again trivial. A frame there is $\pi$-periodic,
and a $\pi$-periodic loop has even winding, which can be removed by $e^{-iw\theta}$ with $w$ even. Every term of the Moyal expansion is a product of factors of
definite parity under $k\mapsto-k$. Each $k$-derivative changes the parity and lowers the degree by one, while $u$-derivatives and powers of $B$ change
neither. Finally, a Hermitian $f$ with $f(\theta+\pi)=-f(\theta)\ne0$ has eigenvalues of both signs.
\end{proof}
When the degree-one flat term does not vanish, the essential spectrum fills large sets.
\begin{remark}\label{s:rem:disp}
Let $m=2$ with a kernel depending on $u$, let $\Meff^0$ be the degree-$0$ part of the flat block, and let $\Sigma$ be the closure of
$\bigcup_{u,\theta,t\ge0}\spec(\Meff^0+tf_1)$. Then interior points of $\Sigma$ lie in $\spec_{\rm ess}H$. If $f_1$ is indefinite somewhere, then $\Sigma=\R$. If
$f_1\ge0$ (resp. $f_1\le0$), then energies below (resp. above) $\Sigma$ are not in $\spec_{\rm ess}H$. Consequently, by Proposition~\ref{s:prop:parity}, a
second-order differential operator with constant kernel rank either has $f_1\equiv0$, and is then in (DN$_0$), or has $\spec_{\rm ess}H=\R$. For the first statement, consider wavepackets at points $u_\rho$ near the zero set of $f_1$, of momentum $\rho\theta$ and width $w$ with $\rho^{-1}\ll w\ll(\rho|\nabla f_1(u_\rho)|)^{-1}$.
They are quasimodes at energy $\mu(u_\rho)+\rho f_1(u_\rho)$, where $\mu$ is an eigenvalue of $\Meff^0$. Such widths exist because $f_1$ vanishes to infinite
order where the strip meets the bulks. The second statement follows from the intermediate value theorem in $\theta$, and the third one from a Feshbach
argument.

For every $m$, when the kernel depends on $u$, the flat block of a Peierls operator contains, already without field, the term
\[
  f_{m-1}=V_\sff^*a_{m-1}V_\sff+\frac i2\sum_j\bigl(\partial_{k_j}V_\sff^*\,a_m\,\partial_{u_j}V_\sff-\partial_{u_j}V_\sff^*\,a_m\,\partial_{k_j}V_\sff\bigr)
\]
of degree $m-1$, which is $f_1$ for $m=2$. For $m=1$, it is part of $\Meff$, and for $m\ge2$ it must vanish for (DN$_0$) to hold in homogeneous frames. For
the twisted symbols $U\rho^{m-2}k^\perp(k^\perp)^TU^*$, $U=e^{i\beta(y)\sigma_z}$, $f_{m-1}=\beta'(y)\rho^{m-1}\cos\theta\sin2\theta$ is indefinite, whereas the
factorized operator $U\curl_B^*\curl_BU^*$ keeps an exact flat band (Example~\ref{s:ex:curl}). The symbols $|k|^{m-2}k^\perp(k^\perp)^T$, $m=6,8$, coupled
through the factors $\lsym=|k|^2(k^\perp)^T$ and $\lsym=(\xi+i\zeta)^3(k^\perp)^T$, satisfy all hypotheses.
\end{remark}
For the Peierls versions of these two symbols, by contrast, (N1) fails at every $B\ne0$ (Proposition~\ref{prop:Meffcomp}(e)), so that the results apply to
them only at $B=0$, and the deformation $t\mapsto tB$ of Corollary~\ref{cor:bulk}(b) is not available. The
explicit field terms for the examples of Proposition~\ref{prop:Meffcomp}(e) are computed in Proposition~\ref{s:prop:exactB}.
In Proposition~\ref{s:prop:exactB}(a), the Schur term of Proposition~\ref{s:prop:hier}(c) alone would give $-\frac{m^2}4B^2=-4B^2$, while the
second-order magnetic terms contribute $+\frac{11}4B^2$, which gives the coefficient $-\frac54$.

\smallskip
\noindent\textbf{Notation for the remaining proofs.} Throughout, $U$ is a frame symbol for which (DN$_0$) holds, with $U^*\sharp^BU=1+\upsilon$, $\tilde a_z=\tilde a-z(1+\upsilon)$, $s=(s_\sff,s_\stt)=(0,\frac m2)$ and
$\mathcal D=\mathrm{diag}(1,\Lm^{m/2})$. A matrix symbol $c$ belongs to the graded class $\mathcal S^\mu_\epsilon$ if on $\{|k|\ge R\}$ its blocks satisfy
$c_{ij}\in S^{\epsilon s_i+\epsilon s_j+\mu}$ for $\epsilon=\pm$ and $c_{ij}\in S^{s_i-s_j+\mu}$ for $\epsilon=0$. For a constant symbol $\ab$ and $B\ne0$,
$K_B(\ab)=\Op^w_y(\ab(By,\zeta))$ is the reduced bulk operator, $\ell(y,\zeta)=(By,\zeta)$, $\Gamma^\mu_{3/4}$ is the isotropic class of symbols $b$ on
$\R^2_X$, $X=(y,\zeta)$, with $|\partial^\beta b|\le C\jp X^{\mu-\frac34|\beta|}$, $\Gamma^\mu$ the class with $\frac34$ replaced by $1$, and $\tau$ is the pairing \eqref{eq:tau}.

\smallskip
\noindent\textbf{The gap of the references.}
\begin{proof}[Proof of Proposition~\ref{prop:gap}]
We give the argument for the reduced bulk references, in a frame for which $\U$ is unitary, so that the reference operator is unitarily equivalent to
$K_B(b_{\rm ref})$; the fibers
of the interface reference are treated in the same way in the fiber calculus of Lemma~\ref{s:lem:admissible}, with $\Lm_{\xi,R}$ in place of $\Lm_R$. Let
$g_R$ be the metric $|dX|^2/\Lm_R(\ell(X))^2$, where $\Lm_R=(R^2+\jp k^2)^{1/2}$. By the argument of Lemma~\ref{s:lem:admissible}, since $\Lm_R\circ\ell$ is
Lipschitz with constant $\max(|B|,1)$ and $\Lm_R\ge R$, the metric $g_R$ is admissible uniformly in $R\ge1$, with Planck function $h_R=(\Lm_R\circ\ell)^{-2}\le
R^{-2}$, and $\Lm_R^{\mu}\circ\ell$ is a $g_R$-admissible weight for every $\mu$.

\emph{Step 1: the symbols.} Since $E^{\rm ext}$ is smooth on the unit disk, $\Sigma_R=1-2\tilde E_R$ satisfies $|\partial_k^\beta\Sigma_R|\le C_\beta R^{-|\beta|}$ on
$\{|k|\le R\}$, while $\Sigma_R=1-2\tilde E^\infty\circ\theta$ has $k$-derivatives of order $\beta$ bounded by $C_\beta|k|^{-|\beta|}$ on $\{|k|\ge R\}$. Hence
$\Sigma_R\circ\ell\in S(1,g_R)$ uniformly in $R$, and the same holds for $\Sigma_R^2=1$. The diagonal weight $D_R=\mathrm{diag}(1,\Lm_R^{m/2})$ has blocks in
$S(1,g_R)$ and $S(\Lm_R^{m/2},g_R)$, uniformly in $R$.

\emph{Step 2: factorization.} Let $\mathcal D_R:=\mathrm{diag}(1,\Op^w(\Lm_R^{m/2}\circ\ell))$. Its transverse block is self-adjoint, and by (T1) and
(T4) it is invertible for $R$ large, with $\mathcal D_R^{-1}=\mathrm{diag}(1,\Op^w(\Lm_R^{-m/2}\circ\ell))(1+O(R^{-1}))$. By (T1), each block of
$b_{\rm ref}\circ\ell-(D_R\circ\ell)\#(\Sigma_R\circ\ell)\#(D_R\circ\ell)$ belongs to $S(d_id_jh_R,g_R)$, where $d_\sff=1$ and $d_\stt=\Lm_R^{m/2}\circ\ell$,
since each correction contains a derivative of $D_R$ or of $\Sigma_R$. Therefore
\[
  T_R:=\mathcal D_R^{-1}K_B(b_{\rm ref})\mathcal D_R^{-1}=K_B(\Sigma_R)+O(R^{-1})
\]
in operator norm, by (T2) applied blockwise after multiplication by the inverse weights.

\emph{Step 3: inversion.} By (T1), $(\Sigma_R\circ\ell)\#(\Sigma_R\circ\ell)-1\in S(h_R,g_R)$ uniformly, so that $K_B(\Sigma_R)^2=1+O(R^{-1})$. The operator
$T_R$ is self-adjoint and bounded, and $T_R^2=1+O(R^{-1})$, so that $\spec T_R\subset\{|\lambda|\ge\frac12\}$ and $\norm{T_R^{-1}}\le2$ for $R$ large. For real
$\lambda$, we write $K_B(b_{\rm ref})-\lambda=\mathcal D_R(T_R-\lambda\mathcal D_R^{-2})\mathcal D_R$. Since $\mathcal D_R\ge\frac12$ for
$R$ large, $\norm{\lambda\mathcal D_R^{-2}}\le4|\lambda|<\frac12\le\norm{T_R^{-1}}^{-1}$ for $|\lambda|<\frac18$, and $T_R-\lambda\mathcal D_R^{-2}$ is then
invertible by a Neumann series. Thus $K_B(b_{\rm ref})-\lambda$ is invertible between the weighted spaces, and $(-\frac18,\frac18)$ does
not meet its spectrum. For $r=0$, $\mathcal D_R=\Op^w(\Lm_R^{m/2}\circ\ell)$ is scalar, and the same argument, with $\mathcal D_R^{-2}\le CR^{-m}$, shows that the gap contains
$(-cR^m,cR^m)$. All constants depend only on finitely many seminorms of $E^{\rm ext}$ and of the frame, and not on $B$ in a compact set of nonzero
fields, nor on $\xi$ for the fibers.
\end{proof}

\smallskip
\noindent\textbf{Large-parameter and functional calculus.} We next construct a parametrix that is uniform in the spectral parameter for one-dimensional operators. Let $c(X)$ be a Hermitian symbol on
$\R^2_X$, $X=(y,\zeta)$, written in a fixed splitting, with $c_{\sff\sff}\in\Gamma^0$, $c_{\sff\stt},c_{\stt\sff}\in\Gamma^{m/2}$, $c_{\stt\stt}\in\Gamma^m$ and
$|c_{\stt\stt}(X)v|\ge c_0\jp X^m|v|$ for $|X|\ge R$. This is the situation of the reduced bulk operators in a unitary frame (Lemma~\ref{lem:globalframe}(b)), and of the reference and deformation symbols of Lemma~\ref{lem:path}. For $\mu\ge1$, let $\Lm_\mu(X)=(\jp X^2+\mu^{2/m})^{1/2}$,
$\omega_\sff=\mu$, $\omega_\stt=\Lm_\mu^m$, and $\hat\omega=\mathrm{diag}(\omega_\sff^{1/2},\omega_\stt^{1/2})$. The entries of $\hat\omega^{\pm1}$ are order functions with
constants independent of $\mu$, they are admissible for the isotropic metric $g=|dX|^2/\jp X^2$ uniformly in $\mu$, and
$|\partial^\beta\hat\omega^{\pm1}|\le C_\beta\hat\omega^{\pm1}\Lm_\mu^{-|\beta|}\le C_\beta\mu^{-|\beta|/m}\hat\omega^{\pm1}$. For a symbol $d$, we write $\hat
d=\hat\omega^{-1}d\hat\omega^{-1}$ pointwise.
\begin{lemma}\label{s:lem:normalized}
Let $\lambda$ be bounded, $t\in\R$ with $\mu:=|t|\ge1$, and $c_t:=c-\lambda-it$. (i) The normalized symbol $\hat c_t$ is bounded in $S(1,g)$ uniformly in
$t$, and its derivatives of order at least one are bounded by $C\mu^{-1/m}$. (ii) For $\mu\ge\mu_0$, $\hat c_t(X)$ is invertible for all $X$, its
off-diagonal blocks and those of its inverse are $O(\mu^{-1/2})$, and $\hat c_t^{-1}$ has the properties stated in (i).
\end{lemma}
\begin{proof}
On $\{|X|\le2R\}$, $c$ is bounded with its derivatives and $\hat\omega^{-1}\le\mu^{-1/2}$, so that the normalized contribution of $c$ there is $O(\mu^{-1})$
with all its derivatives. For $|X|\ge R$, $(c_{\sff\sff}-\lambda)/\mu$ is bounded in $S(1,g)$ with derivatives $O(\mu^{-1})$, and $-it/\mu=\mp i$ is constant.
The entry $c_{\sff\stt}/(\mu\omega_\stt)^{1/2}$ is bounded by $C\jp X^{m/2}\mu^{-1/2}\Lm_\mu^{-m/2}\le C\mu^{-1/2}$, and its derivatives satisfy the same bound
with an extra factor $\jp X^{-|\beta|}$, using $\jp X\le\Lm_\mu$ and the bounds on $\hat\omega^{-1}$. The entry $(c_{\stt\stt}-\lambda-it)/\omega_\stt$ is bounded
since $|c_{\stt\stt}|\le C\jp X^m$ and $|t|\le\Lm_\mu^m$, and its derivatives are $O(\jp X^{m-1}\Lm_\mu^{-m}+\Lm_\mu^{-1})=O(\mu^{-1/m})$, with the extra factor
$\jp X^{-|\beta|}$ for derivatives of higher order. This proves (i). For (ii), the Hermitian matrix $c_{\stt\stt}-\lambda$ satisfies $|(c_{\stt\stt}-\lambda)v|\ge
c\jp X^m|v|$ for $|X|\ge R'$, so that $|(c_{\stt\stt}-\lambda-it)v|^2\ge(c^2\jp X^{2m}+t^2)|v|^2\ge c'\omega_\stt^2|v|^2$; for $|X|\le R'$, the same bound holds for
$\mu\ge\mu_0$ since $c_{\stt\stt}$ is bounded there. The normalized transverse block therefore has a uniformly bounded inverse. Since the normalized
off-diagonal blocks are $O(\mu^{-1/2})$, the normalized Schur complement is $(c_{\sff\sff}-\lambda)/\mu\mp i+O(\mu^{-1})$, which is invertible with inverse of
norm at most $2$ for $\mu\ge\mu_0$. Block inversion gives $\hat c_t^{-1}$, with off-diagonal blocks $O(\mu^{-1/2})$, and the identity
$\partial(\hat c_t^{-1})=-\hat c_t^{-1}(\partial\hat c_t)\hat c_t^{-1}$ gives the bounds on the derivatives.
\end{proof}
Compositions of normalized symbols are pointwise products up to small errors.
\begin{lemma}\label{s:lem:normcomp}
Let $c$ and $d$ be matrix symbols such that $\hat\omega^{-1}c\hat\omega^{-1}$ and $\hat\omega d\hat\omega$ belong to $S(1)$, with derivatives of order at least
one bounded by $\epsilon$. Then $c\#d=cd+r$, where the entries of $r$ satisfy $r_{ij}\in S(\omega_i^{1/2}\omega_j^{-1/2})$ with seminorms
$O(\epsilon+\mu^{-1/m})$. If moreover $\hat\omega^{-1}c\hat\omega^{-1}$ and $\hat\omega d\hat\omega$ are bounded in $S(1,g)$, then also $r_{ij}\in
S(\omega_i^{1/2}\omega_j^{-1/2}\jp X^{-2},g)$ uniformly. The same holds for $d\#c$, with $\omega_i^{-1/2}\omega_j^{1/2}$ in place of $\omega_i^{1/2}\omega_j^{-1/2}$.
The statement also holds with the weights $\mathrm{diag}(1,\Lm_R^{m/2})$, $\Lm_R=(R^2+\jp X^2)^{1/2}$, in place of $\hat\omega$ and $R^{-1}$ in place of
$\mu^{-1/m}$.
\end{lemma}
\begin{proof}
The entries $c_{il}=\omega_i^{1/2}\hat c_{il}\omega_l^{1/2}$ belong to $S(\omega_i^{1/2}\omega_l^{1/2})$, and their first derivatives belong to the same class with
seminorms $O(\epsilon+\mu^{-1/m})$, since a derivative falls either on $\hat c_{il}$ or on a weight. Similarly, $d_{lj}\in S(\omega_l^{-1/2}\omega_j^{-1/2})$ with
small first derivatives. By (C1) with $\hbar=1$, $r_{ij}=\sum_lR_1(c_{il},d_{lj})$ belongs to $S(\omega_i^{1/2}\omega_j^{-1/2})$, with seminorms bounded by
products of seminorms of first derivatives, which are $O(\epsilon+\mu^{-1/m})$. The second statement follows from (C2) with $h_g=\jp X^{-2}$. The last
statement is proved in the same way.
\end{proof}
\begin{proposition}\label{s:prop:largeparam}
There is $\mu_0$ such that, for $|t|\ge\mu_0$, the resolvent $(\Op^w(c)-\lambda-it)^{-1}$ has the symbol $Q_t+e_t$, where $Q_t=(c-\lambda-it)^{-1}$ pointwise,
and where the entries of $e_t$ satisfy, for all $\beta$ and with $\mu=|t|$,
\[
  |\partial^\beta e_{t,ij}|\le C_\beta\,\mu^{-1/m}(\omega_i\omega_j)^{-1/2},\qquad |\partial^\beta e_{t,ij}|\le C_\beta\,(\omega_i\omega_j)^{-1/2}\jp X^{-2-|\beta|},
\]
with constants independent of $t$.
\end{proposition}
\begin{proof}
By Lemma~\ref{s:lem:normalized}, $Q_t=\hat\omega^{-1}\hat c_t^{-1}\hat\omega^{-1}$, with $\hat c_t^{-1}$ bounded in $S(1,g)$ and derivatives $O(\mu^{-1/m})$.
Lemma~\ref{s:lem:normcomp} gives $c_t\#Q_t=1+r$, with $r_{ij}$ in $S(\omega_i^{1/2}\omega_j^{-1/2})$ with seminorms $O(\mu^{-1/m})$ and in
$S(\omega_i^{1/2}\omega_j^{-1/2}\jp X^{-2},g)$ uniformly. Let $\Omega=\Op^w(\hat\omega)$ and $\Omega'=\Op^w(\hat\omega^{-1})$. By (C1) and (C2), $\Omega'\Omega=1+\Op^w(s)$
and $\Omega\Omega'=1+\Op^w(s')$ with $s,s'$ diagonal, of size $O(\mu^{-1/m})$ in $S(1)$ and bounded in $S(\jp X^{-2},g)$, so that $\Omega$ and $\Omega'$ are invertible
for $\mu\ge\mu_0$. Then $\Omega'(1+\Op^w(r))\Omega=1+\Op^w(\rho)$ with $\rho=s+\hat\omega^{-1}\#r\#\hat\omega$, which by (C1) and (C2) is $O(\mu^{-1/m})$ in $S(1)$ and
bounded in $S(\jp X^{-2},g)$. For $\mu\ge\mu_0$, $\norm{\Op^w(\rho)}\le\frac12$ by (C3), and (C4) gives $(1+\Op^w(\rho))^{-1}=1+\Op^w(\tilde\rho)$, where
$\tilde\rho=-\rho\#(1+\tilde\rho)$ has the same properties as $\rho$. Since $\Op^w(c)$ is self-adjoint and $t\ne0$, the right inverse
$\Op^w(Q_t)(1+\Op^w(r))^{-1}=\Op^w(Q_t)\Omega(1+\Op^w(\tilde\rho))\Omega'$ is the resolvent. Its symbol is $Q_t+e_t$ with
$e_t=Q_t\#(\hat\omega\#\hat\omega^{-1}-1)+Q_t\#\hat\omega\#\tilde\rho\#\hat\omega^{-1}$, and the entries of $Q_t$ belong to $S(\omega_i^{-1/2}\omega_j^{-1/2})$ and to
$S(\omega_i^{-1/2}\omega_j^{-1/2},g)$. The composition theorems (C1) and (C2) then give the two bounds.
\end{proof}

\begin{proof}[Proof of Proposition~\ref{prop:gradedFC}]
By Lemma~\ref{lem:globalframe}(b) in the isotropic calculus, we may use a frame for which $\hat\U$ is unitary. Then $\hat P_\alpha=\hat\U P\hat\U^*$
with $P=\mathbf 1_{(-\infty,0)}(\Op^w(c))$, where $c=\tilde a\circ\ell-\alpha$ is of the type considered above, and $\Op^w(c)$ is self-adjoint, by the isotropic
version of Theorem~\ref{thm:analysis}(a), and invertible, since $\alpha\notin\spec K_B(\ab)$. For a self-adjoint invertible operator $T$, the
spectral theorem and $\int_{-T'}^{T'}(\lambda-it)^{-1}dt=2\arctan(T'/\lambda)$ give $\mathbf 1_{(-\infty,0)}(T)=\frac12-\frac1{2\pi}\lim_{T'\to\infty}\int_{-T'}^{T'}(T-it)^{-1}dt$
strongly. We split the integral into $|t|\ge\mu_0$ and $|t|\le\mu_0$.

For $|t|\ge\mu_0$, Proposition~\ref{s:prop:largeparam} gives the resolvent symbol $Q_t+e_t$. Interpolating the two bounds on $e_t$ with exponents $\frac14$ and
$\frac34$, and using $(\omega_i\omega_j)^{-1/2}\le\mu^{-1}$, we obtain $|\partial^\beta e_t|\le C_\beta\mu^{-1-1/(4m)}\jp X^{-\frac32-\frac34|\beta|}$, which is integrable
in $t$. Hence $\int_{|t|\ge\mu_0}e_t\,dt\in\Gamma^{-3/2}_{3/4}$. For $|t|\le\mu_0$, the isotropic version of Proposition~\ref{prop:paramalpha},
whose construction is uniform for $t$ in a compact set, gives $(\Op^w(c)-it)^{-1}=\Op^w(\chi_{R_1}Q_t)+\Op^w(f_t)$, where the entries of $f_t$ belong to
$\Gamma^{-1}$ uniformly in $t$, and the terms $K'_t(\Op^w(c)-it)^{-1}K_t$ have Schwartz kernels and belong to $\Op^w\Gamma^{-\infty}$. Hence
$\int_{|t|\le\mu_0}f_t\,dt\in\Gamma^{-1}$. Finally, the pointwise terms give $\frac12-\frac1{2\pi}{\rm p.v.}\int(c(X)-it)^{-1}dt=\mathbf 1_{(-\infty,0)}(c(X))$ for
$|X|\ge2R_1$, and a smooth compactly supported function for $|X|\le2R_1$. For $|X|$ large, block perturbation theory for the Hermitian matrix $c(X)$, whose
blocks have sizes $O(1)$, $O(\jp X^{\lfloor m/2\rfloor})$ and at least $c\jp X^m$, gives $\mathbf 1_{(-\infty,0)}(c)-\mathrm{diag}(E^\infty_{\rm fl},E^\infty_{\rm
tr})\circ\theta\circ\ell\in\Gamma^{-1}$, since the mixing $c_{\sff\stt}c_{\stt\stt}^{-1}$ has order $\lfloor m/2\rfloor-m\le-1$, the degrees being integers. Altogether
$P=\Op^w(p)$ with $p-\mathrm{diag}(E^\infty_{\rm fl},E^\infty_{\rm tr})\circ\theta\circ\ell\in\Gamma^{-1}_{3/4}$, and conjugation by $\hat\U$, whose symbol is the frame
at infinity up to $\Gamma^{-1}$, gives $\hat P_\alpha\in\cP_\varepsilon(E^\infty_\ell)$ for every $\varepsilon\in(0,\frac13)$.

For a $C^1$ family $c_s$ with fixed $E^\infty$, we differentiate the resolvents,
$\partial_s(\Op^w(c_s)-it)^{-1}=-(\Op^w(c_s)-it)^{-1}\Op^w(\partial_sc_s)(\Op^w(c_s)-it)^{-1}$. The pointwise terms give $-Q_t(\partial_sc_s)Q_t$, whose principal
value integral is $2\pi\partial_s\mathbf 1_{(-\infty,0)}(c_s(X))$, an element of $\Gamma^{-1}$ since $E^\infty$ is fixed. The remaining terms are compositions in
which at least one factor is $e_t$ or a composition error. Since $\hat\omega\partial_sc_s\hat\omega^{-1}$ is bounded in the normalized classes, they satisfy the
two bounds of Proposition~\ref{s:prop:largeparam}, and the same interpolation shows that their integral belongs to $\Gamma^{-3/2}_{3/4}$, continuously in $s$.
Hence $s\mapsto p_s$ is $C^1$ in the topology of $\cP_\varepsilon(E^\infty_\ell)$. The same argument applies to the reference and deformation symbols.
\end{proof}
\smallskip
\noindent\textbf{Semiclassical limit.} In the following proof, the assumption of Proposition~\ref{prop:semiclassical} that the inverses are of
graded type $\mathcal S^0_-$ means that $\hat\omega_1c_h^{-1}\hat\omega_1$ is bounded in $S(1,g)$, where $\hat\omega_1$ is the weight above with $\mu=1$.
\begin{proof}[Proof of Proposition~\ref{prop:semiclassical}]
We show that $\Op^w_h(c_h)$ is invertible for $h\le h_0$, that $P_h=\Op^w_h(\hat u_hQ_h\hat u_h^*+hr_h)$ with $r_h$ bounded in $\Gamma^{-3/2}_{3/4}$, and the
formula for $\tau(P_h)$. We follow the previous proof with the semiclassical products. For every $t\in\R$, with $\mu=\max(|t|,1)$, the pointwise inverse $Q_t=(c_h-it)^{-1}$
satisfies the conclusions of Lemma~\ref{s:lem:normalized}, since $c_h$ is invertible for all $Y$. By (C1) and (C2), $(c_h-it)\#_h Q_t=1+h
r$, where the entries of $r$ satisfy the bounds of Lemma~\ref{s:lem:normcomp} with $\epsilon=\mu^{-1/m}$, for every $t$. The argument of
Proposition~\ref{s:prop:largeparam}, in which the smallness now comes from the factor $h$ for $h\le h_0$, gives that $\Op^w_h(c_h)-it$ is
invertible for all $t$, and that its inverse has the symbol $Q_t+h e_t$, where $e_t$ satisfies the two bounds of Proposition~\ref{s:prop:largeparam}. Since
$c_h(Y)$ is invertible for all $Y$, the pointwise terms integrate to $Q_h$ everywhere, and the interpolation of the previous proof gives
$\mathbf 1_{(-\infty,0)}(\Op^w_h c_h)=\Op^w_h(Q_h+h r'_h)$ with $r'_h$ bounded in $\Gamma^{-3/2}_{3/4}$. Conjugation by
$\hat\U_h$ gives the stated form, since $\hat u\#_h Q\#_h\hat u^*-\hat uQ\hat u^*\in h\Gamma^{-2}$ by (C2).

For the trace, let $q=\hat uQ\hat u^*+h r$ be the symbol of $P=P_h$, acting on $L^2(\R_s)$ with dual variable $\sigma$. Since $s$ and $h D_s$ are
quantizations of linear symbols, $[P,s]=-ih\Op^w_h(\partial_\sigma q)$ and $[P,D_s]=i\Op^w_h(\partial_sq)$ exactly, so that
$P[[P,D_s],[P,s]]=h\Op^w_h\bigl(q\#_h(\partial_sq\#_h\partial_\sigma q-\partial_\sigma q\#_h\partial_sq)\bigr)$. By (C2), this symbol equals
$h\,q[\partial_sq,\partial_\sigma q]+h^2e_h$, where $e_h$ is bounded in $\Gamma^{-7/2}_{3/4}$, using $\partial q\in\Gamma^{-1}_{3/4}$ and the gain of each
composition. By (C5), $\tau(P)=-i\Tr(P[[P,D_s],[P,s]])=-\frac i{2\pi}\int\tr\bigl(q[\partial_sq,\partial_\sigma q]\bigr)\,ds\,d\sigma+O(h)$. Write
$Q'=\hat uQ\hat u^*$, which equals $E(\theta(Y))$ up to $\Gamma^{-1}$. In $q[\partial_sq,\partial_\sigma q]$, the terms containing $h r$ are $O(h)$ in $L^1$,
since $r\in\Gamma^{-3/2}_{3/4}$, $\partial r\in\Gamma^{-9/4}_{3/4}$ and $\partial Q'\in\Gamma^{-1}$. Moreover, $Q'[\partial_sQ',\partial_\sigma Q']\in\Gamma^{-3}$, since the
leading part $E\circ\theta$ depends on a single angle and the commutator of its derivatives vanishes. Since $c_1(Q')=\frac i{2\pi}\tr(Q'[\partial_sQ',\partial_\sigma
Q'])\,ds\wedge d\sigma$, we obtain $\tau(P)=-\int c_1(Q')+O(h)$.
\end{proof}

\section{Proofs for the interface current}\label{s:sec:routine2}
We recall that $a=a_\pm(k)$ and $B=B_\pm$ are constant on the bulks
$\{\pm y\ge y_0\}$, that $\delta=\min\{\dist(\alpha,\Fl),\dist(\alpha,\spec H_+\cup\spec H_-)\}$ is the margin, with $H_\pm$ as in Lemma~\ref{s:lem:bulks}
below, that $\mathcal E_\alpha=(\alpha-\frac\delta2,\alpha+\frac\delta2)$ is the energy window, that $\Wt=(\U^*)^{-1}\mathcal D\U^{-1}$ with
$\mathcal D=\mathrm{diag}(1,\Lm^{m/2})$ is the weight of Theorem~\ref{thm:analysis}, and that, unless stated otherwise, constants depend only on $n$, $m$, $c_0$,
$y_0$, $|\alpha|$, $\delta$, the frame, and finitely many seminorms of the coefficients and of $B$.

\smallskip
\noindent\textbf{Bulk operators and localization.} 
\begin{lemma}\label{s:lem:bulks}
There are potentials $A_\pm$ with $\curl A_\pm=B_\pm$ and $A_\pm=A$ on $\{\pm y\ge y_0\}$, and frames $U_\pm$ with $U_\pm=U$ on $\{\pm y\ge y_0\}$, such that the
operators $H_\pm:=\Op^{A_\pm}(a_\pm)$ satisfy (N1) and (N3), $H\tilde\theta=H_\pm\tilde\theta$ and $[H,\tilde\theta]=[H_\pm,\tilde\theta]$ for $\tilde\theta\in C_b^\infty$
supported in $\{\pm y>y_0\}$, and $\spec H_\pm=\spec K_{B_\pm}(a_\pm)$.
\end{lemma}
\begin{proof}
On the half-plane $\{\pm y\ge y_0\}$, the form $A-(-B_\pm y,0)$ is closed, hence equal to $\nabla\lambda$. We extend $\lambda$ and set $A_\pm=(-B_\pm
y,0)+\nabla\lambda$. The identities hold because $H$ and $H_\pm$ are differential operators with the same coefficients there. The frame
$U_\pm(x,y,\theta):=U(x,\pm y_0,\theta)$ spans $\ker a_{\pm,m}(\theta)$. Since $U$ does not depend on $y$ on $\{\pm y\ge y_0\}$, all its $y$-derivatives vanish
on $\{y=\pm y_0\}$, so that every term of the expansion of the conjugated symbol of $H_\pm$ at $(x,y,k)$ coincides with the corresponding term of the
conjugated symbol of $H$ at $(x,\pm y_0,k)$, the field and the symbol being $B_\pm$ and $a_\pm$ near that point. Hence (DN$_0$) holds for $H_\pm$, with the
same effective flat symbol $\Meff(x,\pm y_0,\theta)$. The spectrum is gauge invariant, and it equals $\spec K_{B_\pm}(a_\pm)$ by Theorem~\ref{thm:bulkspec} if $B_\pm\ne0$, and by
the Fourier transform if $B_\pm=0$.
\end{proof}
Since the frames agree on the bulk half-planes, compositions with $\Wt$ and with $\Wt_\pm$ agree there up to remainders decaying rapidly away from the strip,
by pseudolocality (Lemma~\ref{s:lem:pseudoloc}). In particular, $\Wt^{-1}X\Wt_\pm^{-1}$ is bounded for a differential operator $X$ of Douglis--Nirenberg order
$\le-1$ with coefficients supported in $\{y_0\le\pm y\le y_0+1\}$. This allows us to prove the localization result.
\begin{proof}[Proof of Proposition~\ref{prop:loc}]
The argument is standard \cite{EGS,BEC}, with the weighted estimates of Theorem~\ref{thm:analysis}. We start with the decay in the normal direction. Let $\theta_++\theta_-+\theta_0=1$ be a partition of unity in $y$ with $\supp\theta_\pm\subset\{\pm y\ge
y_0+1\}$, and let $\tilde\theta_\pm=1$ on $\supp\theta_\pm$, supported in $\{\pm y>y_0\}$. With $R=(H-z)^{-1}$ and $R_\pm=(H_\pm-z)^{-1}$, we have $R\theta_\pm=\tilde\theta_\pm
R_\pm\theta_\pm-R[H_\pm,\tilde\theta_\pm]R_\pm\theta_\pm$. The first term is holomorphic near $\mathcal E_\alpha$ and does not contribute to the Helffer--Sj\"ostrand
integral. The second term is $(R\Wt)(\Wt^{-1}[H_\pm,\tilde\theta_\pm]e^{c\jp y}\Wt_\pm^{-1})(\Wt_\pm e^{-c\jp y}R_\pm e^{c\jp y})\theta_\pm$. The middle factor is bounded
by the observation above, the last one by Theorem~\ref{thm:analysis}(b) applied to $H_\pm$, and $\norm{R\Wt}\le C|\Im z|^{-1}$. The general
statement follows from the same argument with $H'$ in place of $H_\pm$. The decay along the wall follows from Theorem~\ref{thm:analysis}(b) with $\phi$ a
regularized distance to $\supp\chi_2$ and $\mu=c|\Im z|$, together with the Helffer--Sj\"ostrand formula split at $|\Im z|=L^{-1/2}$. To combine the two
estimates, we write $\Phi=\Phi_1\Phi_2\Phi_3$, use momentum smoothing for the outer factors and the decay estimates for the inner ones, and move weights past
each other at the cost of bounded errors. The kernel bound then follows from the diamagnetic bound $|K_T(u,u')|\le C\norm{\Lm^2T\Lm^2}$ of Lemma~\ref{s:lem:diamag}.
\end{proof}

\smallskip
\noindent\textbf{Quantization.} We fix $\varphi$ and set $g:=e^{2\pi i\varphi}-1\in C_c^\infty(\mathcal E_\alpha)$, $\Omega:=g(H)$ and
$U:=U_\varphi=1+\Omega$. Let $\cP$ be the set of bounded real functions $P(x)$ equal to $0$ near $-\infty$ and to $1$ near $+\infty$, which contains smooth switches
and the sharp projections $\mathsf P_{x_0}=\mathbf 1_{[x_0,\infty)}(x)$. The argument follows \cite[Sec.~4]{BEC} and uses only Proposition~\ref{prop:loc}.
\begin{lemma}[index identities]\label{s:lem:tc}
Let $K\subset \mathcal E_\alpha$ be compact. There are $C$ and $k$ such that the following holds for $f\in C_c^\infty(K)$. (i) The operators $\vartheta f(H)$ and
$f(H)\vartheta$ are trace-class for $\vartheta\in C_c^\infty(\R_x)$, with norm at most $C\norm f_{C^k}$. (ii) For $P\in\cP$, $[P,f(H)]$ is trace-class with
$\norm{[P,f(H)]}_1\le C\norm f_{C^k}$, and $\Tr[P,f(H)]=0$. (iii) For smooth $P\in\cP$, $i[H,P]f(H)$ is trace-class.
(iv) For $P\in\cP$ and $u,v\in C_c^\infty(\mathcal E_\alpha)$, let $t(u,v):=\Tr\bigl(u(H)[P,v(H)]\bigr)$. Then there is $c\in\C$, independent of $P$, such that $t(u,v)=c\int
uv'\,dE$. Moreover, $\Tr(U^*PU-P)=2\pi ic$.
(v) Let $U$ be unitary and $\Pi$ an orthogonal projection with $[\Pi,U]\in\mathfrak S_1$. Then $\Pi U\Pi|_{\Ran\Pi}$ is Fredholm, and $\Ind(\Pi
U\Pi|_{\Ran\Pi})=-\Tr(U^*\Pi U-\Pi)$.
\end{lemma}
\begin{proof}
Parts (iv) and (v) are standard \cite{ASS,EG,BEC}; we include the short arguments. For (i), we write $f=ff_2$ with $f_2=1$ near $K$, and $\vartheta f(H)=(\vartheta f(H)\jp x)(\jp x^{-1}f_2(H))$. By Proposition~\ref{prop:loc}(c),
both factors have square-integrable kernels, and are therefore Hilbert--Schmidt. For (ii), assume that $P=0$ on $(-\infty,-R]$ and $P=1$ on $[R,\infty)$. The
kernel $(P(x)-P(x'))K_f$ vanishes unless $\max(|x|,|x'|)\le R+|x-x'|$, so that it is bounded by $C\jp x^{-3}\jp{x'}^{-3}e^{-c(|y|+|y'|)}$, and $[P,f]=([P,f]\jp
x)(\jp x^{-1}f_2)+(f\jp x^{-1})(\jp x[P,f_2])$ is a sum of products of Hilbert--Schmidt operators. For the trace, we write $P=\mathsf P_0+Q$ with $Q$ bounded and
compactly supported. The operator $[\mathsf P_0,f(H)]$ has vanishing diagonal blocks with respect to $\mathsf P_0$, so that its trace vanishes, and with $\vartheta Q=Q$,
$\Tr(Qf)=\Tr(\vartheta fQ)=\Tr(f\vartheta Q\vartheta)=\Tr(fQ)$ by cyclicity with one trace-class factor. For (iii), $[H,P]$ is a differential operator of order $m-1$
with coefficients supported in $\supp P'$. Hence $[H,P]=[H,P]\vartheta$ with $\vartheta\in C_c^\infty$ equal to $1$ near $\supp P'$, and
$i[H,P]f(H)=([H,P]\Lm^{-m})(\Lm^m\vartheta f(H))$, where the second factor is trace-class as in (i), using Proposition~\ref{prop:loc}(a).

For (iv), By (i)--(iii) and cyclicity, the form $t$ has the following properties: (a) $t(u,v)+t(v,u)=\Tr[P,uv]=0$; (b) $t(u,vw)=t(uv,w)+t(wu,v)$; (c)
$t(u,v)=0$ if $\supp u\cap\supp v=\emptyset$, by (b) with $w=1$ near $\supp v$ and $w=0$ near $\supp u$; and (d) $t(u,\theta)=0$ if $\theta=1$ near $\supp u$, since
by (b) and (c), $t(u,\theta)=t(u,\theta^2)=2t(u,\theta)$. Fix $e\in C_c^\infty(\mathcal E_\alpha)$ with $e(E)=E$ near a compact set $K$, and set $\lambda(\rho):=t(\rho,e)$ for
$\rho\in C_c^\infty(K)$, so that $|\lambda(\rho)|\le\norm\rho_\infty\norm{[P,e(H)]}_1$. By (b) and induction, $t(u,e^{n+1})=(n+1)t(ue^n,e)$, so that
$t(u,q(e))=\lambda(uq')$ for polynomials $q$ with $q(0)=0$. Approximating $v$ on a neighborhood of $\supp u$ by such polynomials, and using (c), (d) and
(ii), we obtain $t(u,v)=\lambda(uv')$. If $\int\rho=0$, then $\rho=f'$ with $f\in C_c^\infty$, and with $\theta=1$ near $\supp f$, (a) and (d)
give $\lambda(\rho)=t(\theta,f)=-t(f,\theta)=0$. Hence $\lambda=c\int$. To show the independence of $P$, let $P,P'\in\cP$. Then $Q=P-P'$ is compactly supported,
and $t_P-t_{P'}=\Tr(uQv)-\Tr(uvQ)=0$ by cyclicity. Finally, with $\bar g=e^{-2\pi i\varphi}-1$, we have $U^*PU-P=[P,\Omega]+\bar g(H)[P,\Omega]$, whose trace is
$0+t(\bar g,g)=c\int(1+\bar g)g'\,dE=c\int2\pi i\varphi'\,dE=2\pi ic$.

For (v), Let $T=\Pi U\Pi|_{\Ran\Pi}$ and $\Pi^\perp=1-\Pi$. The operators $1-T^*T=\Pi U^*\Pi^\perp U\Pi$ and $1-TT^*=\Pi U\Pi^\perp U^*\Pi$ are trace-class, so that $T$ is
Fredholm with $\Ind T=\Tr(1-T^*T)-\Tr(1-TT^*)$ by Calder\'on's formula. The diagonal blocks of $U^*\Pi U-\Pi$ with respect to $\Pi$ have traces $-\Tr(\Pi
U^*\Pi^\perp U\Pi)$ and $\Tr(\Pi^\perp U^*\Pi U\Pi^\perp)=\Tr(\Pi U\Pi^\perp U^*\Pi)$.
\end{proof}
\begin{proof}[Proof of Theorem~\ref{thm:quant}]
The kernel bound is Proposition~\ref{prop:loc}. By Lemma~\ref{s:lem:tc}(ii), $[\mathsf P_0,U]=[\mathsf P_0,\Omega]\in\mathfrak S_1$, so that
Lemma~\ref{s:lem:tc}(iv),(v) give $\iota:=\Ind(\mathsf P_0U\mathsf P_0)=-\Tr(U^*\mathsf P_0U-\mathsf P_0)=-2\pi ic$, with $c$ independent of $P$ and $\varphi$. In
particular, $-\Tr(U^*PU-P)=\iota$ for every $P\in\cP$, and $\iota$ is the same for all switches $\varphi$ with $\varphi'\in C_c^\infty(\mathcal E_\alpha)$. For smooth $P$, we write
$\varphi'=\varphi_1\varphi_2$. Since $\varphi_1(H)i[H,P]\varphi_2(H)=i\bigl((E\varphi_1)(H)[P,\varphi_2(H)]-\varphi_1(H)[P,(E\varphi_2)(H)]\bigr)$,
Lemma~\ref{s:lem:tc}(iv) gives
\[
  \sigma_I=\Tr\bigl(\varphi_1(H)i[H,P]\varphi_2(H)\bigr)=ic\int\bigl(E\varphi_1\varphi_2'-\varphi_1(E\varphi_2)'\bigr)dE=-ic\int\varphi_1\varphi_2\,dE=\frac \iota{2\pi}.
\]
Gauge transformations commute with $P$. For $H=D_x$ on $L^2(\R)$, one has $2\pi\sigma_I=\int\varphi'=1$, which fixes the orientation.
\end{proof}

\smallskip
\noindent\textbf{Locality and freezing.} 
\begin{proof}[Proof of Theorem~\ref{thm:local}]
The argument is the locality argument of \cite{EGS,BEC}, adapted to the weights of Theorem~\ref{thm:analysis}. For (a), we prove the
following precise form: if $H$ and $H'$ satisfy the hypotheses of Theorem~\ref{thm:quant} with common constants, and
agree on the slab $S_L=\{|x-x_0|<L\}$, with the same coefficients and field, potentials and frame symbols equal on $S_L$ and $A-A'$ bounded, then
$|\sigma_I(H)-\sigma_I(H')|\le C_NL^{-N}$.
We use for both operators a switch $P$ with $\supp P'\subset[x_0-1,x_0+1]$, and we let $\vartheta\in C_c^\infty$ be equal to $1$ near $[x_0-2,x_0+2]$. Then
$[H,P]=[H',P]=\vartheta[H,P]=[H,P]\vartheta$. We write $\varphi'=f_1f_2$ with $f_j\in C_c^\infty(\mathcal E_\alpha)$, and
$\varphi'(H)-\varphi'(H')=f_1(H)(f_2(H)-f_2(H'))+(f_1(H)-f_1(H'))f_2(H')$. The operators $X=i[H,P]f_1(H)=\vartheta X$ and $Y=f_2(H')i[H,P]=Y\vartheta$ are trace
class by Theorem~\ref{thm:quant}, and cyclicity of the trace gives
\[
  |\sigma_I(H)-\sigma_I(H')|\le\norm X_1\norm{(f_2(H)-f_2(H'))\vartheta}+\norm Y_1\norm{\vartheta(f_1(H)-f_1(H'))}.
\]
Let $\chi\in C_c^\infty(S_L)$ be equal to $1$ on $S_{L/2}$, with $|\chi^{(j)}|\le C_jL^{-j}$. Since $\chi H'=\chi H$ and $[H',\chi]=[H,\chi]$, we have $(H'-z)\chi
R=\chi+[H,\chi]R$, so that $\chi R=R'\chi+R'[H,\chi]R$. Since $\vartheta=\vartheta\chi$, this gives $\vartheta(R-R')=\vartheta R'(\chi-1)+\vartheta R'[H,\chi]R$. The first
term is $O(|\Im z|^{-1}e^{-c|\Im z|L})$ by the Combes--Thomas estimate for $H'$. The second term is $(\vartheta R'\Wt')(\Wt'^{-1}[H,\chi]\Wt^{-1})(\Wt R)$. Its middle
factor is bounded, since $[H,\chi]$ has Douglis--Nirenberg order $-1$ and coefficients supported in $S_L$, where the frames and, up to rapidly decaying tails,
the weights agree. Moreover, $\norm{\vartheta R'\Wt'}\le C|\Im z|^{-1}e^{-c|\Im z|L}$ by the weighted Combes--Thomas bound. The Helffer--Sj\"ostrand formula and the
adjoint estimate conclude the proof of (a), and integrality gives its last statement.

For (c), We prove the following precise form. Let $y_1<y_2$ with $y_2-y_1=L$, and let $H$, $H^{(1)}$ and $H^{(2)}$ satisfy the hypotheses of Theorem~\ref{thm:quant} with common constants. Assume that $H^{(1)}=H$ on $\{y<y_2-L/3\}$, that $H^{(2)}=H$ on $\{y>y_1+L/3\}$, and that the three operators are
equal, on $\{y_1+L/4<y<y_2-L/4\}$, to an operator $H_{\rm mid}$ satisfying (N1) and (N3) with $\alpha\notin\spec H_{\rm mid}$. Then
$|\sigma_I(H)-\sigma_I(H^{(1)})-\sigma_I(H^{(2)})|\le C_NL^{-N}$.

Let $\eta_1+\eta_2=1$ in $y$, with $\eta_1=1$ for $y\le y_1+\frac{L}2-\frac L{24}$, $\eta_1=0$ for $y\ge y_1+\frac L2+\frac L{24}$, and $|\eta_j^{(l)}|\le C_lL^{-l}$.
Since $i[H,P]\varphi'(H)$ is trace-class, $\sigma_I(H)=\Tr(i[H,P]\varphi'(H)\eta_1)+\Tr(i[H,P]\varphi'(H)\eta_2)$, and likewise for $H^{(1)}$ and $H^{(2)}$.

We first compare the pieces. The operators $H$ and $H^{(1)}$ agree on $\{y<y_2-L/3\}$, which contains $\supp\eta_1$ at distance at least $L/8$ from its
boundary. The argument of Theorem~\ref{thm:local}(a), with the cut-off $\chi(y)$ equal to $1$ on $\{y<y_2-2L/5\}$ and supported in $\{y<y_2-L/3\}$ in place of
$\chi(x)$, and with $\eta_1$ in place of $\vartheta$, gives $\norm{(f(H)-f(H^{(1)}))\eta_1}\le C_NL^{-N}$ for $f\in C_c^\infty(\mathcal E_\alpha)$. Here the trace-class
factors are localized in $x$ by $i[H,P]=i[H^{(1)},P]$ on $\{y<y_2-L/3\}$, and the contributions of $\{y\ge y_2-L/3\}$ are controlled by the decay of $f(H)$ and
$f(H^{(1)})$ inside the region where they agree with $H_{\rm mid}$ (Proposition~\ref{prop:loc}). Hence
$|\Tr(i[H,P]\varphi'(H)\eta_1)-\Tr(i[H^{(1)},P]\varphi'(H^{(1)})\eta_1)|\le C_NL^{-N}$, and similarly for $\eta_2$ and $H^{(2)}$.

We next show that the cross terms vanish asymptotically. The operator $H^{(1)}$ agrees with $H_{\rm mid}$ on $\{y_1+L/4<y\}$ up to the bulk $\{y\ge y_2-L/3\}$,
where it is a gapped operator as well. Therefore $\varphi'(H^{(1)})$ decays like $e^{-c\,\dist(y,\{y\le y_1+L/4\})}$ on $\supp\eta_2$ by Proposition~\ref{prop:loc},
and $|\Tr(i[H^{(1)},P]\varphi'(H^{(1)})\eta_2)|\le C_NL^{-N}$. The same holds for $H^{(2)}$ and $\eta_1$. Adding the four estimates gives (c).

For (b), We choose the frames of $H^{x_0}_{t,L}$ continuous in $t$, as in (F), and the gauges $A_t=(-\int_0^yB_t\,ds,0)$, for which $A_t-A_s$ is bounded since
the fields agree for $|y|\ge y_0$. By Theorem~\ref{thm:analysis}(c), the resolvents are norm continuous in $t$, so that $U_{\varphi,t}$ is norm continuous,
$[\mathsf P_{x_0},U_{\varphi,t}]$ is trace-class, and the index of $\mathsf P_{x_0}U_{\varphi,t}\mathsf P_{x_0}$ is constant. Hence $\sigma_I(H^{x_0}_{1,L})=\sigma_I(H^{[x_0]})$. Since $H^{x_0}_{1,L}=H$ on
$\{|x-x_0|<L\}$, part (a) gives $\sigma_I(H)=\sigma_I(H^{x_0}_{1,L})$ for $L$ large. Along the path, (N3) holds with a uniform margin, so that
$E^\infty(x_0,\cdot,\cdot)$ depends continuously on $t$ and $\Cy(x_0)$ is the same for $H$ and $H^{[x_0]}$. Finally, $\Cy(x_0)$ does not depend on $x_0$: on
$M=[x_0,x_1]\times[-y_0,y_0]\times S^1$, the form $(E^\infty)^*c_1$ is closed and smooth by (N3), it vanishes on the faces $\{y=\pm y_0\}$, where $E^\infty$
depends only on $\theta$, and Stokes' theorem shows that the integrals over the cylinders $\{x=x_0\}$ and $\{x=x_1\}$ agree.

For (d), the operator $\OpA(\ab)$ with the field of the statement satisfies $\alpha\notin\spec\OpA(\ab)$, together with a neighborhood of fixed size, by
Proposition~\ref{prop:weak}. Since $\sigma_I$ does not depend on $\varphi$ (Theorem~\ref{thm:quant}), we may choose $\varphi'$ supported in that neighborhood,
in which case $\varphi'(H)=0$ and $\sigma_I=0$.
\end{proof}

\smallskip
\noindent\textbf{Translation-invariant walls.} 
\begin{proof}[Proof of Theorem~\ref{thm:TI} and Lemma~\ref{lem:decouple}]
The proof has four steps: discreteness and the large-momentum gap, the decoupling of Lemma~\ref{lem:decouple}, the identification of the current with a
spectral flow, and its dependence on the energy.

\emph{Discreteness and large-momentum gap.} We use the fiber frames $\hat\U(\xi)$ and the conjugated fibers $\tilde A_z(\xi)$ of
section~\ref{sec:pBIC}. We first prove discreteness and analyticity. After weighting, the fiber parametrix has errors in $S(\Lm_\xi^{-1},g_\xi)$. Since $B_\pm\ne0$, we have
$|A_1(y)|\to\infty$, and hence $\Lm_\xi\to\infty$ at infinity in $\R^2_{y,\zeta}$, so that these errors are compact. Hence $\Hh(\xi)-z$ is Fredholm for $z\in Z$,
and $\spec\Hh(\xi)\cap\mathbb J$ is discrete. In the fixed frame, $\partial_\xi\tilde A_z(\xi)=\Op^w_y(\partial_{k_1}\tilde a)$ has Douglis--Nirenberg order $-1$, and
$\hat\U(\xi)$ is analytic in $\xi$ as a bounded operator. Together with the weighted resolvent bound, this gives norm-resolvent analyticity.

We next prove the large-momentum gap. Let $B_\pm>0$ and $\xi\to-\infty$, the other cases being analogous. The guiding center $y_c(\xi)$, defined by
$A_1(y_c)=\xi$, tends to $+\infty$. On $\{y\ge y_0\}$, we have $|\xi-A_1(y)|=B_+|y-y_c|$, and elsewhere $|\xi-A_1(y)|\ge|\xi|-M_0$ for a constant $M_0$ depending only on $B$ on $\{|y|\le y_0\}$. Let $L\le\frac14(y_c-y_0)$,
and let $\chi_{\rm c}^2+\chi_{\rm w}^2=1$ with $\chi_{\rm c}=1$ on $\{|y-y_c|\le L\}$, $\supp \chi_{\rm c}\subset\{|y-y_c|\le2L\}$ and $|\chi^{(l)}_\bullet|\le C_lL^{-l}$. On $\supp
\chi_{\rm w}$, the kinetic momentum is at least $\min(B_+L,|\xi|-M_0)\ge2R$. Let $\tilde A_+$ be the fiber of the conjugated bulk operator $\U^*(H_+-z)\U$, which is invertible for
$z\in \mathcal E_\alpha+i[-1,1]$ with $\norm{\mathcal D_\xi\tilde A_+^{-1}\mathcal D_\xi}\le C/\delta$, and let $G_{\rm w}$ be the fiber parametrix built from
Proposition~\ref{prop:paramalpha}(b) on $\{\Lm_\xi\ge R\}$. With $\mathcal R:=\chi_{\rm c}\tilde A_+^{-1}\chi_{\rm c}+\chi_{\rm w}G_{\rm w}\chi_{\rm w}$, we have
\[
  \tilde A_z\mathcal R=1+E,\quad E=[\tilde A_z,\chi_{\rm c}]\tilde A_+^{-1}\chi_{\rm c}+(\tilde A_z-\tilde A_+)\chi_{\rm c}\tilde A_+^{-1}\chi_{\rm c}+\chi_{\rm w}\Op^w(e)\chi_{\rm w}+[\tilde A_z,\chi_{\rm w}]G_{\rm w}\chi_{\rm w}.
\]
We bound $\mathcal D_\xi^{-1}E\mathcal D_\xi$. The commutators $[\tilde A_z,\chi_\bullet]$ have symbols $-i\chi_\bullet'\partial_\zeta\tilde a+\dots$, which are of
Douglis--Nirenberg order $-1$ with coefficient $O(L^{-1})$, and $\mathcal D_\xi^{-1}\chi_\bullet\mathcal D_\xi$ is bounded since $\chi_\bullet$ is scalar in the fixed
splitting. The symbols of $\tilde A_z$ and $\tilde A_+$ agree on $\{y\ge y_0\}$, which is at distance at least $\frac12(y_c-y_0)$ from $\supp \chi_{\rm c}$, so that the
second term is $O((y_c-y_0)^{-N})$ by pseudolocality. Finally, $e$ has Douglis--Nirenberg order $-1$ and is supported where $\Lm_\xi\ge R$, so that the third
term is $O(R^{-1})$. Altogether,
$\norm{\mathcal D_\xi^{-1}E\mathcal D_\xi}\le C(L^{-1}\delta^{-1}+L^{-1}+(y_c-y_0)^{-N}+R^{-1})\le\frac12$ for $R$ large, $L=C'(1+\delta^{-1})$ and
$|\xi|\ge Z_\alpha:=M_0+\max(2R,4B_+L)$. Then $\mathcal D_\xi^{-1}\tilde A_z\mathcal D_\xi^{-1}$ has a bounded right inverse, and by the same construction a left
inverse, for real $z\in \mathcal E_\alpha$. If the fields have opposite signs, the end with two guiding centers requires two pieces $\chi_{\rm c}^\pm$, while at the other
end the kinetic momentum is at least $|\xi|-M_0$ everywhere and $G_{\rm w}$ alone suffices.

\emph{Decoupling.} For Lemma~\ref{lem:decouple}, we run the proof of the large-momentum gap above with $L=\frac14\min(y_c-y_0)\to\infty$, for $z$ on the contour. Then $(\Hh-z)^{-1}=\hat\U\mathcal
R(1+E)^{-1}\hat\U^*$ with $\norm{\mathcal D_\xi^{-1}E\mathcal D_\xi}\to0$ uniformly on the contour. Moreover, $\oint G_{\rm w}\,dz=0$ since $G_{\rm w}$ is holomorphic
inside the contour, which avoids $\Fl$. Therefore $P(\xi)-\hat\U \chi_{\rm c}P_+'\chi_{\rm c}\hat\U^*\to0$, where $P'_+$ is the corresponding Riesz projection of the conjugated bulk operator.
Finally, $(1-\chi_{\rm c})P_+\to0$ in norm, since $P_+$ has finite rank and $\chi_{\rm c}\to1$ strongly after translation.

\emph{Spectral flow.} We now show that the current observable is a spectral flow. For $f\in C_c^\infty(\mathcal E_\alpha)$, the operator $f(\Hh(\xi))$ has finite rank, vanishes for
$|\xi|\ge Z_\alpha$, and is smooth in $\xi$ with eigenfunctions in $\cS(\R)$, by momentum smoothing and the Combes--Thomas estimate for the fibers. Hence the
kernel of $f(H)$, which is $K_f(x,y;x',y')=\frac1{2\pi}\int_{-Z_\alpha}^{Z_\alpha}e^{i(x-x')\xi}k_\xi(y,y')\,d\xi$, satisfies $(y,y',s)\mapsto
K_f(s,y;0,y')\in\cS(\R^3)$. Choose $\chi\in C_c^\infty(\mathcal E_\alpha)$ with $\chi\varphi'=\varphi'$, and set $\psi(E)=E\chi(E)$ and $\vartheta(E)=E\varphi'(E)$. Since $\chi(H)$ maps into
$H^\infty_A\subset\cD(H)$ and $P$ preserves $H^\infty_A$,
\[
  [H,P]\varphi'(H)=\bigl(H[P,\chi(H)]+[\psi(H),P]\bigr)\varphi'(H),
\]
and both terms have Schwartz kernels. For $x$-translation-invariant operators $A_1,A_2$ with kernels in $\cS(\R^3)$ in the variables $(y,y',x-x')$, the
\emph{shift identity} $\Tr([A_1,P]A_2)=\frac1{2\pi}\int\tr\bigl(-i\partial_\xi\hat A_1(\xi)\hat A_2(\xi)\bigr)d\xi$ holds. Indeed, integrating $P(x-s)-P(x)$ over
$x$ gives $-s$, and multiplication by $x$ acts on fibers as $i\partial_\xi$. We apply this identity to both terms, use cyclicity in the first one, and use
$\partial_\xi\psi(\Hh)=\partial_\xi\Hh\,\chi(\Hh)+\Hh\,\partial_\xi\chi(\Hh)$. The $\vartheta$-terms cancel, and we obtain
\begin{equation*}
  2\pi\sigma_I=\int_\R\tr\bigl(\partial_\xi\Hh(\xi)\,\varphi'(\Hh(\xi))\bigr)\,d\xi .
\end{equation*}
By the windowed Hellmann--Feynman identity $\partial_\xi\tr f(\Hh)=\tr(f'(\Hh)\partial_\xi\Hh)$, which is proved with Kato's transformation function on the
finite-rank Riesz projections, this integral does not depend on $\varphi$. Choose $-Z_\alpha=\xi_0<\dots<\xi_N=Z_\alpha$ and $\delta_j$ such that
$\alpha\pm\delta_j\notin\spec\Hh(\xi)$ on $[\xi_{j-1},\xi_j]$. For $\varphi$ with $\supp\varphi'\subset(\alpha-\eta,\alpha]$ and $\eta<\min\delta_j$, the integral
over $[\xi_{j-1},\xi_j]$ equals $[\sum_{\lambda\in\spec\Hh(\xi)\cap(\alpha-\delta_j,\alpha+\delta_j)}\varphi(\lambda)]_{\xi_{j-1}}^{\xi_j}$. Letting $\supp\varphi'$
shrink to $\alpha$ gives $\sum_j(n_j(\xi_j)-n_j(\xi_{j-1}))$ with $n_j(\xi)=\#(\spec\Hh(\xi)\cap[\alpha,\alpha+\delta_j))$, which is Phillips' formula for
$\SF(\alpha)$ \cite{Phillips}.

\emph{Energy differences.} We apply the homotopy invariance of the spectral flow to $G(\xi,e)=\Hh(\xi)-e$ on $[-Z_\alpha,Z_\alpha]\times[\alpha_1,\alpha_2]$. Going around the boundary counterclockwise,
we find $\SF(\alpha_1)-n_{\rm R}-\SF(\alpha_2)+n_{\rm L}=0$, where $n_{\rm R}$ and $n_{\rm L}$ are the numbers of eigenvalues of $\Hh(\pm Z_\alpha)$ in $(\alpha_1,\alpha_2)$.
By Lemma~\ref{lem:decouple} with a contour around $[\alpha_1,\alpha_2]$, which avoids $\Fl$, these are the counts of the bulks selected at each end. For
$B_\pm>0$, $\xi=+Z_\alpha$ selects $a_-$ and $\xi=-Z_\alpha$ selects $a_+$, and for $B_\pm<0$ the reverse holds. For $B_+<0<B_-$, $A_1\to+\infty$ at both ends, so that
$\xi=+Z_\alpha$ carries both bulks and $\xi=-Z_\alpha$ none, and for $B_-<0<B_+$ the reverse holds. In all four cases, $\SF(\alpha_2)-\SF(\alpha_1)=n_{\rm L}-n_{\rm
R}=\sgn(B_+)n_+-\sgn(B_-)n_-$.
\end{proof}

We will also use the following comparison of spectral flows.
\begin{lemma}\label{s:lem:sfcompare}
Let $\lambda\mapsto T_\lambda$ and $\lambda\mapsto T'_\lambda$ be norm-resolvent continuous families of self-adjoint Fredholm operators near $0$, invertible at
$\lambda=0,1$. Let $0=\lambda_0<\dots<\lambda_N=1$ and $\delta_j>0$ be a Phillips partition for $T$, that is, $\pm\delta_j\notin\spec T_\lambda$ for
$\lambda\in[\lambda_{j-1},\lambda_j]$, and assume that $\pm\delta_j\notin\spec T'_\lambda$ on the same intervals and that $\mathbf 1_{[0,\delta_j)}(T_\lambda)$ and
$\mathbf 1_{[0,\delta_j)}(T'_\lambda)$ have the same rank at $\lambda=\lambda_{j-1}$ and $\lambda=\lambda_j$, for every $j$. Then $\sfl(T)=\sfl(T')$.
\end{lemma}
\begin{proof}
By Phillips' formula \cite{Phillips}, $\sfl(T)=\sum_j\bigl(\operatorname{rank}\mathbf 1_{[0,\delta_j)}(T_{\lambda_j})-\operatorname{rank}\mathbf
1_{[0,\delta_j)}(T_{\lambda_{j-1}})\bigr)$, and the same partition is admissible for $T'$, with the same ranks.
\end{proof}

The fiber analysis of section~\ref{sec:locality} is also used for the pseudo-differential symbols of the deformation paths of
Lemma~\ref{lem:path}.
\begin{lemma}\label{s:lem:pdofibers}
Let $c$ be a Hermitian symbol independent of $x$, of graded type $\mathcal S^0_+$ in a fixed frame, with elliptic transverse block and degree-$0$ Schur
complement at distance at least $c_1>0$ from $0$ at large momentum, which coincides on $\{\pm y\ge y_0\}$ with constant symbols $c_\pm$ of the same type.
Let $B_\pm\ne0$, and let $\Hh_c(\xi)$ be the fibers of $\OpA(c)$ in a gauge $A=(A_1(y),0)$. Then the fibers are self-adjoint, have discrete spectrum
near $0$, depend continuously on $\xi$ in the norm-resolvent sense, and satisfy the conclusion of Lemma~\ref{lem:decouple} for every compact
$K\subset(-c_1/2,c_1/2)+i\R$ that does not meet $\spec K_{B_+}(c_+)\cup\spec K_{B_-}(c_-)$, with constants depending only on finitely many seminorms of $c$,
on $c_1$ and on the distance from $K$ to these spectra. If $0\notin\spec K_{B_\pm}(c_\pm)$, the fibers have, in addition, no spectrum in a fixed neighborhood of
$0$ for $|\xi|\ge Z$. The same holds uniformly, and continuously in the parameter, for continuous families of such symbols, for compact sets $K$ that avoid
the bulk spectra for all values of the parameter.
\end{lemma}
\begin{proof}
By the proof of Lemma~\ref{s:lem:gauge}, applied to the phase $e^{-i(x-x')F_1(y,y')}$ with $F_1(y,y')=\int_0^1A_1(y'+t(y-y'))\,dt$, the fiber of $\OpA(c)$ at $\xi$
is $\Op^w_y$ of the reduction of the amplitude $c(\frac{y+y'}2,\xi-F_1(y,y'),\zeta)$. Since $\partial_yF_1$ is bounded, this reduction equals
$c(y,\xi-A_1(y),\zeta)$ up to a symbol of lower order in the fiber classes of Lemma~\ref{s:lem:admissible}, uniformly in $\xi$. The proofs in
section~\ref{sec:locality} and in Appendix~\ref{s:sec:routine2} then use only the following ingredients: the fiber parametrix built from the
pointwise inverse at large momentum, as in Proposition~\ref{prop:paramalpha}(b); the commutators with cut-offs $\chi(y)$, whose symbols are
$-i\chi'\partial_\zeta c$ plus terms of lower order in the fiber calculus; the comparison with the bulks by pseudolocality (Lemma~\ref{s:lem:pseudoloc}); the
self-adjointness of the fibers, which follows from the fiber version of the proof of Theorem~\ref{thm:analysis}(a); and the norm-resolvent
continuity, which follows from the fiber version of the proof of Theorem~\ref{thm:analysis}(c). None of these uses that the symbol is
polynomial in $k$, and the conclusions follow verbatim. Uniformity for continuous families follows since all constants depend only on finitely many
seminorms.
\end{proof}

\smallskip
\noindent\textbf{Stability.} 
\begin{proof}[Proof of Theorem~\ref{thm:slab}]
We start with (a). For each $s$, the operators $H_{\pm,s}$ agree with $H_s$ on $\{\pm y\ge y_1\}$, satisfy (N1) and (N3), and do not have $\alpha$ in their
spectrum, so that $H_s$ satisfies the hypotheses of Theorem~\ref{thm:quant} in the form (N$'$). The potentials $A+s\beta$ differ by bounded functions with
bounded derivatives, and the coefficients, frame symbols and conjugated symbols depend continuously on $s$. By Theorem~\ref{thm:analysis}(c), applied to
$H_s$ and to $H_{\pm,s}$, the resolvents are norm continuous in $s$. In particular, the distance from $\alpha$ to $\spec H_{\pm,s}$ is bounded below uniformly on
$[0,1]$, so that the margin $\delta$ is uniform, and we may fix $\varphi$ with $\varphi'\in C_c^\infty(\mathcal E_\alpha)$ for all $s$. The Helffer--Sj\"ostrand
formula shows that $s\mapsto U_\varphi(H_s)$ is norm continuous. By Theorem~\ref{thm:quant}, $\mathsf P_{x_0}U_\varphi(H_s)\mathsf P_{x_0}|_{\Ran\mathsf P_{x_0}}$ is then a
norm-continuous family of Fredholm operators, so that its index $2\pi\sigma_I(H_s)$ is constant. Theorem~\ref{thm:BIC} applied to $H$ then gives $2\pi\sigma_I(H_1)=J_+-J_-+\Cy$. When $v$ and $\beta$ are supported in $\{|y|\le y_1-1\}$, it remains to show
that $\Cy_1(x_0)=\Cy(x_0)$. For $y_0\le|y|\le y_1$, the projection $E^\infty$ of $H$ is that of the
constant bulk symbol and depends only on $\theta$, so that $\Cy(x_0)$ is also the integral of $c_1(E^\infty)$ over $[-y_1,y_1]\times S^1$. By structural
admissibility, the projections $E^\infty$ of $H_s$ at $(x_0,y,\theta)$ depend continuously on $s$, and they do not depend on $s$ for $|y|=y_1$. The integral
of the closed form $c_1$ over the cylinder is therefore invariant along $s\in[0,1]$, by Stokes' theorem applied on $[-y_1,y_1]\times S^1\times[0,1]$, since
the boundary contribution involves maps that depend only on $(s,\theta)$, on which $c_1$ vanishes.

We next justify the sufficient conditions stated before the theorem. For $r=0$, (N1) is the uniform ellipticity of the principal symbol, which neither
perturbations of degree at most $m-1$ nor bounded changes of the potential affect. The frame is trivial, and (N3) is void. If the conjugated symbol $w$ of
$v$ belongs to $\mathcal S^{-1}_+$, the blocks of $\tilde a+sw$ have the orders of (DN$_0$), the transverse principal block is unchanged, and the corrections to
the Schur complement, namely $(\tilde a+sw)_{\sff\stt}(\tilde a+sw)_{\stt\stt}^{-1}(\tilde a+sw)_{\stt\sff}-\tilde a_{\sff\stt}\tilde a_{\stt\stt}^{-1}\tilde a_{\stt\sff}$
and $sw_{\sff\sff}$, have degree at most $-1$, so that $\Meff$ and $\Fl$ are unchanged. For $m\le2$, $\Meff$ does not depend on the field
(Proposition~\ref{prop:Meffcomp}(c)). For $m=1$, (DN$_0$) holds for every first-order operator, and Proposition~\ref{prop:Meffcomp}(a) gives $\Meff=V_\sff^*(W+sv)V_\sff+T=V_\sff^*WV_\sff+T$
when $v$ takes values in $\mathcal V_0$, while $\Meff$ does not depend on the field. For $m\ge2$, a perturbation $v$ of degree $0$ with values in
$\mathcal V_0$ has a conjugated symbol in $\mathcal S^{-1}_+$, since its flat block $V_\sff^*vV_\sff$ vanishes and the Moyal corrections have degree at most $-1$,
so that the previous argument applies. For the gap condition with $H_{\pm,s}=\Op^{A_\pm+s\beta}(a_\pm+sv)$, we write $H_{\pm,s}-\alpha=(\U_\pm^*)^{-1}\mathcal
D\bigl(T_\pm+E_s\bigr)\mathcal D\U_\pm^{-1}$, with $T_\pm=\mathcal D^{-1}\U_\pm^*(H_\pm-\alpha)\U_\pm\mathcal D^{-1}$, which has a bounded inverse by
Theorem~\ref{thm:analysis}(b), and $E_s$ the weighted perturbation, given by Lemma~\ref{s:lem:gauge} for the change of potential. A Neumann series applies when
$\sup_s\norm{E_s}<\norm{T_\pm^{-1}}^{-1}$.

We now prove (b). The operators $H_{\pm,1}:=\Op^{A_\pm+\beta}(a_\pm+v)$ agree with $H_1$ on $\{\pm y\ge y_0\}$, satisfy (N1) and (N3), and do not have $\alpha$ in their spectrum, so that
$H_1$ satisfies the hypotheses of Theorem~\ref{thm:quant} in the form (N$'$). For homogeneous frame symbols, let $U$ be the frame of $H$, $E_{\ker}$ the kernel projector of $H_1$, and $\phi\in C_c^\infty(\R)$ equal to $1$ on $[x_1-L,x_1+L]$ and supported in $(x_1-L-1,x_1+L+1)$. Transport in $x$ from $\{x=x_1+L+1\}$ generated by $\phi[\partial_xE_{\ker},E_{\ker}]+(1-\phi)(\partial_xU)U^*$, which preserves $\Ran E_{\ker}$ and $\Ran(1-E_{\ker})$, gives a frame of $H_1$ equal to $U$ on $\{x\ge x_1+L+1\}$ and to $Ug(y,\theta)$, with $g$ block diagonal, on $\{x\le x_1-L-1\}$; for $m\le2$, (DN$_0$)
holds in this frame by Proposition~\ref{prop:flat}(a). For $m\ge3$, we use instead the frame symbol of the assumption, which equals that of $H$ on
$\{x\ge x_1+L+1\}$. By Theorem~\ref{thm:quant}, $\sigma_I(H)$ and $\sigma_I(H_1)$ do not depend on the
switch, and we choose a switch $P$ with $P'$ supported near $x_2$, with $x_2-x_1-L-1\ge2L'$. The operators $H$ and $H_1$ then coincide on $\{|x-x_2|<L'\}$,
with the same potential and frame symbol, and Theorem~\ref{thm:local}(a) gives $\sigma_I(H_1)=\sigma_I(H)$ for $L'$ large.
\end{proof}

\section{Main techniques, roadmap and notation}\label{s:sec:intro}\label{s:sec:notation}
This appendix describes the main techniques of the paper, gives a roadmap of the proofs, and lists the notation used throughout the paper.

\smallskip
\noindent\textbf{The magnetic calculus.} The operators are of the form $H=a(u,\Pi)$, where $u=(x,y)\in\R^2$, $\Pi=D-A$ with $D=-i\nabla$ is the kinetic momentum of a magnetic potential
$A$ with $\curl A=B$, and $a(u,k)$ is a Hermitian matrix symbol, polynomial of degree $m$ in $k=(\xi,\zeta)$. The operator is defined by the magnetic Weyl
quantization \cite{IMP,IMP2}
\[
  (\OpA(a)f)(u)=(2\pi)^{-2}\iint e^{i(u-u')\cdot k}e^{-i\int_{[u,u']}A}\,a\bigl(\tfrac{u+u'}2,k\bigr)f(u')\,du'\,dk ,
\]
which reduces to the Weyl symmetrization of the polynomial $a(u,\Pi)$, namely the Peierls substitution $k\to\Pi$. The phase $e^{-i\int_{[u,u']}A}$ makes the
calculus gauge covariant: replacing $A$ by $A+\nabla\lambda$ conjugates $\OpA(a)$ by $e^{i\lambda}$, so that all symbolic statements depend on $B$ only. When $A$
is linear, as in the Landau gauge $A=(-By,0)$ of a constant field, the line integral is exactly $A(\frac{u+u'}2)\cdot(u'-u)$, and $\OpA(a)$ is the ordinary Weyl
quantization of the symbol $a(u,k-A(u))$. In exchange, the product of symbols $\sharp^B$ acquires corrections involving $B$, each with at least two derivatives in $k$. For instance,
$[\Pi_1,\Pi_2]=iB$. The paper also treats factorized operators $L_B^*L_B+W$, in which the field enters through the factor $L_B$. The two couplings of the same
field-free symbol differ by terms of order $m-2$ multiplied by $B$, which move the generalized Landau levels and, in the presence of flat bands, may change the flat block
(Proposition~\ref{prop:Meffcomp} and Remark~\ref{rem:quant}).

\smallskip
\noindent\textbf{Lack of ellipticity.} The first difficulty addressed in the paper is the lack of ellipticity. Consider the operator $\curl^*\curl$ acting on vector fields in the plane, whose
symbol is $k^\perp(k^\perp)^T$. It vanishes on the vector $\theta=k/|k|$, so that its principal symbol has a kernel of rank one, and the corresponding band is
identically zero. Adding a potential $W$ makes this band bounded but not constant, and at large momentum it converges to the value of $W$ compressed to the
kernel, here $\theta^TW\theta$. In general, the kernel of $a_m(u,\theta)$ has constant rank $r$, and in a frame $U=[V_\sff,V_\stt]$ adapted to it the
conjugated symbol $\tilde a=U^*\sharp^Ba\sharp^BU$ has a flat block $\tilde a_{\sff\sff}$, off-diagonal blocks, and an elliptic transverse block of order $m$. If the
flat block is of order $0$ and the off-diagonal blocks of order at most $\frac m2$, which is the Douglis--Nirenberg condition (DN$_0$) \cite{DN,ADN}, then the
bounded bands converge along each direction $\theta$ to the eigenvalues of the effective flat symbol
\[
  \Meff(u,\theta)=\lim_{\rho\to\infty}\bigl(\tilde a_{\sff\sff}-\tilde a_{\sff\stt}\tilde a_{\stt\stt}^{-1}\tilde a_{\stt\sff}\bigr)(u,\rho\theta),
\]
the limit of the Schur complement at large momentum, which is also its homogeneous part of degree $0$ (Definition~\ref{def:Meff}). Its values
form the flat-band set $\Fl$, which is essential spectrum of $H$ (Proposition~\ref{prop:flat}(b)). The block orders are encoded in graded symbol
classes: with weights $s_\sff=0$ and $s_\stt=\frac m2$, the symbol $\tilde a-z$ has blocks of orders $s_i+s_j$, its pointwise inverse at large momentum has
blocks of orders $-s_i-s_j$, and their composition has blocks of orders $s_i-s_j$, like the identity. The asymptotic Neumann series that turns the pointwise
inverse into a parametrix converges when the composition error has blocks of orders at most $s_i-s_j-1$, that is, $\frac m2-1$ from the flat to the
transverse components. This is why the frame must be fixed: in the moving splitting $\Ran E_{\ker}(u,\theta)\oplus\Ran Q(u,\theta)$, derivatives in $u$ of the
moving projector produce an error of order $m-1$ in that block, whereas in the fixed splitting $\C^r\oplus\C^{n-r}$ every composition error gains one order in
each block, as for scalar symbols (Proposition~\ref{prop:paramalpha}(a)).

\smallskip
\noindent\textbf{Analysis of the operator.} The analysis of $H$ is based on a parametrix near the energy $\alpha$ (Proposition~\ref{prop:paramalpha}). For $\alpha\notin\Fl$ and $z$
near $\alpha$, the conjugated symbol $\tilde a-z$ is invertible at large momentum, with an inverse $q_z$ of the opposite graded type, and an asymptotic sum of the
Neumann series built on $q_z$, which exists by Borel's lemma \cite[Prop.~18.1.3]{Hormander3}, gives a symbol $g_z$ with $(\tilde a-z)\sharp^Bg_z-1$ smoothing. With $\U=\OpA(U)$, the operator
$G_z=\U\OpA(g_z)\U^*$ then satisfies $(H-z)G_z=1+K_z$ with $K_z$ smoothing, and a left parametrix $G'_z$, with $G'_z(H-z)=1+K'_z$ and $K'_z$ smoothing,
is obtained in the same way (Proposition~\ref{prop:paramalpha}(c)). This yields
self-adjointness, since an element of $\ker(H_{\max}-z)$ with $\Im z\ne0$ equals minus a smoothing operator applied to itself, and resolvent estimates, through
$(H-z)^{-1}=G'_z-K'_zG_z+K'_z(H-z)^{-1}K_z$: the resolvent gains $\frac m2$ derivatives on the transverse components and none on the flat ones
(Theorem~\ref{thm:analysis}). Since $G_z$ is holomorphic near $\alpha$, the Helffer--Sj\"ostrand formula
$\Phi(H)=\frac1\pi\int\bar\partial\tilde\Phi(z)(H-z)^{-1}d^2z$, where $\tilde\Phi$ is an almost analytic extension of $\Phi$ \cite{DS99}, gives
$\Phi(H)=-\frac1\pi\int\bar\partial\tilde\Phi(z)(H-z)^{-1}K_z\,d^2z$, so that $\Phi(H)$ is smoothing in momentum for $\Phi$ supported near $\alpha$. Exponential
weights are handled as in the Combes--Thomas argument \cite{CT}: with a frame analytic in $k$ in a strip, conjugation by $e^{\mu\phi}$, $|\nabla\phi|\le1$,
changes the conjugated symbol by $\mu$ times a symbol of the same graded type and one order lower, and a Neumann series in the weighted spaces gives
exponential decay of the resolvent kernel at rate proportional to the distance to the spectrum. The same parametrix gives norm continuity of the resolvent
along deformations, even when the kernel bundle moves.

\smallskip
\noindent\textbf{Functional calculus and localization.} The parametrix of Proposition~\ref{prop:paramalpha} describes the resolvent only for spectral parameters near $\alpha$, which suffices for
functions $\Phi(H)$ with $\Phi$ compactly supported. The bulk invariant, by contrast, involves the spectral projections $\hat P_\alpha$ of the
one-dimensional bulk operators of section~\ref{sec:pbulk}, whose spectra are in general unbounded above and below. These projections are integrals of
resolvents along the vertical line $\Re z=\alpha$, and identifying their symbol class requires the symbol of the resolvent for all $z$ on that line,
uniformly as $|\Im z|\to\infty$ (Proposition~\ref{s:prop:largeparam}). For that purpose one weights the flat components by
$|\Im z|^{1/2}$ and the transverse ones by $(\jp X^2+|\Im z|^{2/m})^{m/4}$. In the weighted picture every block is of order zero, and every composition error
contains a derivative falling either on a symbol or on a weight, which is small when $|\Im z|$ is large. There, the frame is chosen with unitary quantization (Lemma~\ref{lem:globalframe}(b)). Finally, the localization of $\Phi(H)$ near the wall follows from the resolvent bounds of the bulk operators
$H_\pm$ and from the geometric resolvent identity $R\theta_\pm=\tilde\theta_\pm R_\pm\theta_\pm-R[H_\pm,\tilde\theta_\pm]R_\pm\theta_\pm$, for cut-offs
$\theta_\pm$ supported in the bulks and $\tilde\theta_\pm=1$ on their supports, in which the first term is holomorphic in the bulk gap and does not
contribute to the Helffer--Sj\"ostrand integral (Proposition~\ref{prop:loc}). The kernel bounds follow from the diamagnetic inequality
$|e^{-t\Pi^2}g|\le e^{t\Delta}|g|$ \cite{AHS,Simon}, which bounds the kernel of $\Lm^{-2s}$ by that of $(1-\Delta)^{-s}$, and hence the kernel of an
operator $T$ by $\norm{\Lm^{2s}T\Lm^{2s}}$ for $s>\frac12$ (Lemma~\ref{s:lem:diamag}).

\smallskip
\noindent\textbf{The magnetized bulk.} The first effect of a constant field is to replace bands by Landau levels, and this is best understood through a reduction to one dimension. For a constant
symbol $\ab(k)$ and the Landau gauge, $H_B(\ab)$ commutes with translations in $x$, and the partial Fourier transform in $x$ gives the fibers
$\Op^w_y(\ab(\xi+By,\zeta))$. Translating $y$ by the guiding center $y_c=-\xi/B$ shows that all fibers are unitarily equivalent to the reduced bulk operator
$K_B(\ab)=\Op^w_y(\ab(By,\zeta))$ on $L^2(\R_y)$. For $\ab=|k|^2$, this is the harmonic oscillator $B^2y^2+D_y^2$, with eigenvalues $|B|(2j+1)$, $j\ge0$. Each
eigenvalue of $K_B(\ab)$ is therefore an eigenvalue of $H_B(\ab)$ of infinite multiplicity, one eigenfunction for each guiding center, and $\spec
H_B(\ab)=\spec K_B(\ab)$ (Theorem~\ref{thm:bulkspec}). The field enters only through the noncommutativity of $\Pi_1$ and $\Pi_2$, that is, through the Moyal product $\sharp^B$, which creates the Landau levels and, for $m\ge3$, may
modify the effective flat symbol. For a general symbol, $\ab(By,\zeta)$ is a function on
the phase plane $X=(y,\zeta)$, and the calculus adapted to $K_B(\ab)$ is the isotropic calculus \cite{Shubin}, in which symbols are measured in
powers of $\jp X$ and every derivative gains one power. In particular, an elliptic isotropic symbol of positive order has compact resolvent, so that $K_B(\ab)$
has discrete spectrum away from $\Fl_B(\ab)$. Weyl quantization is covariant under linear symplectic changes of variables, through
the metaplectic representation \cite[Ch.~4]{Folland}. The map $(y,\zeta)\mapsto(By,\zeta)$ multiplies the symplectic form by $B$, and composing it with the
symplectic dilation $(y,\zeta)\mapsto(y/|B|,|B|\zeta)$ shows that $K_B(\ab)$ is unitarily equivalent to the semiclassical quantization
$\Op^w_h(\ab\circ\varsigma_B)$ with $h=|B|$, where $\varsigma_B$ is the identity for $B>0$ and exchanges the two variables for $B<0$, which records the
reversal of orientation (Lemma~\ref{lem:reduction}(d)). A further isotropic dilation gives a form in which the direction at infinity
does not depend on $|B|$, which is the form used for the pairing $\tau$ along families in $B$. Weak fields thus correspond to the semiclassical limit $h\to0$, which is how the field-free Chern
integral appears as the weak-field limit of the Hall conductivity (Theorem~\ref{thm:bulk}(c)). This requires the weak-field condition (W), which keeps $E^\infty$, and hence the fractional part of $J$, fixed as the field is switched on.

\smallskip
\noindent\textbf{The Kubo pairing.} The bulk Hall conductivity is defined by the Kubo formula $\sigma_H=i\,\trarea(P[[P,x],[P,y]])$ for the Fermi projector $P$ \cite{ASS,BES,PSB}, where
$\trarea$ is the trace per unit area. For an operator $X$ commuting with magnetic translations and with continuous kernel, $\trarea X=\tr X(u,u)$ for
every $u$, with $\tr$ the matrix trace. In the Landau gauge, multiplication by $x$ acts on fibers as $i\partial_\xi$, and since the fiber at $\xi$ is $K_B(\ab)$ translated by $y_c=-\xi/B$,
$\partial_\xi$ acts on fibers as $iB^{-1}[D_y,\cdot]$. The two-dimensional formula therefore reduces to the one-dimensional pairing \eqref{eq:tau} (Lemma~\ref{lem:kubo})
\[
  \tau(\hat P)=-i\Tr\bigl(\hat P[[\hat P,D_y],[\hat P,y]]\bigr),\qquad J=2\pi\sigma_H=\sgn(B)\,\tau(\hat P),
\]
for the spectral projector $\hat P$ of $K_B(\ab)$. Up to normalization, $\tau(\hat P)$ is the value at $(\hat P,\hat P,\hat P)$ of the cyclic $2$-cocycle
$(a_0,a_1,a_2)\mapsto\Tr\bigl(a_0(\delta_1a_1\,\delta_2a_2-\delta_2a_1\,\delta_1a_2)\bigr)$ built from the commuting derivations $\delta_1=[D_y,\cdot]$ and
$\delta_2=[y,\cdot]$, of the type used in the noncommutative-geometric description of the quantum Hall effect \cite{Connes,BES}, here on operators on the line. The pairing $\tau$ has three properties used in the bulk analysis (Lemma~\ref{lem:tau} and Proposition~\ref{prop:semiclassical}).
First, $\tau(\hat P)=\operatorname{rank}\hat P$ for a finite-rank projector. Hence every isolated eigenvalue of $K_B(\ab)$ of
multiplicity $d$ contributes exactly $d\,\sgn(B)$ to $J$, independently of the symbol: a simple generalized Landau level always carries the Chern number
$\sgn(B)$. Second, $\tau$ is constant along continuous families of projectors with a fixed principal part at infinity in phase space, so that $J$ changes
only when eigenvalues cross the energy, by the spectral flow. Third, in the semiclassical limit, $\tau(\hat P)$ converges to $-\int c_1$ of the principal
symbol of $\hat P$ in the variables of $\varsigma_B$, which for weak fields is $\sgn(B)\,I$, so that $J=\sgn(B)\tau(\hat P)$ tends to the field-free Chern
integral $I=-\int c_1(E_\alpha)$. The first property is the St\v{r}eda formula \cite{Streda} in this setting: each level of $K_B(\ab)$ is an eigenvalue of $H_B(\ab)$ with $\frac{|B|}{2\pi}$
states per unit area, one per guiding center, so that the density of states below $\alpha$ is $n=\frac{|B|}{2\pi}N$ for $N$ levels below $\alpha$, and
$\sigma_H=\frac{\sgn B}{2\pi}N=\partial_Bn$ when $N$ does not change with $B$. The first and second properties give the relative version, in which $N$ is
replaced by the index of a pair of projections (Corollary~\ref{cor:bulk}(a)). The invariant $J$ is first computed at one energy, for instance at weak field in a field-free gap,
where $J=I$, or below the spectrum, where $J=0$. Its value at another energy then differs by $\sgn(B)$ times the number of eigenvalues of $K_B(\ab)$ between
the two energies, and its value at another field along $t\mapsto tB$ differs by $-\sgn(B)$ times the net number of eigenvalues crossing the energy upward.

\smallskip
\noindent\textbf{Topology at infinite momentum.} Topology enters the bulk invariant through the behavior of symbols at infinite momentum. For $\alpha\notin\Fl$, the occupied projector at infinity $E^\infty(u,\theta)$ is a map
from the circle of directions, at each point of the wall, to the Grassmannian $\Gr_q(\C^n)$. For $0<q<n$, the Grassmannian $\Gr_q(\C^n)$ is simply connected,
and its second homotopy group $\pi_2(\Gr_q(\C^n))$ is isomorphic to $\Z$, the isomorphism being given by the integral of the first Chern class
\cite{Hatcher}. Such a loop therefore extends to a map $E^{\rm ext}$ from the disk, and the homotopy classes of extensions are distinguished by their Chern
integral $\kappa(E^{\rm ext})=\int(E^{\rm ext})^*c_1$, whose value modulo $\Z$ is fixed by the loop. Each extension defines a reference symbol, built from $1-2E^{\rm ext}$ rescaled
to a ball of large radius in momentum and equal to $1-2E^\infty$ outside it, whose invariant is computed by the semiclassical limit of $\tau$ and equals
$-\kappa(E^{\rm ext})$. The bulk invariant is then $J=\mathcal I_{E^{\rm ext}}-\kappa(E^{\rm ext})$, where $\mathcal I_{E^{\rm ext}}$ is an integer spectral flow along a
deformation from the reference to the bulk symbol (Theorem~\ref{thm:bulk}(b)). This shows that $J$ is an integer up to the Berry phase
$\Theta$ of the loop $E^\infty$, namely $J+\Theta/2\pi\in\Z$, and that the cylinder term $\Cy$ is the difference of these Berry phases across the wall. For
operators without field, the sum $I_+-I_-+\Cy$ equals $-d_\alpha$, minus the Chern number of the occupied projector on the boundary of $[-y_0,y_0]\times\bar{\mathbb B}$ in
the variables $(y,\xi,\zeta)$, oriented as in Lemma~\ref{lem:degree} (section~\ref{sec:compute}), in agreement with
the Fedosov--H\"ormander formula \cite[Ch.~19]{Hormander3}, \cite{Faure,QB21}. The deformations must be chosen with care, since the straight path from a
symbol to its reference may close the flat-band gap (Lemma~\ref{lem:path}). In particular, $J$ is defined without any regularization at
infinite momentum, including for the unregularized Dirac operator, and a regularization changes $E^\infty$ and shifts $J$ by an integer
(Example~\ref{s:ex:Dirac}).

\smallskip
\noindent\textbf{The interface current.} The interface current is $\sigma_I=\Tr\,i[H,P]\varphi'(H)$, with a switch $P(x)$ in position and a switch $\varphi$ in energy,
$\varphi'$ being supported in a common gap of the two bulks. It is an index \cite{ASS,EG,BEC}. Let
$U_\varphi=e^{2\pi i\varphi(H)}$, so that $U_\varphi-1$ vanishes on spectral subspaces outside $\supp\varphi'$, and $U_\varphi(H_\pm)=1$. The localization of $\varphi'(H)$ near the wall makes $[\mathsf P,U_\varphi]$
trace-class for the sharp switch $\mathsf P=\mathbf 1_{[x_0,\infty)}(x)$, so that $\mathsf PU_\varphi\mathsf P$ is a Fredholm operator on $\Ran\mathsf P$ with
index $-\Tr(U_\varphi^*\mathsf PU_\varphi-\mathsf P)$ (Lemma~\ref{s:lem:tc}(v)). A computation that uses only the trace-class properties of
$[P,f(H)]$ and the cyclicity of the trace shows that $(u,v)\mapsto\Tr(u(H)[P,v(H)])$ is a multiple of $\int uv'$, which identifies $2\pi\sigma_I$ with this index
for every switch $P$ and every $\varphi$ (Lemma~\ref{s:lem:tc}(iv)). This explains the quantization and the independence of $P$ and $\varphi$. For walls that
are invariant along the interface, the same quantity is the spectral flow of the fibers $\xi\mapsto\Hh(\xi)$, that is, the signed number of eigenvalue
branches crossing $\alpha$ as $\xi$ increases \cite{Phillips,QB22}. This follows from the identity
$2\pi\sigma_I=\int\tr(\partial_\xi\Hh(\xi)\varphi'(\Hh(\xi)))\,d\xi$, obtained by writing the trace in the fiber representation (Theorem~\ref{thm:TI}). When both bulk fields are nonzero, the fibers have discrete spectrum near $\alpha$, and at large $|\xi|$ they decouple into translated copies of
the reduced bulk operators, centered at the guiding centers (Lemma~\ref{lem:decouple}). The spectral flow around the boundary of a rectangle in
the variables $(\xi,s)$, where $s$ parametrizes a deformation of the wall to a reference, vanishes, which relates the interface spectral flow to the bulk
spectral flows at the two ends (Theorem~\ref{thm:relcount}).

\smallskip
\noindent\textbf{Cutting and pasting.} General walls are analyzed by cutting and pasting, in the spirit of \cite{EGS,QB21}. Since $\Phi(H)$ decays rapidly away from the wall and $[H,P]$ is
supported near the switch, $i[H,P]\Phi(H)$ is localized near the intersection of the wall and the switch, and $\sigma_I$ depends only on the operator there, which allows one to freeze the coefficients along the interface and
reduce general walls to translation-invariant ones (Theorem~\ref{thm:local}(a),(b)). Likewise, a wall made of two walls separated by a wide
gapped region carries the sum of their currents (Theorem~\ref{thm:local}(c)). These two properties are used to handle bulks without field. A wall across
which only a weak field is switched on carries no current, since weak fields create no level in a field-free gap (Proposition~\ref{prop:weak} and
Theorem~\ref{thm:local}(d)). Inserting a region of weak constant field next to a field-free bulk, which does not change $\sigma_I$ by stability (Theorem~\ref{thm:slab}(a)),
therefore splits the operator into two walls (Theorem~\ref{thm:local}(c)), one of which carries no current, and reduces the problem to an interface
between magnetized bulks, where the spectral flow argument applies (Theorem~\ref{thm:BIC}). The same locality, together
with the norm continuity of the resolvent, gives stability under perturbations of arbitrary size that preserve the flat-band structure and the bulk gaps
(Theorem~\ref{thm:slab}).

\smallskip
\noindent\textbf{Roadmap.} The main results are the description of the magnetized bulk and of its invariant (Theorems~\ref{thm:bulkspec} and
\ref{thm:bulk}), the quantization of the interface current (Theorems~\ref{thm:quant} and \ref{thm:TI}), the bulk--interface correspondence
(Theorem~\ref{thm:BIC}) with Corollaries~\ref{cor:trans} and \ref{cor:main}, and stability (Theorem~\ref{thm:slab}). Their proofs follow three chains of
results. (a) For the operator, the frames of Lemma~\ref{lem:globalframe} lead to the graded parametrix
(Proposition~\ref{prop:paramalpha}), from which Theorem~\ref{thm:analysis} and Proposition~\ref{prop:flat} follow. (b) For the bulk, the reduction to $K_B$
(Lemma~\ref{lem:reduction}), the class of projections (Proposition~\ref{prop:gradedFC}) and the Kubo pairing (Lemmas~\ref{lem:tau} and \ref{lem:kubo}) are
combined with the references, the gapped paths and the semiclassical limit (Lemma~\ref{lem:path}, Propositions~\ref{prop:semiclassical} and \ref{prop:gap})
to give Theorem~\ref{thm:bulk}. (c) For the interface, localization (Proposition~\ref{prop:loc}) gives Theorem~\ref{thm:quant}. Locality and freezing
(Theorem~\ref{thm:local}) reduce general walls to translation-invariant ones, for which Theorem~\ref{thm:TI} and the decoupling of Lemma~\ref{lem:decouple}
lead, through the extensions of Lemma~\ref{lem:degree}, to the level counts of Theorem~\ref{thm:relcount} and to Theorem~\ref{thm:BIC} for $B_\pm\ne0$. The
insertion of a weak field then gives the case of vanishing bulk fields, and the corollaries follow. Stability (Theorem~\ref{thm:slab}) combines locality with the norm continuity of the resolvent
(Theorem~\ref{thm:analysis}(c)). The table below lists the main ingredients of each proof, and results numbered with a letter are in the appendices.

\begin{center}
\small
\begin{tabular}{@{}p{0.27\textwidth}p{0.68\textwidth}@{}}
\toprule
Result & Main ingredients of the proof\\
\midrule
Proposition~\ref{prop:flat} (flat bands) & frames (Lemma~\ref{lem:globalframe}), graded parametrix (Proposition~\ref{prop:paramalpha}), Weyl sequences\\
Proposition~\ref{prop:Meffcomp} ($\Meff$ for the two couplings) & Moyal expansion of the flat block; for factorized operators, the frame built on
$(1+L_BL_B^*)^{-1}$ and the composition rules of Proposition~\ref{prop:paramalpha}(a)\\
Theorem~\ref{thm:analysis} (resolvent) & graded parametrix, Combes--Thomas (Lemma~\ref{s:lem:expconj})\\
Theorem~\ref{thm:bulkspec} (bulk spectrum) & reduction to $K_B$ (Lemma~\ref{lem:reduction}), isotropic parametrix, coherent states\\
Theorem~\ref{thm:bulk}(a) (levels) & Kubo pairing (Lemmas~\ref{lem:kubo} and \ref{lem:tau}), functional calculus (Proposition~\ref{prop:gradedFC})\\
Theorem~\ref{thm:bulk}(b) (fractional part) & references (Lemma~\ref{lem:path}, Proposition~\ref{prop:gap}), semiclassical limit (Proposition~\ref{prop:semiclassical})\\
Theorem~\ref{thm:bulk}(c) (weak field) & no levels at weak field (Proposition~\ref{prop:weak}), semiclassical limit\\
Theorem~\ref{thm:quant} (quantization) & localization (Proposition~\ref{prop:loc}), index identities (Lemma~\ref{s:lem:tc})\\
Theorem~\ref{thm:TI} (spectral flow) & fiber calculus (Lemma~\ref{s:lem:pdofibers}), decoupling (Lemma~\ref{lem:decouple})\\
Theorem~\ref{thm:relcount} (level counts) & Theorem~\ref{thm:TI}, extensions (Lemma~\ref{lem:degree}), interface references, Lemma~\ref{s:lem:sfcompare}\\
Theorem~\ref{thm:BIC} (correspondence) & Theorems~\ref{thm:relcount} and \ref{thm:bulk}(b), locality (Theorem~\ref{thm:local}); for $B_\pm=0$, Theorems~\ref{thm:slab}(a),
\ref{thm:local}(c),(d) and \ref{thm:bulk}(c)\\
Theorem~\ref{thm:slab} (stability) & Theorem~\ref{thm:analysis}(c), Proposition~\ref{prop:loc}, Theorem~\ref{thm:local}\\
Corollaries~\ref{cor:bulk}, \ref{cor:trans}, \ref{cor:main} & Theorems~\ref{thm:bulk} and \ref{thm:BIC}, Lemma~\ref{lem:tau}\\
\bottomrule
\end{tabular}
\end{center}

\smallskip
\noindent\textbf{Notation.} The following list collects the notation of the paper.

{\small
\begin{longtable}{@{}p{0.25\textwidth}p{0.55\textwidth}p{0.14\textwidth}@{}}
\toprule
Symbol & Meaning & Defined in\\
\midrule
\endhead
\multicolumn{3}{@{}l}{\emph{Variables, classes and quantizations}}\\
$u=(x,y)$, $k=(\xi,\zeta)$ & position and momentum; the interface is the $x$-axis & sec.~\ref{sec:intro}\\
$\rho=|k|$, $\theta=k/|k|$ & modulus and direction of the momentum & sec.~\ref{sec:intro}\\
$D=-i\nabla$, $\jp\cdot$ & derivative, $\jp\cdot=(1+|\cdot|^2)^{1/2}$ & sec.~\ref{sec:intro}\\
$S^\nu$, $a=b+S^\nu$ & symbol classes; $a-b\in S^\nu$ & sec.~\ref{sec:intro}\\
$C_b^\infty$, $\mathfrak S_1$, $\mathfrak S_2$, $\Herm(n)$ & bounded smooth functions; trace and Hilbert--Schmidt classes; Hermitian matrices & sec.~\ref{sec:intro}\\
$\Op^w$, $\Op^w_y$, $\Op^w_h$, $\#$, $\#_h$ & Weyl quantizations without field and Moyal products & sec.~\ref{sec:intro}\\
$A$, $B=\curl A$, $\Pi=D-A$ & magnetic potential, field and kinetic momentum & sec.~\ref{sec:operators}\\
$\OpA$, $\sharp^B$ & magnetic Weyl quantization and magnetic Moyal product & sec.~\ref{sec:operators}\\
\multicolumn{3}{@{}l}{\emph{Operators, frames and flat bands}}\\
$H=\OpA(a)$, $a_m$, $W=a_0$ & operator, principal symbol, zero-order term & sec.~\ref{sec:operators}\\
$L_B=\OpA(\lsym)$ & factor of a factorized operator $L_B^*L_B+W$ & sec.~\ref{sec:operators}\\
$r$, $E_{\ker}$, $Q=1-E_{\ker}$ & rank of $\ker a_m$, projector onto it, complementary projector & sec.~\ref{sec:operators}\\
$U=[V_\sff,V_\stt]$, $\U=\OpA(U)$ & frame symbol and its quantization; flat and transverse columns & sec.~\ref{sec:operators}\\
$\tilde a$, $\upsilon$ & conjugated symbol $U^*\sharp^Ba\sharp^BU$; $U^*\sharp^BU=1+\upsilon$ & sec.~\ref{sec:operators}\\
(DN$_0$), $\Meff$ & Douglis--Nirenberg condition; effective flat symbol & Def.~\ref{def:Meff}\\
$\Fl$, $\mathcal V_0(u)$ & flat-band set; elliptic directions & sec.~\ref{sec:operators}\\
$E^\infty$, $E^\infty_{\rm fl}$, $E^\infty_{\rm tr}$, $q$ & occupied projector at infinite momentum, its flat and transverse parts, its rank & sec.~\ref{sec:operators}\\
$\Cy(x_0)$, $\Cy$, $c_1$ & cylinder term; first Chern form & \eqref{eq:cyl}\\
\multicolumn{3}{@{}l}{\emph{Bulks and hypotheses}}\\
$a_\pm$, $B_\pm$, $y_0$ & bulk symbols and fields; half-width of the wall $\{|y|<y_0\}$ & (N2)\\
$\spec H_0(a)$, $\mathcal G(a)$, $E_\alpha$, $I(a;\alpha)$ & field-free spectrum and gaps; occupied projector; Chern integral & \eqref{eq:I}\\
$H_B(a)$, $K_B(a)$ & bulk operator; reduced one-dimensional operator & \eqref{eq:KB}\\
$\Fl_B(a)$ & flat-band set of $H_B(a)$ & sec.~\ref{sec:hypotheses}\\
$P_\alpha$, $\hat P_\alpha$, $\sigma_H$, $J(a,B;\alpha)$ & Fermi projectors; Hall conductivity; bulk invariant & \eqref{eq:sigmaH}\\
$\trarea$ & trace per unit area & \eqref{eq:sigmaH}\\
(N1)--(N4), (N), (N$'$) & standing hypotheses & sec.~\ref{sec:hypotheses}\\
$H_\pm$, $\delta$, $\mathcal E_\alpha$ & bulk operators; margin; energy window & sec.~\ref{sec:hypotheses}\\
(TI), (F), $\chi_L$, $H^{x_0}_{t,L}$, $L_0$ & translation invariance; freezing condition and its data & sec.~\ref{sec:hypotheses}\\
(W1)--(W5), $b_1$, $N_0$, $\ab^{(B)}$ & weak-field condition, its constants, field-dependent bulk symbols & sec.~\ref{sec:hypotheses}\\
\multicolumn{3}{@{}l}{\emph{Bulk invariant}}\\
$\Gr_q(\C^n)$, $\mathbb D$ & Grassmannian; unit disk & sec.~\ref{sec:magbulk}\\
$E^{\rm ext}$, $\kappa(E^{\rm ext})$ & bulk extension of $E^\infty$ and its Chern integral & \eqref{eq:kappa}\\
$\mathcal N_{E^{\rm ext}}$, $\mathcal I_{E^{\rm ext}}$ & relative level count and its signed version & \eqref{eq:Ihat}\\
$\ind(P,Q)$, $\sfl$, $\Fl_{tB}(a)$ & index of a pair; spectral flow; flat-band set at the field $tB$ & sec.~\ref{sec:magbulk}\\
$X=(y,\zeta)$, $\ell$, $\Gamma^\mu_{3/4}$ & phase-plane variable; $\ell(y,\zeta)=(By,\zeta)$; isotropic classes & \eqref{eq:ell}, sec.~\ref{sec:pbulk}\\
$\cP_\varepsilon(E)$, $\tau$, $\delta_1$, $\delta_2$ & class of projections; Kubo pairing; derivations $[D_y,\cdot]$, $[y,\cdot]$ & \eqref{eq:tau}\\
$h=|B|$, $\varsigma_B$ & semiclassical parameter; identity or swap of $(y,\zeta)$ & Lemma~\ref{lem:reduction}(d)\\
$b_{\rm ref}$, $U_R$, $\Sigma_R$ & reference symbol, its frame and sign matrix & sec.~\ref{sec:pbulk}\\
$b_0$ & weak-field threshold & Prop.~\ref{prop:weak}\\
\multicolumn{3}{@{}l}{\emph{Interface current}}\\
$P$, $\mathsf P_{x_0}$, $\varphi$, $U_\varphi$ & switch; sharp switch; energy switch; $e^{2\pi i\varphi(H)}$ & \eqref{eq:sigmaI}, sec.~\ref{sec:interface}\\
$\sigma_I$, $\Ind$ & interface current; Fredholm index & \eqref{eq:sigmaI}, sec.~\ref{sec:interface}\\
$\Hh(\xi)$, $\SF(\alpha)$, $Z_\alpha$ & fibers; their spectral flow; large-momentum threshold & sec.~\ref{sec:interface}\\
$\Theta$, $\tilde J_\pm$, $J_\pm$ & Berry phase of $E^\infty$; integer bulk data; $J(a_\pm,B_\pm;\alpha)$ & Thm.~\ref{thm:BIC}\\
$v$, $\beta$, $H_s$, $y_1$ & perturbation and perturbed path & sec.~\ref{sec:stability}\\
$y_c^\pm(\xi)$, $\hat\U(\xi)$, $\tilde A_z(\xi)$ & guiding centers; fiber frames; conjugated fibers & sec.~\ref{sec:pBIC}\\
\multicolumn{3}{@{}l}{\emph{Analysis}}\\
$\mathcal S^\mu_+$, $\mathcal S^\mu_-$, $\mathcal S^\mu_0$ & graded symbol classes & sec.~\ref{sec:engine}\\
$Z$, $\chi_R$, $q_z$, $g_z$ & spectral parameters; cut-off; pointwise inverse; parametrix & sec.~\ref{sec:engine}\\
$\Lm=(1+\Pi^2)^{1/2}$, $\mathcal D$, $\Wt$ & kinetic weight; graded weight $\mathrm{diag}(1,\Lm^{m/2})$; weight of the resolvent estimates & sec.~\ref{sec:operators}, \ref{sec:engine}\\
$\mathfrak a$, $e_\pm$ & annihilation operator $(\Pi_1+i\Pi_2)/(2|B|)^{1/2}$; $e_\pm=(1,\pm i)/\sqrt2$ & sec.~\ref{s:sec:examples}\\
\bottomrule
\end{longtable}
}


\begin{thebibliography}{99}\small\setlength{\itemsep}{0pt}\setlength{\parsep}{0pt}\setlength{\parskip}{0pt}
\bibitem{ADN} S.~Agmon, A.~Douglis and L.~Nirenberg, Estimates near the boundary for solutions of elliptic partial differential equations satisfying general boundary
conditions II, \emph{Comm.\ Pure Appl.\ Math.} 17 (1964), 35--92.
\bibitem{AC79} Y.~Aharonov and A.~Casher, Ground state of a spin-$\frac12$ charged particle in a two-dimensional magnetic field, \emph{Phys.\ Rev.\ A} 19 (1979), 2461--2462.
\bibitem{AAM96} H.~Aoki, M.~Ando and H.~Matsumura, Hofstadter butterflies for flat bands, \emph{Phys.\ Rev.\ B} 54 (1996), R17296--R17299.
\bibitem{AHS} J.~Avron, I.~Herbst and B.~Simon, Schr\"odinger operators with magnetic fields I. General interactions, \emph{Duke Math.\ J.} 45 (1978), 847--883.
\bibitem{ASS} J.~E.~Avron, R.~Seiler and B.~Simon, Charge deficiency, charge transport and comparison of dimensions, \emph{Comm.\ Math.\ Phys.} 159 (1994), 399--422.
\bibitem{ASS94} J.~E.~Avron, R.~Seiler and B.~Simon, The index of a pair of projections, \emph{J.\ Funct.\ Anal.} 120 (1994), 220--237.
\bibitem{Bal19} G.~Bal, Continuous bulk and interface description of topological insulators, \emph{J.\ Math.\ Phys.} 60 (2019), 081506.
\bibitem{Bal22} G.~Bal, Topological invariants for interface modes, \emph{Comm.\ Partial Differential Equations} 47 (2022), 1636--1679.
\bibitem{Bal23} G.~Bal, Topological charge conservation for continuous insulators, \emph{J.\ Math.\ Phys.} 64 (2023), 031508.
\bibitem{BEC} G.~Bal, A bulk--edge correspondence for two-dimensional self-adjoint boundary and interface problems, arXiv:2610.05616, 2026.
\bibitem{BBD} G.~Bal, S.~Becker and A.~Drouot, Magnetic slowdown of topological edge states, \emph{Comm.\ Pure Appl.\ Math.} 77 (2024), 1235--1277.
\bibitem{BCMQ26} G.~Bal, P.~Cazeaux, D.~Massatt and S.~Quinn, Macroscopic approximation of tight-binding models near spectral degeneracies and validity for
wave packet propagation, arXiv:2602.08073, 2026.
\bibitem{BY24} G.~Bal and J.~Yu, Topological equatorial waves and violation (or not) of the bulk edge correspondence, \emph{J.\ Phys.\ A} 57 (2024), 405204.
\bibitem{BCLRS24} J.-M.~Barbaroux, H.~D.~Cornean, L.~Le~Treust, N.~Raymond and E.~Stockmeyer, Magnetic Dirac systems: violation of bulk-edge correspondence in the zigzag
limit, \emph{Lett.\ Math.\ Phys.} 114 (2024), 93.
\bibitem{BCLRS25} J.-M.~Barbaroux, H.~D.~Cornean, L.~Le~Treust, N.~Raymond and E.~Stockmeyer, Green's functions for magnetic Dirac operators and bulk-edge correspondence,
\emph{Ann.\ Henri Poincar\'e} (2026), doi:10.1007/s00023-026-01759-z.
\bibitem{BGZ24} M.~Barsukova, F.~Gris\'e, Z.~Zhang, S.~Vaidya, J.~Guglielmon, M.~I.~Weinstein, L.~He, B.~Zhen, R.~McEntaffer and M.~C.~Rechtsman, Direct
observation of Landau levels in silicon photonic crystals, \emph{Nat.\ Photon.} 18 (2024), 580--585.
\bibitem{BES} J.~Bellissard, A.~van Elst and H.~Schulz-Baldes, The noncommutative geometry of the quantum Hall effect, \emph{J.\ Math.\ Phys.} 35 (1994), 5373--5451.
\bibitem{BHZ06} B.~A.~Bernevig, T.~L.~Hughes and S.-C.~Zhang, Quantum spin Hall effect and topological phase transition in HgTe quantum wells,
\emph{Science} 314 (2006), 1757--1761.
\bibitem{BonyChemin} J.-M.~Bony and J.-Y.~Chemin, Espaces fonctionnels associ\'es au calcul de Weyl--H\"ormander, \emph{Bull.\ Soc.\ Math.\ France} 122 (1994), 77--118.
\bibitem{CB25} B.~Chen and G.~Bal, Scattering theory of topologically protected edge transport, \emph{Pure Appl.\ Anal.} 7 (2025), 701--731.
\bibitem{CG05} J.-M.~Combes and F.~Germinet, Edge and impurity effects on quantization of Hall currents, \emph{Comm.\ Math.\ Phys.} 256 (2005), 159--180.
\bibitem{CT} J.-M.~Combes and L.~Thomas, Asymptotic behaviour of eigenfunctions for multiparticle Schr\"odinger operators, \emph{Comm.\ Math.\ Phys.} 34 (1973), 251--270.
\bibitem{Connes} A.~Connes, Non-commutative differential geometry, \emph{Publ.\ Math.\ IH\'ES} 62 (1985), 257--360.
\bibitem{CHP18} H.~D.~Cornean, B.~Helffer and R.~Purice, A Beals criterion for magnetic pseudo-differential operators proved with magnetic Gabor frames,
\emph{Comm.\ Partial Differential Equations} 43 (2018), 1196--1204.
\bibitem{CHP24} H.~D.~Cornean, B.~Helffer and R.~Purice, Matrix representation of magnetic pseudo-differential operators via tight Gabor frames,
\emph{J.\ Fourier Anal.\ Appl.} 30 (2024), 21.
\bibitem{CIP19} H.~D.~Cornean, V.~Iftimie and R.~Purice, Peierls' substitution via minimal coupling and magnetic pseudo-differential calculus,
\emph{Rev.\ Math.\ Phys.} 31 (2019), 1950008.
\bibitem{CMS} H.~D.~Cornean, M.~Moscolari and K.~S.~S{\o}rensen, Bulk--edge correspondence for unbounded Dirac--Landau operators, \emph{J.\ Math.\ Phys.} 64 (2023), 021902.
\bibitem{CMT} H.~D.~Cornean, M.~Moscolari and S.~Teufel, From orbital magnetism to bulk-edge correspondence, \emph{Ann.\ Henri Poincar\'e} 26 (2025), 3579--3633.
\bibitem{DMV} P.~Delplace, J.~B.~Marston and A.~Venaille, Topological origin of equatorial waves, \emph{Science} 358 (2017), 1075--1077.
\bibitem{DS99} M.~Dimassi and J.~Sj\"ostrand, \emph{Spectral Asymptotics in the Semi-Classical Limit}, London Math.\ Soc.\ Lecture Note Ser.\ 268, Cambridge
University Press, 1999.
\bibitem{DGR} N.~Dombrowski, F.~Germinet and G.~Raikov, Quantization of edge currents along magnetic barriers and magnetic guides, \emph{Ann.\ Henri Poincar\'e} 12 (2011), 1169--1197.
\bibitem{DN} A.~Douglis and L.~Nirenberg, Interior estimates for elliptic systems of partial differential equations, \emph{Comm.\ Pure Appl.\ Math.} 8 (1955), 503--538.
\bibitem{Drouot} A.~Drouot, Microlocal analysis of the bulk-edge correspondence, \emph{Comm.\ Math.\ Phys.} 383 (2021), 2069--2112.
\bibitem{EG} P.~Elbau and G.-M.~Graf, Equality of bulk and edge Hall conductance revisited, \emph{Comm.\ Math.\ Phys.} 229 (2002), 415--432.
\bibitem{EGS} A.~Elgart, G.-M.~Graf and J.~H.~Schenker, Equality of the bulk and edge Hall conductances in a mobility gap, \emph{Comm.\ Math.\ Phys.} 259 (2005), 185--221.
\bibitem{Faure} F.~Faure, Manifestation of the topological index formula in quantum waves and geophysical waves, \emph{Ann.\ H.\ Lebesgue} 6 (2023),
449--492.
\bibitem{Fed96} B.~V.~Fedosov, Index theorems, in \emph{Partial Differential Equations VIII}, Encyclopaedia Math.\ Sci.\ 65, Springer, Berlin,
1996, 155--251.
\bibitem{Folland} G.~B.~Folland, \emph{Harmonic Analysis in Phase Space}, Annals of Mathematics Studies 122, Princeton University Press, 1989.
\bibitem{FB25} M.~Frazier and G.~Bal, Topological edge states of continuous Hamiltonians, \emph{New J.\ Phys.} 27 (2025), 105001.
\bibitem{FB26} M.~Frazier and G.~Bal, Quantum anomalous Hall phases in gated rhombohedral graphene, arXiv:2509.05439 (2026).
\bibitem{GJT} G.~M.~Graf, H.~Jud and C.~Tauber, Topology in shallow-water waves: a violation of bulk-edge correspondence, \emph{Comm.\ Math.\ Phys.} 383 (2021), 731--761.
\bibitem{GP} G.~M.~Graf and M.~Porta, Bulk-edge correspondence for two-dimensional topological insulators, \emph{Comm.\ Math.\ Phys.} 324 (2013), 851--895.
\bibitem{GRW21} J.~Guglielmon, M.~C.~Rechtsman and M.~I.~Weinstein, Landau levels in strained two-dimensional photonic crystals, \emph{Phys.\ Rev.\ A} 103
(2021), 013505.
\bibitem{Halperin} B.~I.~Halperin, Quantized Hall conductance, current-carrying edge states, and the existence of extended states in a two-dimensional
disordered potential, \emph{Phys.\ Rev.\ B} 25 (1982), 2185--2190.
\bibitem{Hatcher} A.~Hatcher, \emph{Algebraic Topology}, Cambridge University Press, 2002.
\bibitem{Hatsugai93} Y.~Hatsugai, Chern number and edge states in the integer quantum Hall effect, \emph{Phys.\ Rev.\ Lett.} 71 (1993), 3697--3700.
\bibitem{Hormander3} L.~H\"ormander, \emph{The Analysis of Linear Partial Differential Operators III}, Springer, Berlin, 1985.
\bibitem{IMP} V.~Iftimie, M.~M\u{a}ntoiu and R.~Purice, Magnetic pseudodifferential operators, \emph{Publ.\ RIMS} 43 (2007), 585--623.
\bibitem{IMP2} V.~Iftimie, M.~M\u{a}ntoiu and R.~Purice, Commutator criteria for magnetic pseudodifferential operators, \emph{Comm.\ Partial Differential Equations} 35 (2010), 1058--1094.
\bibitem{Kato} T.~Kato, \emph{Perturbation Theory for Linear Operators}, 2nd ed., Springer, 1976.
\bibitem{KRS} J.~Kellendonk, T.~Richter and H.~Schulz-Baldes, Edge current channels and Chern numbers in the integer quantum Hall effect, \emph{Rev.\ Math.\ Phys.} 14 (2002), 87--119.
\bibitem{KS04} J.~Kellendonk and H.~Schulz-Baldes, Quantization of edge currents for continuous magnetic operators, \emph{J.\ Funct.\ Anal.} 209 (2004),
388--413.
\bibitem{KDDC17} \'A.~D.~Kov\'acs, G.~D\'avid, B.~D\'ora and J.~Cserti, Frequency-dependent magneto-optical conductivity in the generalized $\alpha$--$\mathcal T_3$ model, \emph{Phys.\ Rev.\ B} 95 (2017),
035414.
\bibitem{Lerner} N.~Lerner, \emph{Metrics on the Phase Space and Non-Selfadjoint Pseudo-Differential Operators}, Birkh\"auser, 2010.
\bibitem{Lut51} J.~M.~Luttinger, The effect of a magnetic field on electrons in a periodic potential, \emph{Phys.\ Rev.} 84 (1951), 814--817.
\bibitem{Lut56} J.~M.~Luttinger, Quantum theory of cyclotron resonance in semiconductors: general theory, \emph{Phys.\ Rev.} 102 (1956), 1030--1041.
\bibitem{MN16} J.~D.~Malcolm and E.~J.~Nicol, Analytic evaluation of Kane fermion magneto-optics in two and three dimensions, \emph{Phys.\ Rev.\ B} 94 (2016), 224305.
\bibitem{MP04} M.~M\u{a}ntoiu and R.~Purice, The magnetic Weyl calculus, \emph{J.\ Math.\ Phys.} 45 (2004), 1394--1417.
\bibitem{PST03} G.~Panati, H.~Spohn and S.~Teufel, Effective dynamics for Bloch electrons: Peierls substitution and beyond, \emph{Comm.\ Math.\ Phys.}
242 (2003), 547--578.
\bibitem{Peierls33} R.~Peierls, Zur Theorie des Diamagnetismus von Leitungselektronen, \emph{Z.\ Phys.} 80 (1933), 763--791.
\bibitem{Phillips} J.~Phillips, Self-adjoint Fredholm operators and spectral flow, \emph{Canad.\ Math.\ Bull.} 39 (1996), 460--467.
\bibitem{PSB} E.~Prodan and H.~Schulz-Baldes, \emph{Bulk and Boundary Invariants for Complex Topological Insulators}, Springer, 2016.
\bibitem{QB21} S.~Quinn and G.~Bal, Approximations of interface topological invariants, \emph{SIAM J.\ Math.\ Anal.} 56 (2024), 5521--5582.
\bibitem{QB22} S.~Quinn and G.~Bal, Asymmetric transport for magnetic Dirac equations, \emph{Pure Appl.\ Anal.} 6 (2024), 353--377.
\bibitem{RKY20} J.-W.~Rhim, K.~Kim and B.-J.~Yang, Quantum distance and anomalous Landau levels of flat bands, \emph{Nature} 584 (2020), 59--63.
\bibitem{RT24} S.~Rossi and A.~Tarantola, Topology of 2D Dirac operators with variable mass and an application to shallow-water waves, \emph{J.\ Phys.\ A}
57 (2024), 065201.
\bibitem{SKR00} H.~Schulz-Baldes, J.~Kellendonk and T.~Richter, Simultaneous quantization of edge and bulk Hall conductivity, \emph{J.\ Phys.\ A} 33 (2000),
L27--L32.
\bibitem{Shubin} M.~A.~Shubin, \emph{Pseudodifferential Operators and Spectral Theory}, 2nd ed., Springer, 2001.
\bibitem{Silv} M.~G.~Silveirinha, Chern invariants for continuous media, \emph{Phys.\ Rev.\ B} 92 (2015), 125153.
\bibitem{Simon} B.~Simon, \emph{Functional Integration and Quantum Physics}, 2nd ed., AMS Chelsea, 2005.
\bibitem{Streda} P.~St\v{r}eda, Theory of quantised Hall conductivity in two dimensions, \emph{J.\ Phys.\ C} 15 (1982), L717--L721.
\bibitem{Taarabt} A.~Taarabt, Equality of bulk and edge Hall conductances for continuous magnetic random Schr\"odinger operators, arXiv:1403.7767, 2014.
\bibitem{TDV} C.~Tauber, P.~Delplace and A.~Venaille, A bulk-interface correspondence for equatorial waves, \emph{J.\ Fluid Mech.} 868 (2019), R2.
\bibitem{Teppe16} F.~Teppe et al., Temperature-driven massless Kane fermions in HgCdTe crystals, \emph{Nat.\ Commun.} 7 (2016), 12576.
\bibitem{TKNN} D.~J.~Thouless, M.~Kohmoto, M.~P.~Nightingale and M.~den~Nijs, Quantized Hall conductance in a two-dimensional periodic potential,
\emph{Phys.\ Rev.\ Lett.} 49 (1982), 405--408.
\bibitem{Zworski} M.~Zworski, \emph{Semiclassical Analysis}, Graduate Studies in Mathematics 138, American Mathematical Society, 2012.
\end{thebibliography}
\end{document}